\documentclass[reprint,amsmath,amssymb,aps]{revtex4-2}
\usepackage{graphicx}
\usepackage{dcolumn}
\usepackage{bm}
\usepackage{tabstackengine}
\usepackage[dvipsnames]{xcolor}

\begin{document}
\title{Capturing the ground state of uranium dioxide from first principles:\protect\\ crystal distortion, magnetic structure, and phonons}
\author{Shuxiang Zhou$^1$, Hao Ma$^2$, Enda Xiao$^3$, Krzysztof Gofryk$^1$, Chao Jiang$^1$, Michael E. Manley$^2$, David H. Hurley$^1$, and Chris A. Marianetti$^4$}
\affiliation{$^1$Idaho National Laboratory, Idaho Falls, Idaho 83415, USA\\$^2$Oak Ridge National Laboratory, Oak Ridge, Tennessee 37831, USA\\$^3$Department of Chemistry, Columbia University, New York, New York 10027, USA\\$^4$Department of Applied Physics and Applied Mathematics, Columbia University, New York, New York 10027, USA}

\begin{abstract}
Uranium dioxide (UO$_2$) remains a formidable challenge for first-principles approaches due to
the complex interplay among spin-orbit coupling, Mott physics, magnetic ordering,
and crystal distortions. Here we use DFT+$U$ to explore UO$_2$ at zero temperature, incorporating all the aforementioned phenomena. The technical
challenge is to navigate the many metastable electronic states produced by
DFT+$U$, which is accomplished using $f$-orbital
occupation matrix control to search for the ground state. 
We restrict our search to the 
high-symmetry
ferromagnetic phase, including spin-orbit coupling,  
which produces a previously unreported occupation
matrix. 
This newfound occupation matrix is then used as an initialization to explore the broken symmetry
phases. We find the oxygen cage distortion of the 3\textbf{k} antiferromagnetic
state to be in excellent agreement with experiments, and both the
spin-orbit coupling and the Hubbard $U$ are critical ingredients. We
demonstrate that only select phonon modes have a strong dependence on the
Hubbard $U$, whereas magnetic ordering has only a small influence overall. 
We perform measurements of the phonon dispersion curves using inelastic neutron scattering, 
and our calculations show good agreement when using reasonable values of $U$.
The
quantitative success of DFT+$U$ warrants exploration of thermal transport
and other observables within this level of theory.

\end{abstract}

\maketitle

\section{\label{sec:intro}Introduction}

Uranium dioxide (UO$_2$) has attracted a great deal of research interest ever since the
1950s, given its use as a standard nuclear fuel. The partially filled $f$
shell in uranium sets the stage for rich physics, and extensive experiments have
characterized the behavior of UO$_2$. At ambient temperature and
pressure, UO$_2$ crystallizes in the flourite structure with a lattice parameter of
5.47  \AA\,   \cite{idiri_behavior_2004}, and exhibits
paramagnetism  \cite{lander_neutron-diffraction_1976,santini_multipolar_2009}.
Upon cooling, UO$_2$ undergoes an antiferromagnetic (AFM) transition at $T_N=30.8$ K
  \cite{jones_heat_1952}, and there is a concomitant oxygen cage distortion whereby
the oxygen ions move 0.014  \AA\, from their fluorite positions along the $<$111$>$
directions  \cite{faberNeutronDiffractionStudyMathrmO1975} (see Figure 1 of Ref \cite{dorado_stability_2010} for a schematic, in addition to further details). More specifically, the antiferromagnetism is a noncollinear
3\textbf{k} ordering  \cite{burlet_neutron_1986,ikushima_first-order_2001,
blackburn_spherical_2005,wilkins_direct_2006}, which is aligned with the
oxygen cage distortion.

Extensive first-principles calculations were also performed on UO$_2$, and here we focus on
ground state properties. It
is well-known that UO$_2$ is incorrectly predicted to be metallic when using  
density functional theory (DFT) within conventional approximations to the exchange-correlation energy, including the
local density approximation (LDA) and generalized gradient approximation (GGA),
due to their failure to describe the strong correlation of the 5\emph{f}
electrons of uranium. Going beyond conventional DFT, various approaches have
been developed to better capture the effects of strong local interactions; these approaches include hybrid
functionals  \cite{kudin_hybrid_2002,prodan_covalency_2007,becke_new_1993},
self-interaction correction
\cite{petit_electronic_2010,perdew_self-interaction_1981}, 
DFT+$U$  \cite{dudarev_effect_1997,anisimov_band_1991}, and DFT plus
dynamic mean-field theory
(DMFT)  \cite{georges_dynamical_1996,kotliar_electronic_2006}. Although all these
approaches predict UO$_2$ to be an insulator, DFT+$U$ is the most widely used, thanks
to its simplicity and relatively low computational cost. Furthermore, DFT+$U$ shares a degree of overlap with all
the aforementioned techniques. DFT+$U$ is recovered from DFT+DMFT when solving
the DMFT impurity problem within Hartree-Fock  \cite{kotliar_electronic_2006},
DFT+$U$ can be viewed as a local hybrid functional   \cite{agapito_reformulation_2015},
and the DFT+$U$ functional is general enough to account for self-interaction corrections. Given that DFT+$U$ relies on a Hartree-Fock-like approximation, it
must break symmetry and strongly polarize the system in order to offer a reasonable description of energetics
when dealing with strong interactions. 
Therefore, DFT+$U$ brings challenges
to modeling UO$_2$, given the large number of possible spin and orbital orderings in the case
of 5\emph{f} electrons. Indeed, many metastable states are found when performing DFT+$U$
on UO$_2$, and
some calculated properties (e.g., the band gap) are highly dependent on the 
exact ordered state  \cite{dorado_textdfttextu_2009}.

Multiple methods have been designed to search for the ground state in
DFT+$U$, including occupation matrix control
(OMC)  \cite{dorado_textdfttextu_2009,amadon_ensuremathgamma_2008,jomard_structural_2008,zhou_crystal_2011},
the quasi-annealing scheme  \cite{geng_interplay_2010}, and the $U$-ramping
method  \cite{meredig_method_2010}, all of which have been applied to UO$_2$.
None of these schemes can guarantee that the ground state will be found, and different
research efforts using these different methods have, in practice, generated different results. For example,
both Thompson \emph{et al.}  \cite{wang_electronic_2013} and Wang \emph{et al.}  \cite{thompson_first-principles_2011} used the $U$-ramping method and found 1\textbf{k} AFM ground states with different occupation matrices.
Earlier work by Laskowski \emph{et al.}   \cite{laskowski_magnetic_2004} and Gryaznov \emph{et al.}   \cite{gryaznov_density_2010}, which did not elaborate on how the ground state search was performed, found 3\textbf{k} AFM ground states with significantly larger oxygen cage distortions (0.16 and 0.09  \AA, respectively). Dorado \emph{et al.}  \cite{dorado_stability_2010} and Thompson \emph{et al.}  \cite{thompson_first-principles_2011} employed OMC and the $U$-ramping method, respectively, and both found a lowest-energy structure with 1\textbf{k} AFM ordering.
The only study that found a 3\textbf{k} AFM ground state featuring oxygen cage distortion on the same scale as the experiments was Zhou \emph{et al.}   \cite{zhou_crystal_2011} (i.e., 0.024  \AA), who used an OMC technique based on inputs
from crystal field calculations, along with a different type DFT+$U$ functional. 

Phonons may serve as a delicate probe of the electronic structure, and are a
key ingredient in thermal transport. As such, a large number of experimental
studies
  \cite{fritz_elastic_1976,brandt_temperature_1967,dolling_crystal_1965,pang_phonon_2013}
and DFT-based computational studies   \cite{yin_origin_2008, devey_first_2011,
sanati_elastic_2011, yun_phonon_2012, pang_phonon_2013, wang_phonon_2013,
kaur_thermal_2013} have been performed on UO$_2$. While different
first-principles methods have successfully captured various aspects of the
phonon dispersion curves and elastic constants, large discrepancies still exist.
For example, the predicted phonon frequency of the highest longitudinal optical
mode, denoted LO2   \cite{pang_phonon_2013}, at the $L$ point (0.5, 0.5, 0.5)
varies from 50   \cite{yun_phonon_2012} to 70 meV   \cite{wang_phonon_2013}. 
More details on the discrepancies of previous DFT+$U$ calculations of phonons as compared with experiments are reviewed in Supplementary Material (SM)\cite{Supplemental_Material}.
These discrepancies might be attributable to various differences in the
calculations, such as the value of $U$, the magnetic
ordering, and whether or not spin-orbit coupling (SOC) was included.  Furthermore, even for
calculations performed under the exact same theory, there is no
guarantee that the different researchers converged upon the same ground state, which potentially
yields different phonons. Indeed, very few previous studies report
details on their obtained UO$_2$ ground state (e.g., the uranium
$f$-electron occupation matrices), making it extremely challenging to reproduce
prior work and understand the origin of discrepancies.

In this DFT+$U$ study of UO$_2$, we perform a ground state search using OMC
within the ferromagnetic (FM) state, both with and without SOC, yielding a
previously unreported occupation matrix. Our new ground state is then
used as the starting point for exploring states with further broken symmetries, yielding
unprecedented agreement with experiments. Given this new understanding, we then
explore the elastic constants and phonon dispersion curves, comparing the phonons with new inelastic neutron scattering data and untangling the various roles of
$U$, magnetic ordering, and SOC.  

\begin{table*}[t]%
\caption{\label{tab:table2}%
States of FM UO$_2$ with SOC, as found in stage 2 of our ground state search. 
The results are sorted by increasing relative energy, with
respect to the fully relaxed crystal having occupation matrix $\mathbb{S}_0$ and energy  $E_{min}$.}
\begin{ruledtabular}
\begin{tabular}{lcccccr}
\textrm{State}&
\textrm{Degeneracy}&
\textrm{Initialized from States}&
\textrm{\begin{tabular}{@{}c@{}}$E_{und}-E_{min}$ \\ (meV/UO\textsubscript{2})\end{tabular} }&
\textrm{\begin{tabular}{@{}c@{}}$E_{dis}-E_{min}$ \\ (meV/UO\textsubscript{2})\end{tabular} }&
\textrm{{\begin{tabular}{@{}c@{}}Strain Distortion \\ ($\epsilon_{xx},\epsilon_{yy},\epsilon_{zz},\epsilon_{xy},\epsilon_{xz},\epsilon_{yz}$)$\times10^3$\end{tabular} } }&
\textrm{\begin{tabular}{@{}c@{}}Oxygen Cage \\ Distortion\end{tabular} }\\
\colrule
$\mathbb{S}_0$ &1& S$_0$, S$_1$, S$_2$, S$_4$, S$_5$, S$_6$, S$_8$ & 0.3 & 0 & (-1, -1, 2, 0, 0, 0) & -\\
$\mathbb{S}_1$ &2& S$_2$, S$_4$, S$_7$ & 68.9 & 67.7 & (-5, 2, 3, 0, 0, 0)\footnotemark[1]& -\\
$\mathbb{S}_2$ &4& S$_1$, S$_2$ & 72.3 & 70.8 & (-4, 5, 0, 0, 0, 0)\footnotemark[1]& -\\
$\mathbb{S}_3$ &1& S$_3$ & 72.9 & 71.5 & (-3, -3, 5, 0, 0, 0)& -\\
\end{tabular}
\footnotetext[1]{Symmetry-equivalent states are given in the Supplementary Material  \cite{Supplemental_Material}.}
\end{ruledtabular}
\end{table*}

\section{\label{sec:compute}Computational Details}

Our DFT calculations were carried out using the projector augmented-wave (PAW) method   \cite{blochl_projector_1994, kresse_ultrasoft_1999}, as implemented in the Vienna ab initio Simulation Package (VASP) code   \cite{kresse_ab_1993, kresse_efficient_1996}. Two exchange correlation functionals were employed: LDA and GGA as formulated by Perdew, Burke, and Ernzerhof (PBE)   \cite{perdew_generalized_1996}. A plane-wave cutoff energy of 550 eV was used, and the energy convergence criterion was $10^{-8}$ eV. The Hubbard correction was included in the LDA+$U$ or GGA+$U$ approximation to account for the strong local interactions of the uranium \emph{f} electrons. Specifically, we used the simplified rotationally invariant DFT+$U$ approach from Dudarev \emph{et al.}   \cite{dudarev_electron-energy-loss_1998}, which only employs a single effective interaction.  We customized the VASP code to monitor and impose the occupation matrices during the calculation \cite{Supplemental_Material}. The calculated states were controlled via two methods. First, specific occupation matrices were imposed during the first several electronic steps, and then the calculation was allowed to proceed self-consistently without constraint (i.e., OMC). Second, a specific charge density was used to initialize a calculation---a process we refer to as charge density initialization.
Symmetry was turned off
for all calculations. 

Above $T_N$, the space group of UO$_2$ is $Fm\bar{3}m$ and the point group
symmetry of the uranium site is $O_h$. For FM calculations, the primitive cell
of UO$_2$ was used and a 13$\times$13$\times$13 Monkhorst-Pack \emph{k}-point mesh
  \cite{monkhorst_special_1976} was applied; for AFM calculations, the
conventional cubic cell was used and a 7$\times$7$\times$7 Monkhorst-Pack
\emph{k}-point mesh was applied.
The phonons and elastic constants were calculated via the lone irreducible
derivative (LID) approach  \cite{fu_group_2019}. 
LID uses central finite difference to individually compute each group
theoretically irreducible derivative in the smallest possible supercell, minimizing the possibility
of numerical inadequacies. 
Finite difference displacement amplitudes of 
0.01, 0.02, 0.03, and 0.04  \AA\, were used to construct quadratic error tails, ensuring that
the discretization error is properly accounted for.
The face-centered cubic lattice vectors are encoded in a $3\times3$ row stacked
matrix $\hat{\mathbf{a}}=\frac{a_o}{2}(\hat{\mathbf{J}}-\hat{\mathbf{1}})$,
where $\hat{\mathbf{1}}$ is the identity matrix and
$\hat{\mathbf{J}}$ is a matrix in which each element is 1.
The Brillouin zone is discretized using a real space supercell
$\hat{\mathbf{S}}_{BZ}\hat{\mathbf{a}}$, where $\hat{\mathbf{S}}_{BZ}$ is an
invertible matrix of integers which produces superlattice vectors that satisfy
the point group \cite{fu_group_2019}. In this study, all irreducible
derivatives corresponding to $\hat{\mathbf{S}}_{BZ}=2\hat{\mathbf{S}}_{C}$ were
obtained, where $\hat{\mathbf{S}}_{C}$ generates the conventional cubic cell
and is defined as $\hat{\mathbf{S}}_{C}=(\hat{\mathbf{J}}-2\hat{\mathbf{1}})$. The
supercell $2\hat{\mathbf{S}}_{C}$ has
a multiplicity of 32 and therefore contains 96 atoms; though the LID
approach allows
one to extract all of the irreducible derivatives from supercells that have
multiplicity 4 (i.e. 12 atoms) or less. In the case of 1\textbf{k} and 3\textbf{k} AFM,
the magnetism breaks the symmetry of the $Fm\bar{3}m$ space group, and therefore we
use a primitive unit cell of $\hat{\mathbf{S}}_{C}\hat{\mathbf{a}}$, which is commensurate
with both 1\textbf{k} and 3\textbf{k}. When computing phonons for 1\textbf{k} and 3\textbf{k},
we discretize the Brillouin zone using $\hat{\mathbf{S}}_{BZ}=2\hat{\mathbf{1}}$, meaning that
all irreducible derivatives consistent with the AFM supercell $2\hat{\mathbf{S}}_{C}$ will be extracted.
In order to best compare with experiment, we also unfold the AFM phonon band structure back to the 
original primitive unit cell $\hat{\mathbf{a}}$, averaging any translational symmetry breaking. 
The
Born effective charges ($Z_U^\star=5.54$ and $Z_O^\star=-2.77$) of the U and O ions
and the dielectric constant ($\epsilon=5.69$) 
were used to account for LO-TO splitting   \cite{Mark_arXiv,gonzeDynamicalMatricesBorn1997a}.

\section{\label{sec:ground}Ground state search for UO\textsubscript{2}}

Determining the ground state of UO$_2$ within DFT+$U$ is very complex, and exploring all of the phase space is not tractable. Our approach consists of three different stages, all executed using GGA. For stage 1, we used OMC for a given trial occupation matrix and performed computations in the FM state without SOC. For stage 2, the charge density from the stage 1 results was used to initialize a FM calculation with SOC. For stage 3, we used OMC with the resulting stage 2 occupation matrix, and performed an AFM calculation with SOC. Stage 1 was executed for a large number of trial occupation matrices (see below for details), while stage 2 was only executed for the low-energy subset of the stage 1 results. Stage 3 was only executed for the lowest-energy state of stage 2. All the calculations evaluated two types of structural relaxations: relaxing only the volume of the structural cubic crystal (denoted as the undistorted crystal), and full structural relaxations that may break the symmetry (denoted as the distorted crystal).

For stage 1, we followed Dorado \emph{et
al.}'s works   \cite{dorado_textdfttextu_2009, dorado_stability_2010} and used
randomly generated diagonal and non-diagonal occupation matrices
(221 and 900 matrices, respectively) as trials. The nine lowest energies we found (denoted as $S_0$--$S_8$, with $S_0$ having the lowest
energy) consisted of 25 distinct occupation matrices, some of which were related by symmetry.
Details on this search process and the selected properties of the 25 states,
including occupation matrices, can be found in SM\cite{Supplemental_Material}.
Our lowest-energy class of states, $S_0$, contains an orbital ordering that induces structural symmetry breaking
via a pure shear strain ($\pm$0.017 T$_{2g}$ mode), further lowering the energy
by 10 meV relative to the relaxed cubic structure. The $S_1$ class is 13 meV higher in energy
than the $S_0$ class, and one occupation matrix within $S_1$ is what Dorado \emph{et al.} found to be
the lowest-energy state in their search. Therefore, we have found a lower energy state within our search,
though there is no guarantee that even lower-energy states do not exist.
However, the shear strain distortion associated with the $S_0$ class of states is not observed in the experiments.
This discrepancy is not particularly problematic, given that SOC will favor a different
occupation matrix.

\subsection{\label{sec:fmsoc}Ferromagnetism with SOC}

Previous work reported that SOC may have important impacts on phonons   \cite{dorado_advances_2013}. However, the role of SOC in influencing the ground
state occupation matrix of UO$_2$ has not been carefully explored. 
Therefore, we now proceed to our stage 2, in which the low-energy subset of results from stage 1
are used to initialize FM calculations that include SOC. 
The results of stage 2 are given in Table~\ref{tab:table2}, comparing both
the relative energy and the distortion. 
In regard to the energy, the nine lowest-energy classes
found without SOC converge into four lowest-energy classes when using SOC, and the relative
energies between the classes are significantly changed. 
For example, the calculation
initialized by one state in $S_2$ ended up in $\mathbb{S}_3$ with the highest
energy, while the calculation initialized by higher-energy state
$S_8$ ended up in $\mathbb{S}_0$. Furthermore, calculations
initialized from degenerate states within the same class sometimes converged to different classes
when using SOC: the
six degenerate states in $S_2$ converged to $\mathbb{S}_0$, $\mathbb{S}_1$, and
$\mathbb{S}_2$. It is clear that SOC significantly perturbs the system.
SOC also changes the strain distortions. For example, a $T_{2g}$ strain exists
in $S_0$, $S_3$, $S_6$, and $S_8$, but not in any of
the reported SOC states. The same trend holds for the oxygen cage distortion
 (compare Table~\ref{tab:table2} to Table S1 in SM \cite{Supplemental_Material}).

Stage 2 of our ground state search resulted in a lowest-energy state $\mathbb{S}_0$ (see SM \cite{Supplemental_Material}, Section I.D.1) that was substantially lower than all the others.
Interestingly, the $\mathbb{S}_0$ state is also found without using OMC (i.e. initializing the calculation in the default manner), which might be reasonable given that it
is so much lower in energy than $\mathbb{S}_1$.
It should be noted that $\mathbb{S}_0$ has no appreciable strain, thus there
are no immediate inconsistencies with experiments. Although there is no oxygen cage distortion
for $\mathbb{S}_0$, we will demonstrate that magnetic ordering will generate such distortions.

\begin{table*}[t]
\caption{\label{tab:table3}%
Energies and distortions calculated for different magnetic structures and functionals at $U=4$ eV with SOC: FM, 1\textbf{k} AFM, and 3\textbf{k} AFM, as computed using LDA+$U$ and GGA+$U$. 
}
\begin{ruledtabular}
\begin{tabular}{lcccccr}
\textrm{Potential}&
\textrm{\begin{tabular}{@{}c@{}}Magnetic \\ Structure\end{tabular}}&
\textrm{\begin{tabular}{@{}c@{}}$E_{und}-E_{dis}^{3\textbf{k}}$ \\ (meV/UO\textsubscript{2})\end{tabular}}  &
\textrm{\begin{tabular}{@{}c@{}}$E_{dis}-E_{dis}^{3\textbf{k}}$ \\ (meV/UO\textsubscript{2})\end{tabular}}  &
\textrm{\begin{tabular}{@{}c@{}}Lattice Parameters \\ (  \AA)\end{tabular} } &
\textrm{{\begin{tabular}{@{}c@{}}Strain Distortion \\ ($\epsilon_{xx},\epsilon_{yy},\epsilon_{zz},\epsilon_{xy},\epsilon_{xz},\epsilon_{yz}$)$\times10^3$\end{tabular} } }&
\textrm{\begin{tabular}{@{}c@{}}Oxygen Cage \\Distortion  \end{tabular} }\\
\colrule
LDA+$U$ & FM             & 20.0 & 17.7 & 5.450 & (-1, -1, 2, 0, 0, 0) & -\\
LDA+$U$ & 1\textbf{k} AFM & 8.6 & 8.4  & 5.450 & (0, -1, 1, 0, 0, 0) & -\\
LDA+$U$ & 3\textbf{k} AFM & 1.3 & 0  & 5.450 & - & $<$111$>$ 0.016  \AA\\
GGA+$U$ & FM             & 0.4 & 0.1  & 5.546 & (-1, -1, 2, 0, 0, 0) & -\\
GGA+$U$ & 1\textbf{k} AFM & -3.7 & -3.9 & 5.546 & (0, -1, 1, 0, 0, 0) & -\\
GGA+$U$ & 3\textbf{k} AFM & 1.0 &  0 & 5.547 & - & $<$111$>$ 0.016  \AA\\

\end{tabular}
\end{ruledtabular}
\end{table*}

\begin{figure}[t]
\includegraphics[width=0.825\columnwidth]{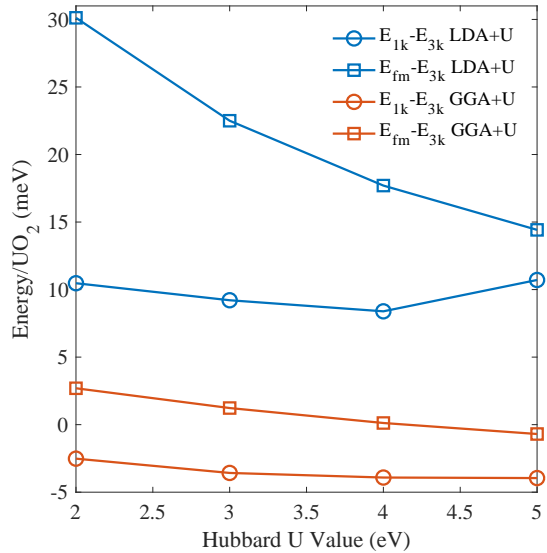}
\caption{\label{fig:figure1} Calculated energy difference of distorted UO$_2$ for
FM and 1\textbf{k} AFM relative to 3\textbf{k} AFM using LDA+$U$ and GGA+$U$ with SOC, as a 
function of the Hubbard $U$.}
\end{figure}

\begin{figure}[t]
\includegraphics[width=0.825\columnwidth]{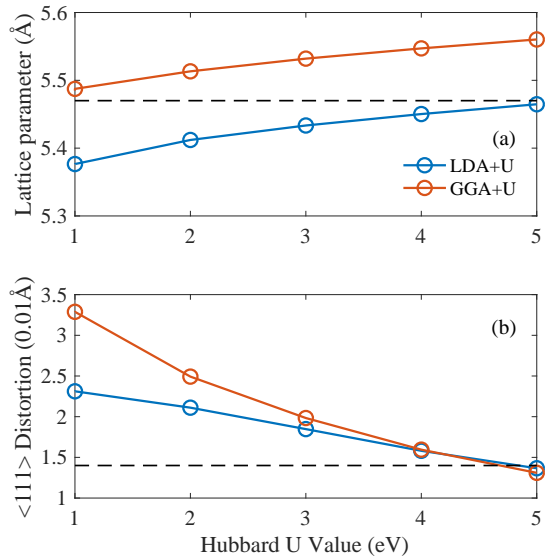}
\caption{\label{fig:figure2} Calculated (a) lattice parameter and (b) $<$111$>$ oxygen cage distortion in
the 3\textbf{k} AFM state, using LDA+$U$+SOC and GGA+$U$+SOC, as a function of the Hubbard $U$. The horizontal dashed lines represent the experimental values of the lattice parameter from Idiri \emph{et al.}   \cite{idiri_behavior_2004} and the distortion from Santini \emph{et al.}   \cite{santini_multipolar_2009}.}
\end{figure}

\subsection{\label{sec:afm} AFM UO$_2$ with SOC}

In stage 3, we used $\mathbb{S}_0$ to initialize the 1\textbf{k} and
3\textbf{k} AFM calculations performed using GGA+$U$ with $U=4$ eV. These results
were then used to initialize GGA+$U$ for other $U$ values in
addition to LDA+$U$ calculations. We found that, for each magnetic order, the
converged occupation matrix of LDA+$U$ was approximately the same as the
occupation matrix of GGA+$U$ (see SM \cite{Supplemental_Material} Section I.E). 

For $U=4$ eV, the energies, lattice parameters, and distortions of UO$_2$ for
each magnetic order are given in Table~\ref{tab:table3}. GGA+$U$ predicts a
lattice parameter 0.08 \AA\ larger than experiment, 
while LDA+$U$ is 0.02 \AA\, smaller. For both LDA+$U$ and GGA+$U$,
the FM and 1\textbf{k} AFM order have a non-zero $E_g$ strain mode, while
the 3\textbf{k} AFM order has no non-zero strains; and the latter is consistent with experiments.
Alternatively, for both LDA+$U$ and GGA+$U$, neither the FM nor 1\textbf{k} AFM order
has an oxygen cage distortion, whereas the 3\textbf{k} AFM
has an oxygen cage distortion with an amplitude of 0.016  \AA\, in the
$<$111$>$-type direction; which is in good agreement with the experimental
result of 0.014  \AA.
Therefore, we see that the 3\textbf{k} AFM order is a necessary condition for
achieving the oxygen cage distortion along the $<$111$>$ direction. 
Furthermore, the 3\textbf{k} AFM order is stable even in the absence of the 
oxygen cage distortion, indicating magnetism to be the dominant energy
scale and the oxygen cage distortion acts cooperatively.
LDA+$U$ predicts the 3\textbf{k} AFM order to be the lowest in energy, while
GGA+$U$ predicts the 1\textbf{k} AFM to be lowest and the 3\textbf{k} AFM and FM are essentially
degenerate. Overall, for the above set of experimental observables, LDA+$U$ appears to be in better agreement with the experiments
than is GGA+$U$.

We now study how the results depend on $U$.
Figure~\ref{fig:figure1} shows the energies of FM and 1\textbf{k} AFM relative to the 3\textbf{k} AFM, including oxygen cage distortions. Neither LDA+$U$ nor GGA+$U$ show any
qualitative changes over this range. The GGA+$U$ results are rather insensitive
to $U$, while the FM state in LDA+$U$ changes rather strongly. The
lattice parameter and oxygen cage distortion are shown in Figure~\ref{fig:figure2}.
The lattice parameter increases with $U$ in both cases, and LDA+$U$ agrees better
with experimental results at $U=4$ eV, given that it generated a large underprediction at $U=0$, as expected.
The oxygen cage distortion is largest at $U=1$ eV and decreases with increasing $U$.
Coincidentally, the LDA+$U$ and GGA+$U$ results happen to cross at approximately $U=4$ eV.

It is important to make comparisons with previous calculations whenever possible,
and we attempted to reproduce the 3\textbf{k} AFM states and
corresponding large oxygen cage distortions reported by Laskowski \emph{et al.} (0.16   \AA)   \cite{laskowski_magnetic_2004}
and Gryaznov \emph{et al.} (0.09   \AA)   \cite{gryaznov_density_2010} with GGA+$U$+SOC. 
However,
we were unable to reproduce any of these findings, highlighting the complexity of
performing DFT+$U$ in correlated $f$ electron systems, and illustrating why
explicit reporting of the occupation matrix is critical. 
Interestingly, the $\mathbb{S}_0$ state in the 1\textbf{k} AFM structure is found without using OMC
(i.e. initializing the calculation in the default manner), while this was not the case for the 3\textbf{k} AFM structure.

\section{\label{sec:phonon}Phonons }

\begin{table}[b]
\caption{\label{tab:table4}%
Calculated elastic constants based on the lowest-energy FM and AFM states, using DFT+$U$ ($U=4$ eV) with SOC,
in addition to 
experimental results at 10 K from Brandt \emph{et al.}  \cite{brandt_temperature_1967}.
}
\begin{ruledtabular}
\begin{tabular}{lccc}
\textrm{ }&
\textrm{$C_{11}$ (GPa)}&
\textrm{$C_{12}$ (GPa)}&
\textrm{$C_{44}$ (GPa)}\\
\colrule
Brandt \emph{et al.}  \cite{brandt_temperature_1967} & 400 & 125 & 59 \\
LDA+$U$: FM & 394.8 & 136.2 & 78.7 \\
LDA+$U$: 1\textbf{k} AFM & 398.8 & 131.8 & 78.5 \\
LDA+$U$: 3\textbf{k} AFM & 394.4 & 132.0 & 77.8 \\
GGA+$U$: FM  & 358.4 & 113.0 & 60.8 \\
GGA+$U$: 1\textbf{k} AFM  & 379.8 & 120.1 & 62.0 \\
GGA+$U$: 3\textbf{k} AFM  & 380.1 & 119.7 & 62.6 
\end{tabular}
\end{ruledtabular}
\end{table}

Based on the lowest-energy state $\mathbb{S}_0$, the elastic
constants and phonon dispersion curves were calculated via the
LID approach  \cite{fu_group_2019} using DFT+$U$+SOC calculations. The elastic constants calculated
with $U=4$ eV for the FM, 1\textbf{k} AFM, and 3\textbf{k} AFM states are
presented in Table~\ref{tab:table4}, in addition to Brandt
\emph{et al.}'s low temperature results for the 3\textbf{k} AFM state \cite{brandt_temperature_1967}. 
All of the magnetic states have small structural distortions which break
$O_h$ symmetry (see Table~\ref{tab:table3}), but the deviation is very small in all cases.
Therefore, we use $C_1$ symmetry and average the elastic constants to restore $O_h$ symmetry (see Table~\ref{tab:table4}), and the unaveraged results are in SM (see SM \cite{Supplemental_Material} Section II G).
We find that magnetic ordering has essentially
no influence on the LDA+$U$ results, and only a small influence 
in the case of GGA+$U$.
Neither LDA+$U$ nor GGA+$U$ is in perfect agreement with the experiments, though
the GGA+$U$ results fall within a 10\% margin of error. 
Furthermore,
by using $U=5$ eV in LDA+$U$ for the 3\textbf{k} AFM state, the calculated elastic constants are
$(C_{11},C_{12},C_{44})=(387,138,78)$ GPa, which are very close to the $U=4$ eV
values, meaning that the acoustic phonon properties are approximately saturated
at $U=4$ eV. 

\begin{figure}[t]
\includegraphics[width=1.0\columnwidth]{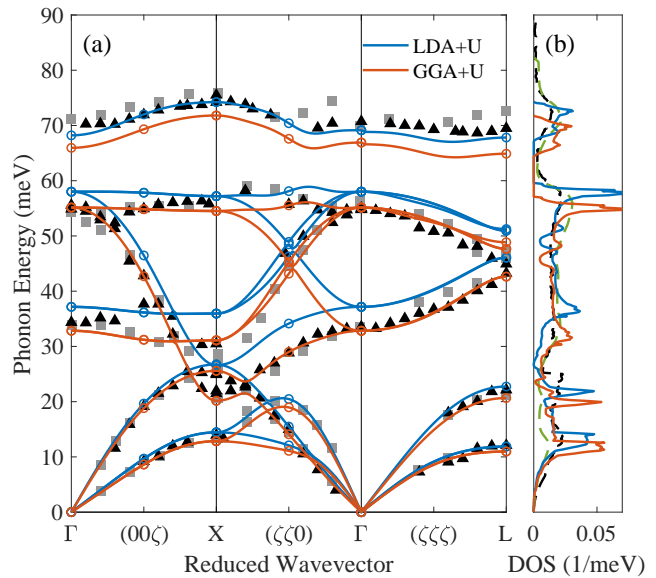}
\caption{
Phonons of  3\textbf{k} AFM UO$_2$. (a) The unfolded phonon dispersion curves and (b) density of states of 
LDA+$U$+SOC and GGA+$U$+SOC ($U=4$ eV) (solid lines), compared with inelastic neutron
scattering data from Pang \emph{et al.}  \cite{pang_phonon_2013} at 300 K (grey squares), this work at 600 K (black triangles), this work at 77 K (dashed black curves), and Bryan \emph{et al.}  \cite{bryan_impact_2019} at 10 K (dashed green
curves). The hollow points were directly computed using DFT+$U$, while the 
corresponding lines are Fourier interpolations. 
}
\label{fig:figure4}
\end{figure}

\begin{figure}[t]
\includegraphics[width=1.0\columnwidth]{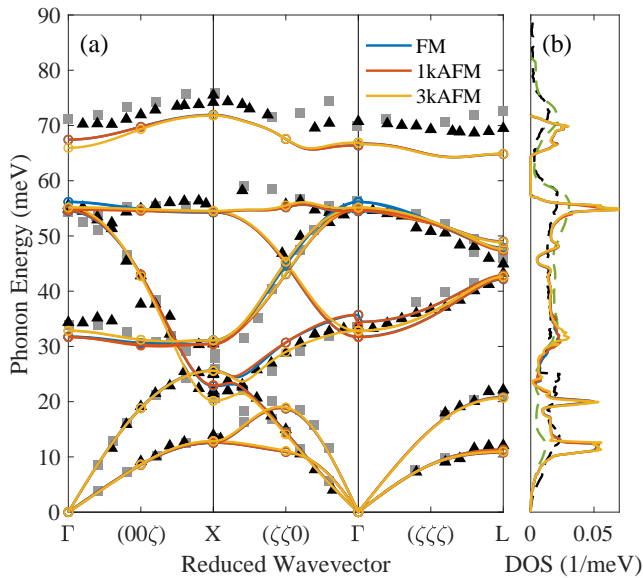}
\caption{Phonons of UO$_2$ in different magnetic states. (a) The GGA+$U$+SOC ($U=4$ eV)  phonon dispersion of FM and unfolded dispersion of 1\textbf{k} AFM and 3\textbf{k} AFM, and (b) density of states, compared with inelastic neutron
scattering data from Pang \emph{et al.}  \cite{pang_phonon_2013} at 300 K (grey squares), this work at 600 K (black triangles), this work at 77 K (dashed black curves), and Bryan \emph{et al.}  \cite{bryan_impact_2019} at 10 K (dashed green
curves). The hollow points were directly computed using DFT+$U$, while the 
corresponding lines are Fourier interpolations. }
\label{fig:figure3}
\end{figure}

We now consider the unfolded phonon dispersion curves of $\mathbb{S}_0$ in the 3\textbf{k} AFM state 
(see Section IIE in SM\cite{Supplemental_Material} for additional details), using
LDA+$U$+SOC and GGA+$U$+SOC ($U=4$ eV), and compare against the experimental
phonon dispersion curves from Pang \emph{et al.} at 300 K   \cite{pang_phonon_2013},
the density of states from Bryan \emph{et al.} at 10 K   \cite{bryan_impact_2019}, the phonon dispersion curves at 600 K from this work, and the density of states at 77 K from this work 
(see Figure~\ref{fig:figure4}).
Generally, the GGA+$U$ results are in better agreement with experiments for all
phonon branches except the highest frequency one (LO2). In the original work by Pang \emph{et al.}   \cite{pang_phonon_2013}, the LO2 mode fit could not have accounted for Bryan \emph{et al.}'s later discovery of an anharmonic feature appearing just above the phonons  \cite{bryanNonlinearPropagatingModes2020}, which probably explains the additional scatter toward higher frequencies in their fit points as compared to ours. 
Interestingly, many of the differences between 
LDA+$U$ and GGA+$U$ are attributable to the difference in lattice parameter (see in SM \cite{Supplemental_Material} Section II B).

\begin{figure}[t]
\includegraphics[width=0.941\columnwidth]{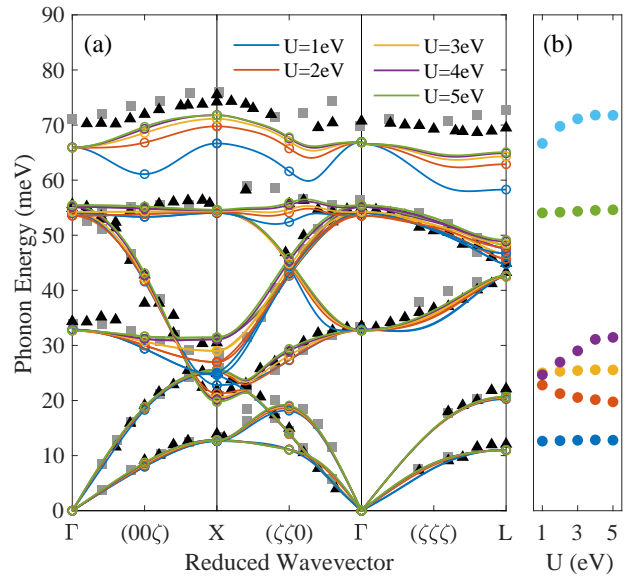}
\caption{ (a) Unfolded phonon dispersion (solid curves) and (b) phonon frequencies at the $X$ point (filled points), calculated for 3\textbf{k} AFM UO$_2$ using GGA+$U$+SOC, as functions of the Hubbard $U$.
In panel (a), the phonon dispersion are compared with inelastic neutron
scattering data from Pang \emph{et al.}  \cite{pang_phonon_2013} at 300 K (grey squares), and from this work at 600 K (black triangles). The hollow points were directly computed using DFT+$U$, while the 
corresponding lines are Fourier interpolations. 
In panel (b), the different colors represent different phonon modes at $X$ point.
}
\label{fig:figure6}
\end{figure}

We now explore the effect of magnetic ordering and the Hubbard $U$ on the unfolded phonon dispersion curves.
Comparing the phonons for
the FM, 1\textbf{k} AFM, and 3\textbf{k} AFM states when 
using 
GGA+$U$+SOC ($U=4$ eV) (see Figure~\ref{fig:figure3}), 
there are only small differences,
demonstrating that magnetic ordering only
has a minor effect on the quadratic portion of the vibrational Hamiltonian.  
The symmetry breaking in the FM and 1\textbf{k} AFM states is large enough to be noticed visually, but still very
small (though see Figure S7 \cite{Supplemental_Material} for plots along additional directions where 
splittings are more pronounced).
Alternatively, the Hubbard $U$ does have a strong effect on select phonon branches,
as demonstrated by computing the phonon 
dispersion curves of
$\mathbb{S}_0$ in the 3\textbf{k} AFM state,
using GGA+$U$+SOC (see Figure~\ref{fig:figure6}(a)). All acoustic phonon branches
are approximately independent of the Hubbard $U$, while two optical phonon
branches, TO1 and LO2, are significantly affected by the Hubbard $U$. More
specifically, the phonon frequencies of the TO1 and LO2 branches at $X$ points (see Figure~\ref{fig:figure6}(b)), along with
the LO2 branch at $L$ points, increase with increasing $U$. However, the
$U$ dependence eventually saturates, as indicated by the small difference 
between $U=4$ and $U=5$ eV.

\section{\label{sec:conclude}Conclusion}

In summary, we investigated the ground state properties of UO$_2$ by using DFT+$U$, including SOC.
Particular emphasis was placed on searching for the ground state occupation matrix,
and this was executed in the FM state using $U=4$ eV. When incorporating SOC, we found a previously
unreported occupation matrix $\mathbb{S}_0$ that we then used to extensively explore the symmetry-broken phases,
where the resulting occupation matrices are closely related to $\mathbb{S}_0$. When exploring the 3\textbf{k} AFM state, we found an oxygen cage distortion that was 
in excellent agreement with experimental results. Both the Hubbard $U$ and SOC, along with the
concomitant occupation matrix $\mathbb{S}_0$, are necessary for obtaining the
3\textbf{k} AFM state and the oxygen cage distortion. Furthermore, including the Hubbard $U$ 
in the absence of SOC results in a spurious 
shear strain distortion that is not observed in experiment. 

Both LDA and GGA were explored in our DFT+$U$ calculations, and each appears to
offer some particular advantage in describing the known experimental outcomes. As expected, at
$U=0$, LDA underpredicts and GGA overpredicts the lattice constant,
and an increasing $U$ increases the lattice parameter in both cases,
yielding better agreement for LDA+$U$ at $U=4$ eV. Both functionals accurately
predict the oxygen cage distortion for a sufficiently large value of $U$ (i.e., $U=4$ eV).
For the magnetic structure, LDA+$U$ correctly predicts the 3\textbf{k} AFM state to be lower
in energy than the 1\textbf{k} AFM and FM states, while GGA+$U$ incorrectly predicts 
the 1\textbf{k} AFM to be lower in energy than the 3\textbf{k} AFM state. That being said, these energy differences
are all on a small scale and should be treated with caution. 
Phonons were extensively explored: only selected modes have a nontrivial dependence on $U$, 
and magnetic ordering only has a small influence. At $U=4$ eV, both LDA and GGA give reasonable results,
though LDA better describes the LO2 mode (albeit showing some deficiency with the TO1/LO1 modes), whereas GGA gives the opposite
result. Most of these differences are attributable to the difference in lattice constant between LDA and GGA.
In summary, DFT+$U$ offers a quantitatively reasonable picture of UO$_2$, as compared to experiments. 
The success of DFT+$U$ in describing the ground state properties of UO$_2$ warrants exploration of finite
temperature properties, including thermal conductivity.

\section{Acknowledgements}

This work is supported by the Center for Thermal Energy Transport under Irradiation, an Energy Frontier Research Center funded by the U.S. Department of Energy (DOE) Office of Basic Energy Sciences. This research used resources at the Spallation Neutron Source, a DOE Office of Science User Facility operated by the ORNL. This research made use of Idaho National Laboratory computing resources, which are supported by the DOE Office of Nuclear Energy and the Nuclear Science User Facilities under contract no. DE-AC07-05ID14517.
This research also used resources of the National Energy Research Scientific Computing Center, a DOE Office of Science User Facility supported by the Office of Science of the U.S. Department of Energy under Contract No. DE-AC02-05CH11231.
The phonon unfolding was supported by grant DE-SC0016507 funded by the U.S. Department of Energy, Office of Science.

\bibliographystyle{apsrev4-1}

\begin{thebibliography}{54}%
\makeatletter
\providecommand \@ifxundefined [1]{%
 \@ifx{#1\undefined}
}%
\providecommand \@ifnum [1]{%
 \ifnum #1\expandafter \@firstoftwo
 \else \expandafter \@secondoftwo
 \fi
}%
\providecommand \@ifx [1]{%
 \ifx #1\expandafter \@firstoftwo
 \else \expandafter \@secondoftwo
 \fi
}%
\providecommand \natexlab [1]{#1}%
\providecommand \enquote  [1]{``#1''}%
\providecommand \bibnamefont  [1]{#1}%
\providecommand \bibfnamefont [1]{#1}%
\providecommand \citenamefont [1]{#1}%
\providecommand \href@noop [0]{\@secondoftwo}%
\providecommand \href [0]{\begingroup \@sanitize@url \@href}%
\providecommand \@href[1]{\@@startlink{#1}\@@href}%
\providecommand \@@href[1]{\endgroup#1\@@endlink}%
\providecommand \@sanitize@url [0]{\catcode `\\12\catcode `\$12\catcode
  `\&12\catcode `\#12\catcode `\^12\catcode `\_12\catcode `\%12\relax}%
\providecommand \@@startlink[1]{}%
\providecommand \@@endlink[0]{}%
\providecommand \url  [0]{\begingroup\@sanitize@url \@url }%
\providecommand \@url [1]{\endgroup\@href {#1}{\urlprefix }}%
\providecommand \urlprefix  [0]{URL }%
\providecommand \Eprint [0]{\href }%
\providecommand \doibase [0]{http://dx.doi.org/}%
\providecommand \selectlanguage [0]{\@gobble}%
\providecommand \bibinfo  [0]{\@secondoftwo}%
\providecommand \bibfield  [0]{\@secondoftwo}%
\providecommand \translation [1]{[#1]}%
\providecommand \BibitemOpen [0]{}%
\providecommand \bibitemStop [0]{}%
\providecommand \bibitemNoStop [0]{.\EOS\space}%
\providecommand \EOS [0]{\spacefactor3000\relax}%
\providecommand \BibitemShut  [1]{\csname bibitem#1\endcsname}%
\let\auto@bib@innerbib\@empty
\bibitem [{\citenamefont {Idiri}\ \emph {et~al.}(2004)\citenamefont {Idiri},
  \citenamefont {Le~Bihan}, \citenamefont {Heathman},\ and\ \citenamefont
  {Rebizant}}]{idiri_behavior_2004}%
  \BibitemOpen
  \bibfield  {author} {\bibinfo {author} {\bibfnamefont {M.}~\bibnamefont
  {Idiri}}, \bibinfo {author} {\bibfnamefont {T.}~\bibnamefont {Le~Bihan}},
  \bibinfo {author} {\bibfnamefont {S.}~\bibnamefont {Heathman}}, \ and\
  \bibinfo {author} {\bibfnamefont {J.}~\bibnamefont {Rebizant}},\ }\href
  {\doibase 10.1103/PhysRevB.70.014113} {\bibfield  {journal} {\bibinfo
  {journal} {Physical Review B}\ }\textbf {\bibinfo {volume} {70}},\ \bibinfo
  {pages} {014113} (\bibinfo {year} {2004})}\BibitemShut {NoStop}%
\bibitem [{\citenamefont {Lander}\ \emph {et~al.}(1976)\citenamefont {Lander},
  \citenamefont {Faber}, \citenamefont {Freeman},\ and\ \citenamefont
  {Desclaux}}]{lander_neutron-diffraction_1976}%
  \BibitemOpen
  \bibfield  {author} {\bibinfo {author} {\bibfnamefont {G.~H.}\ \bibnamefont
  {Lander}}, \bibinfo {author} {\bibfnamefont {J.}~\bibnamefont {Faber}},
  \bibinfo {author} {\bibfnamefont {A.~J.}\ \bibnamefont {Freeman}}, \ and\
  \bibinfo {author} {\bibfnamefont {J.~P.}\ \bibnamefont {Desclaux}},\ }\href
  {\doibase 10.1103/PhysRevB.13.1177} {\bibfield  {journal} {\bibinfo
  {journal} {Physical Review B}\ }\textbf {\bibinfo {volume} {13}},\ \bibinfo
  {pages} {1177} (\bibinfo {year} {1976})}\BibitemShut {NoStop}%
\bibitem [{\citenamefont {Santini}\ \emph {et~al.}(2009)\citenamefont
  {Santini}, \citenamefont {Carretta}, \citenamefont {Amoretti}, \citenamefont
  {Caciuffo}, \citenamefont {Magnani},\ and\ \citenamefont
  {Lander}}]{santini_multipolar_2009}%
  \BibitemOpen
  \bibfield  {author} {\bibinfo {author} {\bibfnamefont {P.}~\bibnamefont
  {Santini}}, \bibinfo {author} {\bibfnamefont {S.}~\bibnamefont {Carretta}},
  \bibinfo {author} {\bibfnamefont {G.}~\bibnamefont {Amoretti}}, \bibinfo
  {author} {\bibfnamefont {R.}~\bibnamefont {Caciuffo}}, \bibinfo {author}
  {\bibfnamefont {N.}~\bibnamefont {Magnani}}, \ and\ \bibinfo {author}
  {\bibfnamefont {G.~H.}\ \bibnamefont {Lander}},\ }\href {\doibase
  10.1103/RevModPhys.81.807} {\bibfield  {journal} {\bibinfo  {journal}
  {Reviews of Modern Physics}\ }\textbf {\bibinfo {volume} {81}},\ \bibinfo
  {pages} {807} (\bibinfo {year} {2009})}\BibitemShut {NoStop}%
\bibitem [{\citenamefont {Jones}\ \emph {et~al.}(1952)\citenamefont {Jones},
  \citenamefont {Gordon},\ and\ \citenamefont {Long}}]{jones_heat_1952}%
  \BibitemOpen
  \bibfield  {author} {\bibinfo {author} {\bibfnamefont {W.~M.}\ \bibnamefont
  {Jones}}, \bibinfo {author} {\bibfnamefont {J.}~\bibnamefont {Gordon}}, \
  and\ \bibinfo {author} {\bibfnamefont {E.~A.}\ \bibnamefont {Long}},\ }\href
  {\doibase 10.1063/1.1700518} {\bibfield  {journal} {\bibinfo  {journal} {The
  Journal of Chemical Physics}\ }\textbf {\bibinfo {volume} {20}},\ \bibinfo
  {pages} {695} (\bibinfo {year} {1952})}\BibitemShut {NoStop}%
\bibitem [{\citenamefont {Faber}\ \emph {et~al.}(1975)\citenamefont {Faber},
  \citenamefont {Lander},\ and\ \citenamefont
  {Cooper}}]{faberNeutronDiffractionStudyMathrmO1975}%
  \BibitemOpen
  \bibfield  {author} {\bibinfo {author} {\bibfnamefont {J.}~\bibnamefont
  {Faber}}, \bibinfo {author} {\bibfnamefont {G.~H.}\ \bibnamefont {Lander}}, \
  and\ \bibinfo {author} {\bibfnamefont {B.~R.}\ \bibnamefont {Cooper}},\
  }\href {\doibase 10.1103/PhysRevLett.35.1770} {\bibfield  {journal} {\bibinfo
   {journal} {Physical Review Letters}\ }\textbf {\bibinfo {volume} {35}},\
  \bibinfo {pages} {1770} (\bibinfo {year} {1975})}\BibitemShut {NoStop}%
\bibitem [{\citenamefont {Dorado}\ \emph {et~al.}(2010)\citenamefont {Dorado},
  \citenamefont {Jomard}, \citenamefont {Freyss},\ and\ \citenamefont
  {Bertolus}}]{dorado_stability_2010}%
  \BibitemOpen
  \bibfield  {author} {\bibinfo {author} {\bibfnamefont {B.}~\bibnamefont
  {Dorado}}, \bibinfo {author} {\bibfnamefont {G.}~\bibnamefont {Jomard}},
  \bibinfo {author} {\bibfnamefont {M.}~\bibnamefont {Freyss}}, \ and\ \bibinfo
  {author} {\bibfnamefont {M.}~\bibnamefont {Bertolus}},\ }\href {\doibase
  10.1103/PhysRevB.82.035114} {\bibfield  {journal} {\bibinfo  {journal}
  {Physical Review B}\ }\textbf {\bibinfo {volume} {82}},\ \bibinfo {pages}
  {035114} (\bibinfo {year} {2010})}\BibitemShut {NoStop}%
\bibitem [{\citenamefont {Burlet}\ \emph {et~al.}(1986)\citenamefont {Burlet},
  \citenamefont {Rossat-Mignod}, \citenamefont {vuevel}, \citenamefont {Vogt},
  \citenamefont {Spirlet},\ and\ \citenamefont
  {Rebivant}}]{burlet_neutron_1986}%
  \BibitemOpen
  \bibfield  {author} {\bibinfo {author} {\bibfnamefont {P.}~\bibnamefont
  {Burlet}}, \bibinfo {author} {\bibfnamefont {J.}~\bibnamefont
  {Rossat-Mignod}}, \bibinfo {author} {\bibfnamefont {S.}~\bibnamefont
  {vuevel}}, \bibinfo {author} {\bibfnamefont {O.}~\bibnamefont {Vogt}},
  \bibinfo {author} {\bibfnamefont {J.~C.}\ \bibnamefont {Spirlet}}, \ and\
  \bibinfo {author} {\bibfnamefont {J.}~\bibnamefont {Rebivant}},\ }\href
  {\doibase 10.1016/0022-5088(86)90521-7} {\bibfield  {journal} {\bibinfo
  {journal} {Journal of the Less Common Metals}\ }\bibinfo {series}
  {Proceedings of {Actinides} 85, {Aix} en {Provence} - {Part} {I}},\ \textbf
  {\bibinfo {volume} {121}},\ \bibinfo {pages} {121} (\bibinfo {year}
  {1986})}\BibitemShut {NoStop}%
\bibitem [{\citenamefont {Ikushima}\ \emph {et~al.}(2001)\citenamefont
  {Ikushima}, \citenamefont {Tsutsui}, \citenamefont {Haga}, \citenamefont
  {Yasuoka}, \citenamefont {Walstedt}, \citenamefont {Masaki}, \citenamefont
  {Nakamura}, \citenamefont {Nasu},\ and\ \citenamefont
  {Ōnuki}}]{ikushima_first-order_2001}%
  \BibitemOpen
  \bibfield  {author} {\bibinfo {author} {\bibfnamefont {K.}~\bibnamefont
  {Ikushima}}, \bibinfo {author} {\bibfnamefont {S.}~\bibnamefont {Tsutsui}},
  \bibinfo {author} {\bibfnamefont {Y.}~\bibnamefont {Haga}}, \bibinfo {author}
  {\bibfnamefont {H.}~\bibnamefont {Yasuoka}}, \bibinfo {author} {\bibfnamefont
  {R.~E.}\ \bibnamefont {Walstedt}}, \bibinfo {author} {\bibfnamefont {N.~M.}\
  \bibnamefont {Masaki}}, \bibinfo {author} {\bibfnamefont {A.}~\bibnamefont
  {Nakamura}}, \bibinfo {author} {\bibfnamefont {S.}~\bibnamefont {Nasu}}, \
  and\ \bibinfo {author} {\bibfnamefont {Y.}~\bibnamefont {Ōnuki}},\ }\href
  {\doibase 10.1103/PhysRevB.63.104404} {\bibfield  {journal} {\bibinfo
  {journal} {Physical Review B}\ }\textbf {\bibinfo {volume} {63}},\ \bibinfo
  {pages} {104404} (\bibinfo {year} {2001})}\BibitemShut {NoStop}%
\bibitem [{\citenamefont {Blackburn}\ \emph {et~al.}(2005)\citenamefont
  {Blackburn}, \citenamefont {Caciuffo}, \citenamefont {Magnani}, \citenamefont
  {Santini}, \citenamefont {Brown}, \citenamefont {Enderle},\ and\
  \citenamefont {Lander}}]{blackburn_spherical_2005}%
  \BibitemOpen
  \bibfield  {author} {\bibinfo {author} {\bibfnamefont {E.}~\bibnamefont
  {Blackburn}}, \bibinfo {author} {\bibfnamefont {R.}~\bibnamefont {Caciuffo}},
  \bibinfo {author} {\bibfnamefont {N.}~\bibnamefont {Magnani}}, \bibinfo
  {author} {\bibfnamefont {P.}~\bibnamefont {Santini}}, \bibinfo {author}
  {\bibfnamefont {P.~J.}\ \bibnamefont {Brown}}, \bibinfo {author}
  {\bibfnamefont {M.}~\bibnamefont {Enderle}}, \ and\ \bibinfo {author}
  {\bibfnamefont {G.~H.}\ \bibnamefont {Lander}},\ }\href {\doibase
  10.1103/PhysRevB.72.184411} {\bibfield  {journal} {\bibinfo  {journal}
  {Physical Review B}\ }\textbf {\bibinfo {volume} {72}},\ \bibinfo {pages}
  {184411} (\bibinfo {year} {2005})}\BibitemShut {NoStop}%
\bibitem [{\citenamefont {Wilkins}\ \emph {et~al.}(2006)\citenamefont
  {Wilkins}, \citenamefont {Caciuffo}, \citenamefont {Detlefs}, \citenamefont
  {Rebizant}, \citenamefont {Colineau}, \citenamefont {Wastin},\ and\
  \citenamefont {Lander}}]{wilkins_direct_2006}%
  \BibitemOpen
  \bibfield  {author} {\bibinfo {author} {\bibfnamefont {S.~B.}\ \bibnamefont
  {Wilkins}}, \bibinfo {author} {\bibfnamefont {R.}~\bibnamefont {Caciuffo}},
  \bibinfo {author} {\bibfnamefont {C.}~\bibnamefont {Detlefs}}, \bibinfo
  {author} {\bibfnamefont {J.}~\bibnamefont {Rebizant}}, \bibinfo {author}
  {\bibfnamefont {E.}~\bibnamefont {Colineau}}, \bibinfo {author}
  {\bibfnamefont {F.}~\bibnamefont {Wastin}}, \ and\ \bibinfo {author}
  {\bibfnamefont {G.~H.}\ \bibnamefont {Lander}},\ }\href {\doibase
  10.1103/PhysRevB.73.060406} {\bibfield  {journal} {\bibinfo  {journal}
  {Physical Review B}\ }\textbf {\bibinfo {volume} {73}},\ \bibinfo {pages}
  {060406} (\bibinfo {year} {2006})}\BibitemShut {NoStop}%
\bibitem [{\citenamefont {Kudin}\ \emph {et~al.}(2002)\citenamefont {Kudin},
  \citenamefont {Scuseria},\ and\ \citenamefont {Martin}}]{kudin_hybrid_2002}%
  \BibitemOpen
  \bibfield  {author} {\bibinfo {author} {\bibfnamefont {K.~N.}\ \bibnamefont
  {Kudin}}, \bibinfo {author} {\bibfnamefont {G.~E.}\ \bibnamefont {Scuseria}},
  \ and\ \bibinfo {author} {\bibfnamefont {R.~L.}\ \bibnamefont {Martin}},\
  }\href {\doibase 10.1103/PhysRevLett.89.266402} {\bibfield  {journal}
  {\bibinfo  {journal} {Physical Review Letters}\ }\textbf {\bibinfo {volume}
  {89}},\ \bibinfo {pages} {266402} (\bibinfo {year} {2002})}\BibitemShut
  {NoStop}%
\bibitem [{\citenamefont {Prodan}\ \emph {et~al.}(2007)\citenamefont {Prodan},
  \citenamefont {Scuseria},\ and\ \citenamefont
  {Martin}}]{prodan_covalency_2007}%
  \BibitemOpen
  \bibfield  {author} {\bibinfo {author} {\bibfnamefont {I.~D.}\ \bibnamefont
  {Prodan}}, \bibinfo {author} {\bibfnamefont {G.~E.}\ \bibnamefont
  {Scuseria}}, \ and\ \bibinfo {author} {\bibfnamefont {R.~L.}\ \bibnamefont
  {Martin}},\ }\href {\doibase 10.1103/PhysRevB.76.033101} {\bibfield
  {journal} {\bibinfo  {journal} {Physical Review B}\ }\textbf {\bibinfo
  {volume} {76}},\ \bibinfo {pages} {033101} (\bibinfo {year}
  {2007})}\BibitemShut {NoStop}%
\bibitem [{\citenamefont {Becke}(1993)}]{becke_new_1993}%
  \BibitemOpen
  \bibfield  {author} {\bibinfo {author} {\bibfnamefont {A.~D.}\ \bibnamefont
  {Becke}},\ }\href {\doibase 10.1063/1.464304} {\bibfield  {journal} {\bibinfo
   {journal} {The Journal of Chemical Physics}\ }\textbf {\bibinfo {volume}
  {98}},\ \bibinfo {pages} {1372} (\bibinfo {year} {1993})}\BibitemShut
  {NoStop}%
\bibitem [{\citenamefont {Petit}\ \emph {et~al.}(2010)\citenamefont {Petit},
  \citenamefont {Svane}, \citenamefont {Szotek}, \citenamefont {Temmerman},\
  and\ \citenamefont {Stocks}}]{petit_electronic_2010}%
  \BibitemOpen
  \bibfield  {author} {\bibinfo {author} {\bibfnamefont {L.}~\bibnamefont
  {Petit}}, \bibinfo {author} {\bibfnamefont {A.}~\bibnamefont {Svane}},
  \bibinfo {author} {\bibfnamefont {Z.}~\bibnamefont {Szotek}}, \bibinfo
  {author} {\bibfnamefont {W.~M.}\ \bibnamefont {Temmerman}}, \ and\ \bibinfo
  {author} {\bibfnamefont {G.~M.}\ \bibnamefont {Stocks}},\ }\href {\doibase
  10.1103/PhysRevB.81.045108} {\bibfield  {journal} {\bibinfo  {journal}
  {Physical Review B}\ }\textbf {\bibinfo {volume} {81}},\ \bibinfo {pages}
  {045108} (\bibinfo {year} {2010})}\BibitemShut {NoStop}%
\bibitem [{\citenamefont {Perdew}\ and\ \citenamefont
  {Zunger}(1981)}]{perdew_self-interaction_1981}%
  \BibitemOpen
  \bibfield  {author} {\bibinfo {author} {\bibfnamefont {J.~P.}\ \bibnamefont
  {Perdew}}\ and\ \bibinfo {author} {\bibfnamefont {A.}~\bibnamefont
  {Zunger}},\ }\href {\doibase 10.1103/PhysRevB.23.5048} {\bibfield  {journal}
  {\bibinfo  {journal} {Physical Review B}\ }\textbf {\bibinfo {volume} {23}},\
  \bibinfo {pages} {5048} (\bibinfo {year} {1981})}\BibitemShut {NoStop}%
\bibitem [{\citenamefont {Dudarev}\ \emph {et~al.}(1997)\citenamefont
  {Dudarev}, \citenamefont {Manh},\ and\ \citenamefont
  {Sutton}}]{dudarev_effect_1997}%
  \BibitemOpen
  \bibfield  {author} {\bibinfo {author} {\bibfnamefont {S.~L.}\ \bibnamefont
  {Dudarev}}, \bibinfo {author} {\bibfnamefont {D.~N.}\ \bibnamefont {Manh}}, \
  and\ \bibinfo {author} {\bibfnamefont {A.~P.}\ \bibnamefont {Sutton}},\
  }\href {\doibase 10.1080/13642819708202343} {\bibfield  {journal} {\bibinfo
  {journal} {Philosophical Magazine B}\ }\textbf {\bibinfo {volume} {75}},\
  \bibinfo {pages} {613} (\bibinfo {year} {1997})}\BibitemShut {NoStop}%
\bibitem [{\citenamefont {Anisimov}\ \emph {et~al.}(1991)\citenamefont
  {Anisimov}, \citenamefont {Zaanen},\ and\ \citenamefont
  {Andersen}}]{anisimov_band_1991}%
  \BibitemOpen
  \bibfield  {author} {\bibinfo {author} {\bibfnamefont {V.~I.}\ \bibnamefont
  {Anisimov}}, \bibinfo {author} {\bibfnamefont {J.}~\bibnamefont {Zaanen}}, \
  and\ \bibinfo {author} {\bibfnamefont {O.~K.}\ \bibnamefont {Andersen}},\
  }\href {\doibase 10.1103/PhysRevB.44.943} {\bibfield  {journal} {\bibinfo
  {journal} {Physical Review B}\ }\textbf {\bibinfo {volume} {44}},\ \bibinfo
  {pages} {943} (\bibinfo {year} {1991})}\BibitemShut {NoStop}%
\bibitem [{\citenamefont {Georges}\ \emph {et~al.}(1996)\citenamefont
  {Georges}, \citenamefont {Kotliar}, \citenamefont {Krauth},\ and\
  \citenamefont {Rozenberg}}]{georges_dynamical_1996}%
  \BibitemOpen
  \bibfield  {author} {\bibinfo {author} {\bibfnamefont {A.}~\bibnamefont
  {Georges}}, \bibinfo {author} {\bibfnamefont {G.}~\bibnamefont {Kotliar}},
  \bibinfo {author} {\bibfnamefont {W.}~\bibnamefont {Krauth}}, \ and\ \bibinfo
  {author} {\bibfnamefont {M.~J.}\ \bibnamefont {Rozenberg}},\ }\href {\doibase
  10.1103/RevModPhys.68.13} {\bibfield  {journal} {\bibinfo  {journal} {Reviews
  of Modern Physics}\ }\textbf {\bibinfo {volume} {68}},\ \bibinfo {pages} {13}
  (\bibinfo {year} {1996})}\BibitemShut {NoStop}%
\bibitem [{\citenamefont {Kotliar}\ \emph {et~al.}(2006)\citenamefont
  {Kotliar}, \citenamefont {Savrasov}, \citenamefont {Haule}, \citenamefont
  {Oudovenko}, \citenamefont {Parcollet},\ and\ \citenamefont
  {Marianetti}}]{kotliar_electronic_2006}%
  \BibitemOpen
  \bibfield  {author} {\bibinfo {author} {\bibfnamefont {G.}~\bibnamefont
  {Kotliar}}, \bibinfo {author} {\bibfnamefont {S.~Y.}\ \bibnamefont
  {Savrasov}}, \bibinfo {author} {\bibfnamefont {K.}~\bibnamefont {Haule}},
  \bibinfo {author} {\bibfnamefont {V.~S.}\ \bibnamefont {Oudovenko}}, \bibinfo
  {author} {\bibfnamefont {O.}~\bibnamefont {Parcollet}}, \ and\ \bibinfo
  {author} {\bibfnamefont {C.~A.}\ \bibnamefont {Marianetti}},\ }\href
  {\doibase 10.1103/RevModPhys.78.865} {\bibfield  {journal} {\bibinfo
  {journal} {Reviews of Modern Physics}\ }\textbf {\bibinfo {volume} {78}},\
  \bibinfo {pages} {865} (\bibinfo {year} {2006})}\BibitemShut {NoStop}%
\bibitem [{\citenamefont {Agapito}\ \emph {et~al.}(2015)\citenamefont
  {Agapito}, \citenamefont {Curtarolo},\ and\ \citenamefont
  {Buongiorno~Nardelli}}]{agapito_reformulation_2015}%
  \BibitemOpen
  \bibfield  {author} {\bibinfo {author} {\bibfnamefont {L.~A.}\ \bibnamefont
  {Agapito}}, \bibinfo {author} {\bibfnamefont {S.}~\bibnamefont {Curtarolo}},
  \ and\ \bibinfo {author} {\bibfnamefont {M.}~\bibnamefont
  {Buongiorno~Nardelli}},\ }\href {\doibase 10.1103/PhysRevX.5.011006}
  {\bibfield  {journal} {\bibinfo  {journal} {Physical Review X}\ }\textbf
  {\bibinfo {volume} {5}},\ \bibinfo {pages} {011006} (\bibinfo {year}
  {2015})}\BibitemShut {NoStop}%
\bibitem [{\citenamefont {Dorado}\ \emph {et~al.}(2009)\citenamefont {Dorado},
  \citenamefont {Amadon}, \citenamefont {Freyss},\ and\ \citenamefont
  {Bertolus}}]{dorado_textdfttextu_2009}%
  \BibitemOpen
  \bibfield  {author} {\bibinfo {author} {\bibfnamefont {B.}~\bibnamefont
  {Dorado}}, \bibinfo {author} {\bibfnamefont {B.}~\bibnamefont {Amadon}},
  \bibinfo {author} {\bibfnamefont {M.}~\bibnamefont {Freyss}}, \ and\ \bibinfo
  {author} {\bibfnamefont {M.}~\bibnamefont {Bertolus}},\ }\href {\doibase
  10.1103/PhysRevB.79.235125} {\bibfield  {journal} {\bibinfo  {journal}
  {Physical Review B}\ }\textbf {\bibinfo {volume} {79}},\ \bibinfo {pages}
  {235125} (\bibinfo {year} {2009})}\BibitemShut {NoStop}%
\bibitem [{\citenamefont {Amadon}\ \emph {et~al.}(2008)\citenamefont {Amadon},
  \citenamefont {Jollet},\ and\ \citenamefont
  {Torrent}}]{amadon_ensuremathgamma_2008}%
  \BibitemOpen
  \bibfield  {author} {\bibinfo {author} {\bibfnamefont {B.}~\bibnamefont
  {Amadon}}, \bibinfo {author} {\bibfnamefont {F.}~\bibnamefont {Jollet}}, \
  and\ \bibinfo {author} {\bibfnamefont {M.}~\bibnamefont {Torrent}},\ }\href
  {\doibase 10.1103/PhysRevB.77.155104} {\bibfield  {journal} {\bibinfo
  {journal} {Physical Review B}\ }\textbf {\bibinfo {volume} {77}},\ \bibinfo
  {pages} {155104} (\bibinfo {year} {2008})}\BibitemShut {NoStop}%
\bibitem [{\citenamefont {Jomard}\ \emph {et~al.}(2008)\citenamefont {Jomard},
  \citenamefont {Amadon}, \citenamefont {Bottin},\ and\ \citenamefont
  {Torrent}}]{jomard_structural_2008}%
  \BibitemOpen
  \bibfield  {author} {\bibinfo {author} {\bibfnamefont {G.}~\bibnamefont
  {Jomard}}, \bibinfo {author} {\bibfnamefont {B.}~\bibnamefont {Amadon}},
  \bibinfo {author} {\bibfnamefont {F.}~\bibnamefont {Bottin}}, \ and\ \bibinfo
  {author} {\bibfnamefont {M.}~\bibnamefont {Torrent}},\ }\href {\doibase
  10.1103/PhysRevB.78.075125} {\bibfield  {journal} {\bibinfo  {journal}
  {Physical Review B}\ }\textbf {\bibinfo {volume} {78}},\ \bibinfo {pages}
  {075125} (\bibinfo {year} {2008})}\BibitemShut {NoStop}%
\bibitem [{\citenamefont {Zhou}\ and\ \citenamefont
  {Ozoliņš}(2011)}]{zhou_crystal_2011}%
  \BibitemOpen
  \bibfield  {author} {\bibinfo {author} {\bibfnamefont {F.}~\bibnamefont
  {Zhou}}\ and\ \bibinfo {author} {\bibfnamefont {V.}~\bibnamefont
  {Ozoliņš}},\ }\href {\doibase 10.1103/PhysRevB.83.085106} {\bibfield
  {journal} {\bibinfo  {journal} {Physical Review B}\ }\textbf {\bibinfo
  {volume} {83}},\ \bibinfo {pages} {085106} (\bibinfo {year}
  {2011})}\BibitemShut {NoStop}%
\bibitem [{\citenamefont {Geng}\ \emph {et~al.}(2010)\citenamefont {Geng},
  \citenamefont {Chen}, \citenamefont {Kaneta}, \citenamefont {Kinoshita},\
  and\ \citenamefont {Wu}}]{geng_interplay_2010}%
  \BibitemOpen
  \bibfield  {author} {\bibinfo {author} {\bibfnamefont {H.~Y.}\ \bibnamefont
  {Geng}}, \bibinfo {author} {\bibfnamefont {Y.}~\bibnamefont {Chen}}, \bibinfo
  {author} {\bibfnamefont {Y.}~\bibnamefont {Kaneta}}, \bibinfo {author}
  {\bibfnamefont {M.}~\bibnamefont {Kinoshita}}, \ and\ \bibinfo {author}
  {\bibfnamefont {Q.}~\bibnamefont {Wu}},\ }\href {\doibase
  10.1103/PhysRevB.82.094106} {\bibfield  {journal} {\bibinfo  {journal}
  {Physical Review B}\ }\textbf {\bibinfo {volume} {82}},\ \bibinfo {pages}
  {094106} (\bibinfo {year} {2010})}\BibitemShut {NoStop}%
\bibitem [{\citenamefont {Meredig}\ \emph {et~al.}(2010)\citenamefont
  {Meredig}, \citenamefont {Thompson}, \citenamefont {Hansen}, \citenamefont
  {Wolverton},\ and\ \citenamefont {van~de Walle}}]{meredig_method_2010}%
  \BibitemOpen
  \bibfield  {author} {\bibinfo {author} {\bibfnamefont {B.}~\bibnamefont
  {Meredig}}, \bibinfo {author} {\bibfnamefont {A.}~\bibnamefont {Thompson}},
  \bibinfo {author} {\bibfnamefont {H.~A.}\ \bibnamefont {Hansen}}, \bibinfo
  {author} {\bibfnamefont {C.}~\bibnamefont {Wolverton}}, \ and\ \bibinfo
  {author} {\bibfnamefont {A.}~\bibnamefont {van~de Walle}},\ }\href {\doibase
  10.1103/PhysRevB.82.195128} {\bibfield  {journal} {\bibinfo  {journal}
  {Physical Review B}\ }\textbf {\bibinfo {volume} {82}},\ \bibinfo {pages}
  {195128} (\bibinfo {year} {2010})}\BibitemShut {NoStop}%
\bibitem [{\citenamefont {Wang}\ \emph
  {et~al.}(2013{\natexlab{a}})\citenamefont {Wang}, \citenamefont {Ewing},\
  and\ \citenamefont {Becker}}]{wang_electronic_2013}%
  \BibitemOpen
  \bibfield  {author} {\bibinfo {author} {\bibfnamefont {J.}~\bibnamefont
  {Wang}}, \bibinfo {author} {\bibfnamefont {R.~C.}\ \bibnamefont {Ewing}}, \
  and\ \bibinfo {author} {\bibfnamefont {U.}~\bibnamefont {Becker}},\ }\href
  {\doibase 10.1103/PhysRevB.88.024109} {\bibfield  {journal} {\bibinfo
  {journal} {Physical Review B}\ }\textbf {\bibinfo {volume} {88}},\ \bibinfo
  {pages} {024109} (\bibinfo {year} {2013}{\natexlab{a}})}\BibitemShut
  {NoStop}%
\bibitem [{\citenamefont {Thompson}\ and\ \citenamefont
  {Wolverton}(2011)}]{thompson_first-principles_2011}%
  \BibitemOpen
  \bibfield  {author} {\bibinfo {author} {\bibfnamefont {A.~E.}\ \bibnamefont
  {Thompson}}\ and\ \bibinfo {author} {\bibfnamefont {C.}~\bibnamefont
  {Wolverton}},\ }\href {\doibase 10.1103/PhysRevB.84.134111} {\bibfield
  {journal} {\bibinfo  {journal} {Physical Review B}\ }\textbf {\bibinfo
  {volume} {84}},\ \bibinfo {pages} {134111} (\bibinfo {year}
  {2011})}\BibitemShut {NoStop}%
\bibitem [{\citenamefont {Laskowski}\ \emph {et~al.}(2004)\citenamefont
  {Laskowski}, \citenamefont {Madsen}, \citenamefont {Blaha},\ and\
  \citenamefont {Schwarz}}]{laskowski_magnetic_2004}%
  \BibitemOpen
  \bibfield  {author} {\bibinfo {author} {\bibfnamefont {R.}~\bibnamefont
  {Laskowski}}, \bibinfo {author} {\bibfnamefont {G.~K.~H.}\ \bibnamefont
  {Madsen}}, \bibinfo {author} {\bibfnamefont {P.}~\bibnamefont {Blaha}}, \
  and\ \bibinfo {author} {\bibfnamefont {K.}~\bibnamefont {Schwarz}},\ }\href
  {\doibase 10.1103/PhysRevB.69.140408} {\bibfield  {journal} {\bibinfo
  {journal} {Physical Review B}\ }\textbf {\bibinfo {volume} {69}},\ \bibinfo
  {pages} {140408} (\bibinfo {year} {2004})}\BibitemShut {NoStop}%
\bibitem [{\citenamefont {Gryaznov}\ \emph {et~al.}(2010)\citenamefont
  {Gryaznov}, \citenamefont {Heifets},\ and\ \citenamefont
  {Sedmidubsky}}]{gryaznov_density_2010}%
  \BibitemOpen
  \bibfield  {author} {\bibinfo {author} {\bibfnamefont {D.}~\bibnamefont
  {Gryaznov}}, \bibinfo {author} {\bibfnamefont {E.}~\bibnamefont {Heifets}}, \
  and\ \bibinfo {author} {\bibfnamefont {D.}~\bibnamefont {Sedmidubsky}},\
  }\href {\doibase 10.1039/C0CP00372G} {\bibfield  {journal} {\bibinfo
  {journal} {Physical Chemistry Chemical Physics}\ }\textbf {\bibinfo {volume}
  {12}},\ \bibinfo {pages} {12273} (\bibinfo {year} {2010})}\BibitemShut
  {NoStop}%
\bibitem [{\citenamefont {Fritz}(1976)}]{fritz_elastic_1976}%
  \BibitemOpen
  \bibfield  {author} {\bibinfo {author} {\bibfnamefont {I.~J.}\ \bibnamefont
  {Fritz}},\ }\href {\doibase 10.1063/1.322438} {\bibfield  {journal} {\bibinfo
   {journal} {Journal of Applied Physics}\ }\textbf {\bibinfo {volume} {47}},\
  \bibinfo {pages} {4353} (\bibinfo {year} {1976})}\BibitemShut {NoStop}%
\bibitem [{\citenamefont {Brandt}\ and\ \citenamefont
  {Walker}(1967)}]{brandt_temperature_1967}%
  \BibitemOpen
  \bibfield  {author} {\bibinfo {author} {\bibfnamefont {O.~G.}\ \bibnamefont
  {Brandt}}\ and\ \bibinfo {author} {\bibfnamefont {C.~T.}\ \bibnamefont
  {Walker}},\ }\href {\doibase 10.1103/PhysRevLett.18.11} {\bibfield  {journal}
  {\bibinfo  {journal} {Physical Review Letters}\ }\textbf {\bibinfo {volume}
  {18}},\ \bibinfo {pages} {11} (\bibinfo {year} {1967})}\BibitemShut {NoStop}%
\bibitem [{\citenamefont {Dolling}\ \emph {et~al.}(1965)\citenamefont
  {Dolling}, \citenamefont {Cowley},\ and\ \citenamefont
  {Woods}}]{dolling_crystal_1965}%
  \BibitemOpen
  \bibfield  {author} {\bibinfo {author} {\bibfnamefont {G.}~\bibnamefont
  {Dolling}}, \bibinfo {author} {\bibfnamefont {R.~A.}\ \bibnamefont {Cowley}},
  \ and\ \bibinfo {author} {\bibfnamefont {A.~D.~B.}\ \bibnamefont {Woods}},\
  }\href {\doibase 10.1139/p65-135} {\bibfield  {journal} {\bibinfo  {journal}
  {Canadian Journal of Physics}\ } (\bibinfo {year} {1965}),\
  10.1139/p65-135}\BibitemShut {NoStop}%
\bibitem [{\citenamefont {Pang}\ \emph {et~al.}(2013)\citenamefont {Pang},
  \citenamefont {Buyers}, \citenamefont {Chernatynskiy}, \citenamefont
  {Lumsden}, \citenamefont {Larson},\ and\ \citenamefont
  {Phillpot}}]{pang_phonon_2013}%
  \BibitemOpen
  \bibfield  {author} {\bibinfo {author} {\bibfnamefont {J.~W.~L.}\
  \bibnamefont {Pang}}, \bibinfo {author} {\bibfnamefont {W.~J.~L.}\
  \bibnamefont {Buyers}}, \bibinfo {author} {\bibfnamefont {A.}~\bibnamefont
  {Chernatynskiy}}, \bibinfo {author} {\bibfnamefont {M.~D.}\ \bibnamefont
  {Lumsden}}, \bibinfo {author} {\bibfnamefont {B.~C.}\ \bibnamefont {Larson}},
  \ and\ \bibinfo {author} {\bibfnamefont {S.~R.}\ \bibnamefont {Phillpot}},\
  }\href {\doibase 10.1103/PhysRevLett.110.157401} {\bibfield  {journal}
  {\bibinfo  {journal} {Physical Review Letters}\ }\textbf {\bibinfo {volume}
  {110}},\ \bibinfo {pages} {157401} (\bibinfo {year} {2013})}\BibitemShut
  {NoStop}%
\bibitem [{\citenamefont {Yin}\ and\ \citenamefont
  {Savrasov}(2008)}]{yin_origin_2008}%
  \BibitemOpen
  \bibfield  {author} {\bibinfo {author} {\bibfnamefont {Q.}~\bibnamefont
  {Yin}}\ and\ \bibinfo {author} {\bibfnamefont {S.~Y.}\ \bibnamefont
  {Savrasov}},\ }\href {\doibase 10.1103/PhysRevLett.100.225504} {\bibfield
  {journal} {\bibinfo  {journal} {Physical Review Letters}\ }\textbf {\bibinfo
  {volume} {100}},\ \bibinfo {pages} {225504} (\bibinfo {year}
  {2008})}\BibitemShut {NoStop}%
\bibitem [{\citenamefont {Devey}(2011)}]{devey_first_2011}%
  \BibitemOpen
  \bibfield  {author} {\bibinfo {author} {\bibfnamefont {A.~J.}\ \bibnamefont
  {Devey}},\ }\href {\doibase 10.1016/j.jnucmat.2011.03.012} {\bibfield
  {journal} {\bibinfo  {journal} {Journal of Nuclear Materials}\ }\textbf
  {\bibinfo {volume} {412}},\ \bibinfo {pages} {301} (\bibinfo {year}
  {2011})}\BibitemShut {NoStop}%
\bibitem [{\citenamefont {Sanati}\ \emph {et~al.}(2011)\citenamefont {Sanati},
  \citenamefont {Albers}, \citenamefont {Lookman},\ and\ \citenamefont
  {Saxena}}]{sanati_elastic_2011}%
  \BibitemOpen
  \bibfield  {author} {\bibinfo {author} {\bibfnamefont {M.}~\bibnamefont
  {Sanati}}, \bibinfo {author} {\bibfnamefont {R.~C.}\ \bibnamefont {Albers}},
  \bibinfo {author} {\bibfnamefont {T.}~\bibnamefont {Lookman}}, \ and\
  \bibinfo {author} {\bibfnamefont {A.}~\bibnamefont {Saxena}},\ }\href
  {\doibase 10.1103/PhysRevB.84.014116} {\bibfield  {journal} {\bibinfo
  {journal} {Physical Review B}\ }\textbf {\bibinfo {volume} {84}},\ \bibinfo
  {pages} {014116} (\bibinfo {year} {2011})}\BibitemShut {NoStop}%
\bibitem [{\citenamefont {Yun}\ \emph {et~al.}(2012)\citenamefont {Yun},
  \citenamefont {Legut},\ and\ \citenamefont {Oppeneer}}]{yun_phonon_2012}%
  \BibitemOpen
  \bibfield  {author} {\bibinfo {author} {\bibfnamefont {Y.}~\bibnamefont
  {Yun}}, \bibinfo {author} {\bibfnamefont {D.}~\bibnamefont {Legut}}, \ and\
  \bibinfo {author} {\bibfnamefont {P.~M.}\ \bibnamefont {Oppeneer}},\ }\href
  {\doibase 10.1016/j.jnucmat.2012.03.017} {\bibfield  {journal} {\bibinfo
  {journal} {Journal of Nuclear Materials}\ }\textbf {\bibinfo {volume}
  {426}},\ \bibinfo {pages} {109} (\bibinfo {year} {2012})}\BibitemShut
  {NoStop}%
\bibitem [{\citenamefont {Wang}\ \emph
  {et~al.}(2013{\natexlab{b}})\citenamefont {Wang}, \citenamefont {Zhang},
  \citenamefont {Lizárraga}, \citenamefont {Di~Marco},\ and\ \citenamefont
  {Eriksson}}]{wang_phonon_2013}%
  \BibitemOpen
  \bibfield  {author} {\bibinfo {author} {\bibfnamefont {B.-T.}\ \bibnamefont
  {Wang}}, \bibinfo {author} {\bibfnamefont {P.}~\bibnamefont {Zhang}},
  \bibinfo {author} {\bibfnamefont {R.}~\bibnamefont {Lizárraga}}, \bibinfo
  {author} {\bibfnamefont {I.}~\bibnamefont {Di~Marco}}, \ and\ \bibinfo
  {author} {\bibfnamefont {O.}~\bibnamefont {Eriksson}},\ }\href {\doibase
  10.1103/PhysRevB.88.104107} {\bibfield  {journal} {\bibinfo  {journal}
  {Physical Review B}\ }\textbf {\bibinfo {volume} {88}},\ \bibinfo {pages}
  {104107} (\bibinfo {year} {2013}{\natexlab{b}})}\BibitemShut {NoStop}%
\bibitem [{\citenamefont {Kaur}\ \emph {et~al.}(2013)\citenamefont {Kaur},
  \citenamefont {Panigrahi},\ and\ \citenamefont
  {Valsakumar}}]{kaur_thermal_2013}%
  \BibitemOpen
  \bibfield  {author} {\bibinfo {author} {\bibfnamefont {G.}~\bibnamefont
  {Kaur}}, \bibinfo {author} {\bibfnamefont {P.}~\bibnamefont {Panigrahi}}, \
  and\ \bibinfo {author} {\bibfnamefont {M.~C.}\ \bibnamefont {Valsakumar}},\
  }\href {\doibase 10.1088/0965-0393/21/6/065014} {\bibfield  {journal}
  {\bibinfo  {journal} {Modelling and Simulation in Materials Science and
  Engineering}\ }\textbf {\bibinfo {volume} {21}},\ \bibinfo {pages} {065014}
  (\bibinfo {year} {2013})}\BibitemShut {NoStop}%
\bibitem [{Sup()}]{Supplemental_Material}%
  \BibitemOpen
  \href@noop {} {}\bibinfo {note} {See Supplemental Material at [link] for
  information about UO$_2$ states found in our ground state search, phonon
  spectra of selected states, VASP patch files for OMC, and experimental
  details about inelastic neutron scattering.}\BibitemShut {Stop}%
\bibitem [{\citenamefont {Blöchl}(1994)}]{blochl_projector_1994}%
  \BibitemOpen
  \bibfield  {author} {\bibinfo {author} {\bibfnamefont {P.~E.}\ \bibnamefont
  {Blöchl}},\ }\href {\doibase 10.1103/PhysRevB.50.17953} {\bibfield
  {journal} {\bibinfo  {journal} {Physical Review B}\ }\textbf {\bibinfo
  {volume} {50}},\ \bibinfo {pages} {17953} (\bibinfo {year}
  {1994})}\BibitemShut {NoStop}%
\bibitem [{\citenamefont {Kresse}\ and\ \citenamefont
  {Joubert}(1999)}]{kresse_ultrasoft_1999}%
  \BibitemOpen
  \bibfield  {author} {\bibinfo {author} {\bibfnamefont {G.}~\bibnamefont
  {Kresse}}\ and\ \bibinfo {author} {\bibfnamefont {D.}~\bibnamefont
  {Joubert}},\ }\href {\doibase 10.1103/PhysRevB.59.1758} {\bibfield  {journal}
  {\bibinfo  {journal} {Physical Review B}\ }\textbf {\bibinfo {volume} {59}},\
  \bibinfo {pages} {1758} (\bibinfo {year} {1999})}\BibitemShut {NoStop}%
\bibitem [{\citenamefont {Kresse}\ and\ \citenamefont
  {Hafner}(1993)}]{kresse_ab_1993}%
  \BibitemOpen
  \bibfield  {author} {\bibinfo {author} {\bibfnamefont {G.}~\bibnamefont
  {Kresse}}\ and\ \bibinfo {author} {\bibfnamefont {J.}~\bibnamefont
  {Hafner}},\ }\href {http://link.aps.org/doi/10.1103/PhysRevB.47.558}
  {\bibfield  {journal} {\bibinfo  {journal} {Physical Review B}\ }\textbf
  {\bibinfo {volume} {47}},\ \bibinfo {pages} {558} (\bibinfo {year}
  {1993})}\BibitemShut {NoStop}%
\bibitem [{\citenamefont {Kresse}\ and\ \citenamefont
  {Furthmüller}(1996)}]{kresse_efficient_1996}%
  \BibitemOpen
  \bibfield  {author} {\bibinfo {author} {\bibfnamefont {G.}~\bibnamefont
  {Kresse}}\ and\ \bibinfo {author} {\bibfnamefont {J.}~\bibnamefont
  {Furthmüller}},\ }\href {\doibase 10.1103/PhysRevB.54.11169} {\bibfield
  {journal} {\bibinfo  {journal} {Physical Review B}\ }\textbf {\bibinfo
  {volume} {54}},\ \bibinfo {pages} {11169} (\bibinfo {year}
  {1996})}\BibitemShut {NoStop}%
\bibitem [{\citenamefont {Perdew}\ \emph {et~al.}(1996)\citenamefont {Perdew},
  \citenamefont {Burke},\ and\ \citenamefont
  {Ernzerhof}}]{perdew_generalized_1996}%
  \BibitemOpen
  \bibfield  {author} {\bibinfo {author} {\bibfnamefont {J.~P.}\ \bibnamefont
  {Perdew}}, \bibinfo {author} {\bibfnamefont {K.}~\bibnamefont {Burke}}, \
  and\ \bibinfo {author} {\bibfnamefont {M.}~\bibnamefont {Ernzerhof}},\ }\href
  {\doibase 10.1103/PhysRevLett.77.3865} {\bibfield  {journal} {\bibinfo
  {journal} {Physical Review Letters}\ }\textbf {\bibinfo {volume} {77}},\
  \bibinfo {pages} {3865} (\bibinfo {year} {1996})}\BibitemShut {NoStop}%
\bibitem [{\citenamefont {Dudarev}\ \emph {et~al.}(1998)\citenamefont
  {Dudarev}, \citenamefont {Botton}, \citenamefont {Savrasov}, \citenamefont
  {Humphreys},\ and\ \citenamefont
  {Sutton}}]{dudarev_electron-energy-loss_1998}%
  \BibitemOpen
  \bibfield  {author} {\bibinfo {author} {\bibfnamefont {S.~L.}\ \bibnamefont
  {Dudarev}}, \bibinfo {author} {\bibfnamefont {G.~A.}\ \bibnamefont {Botton}},
  \bibinfo {author} {\bibfnamefont {S.~Y.}\ \bibnamefont {Savrasov}}, \bibinfo
  {author} {\bibfnamefont {C.~J.}\ \bibnamefont {Humphreys}}, \ and\ \bibinfo
  {author} {\bibfnamefont {A.~P.}\ \bibnamefont {Sutton}},\ }\href {\doibase
  10.1103/PhysRevB.57.1505} {\bibfield  {journal} {\bibinfo  {journal}
  {Physical Review B}\ }\textbf {\bibinfo {volume} {57}},\ \bibinfo {pages}
  {1505} (\bibinfo {year} {1998})}\BibitemShut {NoStop}%
\bibitem [{\citenamefont {Monkhorst}\ and\ \citenamefont
  {Pack}(1976)}]{monkhorst_special_1976}%
  \BibitemOpen
  \bibfield  {author} {\bibinfo {author} {\bibfnamefont {H.~J.}\ \bibnamefont
  {Monkhorst}}\ and\ \bibinfo {author} {\bibfnamefont {J.~D.}\ \bibnamefont
  {Pack}},\ }\href {\doibase 10.1103/PhysRevB.13.5188} {\bibfield  {journal}
  {\bibinfo  {journal} {Physical Review B}\ }\textbf {\bibinfo {volume} {13}},\
  \bibinfo {pages} {5188} (\bibinfo {year} {1976})}\BibitemShut {NoStop}%
\bibitem [{\citenamefont {Fu}\ \emph {et~al.}(2019)\citenamefont {Fu},
  \citenamefont {Kornbluth}, \citenamefont {Cheng},\ and\ \citenamefont
  {Marianetti}}]{fu_group_2019}%
  \BibitemOpen
  \bibfield  {author} {\bibinfo {author} {\bibfnamefont {L.}~\bibnamefont
  {Fu}}, \bibinfo {author} {\bibfnamefont {M.}~\bibnamefont {Kornbluth}},
  \bibinfo {author} {\bibfnamefont {Z.}~\bibnamefont {Cheng}}, \ and\ \bibinfo
  {author} {\bibfnamefont {C.~A.}\ \bibnamefont {Marianetti}},\ }\href
  {\doibase 10.1103/PhysRevB.100.014303} {\bibfield  {journal} {\bibinfo
  {journal} {Physical Review B}\ }\textbf {\bibinfo {volume} {100}},\ \bibinfo
  {pages} {014303} (\bibinfo {year} {2019})}\BibitemShut {NoStop}%
\bibitem [{\citenamefont {Mathis}\ \emph {et~al.}(2022)\citenamefont {Mathis},
  \citenamefont {Khanolkar}, \citenamefont {Fu}, \citenamefont {Bryan},
  \citenamefont {Dennett}, \citenamefont {Rickert}, \citenamefont {Mann},
  \citenamefont {Winn}, \citenamefont {Abernathy}, \citenamefont {Manley},
  \citenamefont {Hurley},\ and\ \citenamefont {Marianetti}}]{Mark_arXiv}%
  \BibitemOpen
  \bibfield  {author} {\bibinfo {author} {\bibfnamefont {M.~A.}\ \bibnamefont
  {Mathis}}, \bibinfo {author} {\bibfnamefont {A.}~\bibnamefont {Khanolkar}},
  \bibinfo {author} {\bibfnamefont {L.}~\bibnamefont {Fu}}, \bibinfo {author}
  {\bibfnamefont {M.~S.}\ \bibnamefont {Bryan}}, \bibinfo {author}
  {\bibfnamefont {C.~A.}\ \bibnamefont {Dennett}}, \bibinfo {author}
  {\bibfnamefont {K.}~\bibnamefont {Rickert}}, \bibinfo {author} {\bibfnamefont
  {J.~M.}\ \bibnamefont {Mann}}, \bibinfo {author} {\bibfnamefont
  {B.}~\bibnamefont {Winn}}, \bibinfo {author} {\bibfnamefont {D.~L.}\
  \bibnamefont {Abernathy}}, \bibinfo {author} {\bibfnamefont {M.~E.}\
  \bibnamefont {Manley}}, \bibinfo {author} {\bibfnamefont {D.~H.}\
  \bibnamefont {Hurley}}, \ and\ \bibinfo {author} {\bibfnamefont {C.~A.}\
  \bibnamefont {Marianetti}},\ }\href@noop {} {\enquote {\bibinfo {title} {The
  generalized quasiharmonic approximation via space group irreducible
  derivatives},}\ } (\bibinfo {year} {2022}),\ \Eprint
  {http://arxiv.org/abs/arXiv:2202.14016} {arXiv:2202.14016} \BibitemShut
  {NoStop}%
\bibitem [{\citenamefont {Gonze}\ and\ \citenamefont
  {Lee}(1997)}]{gonzeDynamicalMatricesBorn1997a}%
  \BibitemOpen
  \bibfield  {author} {\bibinfo {author} {\bibfnamefont {X.}~\bibnamefont
  {Gonze}}\ and\ \bibinfo {author} {\bibfnamefont {C.}~\bibnamefont {Lee}},\
  }\href {\doibase 10.1103/PhysRevB.55.10355} {\bibfield  {journal} {\bibinfo
  {journal} {Physical Review B}\ }\textbf {\bibinfo {volume} {55}},\ \bibinfo
  {pages} {10355} (\bibinfo {year} {1997})}\BibitemShut {NoStop}%
\bibitem [{\citenamefont {Dorado}\ \emph {et~al.}(2013)\citenamefont {Dorado},
  \citenamefont {Freyss}, \citenamefont {Amadon}, \citenamefont {Bertolus},
  \citenamefont {Jomard},\ and\ \citenamefont {Garcia}}]{dorado_advances_2013}%
  \BibitemOpen
  \bibfield  {author} {\bibinfo {author} {\bibfnamefont {B.}~\bibnamefont
  {Dorado}}, \bibinfo {author} {\bibfnamefont {M.}~\bibnamefont {Freyss}},
  \bibinfo {author} {\bibfnamefont {B.}~\bibnamefont {Amadon}}, \bibinfo
  {author} {\bibfnamefont {M.}~\bibnamefont {Bertolus}}, \bibinfo {author}
  {\bibfnamefont {G.}~\bibnamefont {Jomard}}, \ and\ \bibinfo {author}
  {\bibfnamefont {P.}~\bibnamefont {Garcia}},\ }\href {\doibase
  10.1088/0953-8984/25/33/333201} {\bibfield  {journal} {\bibinfo  {journal}
  {Journal of Physics: Condensed Matter}\ }\textbf {\bibinfo {volume} {25}},\
  \bibinfo {pages} {333201} (\bibinfo {year} {2013})}\BibitemShut {NoStop}%
\bibitem [{\citenamefont {Bryan}\ \emph {et~al.}(2019)\citenamefont {Bryan},
  \citenamefont {Pang}, \citenamefont {Larson}, \citenamefont {Chernatynskiy},
  \citenamefont {Abernathy}, \citenamefont {Gofryk},\ and\ \citenamefont
  {Manley}}]{bryan_impact_2019}%
  \BibitemOpen
  \bibfield  {author} {\bibinfo {author} {\bibfnamefont {M.~S.}\ \bibnamefont
  {Bryan}}, \bibinfo {author} {\bibfnamefont {J.~W.~L.}\ \bibnamefont {Pang}},
  \bibinfo {author} {\bibfnamefont {B.~C.}\ \bibnamefont {Larson}}, \bibinfo
  {author} {\bibfnamefont {A.}~\bibnamefont {Chernatynskiy}}, \bibinfo {author}
  {\bibfnamefont {D.~L.}\ \bibnamefont {Abernathy}}, \bibinfo {author}
  {\bibfnamefont {K.}~\bibnamefont {Gofryk}}, \ and\ \bibinfo {author}
  {\bibfnamefont {M.~E.}\ \bibnamefont {Manley}},\ }\href {\doibase
  10.1103/PhysRevMaterials.3.065405} {\bibfield  {journal} {\bibinfo  {journal}
  {Physical Review Materials}\ }\textbf {\bibinfo {volume} {3}},\ \bibinfo
  {pages} {065405} (\bibinfo {year} {2019})}\BibitemShut {NoStop}%
\bibitem [{\citenamefont {Bryan}\ \emph {et~al.}(2020)\citenamefont {Bryan},
  \citenamefont {Fu}, \citenamefont {Rickert}, \citenamefont {Turner},
  \citenamefont {Prusnick}, \citenamefont {Mann}, \citenamefont {Abernathy},
  \citenamefont {Marianetti},\ and\ \citenamefont
  {Manley}}]{bryanNonlinearPropagatingModes2020}%
  \BibitemOpen
  \bibfield  {author} {\bibinfo {author} {\bibfnamefont {M.~S.}\ \bibnamefont
  {Bryan}}, \bibinfo {author} {\bibfnamefont {L.}~\bibnamefont {Fu}}, \bibinfo
  {author} {\bibfnamefont {K.}~\bibnamefont {Rickert}}, \bibinfo {author}
  {\bibfnamefont {D.}~\bibnamefont {Turner}}, \bibinfo {author} {\bibfnamefont
  {T.~A.}\ \bibnamefont {Prusnick}}, \bibinfo {author} {\bibfnamefont {J.~M.}\
  \bibnamefont {Mann}}, \bibinfo {author} {\bibfnamefont {D.~L.}\ \bibnamefont
  {Abernathy}}, \bibinfo {author} {\bibfnamefont {C.~A.}\ \bibnamefont
  {Marianetti}}, \ and\ \bibinfo {author} {\bibfnamefont {M.~E.}\ \bibnamefont
  {Manley}},\ }\href {\doibase 10.1038/s42005-020-00483-2} {\bibfield
  {journal} {\bibinfo  {journal} {Communications Physics}\ }\textbf {\bibinfo
  {volume} {3}},\ \bibinfo {pages} {1} (\bibinfo {year} {2020})}\BibitemShut
  {NoStop}%
\bibitem [{\citenamefont {Abernathy}\ \emph {et~al.}(2012)\citenamefont
  {Abernathy}, \citenamefont {Stone}, \citenamefont {Loguillo}, \citenamefont
  {Lucas}, \citenamefont {Delaire}, \citenamefont {Tang}, \citenamefont {Lin},\
  and\ \citenamefont {Fultz}}]{abernathyDesignOperationWide2012}%
  \BibitemOpen
  \bibfield  {author} {\bibinfo {author} {\bibfnamefont {D.~L.}\ \bibnamefont
  {Abernathy}}, \bibinfo {author} {\bibfnamefont {M.~B.}\ \bibnamefont
  {Stone}}, \bibinfo {author} {\bibfnamefont {M.~J.}\ \bibnamefont {Loguillo}},
  \bibinfo {author} {\bibfnamefont {M.~S.}\ \bibnamefont {Lucas}}, \bibinfo
  {author} {\bibfnamefont {O.}~\bibnamefont {Delaire}}, \bibinfo {author}
  {\bibfnamefont {X.}~\bibnamefont {Tang}}, \bibinfo {author} {\bibfnamefont
  {J.~Y.~Y.}\ \bibnamefont {Lin}}, \ and\ \bibinfo {author} {\bibfnamefont
  {B.}~\bibnamefont {Fultz}},\ }\href {\doibase 10.1063/1.3680104} {\bibfield
  {journal} {\bibinfo  {journal} {Review of Scientific Instruments}\ }\textbf
  {\bibinfo {volume} {83}},\ \bibinfo {pages} {015114} (\bibinfo {year}
  {2012})}\BibitemShut {NoStop}%
\bibitem [{\citenamefont {Pang}\ \emph {et~al.}(2014)\citenamefont {Pang},
  \citenamefont {Chernatynskiy}, \citenamefont {Larson}, \citenamefont
  {Buyers}, \citenamefont {Abernathy}, \citenamefont {McClellan},\ and\
  \citenamefont {Phillpot}}]{pangPhononDensityStates2014}%
  \BibitemOpen
  \bibfield  {author} {\bibinfo {author} {\bibfnamefont {J.~W.~L.}\
  \bibnamefont {Pang}}, \bibinfo {author} {\bibfnamefont {A.}~\bibnamefont
  {Chernatynskiy}}, \bibinfo {author} {\bibfnamefont {B.~C.}\ \bibnamefont
  {Larson}}, \bibinfo {author} {\bibfnamefont {W.~J.~L.}\ \bibnamefont
  {Buyers}}, \bibinfo {author} {\bibfnamefont {D.~L.}\ \bibnamefont
  {Abernathy}}, \bibinfo {author} {\bibfnamefont {K.~J.}\ \bibnamefont
  {McClellan}}, \ and\ \bibinfo {author} {\bibfnamefont {S.~R.}\ \bibnamefont
  {Phillpot}},\ }\href {\doibase 10.1103/PhysRevB.89.115132} {\bibfield
  {journal} {\bibinfo  {journal} {Physical Review B}\ }\textbf {\bibinfo
  {volume} {89}},\ \bibinfo {pages} {115132} (\bibinfo {year}
  {2014})}\BibitemShut {NoStop}%
\bibitem [{\citenamefont {Liechtenstein}\ \emph {et~al.}(1995)\citenamefont
  {Liechtenstein}, \citenamefont {Anisimov},\ and\ \citenamefont
  {Zaanen}}]{liechtenstein_density-functional_1995}%
  \BibitemOpen
  \bibfield  {author} {\bibinfo {author} {\bibfnamefont {A.~I.}\ \bibnamefont
  {Liechtenstein}}, \bibinfo {author} {\bibfnamefont {V.~I.}\ \bibnamefont
  {Anisimov}}, \ and\ \bibinfo {author} {\bibfnamefont {J.}~\bibnamefont
  {Zaanen}},\ }\href {\doibase 10.1103/PhysRevB.52.R5467} {\bibfield  {journal}
  {\bibinfo  {journal} {Physical Review B}\ }\textbf {\bibinfo {volume} {52}},\
  \bibinfo {pages} {R5467} (\bibinfo {year} {1995})}\BibitemShut {NoStop}%
\end{thebibliography}

\end{document}


\renewcommand{\thefigure}{S\arabic{figure}}
\renewcommand{\thetable}{S\arabic{table}}
\raggedright{\textbf{\Large Supplemental materials to}}
\title{Capturing the ground state of uranium dioxide from first principles:\\ crystal distortion, magnetic structure, and phonons}
\author{Shuxiang Zhou$^1$, Hao Ma$^2$, Enda Xiao$^3$, Krzysztof Gofryk$^1$, Chao Jiang$^1$, Michael E. Manley$^2$, David H. Hurley$^1$, and Chris A. Marianetti$^4$}
\affiliation{$^1$Idaho National Laboratory, Idaho Falls, Idaho 83415, USA\\$^2$ Oak Ridge National Laboratory, Oak Ridge, Tennessee 37831, USA\\$^3$Department of Chemistry, Columbia University, New York, New York 10027, USA\\$^4$Department of Applied Physics and Applied Mathematics, Columbia University, New York, New York 10027, USA}

\maketitle
\section{\label{sec:smproperties}Ground state search: occupation matrices, energies, and distortions }
\vspace{5mm}

\justifying
In this section, we report the UO$_2$ states found in our ground state search. 
For each state, the energy, lattice vectors, atomic coordinates, strain, spin moments, orbit moments (if spin-orbit coupling [SOC] is applied), and occupation matrices (denoted as $\occmat$) are reported for the fully relaxed crystal; the energy and lattice parameter of the undistorted crystal (relaxing only the volume of the cubic crystal, denoted by "und") are also presented. 
We first report the ground state search results of GGA+$U$ ($U=4$ eV) without SOC, including ferromagnetic (FM) (Section~\ref{sec:smfmnosoc}), 1\textbf{k} antiferromagnetic (AFM) (Section~\ref{sec:sm1kafmnosoc}), and 3\textbf{k} AFM states (Section~\ref{sec:sm3kafmnosoc}), followed by the DFT+$U$+SOC results of FM (Section~\ref{sec:smfmsoc}), 1\textbf{k} AFM (Section~\ref{sec:sm1kafmsoc}), and 3\textbf{k} AFM states (Section~\ref{sec:sm3kafmsoc}).

\subsection{\label{sec:smfmnosoc} FM without SOC (the stage 1)}

First, we introduce the search procedures for the stage 1, in which we employ occupation matrix control (OMC) in FM UO$_2$, using GGA+$U$ ($U=4$ eV) without SOC.
Occupation matrices are calculated in the basis of real spherical harmonics, e.g., for $f$-orbitals, the order is $f_{y(3x^2-y^2)}, f_{xyz}, f_{yz^2}, f_{z^3}, f_{xz^2}, f_{z(x^2-y^2)}, f_{x(x^2-3y^2)}$. A table for different orbitals are given on VASP wiki: \url{https://www.vasp.at/wiki/index.php/Angular_functions}.
As exploring all of the $7\times7$ occupation matrix phase space is not tractable, we followed Dorado \emph{et al.}'s study\cite{dorado_textdfttextu_2009, dorado_stability_2010} and performed the stage 1 in two steps: first, DFT calculations were initialized from diagonal matrices; then for the lowest-energy state obtained, the resulting occupation matrix was used to construct an off-diagonal occupation matrix that we then used in further DFT calculations. All initial diagonal and off-diagonal occupation matrices have global restrictions: each must be symmetric, have a trace of 2 (dictated by the nominal number of 5$f$ electrons in U), and all diagonal and off-diagonal components must be within the range $[0, 1]$ and $[-1, 1]$, respectively. 
For the diagonal matrices in the first step, we constructed all 21 integer diagonal matrices, whose diagonal components are only 0 or 1 (meaning there are $C_7^2=21$ different ways to fill the seven 5\emph{f} levels with two electrons), along with 200 randomly generated diagonal matrices that satisfy the global restrictions. 
Therefore, a total of 221 occupation matrices were used for initialization in the first step of the stage 1. 
After self-consistent calculations, we found three states with the identical lowest energy, due being related by symmetry. The up-spin occupation matrices for the three lowest-energy states are given by (the down-spin matrix is approximately a zero matrix):

$\occmat_{D_1}^\uparrow=
\setstackgap{L}{1.1\baselineskip}
\fixTABwidth{T}
\parenMatrixstack{
  0.4&      0&   -0.4&      0&      0&      0&      0  \\
    0&    0.1&      0&      0&      0&      0&      0  \\
 -0.4&      0&    0.7&      0&      0&      0&      0  \\
    0&      0&      0&      0&      0&      0&      0  \\
    0&      0&      0&      0&    0.7&      0&    0.4  \\
    0&      0&      0&      0&      0&      0&      0  \\
    0&      0&      0&      0&    0.4&      0&    0.4  
}$,
$\occmat_{D_2}^\uparrow=
\setstackgap{L}{1.1\baselineskip}
\fixTABwidth{T}
\parenMatrixstack{
  0.5&      0&   -0.5&      0&      0&      0&      0  \\
    0&    0.1&      0&      0&      0&      0&      0  \\
 -0.5&      0&    0.6&      0&      0&      0&      0  \\
    0&      0&      0&    0.1&      0&   -0.1&      0  \\
    0&      0&      0&      0&      0&      0&      0  \\
    0&      0&      0&   -0.1&      0&    1.0&      0  \\
    0&      0&      0&      0&      0&      0&      0 
}$,\\

$\occmat_{D_3}^\uparrow=
\setstackgap{L}{1.1\baselineskip}
\setstacktabbedgap{9pt}
\fixTABwidth{T}
\parenMatrixstack{
    0&      0&      0&      0&      0&      0&      0  \\
    0&    0.1&      0&      0&      0&      0&      0  \\
    0&      0&      0&      0&      0&      0&      0  \\
    0&      0&      0&    0.1&      0&    0.1&      0  \\
    0&      0&      0&      0&    0.6&      0&    0.5  \\
    0&      0&      0&    0.1&      0&    1.0&      0  \\
    0&      0&      0&      0&    0.5&      0&    0.5 
}$.

In the second step, based on $\occmat_{D_1}^\uparrow$, $\occmat_{D_2}^\uparrow$, and $\occmat_{D_3}^\uparrow$, we built off-diagonal matrices ($\occmat_{D_1^\star}^\uparrow$, $\occmat_{D_2^\star}^\uparrow$, and $\occmat_{D_3^\star}^\uparrow$) that were then used to initialize further calculations. 
Based on a matrix $A$ (whose $i$-th row $j$-th column component is denoted as $A_{ij}$), if $A$ has $n$ diagonal components larger than 0.3 (i.e., there are $n$ different integers $a_1, ..., a_n$ satisfying $A_{a_i a_i}>0.3$), we build an off-diagonal matrix $A^\star$ whose components are all zero except $A_{a_i a_j}$ for $i,j=1,2,...n$:

$\occmat_{D_1^\star}^\uparrow=
\setstackgap{L}{1.1\baselineskip}
\fixTABwidth{T}
\parenMatrixstack{
    a&      0&      b&      0&      c&      0&      e  \\
    0&      0&      0&      0&      0&      0&      0  \\
    b&      0&    1-a&      0&      d&      0&     -c  \\
    0&      0&      0&      0&      0&      0&      0  \\
    c&      0&      d&      0&    1-a&      0&     -b \\
    0&      0&      0&      0&      0&      0&      0  \\
    e&      0&     -c&      0&     -b&      0&      a 
}, \occmat_{D_2^\star}^\uparrow=
\setstackgap{L}{1.1\baselineskip}
\fixTABwidth{T}
\parenMatrixstack{
    a&      0&      c&      0&      0&      d&      0  \\
    0&      0&      0&      0&      0&      0&      0  \\
    c&      0&      b&      0&      0&      e&      0  \\
    0&      0&      0&      0&      0&      0&      0  \\
    0&      0&      0&      0&      0&      0&      0  \\
    d&      0&      e&      0&      0&  2\emph{-a-b}&    0  \\
    0&      0&      0&      0&      0&      0&      0  
}$, 

$\occmat_{D_3^\star}^\uparrow=
\setstackgap{L}{1.1\baselineskip}
\fixTABwidth{T}
\parenMatrixstack{
    0&      0&      0&      0&      0&      0&      0  \\
    0&      0&      0&      0&      0&      0&      0  \\
    0&      0&      0&      0&      0&      0&      0  \\
    0&      0&      0&      0&      0&      0&      0  \\
    0&      0&      0&      0&      a&      c&      d  \\
    0&      0&      0&      0&      c&      b&      e  \\
    0&      0&      0&      0&      d&      e&  2\emph{-a-b}  
}$, \\
where $a$, $b$, $c$, $d$, and $e$ are randomly generated numbers, fulfilling the global restrictions. 

It should be noted that Dorado \emph{et al.} \cite{dorado_textdfttextu_2009} only explored $\occmat_{D_1^\star}$, without off-diagonal components $c$, $d$, and $e$ in their ground state search (i.e., they only included $a$ and $b$). Therefore, we explored a larger phase space of occupation matrices than what was covered in their study. 
For each case of $\occmat_{D_1^\star}$, $\occmat_{D_2^\star}$, and $\occmat_{D_3^\star}$, 300 occupation matrices were randomly generated. In addition, we used the reported occupation matrices of the lowest-energy states in Dorado et al. \cite{dorado_stability_2010}, Thompson et al. \cite{thompson_first-principles_2011}, and Wang et al. \cite{wang_electronic_2013}. Therefore, a total of 903 occupation matrices were used to initialize calculations in the second step. After self-consistent calculations, we found six states with the identical lowest energy:

$\occmat_{S_0^1}^\uparrow=
\setstackgap{L}{1.1\baselineskip}
\fixTABwidth{T}
\parenMatrixstack{
  0.4&      0&   -0.3&      0&    0.1&      0&    0.3  \\
    0&    0.1&      0&      0&      0&      0&      0  \\
 -0.3&      0&    0.7&      0&    0.3&      0&   -0.1  \\
    0&      0&      0&      0&      0&      0&      0  \\
  0.1&      0&    0.3&      0&    0.7&      0&    0.3  \\
    0&      0&      0&      0&      0&      0&      0  \\
  0.3&      0&   -0.1&      0&    0.3&      0&    0.4 
}, \occmat_{S_0^2}^\uparrow=
\setstackgap{L}{1.1\baselineskip}
\fixTABwidth{T}
\parenMatrixstack{
  0.5&      0&   -0.4&   -0.2&      0&    0.2&      0  \\
    0&    0.1&      0&      0&      0&      0&      0  \\
 -0.4&      0&    0.6&    0.2&      0&    0.2&      0  \\
 -0.2&      0&    0.2&    0.2&      0&   -0.1&      0  \\
    0&      0&      0&      0&      0&      0&      0  \\
  0.2&      0&    0.2&   -0.1&      0&    0.9&      0  \\
    0&      0&      0&      0&      0&      0&      0
}$,

$\occmat_{S_0^3}^\uparrow=
\setstackgap{L}{1.1\baselineskip}
\fixTABwidth{T}
\parenMatrixstack{
    0&      0&      0&      0&      0&      0&      0  \\
    0&    0.1&      0&      0&      0&      0&      0  \\
    0&      0&      0&      0&      0&      0&      0  \\
    0&      0&      0&    0.2&    0.2&    0.1&    0.2  \\
    0&      0&      0&    0.2&    0.6&   -0.2&    0.4  \\
    0&      0&      0&    0.1&   -0.2&    0.9&    0.2  \\
    0&      0&      0&    0.2&    0.4&    0.2&    0.5
}, \occmat_{S_0^4}^\uparrow=
\setstackgap{L}{1.1\baselineskip}
\fixTABwidth{T}
\parenMatrixstack{
  0.4&      0&   -0.3&      0&   -0.1&      0&   -0.3  \\
    0&    0.1&      0&      0&      0&      0&      0  \\
 -0.3&      0&    0.7&      0&   -0.3&      0&    0.1  \\
    0&      0&      0&      0&      0&      0&      0  \\
 -0.1&      0&   -0.3&      0&    0.7&      0&    0.3  \\
    0&      0&      0&      0&      0&      0&      0  \\
 -0.3&      0&    0.1&      0&    0.3&      0&    0.4
}$,

$\occmat_{S_0^5}^\uparrow=
\setstackgap{L}{1.1\baselineskip}
\fixTABwidth{T}
\parenMatrixstack{
  0.5&      0&   -0.4&    0.2&      0&   -0.2&      0  \\
    0&    0.1&      0&      0&      0&      0&      0  \\
 -0.4&      0&    0.6&   -0.2&      0&   -0.2&      0  \\
  0.2&      0&   -0.2&    0.2&      0&   -0.1&      0  \\
    0&      0&      0&      0&      0&      0&      0  \\
 -0.2&      0&   -0.2&   -0.1&      0&    0.9&      0  \\
    0&      0&      0&      0&      0&      0&      0
}, \occmat_{S_0^6}^\uparrow=
\setstackgap{L}{1.1\baselineskip}
\fixTABwidth{T}
\parenMatrixstack{
    0&      0&      0&      0&      0&      0&      0  \\
    0&    0.1&      0&      0&      0&      0&      0  \\
    0&      0&      0&      0&      0&      0&      0  \\
    0&      0&      0&    0.2&   -0.2&    0.1&   -0.2  \\
    0&      0&      0&   -0.2&    0.6&    0.2&    0.4  \\
    0&      0&      0&    0.1&    0.2&    0.9&   -0.2  \\
    0&      0&      0&   -0.2&    0.4&   -0.2&    0.5
}$,\\
where degenerate clasess of states are denoted $S_i$, with $i=0$ being the lowest energy,
and $S_i^j$ is the $j$-th member of the class. The properties of each state will be subsequently described.

The nine lowest-energy classes we found (denoted as $S_0$--$S_8$, with $S_0$ having the lowest
energy) contained 25 distinct states with different occupation matrices, with each state
in a given class 
being related by symmetry (note that we may not have found all the states for each energy class in our search). 
Table~\ref{tab:tables1} summarizes the energy and distortion of each $S_i^j$, followed by the occupancy matrices of each state. Note that the lattice parameter of undistorted crystal $x$ represents a face-centred cubic crystal with basis vectors $a=(x, x, 0), b=(0, x, x), c=(x, 0, x)$.

In comparing with previous studies, the lowest-energy state reported by Dorado \emph{et al.} \cite{dorado_stability_2010} is $S_1^2$, and their second-lowest-energy state is $S_4^2$. Dorado \emph{et al.} found the energy difference between the undistorted $S_1^2$ and $S_4^2$ states to be 43 meV per primitive unit cell, whereas we found it to be 39.7 meV in this work, which is a reasonable agreement. 
The lowest-energy states reported by Wang \emph{et al.}\cite{wang_electronic_2013} and Thompson \emph{et al.}\cite{thompson_first-principles_2011} were not found in our search, and the calculations initialized by their occupation matrices end in $S_0$. 

Several different types of distortions were found, including a $E_g$ strain mode, a $T_{2g}$ strain mode, and an oxygen cage distortion. Notably, $S_0$ has a pure shear strain distortion ($\pm$0.017 $T_{2g}$ strain mode). 
Both $S_0^1$ and $S_0^4$ are initialized from $\occmat_{D_1^\star}$, resulting in identical diagonal components in their occupation matrices, but different off-diagonal components; and the resulting distortions are opposite (0.017 and -0.017 of $T_{2g}$ mode shear strain, respectively). 

\begin{table*}[h]
\caption{\label{tab:tables1}%
Results for FM UO$_2$ without SOC from the stage 1 of our ground state search. 
The results are sorted by increasing relative energy with
respect to the fully relaxed crystal having occupation matrix $S_0$ and energy $E_{min}$.
}
\begin{ruledtabular}
\begin{tabular}{lcccr}
\textrm{State}&
\textrm{\begin{tabular}{@{}c@{}}$E_{und}-E_{min}$ \\ (meV/UO\textsubscript{2})\end{tabular} }&
\textrm{\begin{tabular}{@{}c@{}}$E_{dis}-E_{min}$ \\ (meV/UO\textsubscript{2})\end{tabular} }&
\textrm{{\begin{tabular}{@{}c@{}}Strain distortion\\ ($\epsilon_{xx},\epsilon_{yy},\epsilon_{zz},\epsilon_{xy},\epsilon_{xz},\epsilon_{yz}$)$\times10^3$\end{tabular} } }&
\textrm{\begin{tabular}{@{}c@{}}Oxygen cage  \\ distortion\end{tabular} }\\
\colrule
S$_0^1$ & \multirow{6}{*}{9.9} & \multirow{6}{*}{0} & (0, 0, 0, 17, 0, 0) & - \\
S$_0^2$ & & & (0, 0, 0, 0, 0, 17) & - \\
S$_0^3$ & & & (0, 0, 0, 0, 17, 0) & - \\
S$_0^4$ & & & (0, 0, 0, -17, 0, 0) & - \\
S$_0^5$ & & & (0, 0, 0, 0, 0, -17) & - \\
S$_0^6$ & & & (0, 0, 0, 0, -17, 0) & - \\
\hline
S$_1^1$ & \multirow{3}{*}{13.0} & \multirow{3}{*}{12.9}  & (0, 1, 0, 0, 0, 0) & - \\
S$_1^2$ & & & (0, 0, 1, 0, 0, 0) & - \\
S$_1^3$ & & & (1, 0, 0, 0, 0, 0) & - \\
\hline
S$_2^1$ & \multirow{6}{*}{42.1} & \multirow{6}{*}{38.9}  & (-3, -5, 8, 0, 0, 0) & - \\
S$_2^2$ & & & (-5, 8, -3, 0, 0, 0) & - \\
S$_2^3$ & & & (8, -5, -3, 0, 0, 0) & - \\
S$_2^4$ & & & (8, -3, -5, 0, 0, 0) & - \\
S$_2^5$ & & & (-5, -3, 8, 0, 0, 0) & - \\
S$_2^6$ & & & (-3, 8, -5, 0, 0, 0) & - \\
\hline
S$_3^1$ & 43.8 & 40.4 & (-4, -4, 8, 2, 0, 0) & $<$001$>$ 0.003\AA\\
\hline
S$_4^1$ & \multirow{2}{*}{52.7} & \multirow{2}{*}{49.2} & (4, -8, 4, 0, 0, 0)& - \\
S$_4^2$ & & & (4, 4, -8, 0, 0, 0) & - \\
\hline
S$_5^1$ & \multirow{3}{*}{100} & \multirow{3}{*}{99.5} & (-3, 2, 2, 0, 0, 0) & -\\
S$_5^2$ & & & (2, 2, -3, 0, 0, 0) & - \\
S$_5^3$ & & & (2, -3, 2, 0, 0, 0) & - \\
\hline
S$_6^1$ & 114 & 95.6 & (2, 2, -3, 18, 0, 0) & $<$001$>$ 0.010\AA \\
\hline
S$_7^1$ & \multirow{2}{*}{121} & \multirow{2}{*}{117} & (4, 4, -8, 0, 0, 0) & -\\
S$_7^2$ & & & (4, 4, -8, 0, 0, 0) & - \\
\hline
S$_8^1$ & 131 & 108 & (1, 1, 1, -21, 0, 0)& $<$001$>$ 0.009\AA \\

\end{tabular}
\end{ruledtabular}
\end{table*}
\clearpage

\subsubsection{Properties of S$_0^1$}

\begin{table*}[h]%
\begin{ruledtabular}
\begin{tabular}{ll}
\textrm{Properties (unit)}&
\textrm{Values}\\
\colrule
Energy (eV) & -29.137 \\
Basis vectors (\AA)& $a=(2.7799, 2.7799, 0.0001), b=(0.0453, 2.7345, 2.7360), c=(2.7345, 0.0453, 2.7360)$\\
\begin{tabular}{@{}l@{}}Atom coordinates \\ (in basis vectors)\end{tabular} &
{\begin{tabular}{@{}l@{}}$U1=(0.5000, 0.5000, 0.5000)$ \\$O1=(0.2500, 0.2500, 0.2500)$, $O2=(0.7500, 0.7500, 0.7500)$\end{tabular}} \\ 
Strain & \begin{tabular}{@{}l@{}}($\epsilon_{xx},\epsilon_{yy},\epsilon_{zz},\epsilon_{xy},\epsilon_{xz},\epsilon_{yz}$)=(0.0001, 0.0001, 0.0006, 0.0166, 0, 0), \\($\epsilon_{A1g},\epsilon_{Eg.0},\epsilon_{Eg.1},\epsilon_{T2g.0},\epsilon_{T2g.1},\epsilon_{T2g.2}$)=(0.0005, 0, 0.0004, 0.0166, 0, 0).\end{tabular}\\
Spin moments ($\mu_B$)& (2.028, 0.000, 0.000)\\
Energy (eV) and lattice (\AA) of und & -29.127, 2.7731
\end{tabular}
\end{ruledtabular}
\end{table*}

$\occmat^\uparrow=
\setstackgap{L}{1.1\baselineskip}
\fixTABwidth{T}
\parenMatrixstack{
0.4009 & 0 & -0.2737 & 0 & 0.1436 & 0 & 0.3428 \\
0 & 0.1245 & 0 & 0.0105 & 0 & 0 & 0 \\
-0.2737 & 0 & 0.6411 & 0 & 0.3372 & 0 & -0.1436 \\
0 & 0.0105 & 0 & 0.0453 & 0 & 0 & 0 \\
0.1436 & 0 & 0.3372 & 0 & 0.6411 & 0 & 0.2737 \\
0 & 0 & 0 & 0 & 0 & 0.0297 & 0 \\
0.3428 & 0 & -0.1436 & 0 & 0.2737 & 0 & 0.4009
}$,

$\occmat^\downarrow=
\setstackgap{L}{1.1\baselineskip}
\fixTABwidth{T}
\parenMatrixstack{
0.0324 & 0 & 0.0073 & 0 & -0.0014 & 0 & -0.0010 \\
0 & 0.1087 & 0 & 0.0059 & 0 & 0 & 0 \\
0.0073 & 0 & 0.0270 & 0 & -0.0050 & 0 & 0.0014 \\
0 & 0.0059 & 0 & 0.0357 & 0 & 0 & 0 \\
-0.0014 & 0 & -0.0050 & 0 & 0.0270 & 0 & -0.0073 \\
0 & 0 & 0 & 0 & 0 & 0.0246 & 0 \\
-0.0010 & 0 & 0.0014 & 0 & -0.0073 & 0 & 0.0324
}$.

\subsubsection{Properties of S$_0^2$}

\begin{table*}[h]%
\begin{ruledtabular}
\begin{tabular}{ll}
\textrm{Properties (unit)}&
\textrm{Values}\\
\colrule
Energy (eV) & -29.137 \\
Basis vectors (\AA)& $a=(2.7360, 2.7345, 0.0453), b=(0.0001, 2.7799, 2.7799), c=(2.7360, 0.0453, 2.7345)$\\
\begin{tabular}{@{}l@{}}Atom coordinates \\ (in basis vectors)\end{tabular} &
{\begin{tabular}{@{}l@{}}$U1=(0.5000, 0.5000, 0.5000)$ \\$O1=(0.2500, 0.2500, 0.2500)$, $O2=(0.7500, 0.7500, 0.7500)$\end{tabular}} \\ 
Strain & \begin{tabular}{@{}l@{}}($\epsilon_{xx},\epsilon_{yy},\epsilon_{zz},\epsilon_{xy},\epsilon_{xz},\epsilon_{yz}$)=(0.0006, 0.0001, 0.0001, 0, 0, 0.0166), \\($\epsilon_{A1g},\epsilon_{Eg.0},\epsilon_{Eg.1},\epsilon_{T2g.0},\epsilon_{T2g.1},\epsilon_{T2g.2}$)=(0.0005, 0.0004, -0.0002, 0, 0.0166, 0).\end{tabular}\\
Spin moments ($\mu_B$)& (2.028, 0.000, 0.000)\\
Energy (eV) and lattice (\AA) of und & -29.127, 2.7731
\end{tabular}
\end{ruledtabular}
\end{table*}

$\occmat^\uparrow=
\setstackgap{L}{1.1\baselineskip}
\fixTABwidth{T}
\parenMatrixstack{
0.4846 & 0 & -0.3626 & -0.2167 & 0 & 0.2129 & -0.0001 \\
0 & 0.1300 & 0 & 0 & -0.0022 & 0 & 0.0029 \\
-0.3626 & 0 & 0.5635 & 0.2115 & 0 & 0.2188 & 0.0001 \\
-0.2167 & 0 & 0.2115 & 0.1612 & 0 & -0.0534 & 0.0001 \\
0 & -0.0022 & 0 & 0 & 0.0376 & 0.0001 & -0.0078 \\
0.2129 & 0 & 0.2188 & -0.0534 & 0.0001 & 0.8828 & -0.0001 \\
-0.0001 & 0.0029 & 0.0001 & 0.0001 & -0.0078 & -0.0001 & 0.0416
}$,

$\occmat^\downarrow=
\setstackgap{L}{1.1\baselineskip}
\fixTABwidth{T}
\parenMatrixstack{
0.0333 & 0 & 0.0089 & 0.0003 & 0 & -0.0011 & 0 \\
0 & 0.1163 & 0 & 0 & -0.0004 & 0 & 0.0005 \\
0.0089 & 0 & 0.0313 & -0.0015 & 0 & -0.0008 & 0 \\
0.0003 & 0 & -0.0015 & 0.0412 & 0 & 0.0013 & 0 \\
0 & -0.0004 & 0 & 0 & 0.0308 & 0 & -0.0055 \\
-0.0011 & 0 & -0.0008 & 0.0013 & 0 & 0.0235 & 0 \\
0 & 0.0005 & 0 & 0 & -0.0055 & 0 & 0.0337
}$.

\subsubsection{Properties of S$_0^3$}

\begin{table*}[h]%
\begin{ruledtabular}
\begin{tabular}{ll}
\textrm{Properties (unit)}&
\textrm{Values}\\
\colrule
Energy (eV) & -29.137 \\
Basis vectors (\AA)& $a=(2.7345, 2.7360, 0.0453), b=(0.0453, 2.7360, 2.7345), c=(2.7799, 09, 2.7799)$\\
\begin{tabular}{@{}l@{}}Atom coordinates \\ (in basis vectors)\end{tabular} &
{\begin{tabular}{@{}l@{}}$U1=(0.5000, 0.5000, 0.5000)$ \\$O1=(0.2500, 0.2500, 0.2500)$, $O2=(0.7500, 0.7500, 0.7500)$\end{tabular}} \\ 
Strain & \begin{tabular}{@{}l@{}}($\epsilon_{xx},\epsilon_{yy},\epsilon_{zz},\epsilon_{xy},\epsilon_{xz},\epsilon_{yz}$)=(0.0001, 0.0006, 0.0001, 0, 0.0166, 0), \\($\epsilon_{A1g},\epsilon_{Eg.0},\epsilon_{Eg.1},\epsilon_{T2g.0},\epsilon_{T2g.1},\epsilon_{T2g.2}$)=(0.0005, -0.0004, -0.0002, 0, 0, 0.0166).\end{tabular}\\
Spin moments ($\mu_B$)& (2.028, 0.000, 0.000)\\
Energy (eV) and lattice (\AA) of und & -29.127, 2.7731
\end{tabular}
\end{ruledtabular}
\end{table*}

$\occmat^\uparrow=
\setstackgap{L}{1.1\baselineskip}
\fixTABwidth{T}
\parenMatrixstack{
0.0416 & -0.0029 & 0.0078 & -0.0001 & -0.0001 & -0.0001 & -0.0001 \\
-0.0029 & 0.1300 & -0.0022 & 0 & 0 & 0 & 0 \\
0.0078 & -0.0022 & 0.0376 & 0 & 0 & -0.0002 & -0.0001 \\
-0.0001 & 0 & 0 & 0.1622 & 0.2125 & 0.0530 & 0.2174 \\
-0.0001 & 0 & 0 & 0.2125 & 0.5620 & -0.2181 & 0.3626 \\
-0.0001 & 0 & -0.0002 & 0.0530 & -0.2181 & 0.8839 & 0.2119 \\
-0.0001 & 0 & -0.0001 & 0.2174 & 0.3626 & 0.2119 & 0.4841
}$,

$\occmat^\downarrow=
\setstackgap{L}{1.1\baselineskip}
\fixTABwidth{T}
\parenMatrixstack{
0.0337 & -0.0005 & 0.0055 & 0 & 0 & 0 & 0 \\
-0.0005 & 0.1163 & -0.0004 & 0 & 0 & 0 & 0 \\
0.0055 & -0.0004 & 0.0308 & 0 & 0 & 0 & 0 \\
0 & 0 & 0 & 0.0412 & -0.0015 & -0.0013 & -0.0003 \\
0 & 0 & 0 & -0.0015 & 0.0313 & 0.0008 & -0.0089 \\
0 & 0 & 0 & -0.0013 & 0.0008 & 0.0235 & -0.0011 \\
0 & 0 & 0 & -0.0003 & -0.0089 & -0.0011 & 0.0333
}$.

\subsubsection{Properties of S$_0^4$}

\begin{table*}[h]%
\begin{ruledtabular}
\begin{tabular}{ll}
\textrm{Properties (unit)}&
\textrm{Values}\\
\colrule
Energy (eV) & -29.137 \\
Basis vectors (\AA)& $a=(2.7274, 2.7274,     0), b=(-0.0460, 2.7734, 2.7749), c=(2.7734, -0.0460, 2.7749)$\\
\begin{tabular}{@{}l@{}}Atom coordinates \\ (in basis vectors)\end{tabular} &
{\begin{tabular}{@{}l@{}}$U1=(0.5000, 0.5000, 0.5000)$ \\$O1=(0.2500, 0.2500, 0.2500)$, $O2=(0.7500, 0.7500, 0.7500)$\end{tabular}} \\ 
Strain & \begin{tabular}{@{}l@{}}($\epsilon_{xx},\epsilon_{yy},\epsilon_{zz},\epsilon_{xy},\epsilon_{xz},\epsilon_{yz}$)=(0.0001, 0.0001, 0.0007, -0.0166,     0,     0), \\($\epsilon_{A1g},\epsilon_{Eg.0},\epsilon_{Eg.1},\epsilon_{T2g.0},\epsilon_{T2g.1},\epsilon_{T2g.2}$)=(0.0005,     0, 0.0004, -0.0166,     0,     0).\end{tabular}\\
Spin moments ($\mu_B$)& (2.028, 0.000, 0.000)\\
Energy (eV) and lattice (\AA) of und & -29.127, 2.7731
\end{tabular}
\end{ruledtabular}
\end{table*}

$\occmat^\uparrow=
\setstackgap{L}{1.1\baselineskip}
\fixTABwidth{T}
\parenMatrixstack{
0.4009 &     0 & -0.2738 &     0 & -0.1435 &     0 & -0.3428 \\
    0 & 0.1245 &     0 & -0.0105 &     0 &     0 &     0 \\
-0.2738 &     0 & 0.6411 &     0 & -0.3371 &     0 & 0.1435 \\
    0 & -0.0105 &     0 & 0.0453 &     0 &     0 &     0 \\
-0.1435 &     0 & -0.3371 &     0 & 0.6411 &     0 & 0.2738 \\
    0 &     0 &     0 &     0 &     0 & 0.0297 &     0 \\
-0.3428 &     0 & 0.1435 &     0 & 0.2738 &     0 & 0.4009
}$,

$\occmat^\downarrow=
\setstackgap{L}{1.1\baselineskip}
\fixTABwidth{T}
\parenMatrixstack{
0.0324 &     0 & 0.0073 &     0 & 0.0014 &     0 & 0.0010 \\
    0 & 0.1087 &     0 & -0.0059 &     0 &     0 &     0 \\
0.0073 &     0 & 0.0270 &     0 & 0.0050 &     0 & -0.0014 \\
    0 & -0.0059 &     0 & 0.0357 &     0 &     0 &     0 \\
0.0014 &     0 & 0.0050 &     0 & 0.0270 &     0 & -0.0073 \\
    0 &     0 &     0 &     0 &     0 & 0.0246 &     0 \\
0.0010 &     0 & -0.0014 &     0 & -0.0073 &     0 & 0.0324
}$.

\subsubsection{Properties of S$_0^5$}

\begin{table*}[h]%
\begin{ruledtabular}
\begin{tabular}{ll}
\textrm{Properties (unit)}&
\textrm{Values}\\
\colrule
Energy (eV) & -29.137 \\
Basis vectors (\AA)& $a=(2.7749, 2.7734, -0.0460), b=(0, 2.7274, 2.7274), c=(2.7749, -0.0460, 2.7734)$\\
\begin{tabular}{@{}l@{}}Atom coordinates \\ (in basis vectors)\end{tabular} &
{\begin{tabular}{@{}l@{}}$U1=(0.5000, 0.5000, 0.5000)$ \\$O1=(0.2500, 0.2500, 0.2500)$, $O2=(0.7500, 0.7500, 0.7500)$\end{tabular}} \\ 
Strain & \begin{tabular}{@{}l@{}}($\epsilon_{xx},\epsilon_{yy},\epsilon_{zz},\epsilon_{xy},\epsilon_{xz},\epsilon_{yz}$)=(0.0007, 0.0001, 0.0001, 0, 0, -0.0166), \\($\epsilon_{A1g},\epsilon_{Eg.0},\epsilon_{Eg.1},\epsilon_{T2g.0},\epsilon_{T2g.1},\epsilon_{T2g.2}$)=(0.0005, -0.0004, -0.0002, 0, 0, -0.0166).\end{tabular}\\
Spin moments ($\mu_B$)& (2.028, 0.000, 0.000)\\
Energy (eV) and lattice (\AA) of und & -29.127, 2.7731
\end{tabular}
\end{ruledtabular}
\end{table*}

$\occmat^\uparrow=
\setstackgap{L}{1.1\baselineskip}
\fixTABwidth{T}
\parenMatrixstack{
0.0394 & 0.0083 & 0.0075 & 0 & 0 & 0 & 0 \\
0.0083 & 0.1245 & 0.0065 & 0 & 0 & 0 & 0 \\
0.0075 & 0.0065 & 0.0355 & 0 & 0 & 0 & 0 \\
0 & 0 & 0 & 0.2259 & -0.2572 & 0.0478 & -0.2640 \\
0 & 0 & 0 & -0.2572 & 0.5484 & 0.2685 & 0.2977 \\
0 & 0 & 0 & 0.0478 & 0.2685 & 0.8161 & -0.2538 \\
0 & 0 & 0 & -0.2640 & 0.2977 & -0.2538 & 0.4935
}$,

$\occmat^\downarrow=
\setstackgap{L}{1.1\baselineskip}
\fixTABwidth{T}
\parenMatrixstack{
0.0316 & 0.0047 & 0.0054 & 0 & 0 & 0 & 0 \\
0.0047 & 0.1087 & 0.0036 & 0 & 0 & 0 & 0 \\
0.0054 & 0.0036 & 0.0288 & 0 & 0 & 0 & 0 \\
0 & 0 & 0 & 0.0374 & 0.0010 & -0.0008 & 0.0040 \\
0 & 0 & 0 & 0.0010 & 0.0285 & -0.0009 & -0.0077 \\
0 & 0 & 0 & -0.0008 & -0.0009 & 0.0219 & 0.0035 \\
0 & 0 & 0 & 0.0040 & -0.0077 & 0.0035 & 0.0309
}$.

\subsubsection{Properties of S$_0^6$}

\begin{table*}[h]%
\begin{ruledtabular}
\begin{tabular}{ll}
\textrm{Properties (unit)}&
\textrm{Values}\\
\colrule
Energy (eV) & -29.137 \\
Basis vectors (\AA)& $a=(2.7734, 2.7749, -0.0460), b=(-0.0460, 2.7749, 2.7734), c=(2.7274, 0, 2.7274)$\\
\begin{tabular}{@{}l@{}}Atom coordinates \\ (in basis vectors)\end{tabular} &
{\begin{tabular}{@{}l@{}}$U1=(0.5000, 0.5000, 0.5000)$ \\$O1=(0.2500, 0.2500, 0.2500)$, $O2=(0.7500, 0.7500, 0.7500)$\end{tabular}} \\ 
Strain & \begin{tabular}{@{}l@{}}($\epsilon_{xx},\epsilon_{yy},\epsilon_{zz},\epsilon_{xy},\epsilon_{xz},\epsilon_{yz}$)=(0.0001, 0.0007, 0.0001, 0, -0.0166, 0), \\($\epsilon_{A1g},\epsilon_{Eg.0},\epsilon_{Eg.1},\epsilon_{T2g.0},\epsilon_{T2g.1},\epsilon_{T2g.2}$)=(0.0005, 0.0004, -0.0002, 0, -0.0166, 0).\end{tabular}\\
Spin moments ($\mu_B$)& (2.028, 0.000, 0.000)\\
Energy (eV) and lattice (\AA) of und & -29.127, 2.7731
\end{tabular}
\end{ruledtabular}
\end{table*}

$\occmat^\uparrow=
\setstackgap{L}{1.1\baselineskip}
\fixTABwidth{T}
\parenMatrixstack{
0.4935 & 0 & -0.2977 & 0.2640 & 0 & -0.2538 & 0 \\
0 & 0.1245 & 0 & 0 & 0.0065 & 0 & -0.0083 \\
-0.2977 & 0 & 0.5484 & -0.2572 & 0 & -0.2685 & 0 \\
0.2640 & 0 & -0.2572 & 0.2259 & 0 & -0.0478 & 0 \\
0 & 0.0065 & 0 & 0 & 0.0355 & 0 & -0.0075 \\
-0.2538 & 0 & -0.2685 & -0.0478 & 0 & 0.8161 & 0 \\
0 & -0.0083 & 0 & 0 & -0.0075 & 0 & 0.0394
}$,

$\occmat^\downarrow=
\setstackgap{L}{1.1\baselineskip}
\fixTABwidth{T}
\parenMatrixstack{
0.0309 & 0 & 0.0077 & -0.0040 & 0 & 0.0035 & 0 \\
0 & 0.1087 & 0 & 0 & 0.0036 & 0 & -0.0047 \\
0.0077 & 0 & 0.0285 & 0.0010 & 0 & 0.0009 & 0 \\
-0.0040 & 0 & 0.0010 & 0.0374 & 0 & 0.0008 & 0 \\
0 & 0.0036 & 0 & 0 & 0.0288 & 0 & -0.0054 \\
0.0035 & 0 & 0.0009 & 0.0008 & 0 & 0.0219 & 0 \\
0 & -0.0047 & 0 & 0 & -0.0054 & 0 & 0.0316
}$.

\subsubsection{Properties of S$_1^1$}

\begin{table*}[h]%
\begin{ruledtabular}
\begin{tabular}{ll}
\textrm{Properties (unit)}&
\textrm{Values}\\
\colrule
Energy (eV) & -29.124 \\
Basis vectors (\AA)& $a=(2.7722, 2.7756, -0.0002), b=(-0.0002, 2.7756, 2.7722), c=(2.7720, 0, 2.7720)$\\
\begin{tabular}{@{}l@{}}Atom coordinates \\ (in basis vectors)\end{tabular} &
{\begin{tabular}{@{}l@{}}$U1=(0.5000, 0.5000, 0.5000)$ \\$O1=(0.2500, 0.2500, 0.2500)$, $O2=(0.7500, 0.7500, 0.7500)$\end{tabular}} \\ 
Strain & \begin{tabular}{@{}l@{}}($\epsilon_{xx},\epsilon_{yy},\epsilon_{zz},\epsilon_{xy},\epsilon_{xz},\epsilon_{yz}$)=(-0.0004, 0.0009, -0.0004, 0, -0.0001, 0), \\($\epsilon_{A1g},\epsilon_{Eg.0},\epsilon_{Eg.1},\epsilon_{T2g.0},\epsilon_{T2g.1},\epsilon_{T2g.2}$)=(0, -0.0009, -0.0005, 0, 0, -0.0001).\end{tabular}\\
Spin moments ($\mu_B$)& (2.028, 0.000, 0.000)\\
Energy (eV) and lattice (\AA) of und & -29.124, 2.7733
\end{tabular}
\end{ruledtabular}
\end{table*}

$\occmat^\uparrow=
\setstackgap{L}{1.1\baselineskip}
\fixTABwidth{T}
\parenMatrixstack{
0.0382 & 0.0001 & 0.0068 & 0 & 0 & 0 & 0 \\
0.0001 & 0.1213 & 0 & 0 & 0 & 0 & 0 \\
0.0068 & 0 & 0.0347 & 0 & 0 & 0 & 0 \\
0 & 0 & 0 & 0.0502 & 0.0043 & 0.0555 & 0.0031 \\
0 & 0 & 0 & 0.0043 & 0.5865 & 0.0055 & 0.4709 \\
0 & 0 & 0 & 0.0555 & 0.0055 & 0.9942 & -0.0067 \\
0 & 0 & 0 & 0.0031 & 0.4709 & -0.0067 & 0.4580
}$,

$\occmat^\downarrow=
\setstackgap{L}{1.1\baselineskip}
\fixTABwidth{T}
\parenMatrixstack{
0.0307 & 0 & 0.0048 & 0 & 0 & 0 & 0 \\
0 & 0.1081 & 0 & 0 & 0 & 0 & 0 \\
0.0048 & 0 & 0.0282 & 0 & 0 & 0 & 0 \\
0 & 0 & 0 & 0.0400 & 0 & -0.0016 & 0 \\
0 & 0 & 0 & 0 & 0.0297 & 0 & -0.0096 \\
0 & 0 & 0 & -0.0016 & 0 & 0.0211 & 0.0001 \\
0 & 0 & 0 & 0 & -0.0096 & 0.0001 & 0.0314
}$.
\clearpage

\subsubsection{Properties of S$_1^2$}

\begin{table*}[h]%
\begin{ruledtabular}
\begin{tabular}{ll}
\textrm{Properties (unit)}&
\textrm{Values}\\
\colrule
Energy (eV) & -29.124 \\
Basis vectors (\AA)& $a=(2.7722, 2.7722,     0), b=(    0, 2.7722, 2.7756), c=(2.7722,     0, 2.7756)$\\
\begin{tabular}{@{}l@{}}Atom coordinates \\ (in basis vectors)\end{tabular} &
{\begin{tabular}{@{}l@{}}$U1=(0.5000, 0.5000, 0.5000)$ \\$O1=(0.2500, 0.2500, 0.2500)$, $O2=(0.7500, 0.7500, 0.7500)$\end{tabular}} \\ 
Strain & \begin{tabular}{@{}l@{}}($\epsilon_{xx},\epsilon_{yy},\epsilon_{zz},\epsilon_{xy},\epsilon_{xz},\epsilon_{yz}$)=(-0.0004, -0.0004, 0.0008,     0,     0,     0), \\($\epsilon_{A1g},\epsilon_{Eg.0},\epsilon_{Eg.1},\epsilon_{T2g.0},\epsilon_{T2g.1},\epsilon_{T2g.2}$)=(    0,     0, 0.0010,     0,     0,     0).\end{tabular}\\
Spin moments ($\mu_B$)& (2.028, 0.000, 0.000)\\
Energy (eV) and lattice (\AA) of und & -29.124, 2.7733
\end{tabular}
\end{ruledtabular}
\end{table*}

$\occmat^\uparrow=
\setstackgap{L}{1.1\baselineskip}
\fixTABwidth{T}
\parenMatrixstack{
0.3505 &     0 & -0.4432 &     0 &     0 &     0 & 0.0001 \\
    0 & 0.1213 &     0 &     0 &     0 &     0 &     0 \\
-0.4432 &     0 & 0.6940 &     0 & 0.0001 &     0 & -0.0001 \\
    0 &     0 &     0 & 0.0435 &     0 &     0 &     0 \\
    0 &     0 & 0.0001 &     0 & 0.6940 &     0 & 0.4432 \\
    0 &     0 &     0 &     0 &     0 & 0.0294 &     0 \\
0.0001 &     0 & -0.0001 &     0 & 0.4432 &     0 & 0.3505
}$,

$\occmat^\downarrow=
\setstackgap{L}{1.1\baselineskip}
\fixTABwidth{T}
\parenMatrixstack{
0.0344 &     0 & 0.0088 &     0 &     0 &     0 &     0 \\
    0 & 0.1082 &     0 &     0 &     0 &     0 &     0 \\
0.0088 &     0 & 0.0267 &     0 &     0 &     0 &     0 \\
    0 &     0 &     0 & 0.0344 &     0 &     0 &     0 \\
    0 &     0 &     0 &     0 & 0.0267 &     0 & -0.0088 \\
    0 &     0 &     0 &     0 &     0 & 0.0246 &     0 \\
    0 &     0 &     0 &     0 & -0.0088 &     0 & 0.0344
}$.

\subsubsection{Properties of S$_1^3$}

\begin{table*}[h]%
\begin{ruledtabular}
\begin{tabular}{ll}
\textrm{Properties (unit)}&
\textrm{Values}\\
\colrule
Energy (eV) & -29.124 \\
Basis vectors (\AA)& $a=(2.7756, 2.7722,     0), b=(    0, 2.7722, 2.7722), c=(2.7756,     0, 2.7722)$\\
\begin{tabular}{@{}l@{}}Atom coordinates \\ (in basis vectors)\end{tabular} &
{\begin{tabular}{@{}l@{}}$U1=(0.5000, 0.5000, 0.5000)$ \\$O1=(0.2500, 0.2500, 0.2500)$, $O2=(0.7500, 0.7500, 0.7500)$\end{tabular}} \\ 
Strain & \begin{tabular}{@{}l@{}}($\epsilon_{xx},\epsilon_{yy},\epsilon_{zz},\epsilon_{xy},\epsilon_{xz},\epsilon_{yz}$)=(0.0008, -0.0004, -0.0004,     0,     0,     0), \\($\epsilon_{A1g},\epsilon_{Eg.0},\epsilon_{Eg.1},\epsilon_{T2g.0},\epsilon_{T2g.1},\epsilon_{T2g.2}$)=(     0, 0.0009, -0.0005,     0,     0,     0).\end{tabular}\\
Spin moments ($\mu_B$)& (2.028, 0.000, 0.000)\\
Energy (eV) and lattice (\AA) of und & -29.124, 2.7733
\end{tabular}
\end{ruledtabular}
\end{table*}

$\occmat^\uparrow=
\setstackgap{L}{1.1\baselineskip}
\fixTABwidth{T}
\parenMatrixstack{
0.4580 &     0 & -0.4709 &     0 &     0 &     0 &     0 \\
    0 & 0.1213 &     0 &     0 &     0 &     0 &     0 \\
-0.4709 &     0 & 0.5865 &     0 &     0 &     0 &     0 \\
    0 &     0 &     0 & 0.0502 &     0 & -0.0555 &     0 \\
    0 &     0 &     0 &     0 & 0.0347 &     0 & -0.0068 \\
    0 &     0 &     0 & -0.0555 &     0 & 0.9943 &     0 \\
    0 &     0 &     0 &     0 & -0.0068 &     0 & 0.0382
}$,

$\occmat^\downarrow=
\setstackgap{L}{1.1\baselineskip}
\fixTABwidth{T}
\parenMatrixstack{
0.0314 &     0 & 0.0096 &     0 &     0 &     0 &     0 \\
    0 & 0.1082 &     0 &     0 &     0 &     0 &     0 \\
0.0096 &     0 & 0.0297 &     0 &     0 &     0 &     0 \\
    0 &     0 &     0 & 0.0400 &     0 & 0.0016 &     0 \\
    0 &     0 &     0 &     0 & 0.0282 &     0 & -0.0048 \\
    0 &     0 &     0 & 0.0016 &     0 & 0.0211 &     0 \\
    0 &     0 &     0 &     0 & -0.0048 &     0 & 0.0307
}$.

\subsubsection{Properties of S$_2^1$}

\begin{table*}[h]%
\begin{ruledtabular}
\begin{tabular}{ll}
\textrm{Properties (unit)}&
\textrm{Values}\\
\colrule
Energy (eV) & -29.098 \\
Basis vectors (\AA)& $a=(2.7648, 2.7600,     0), b=(  0, 2.7600, 2.7952), c=(2.7648,     0, 2.7952)$\\
\begin{tabular}{@{}l@{}}Atom coordinates \\ (in basis vectors)\end{tabular} &
{\begin{tabular}{@{}l@{}}$U1=(0.5000, 0.5000, 0.5000)$ \\$O1=(0.2500, 0.2500, 0.2500)$, $O2=(0.7500, 0.7500, 0.7500)$\end{tabular}} \\ 
Strain & \begin{tabular}{@{}l@{}}($\epsilon_{xx},\epsilon_{yy},\epsilon_{zz},\epsilon_{xy},\epsilon_{xz},\epsilon_{yz}$)=(-0.0031, -0.0048, 0.0079,     0,     0,     0), \\($\epsilon_{A1g},\epsilon_{Eg.0},\epsilon_{Eg.1},\epsilon_{T2g.0},\epsilon_{T2g.1},\epsilon_{T2g.2}$)=(    0, 0.0012, 0.0097,     0,     0,     0).\end{tabular}\\
Spin moments ($\mu_B$)& (2.018, 0.000, 0.000)\\
Energy (eV) and lattice (\AA) of und & -29.095, 2.7733
\end{tabular}
\end{ruledtabular}
\end{table*}

$\occmat^\uparrow=
\setstackgap{L}{1.1\baselineskip}
\fixTABwidth{T}
\parenMatrixstack{
0.0383 &     0 & 0.0068 &     0 &     0 &     0 &     0 \\
    0 & 0.1346 &     0 &     0 &     0 &     0 &     0 \\
0.0068 &     0 & 0.0357 &     0 &     0 &     0 &     0 \\
    0 &     0 &     0 & 0.9793 &     0 & -0.1152 &     0 \\
    0 &     0 &     0 &     0 & 0.0525 &     0 & 0.1170 \\
    0 &     0 &     0 & -0.1152 &     0 & 0.0466 &     0 \\
    0 &     0 &     0 &     0 & 0.1170 &     0 & 0.9806
}$,

$\occmat^\downarrow=
\setstackgap{L}{1.1\baselineskip}
\fixTABwidth{T}
\parenMatrixstack{
0.0316 &     0 & 0.0054 &     0 &     0 &     0 &     0 \\
    0 & 0.1102 &     0 &     0 &     0 &     0 &     0 \\
0.0054 &     0 & 0.0291 &     0 &     0 &     0 &     0 \\
    0 &     0 &     0 & 0.0316 &     0 & 0.0003 &     0 \\
    0 &     0 &     0 &     0 & 0.0290 &     0 & -0.0056 \\
    0 &     0 &     0 & 0.0003 &     0 & 0.0246 &     0 \\
    0 &     0 &     0 &     0 & -0.0056 &     0 & 0.0274
}$.

\subsubsection{Properties of S$_2^2$}

\begin{table*}[h]%
\begin{ruledtabular}
\begin{tabular}{ll}
\textrm{Properties (unit)}&
\textrm{Values}\\
\colrule
Energy (eV) & -29.098 \\
Basis vectors (\AA)& $a=(2.7600, 2.7952,     0), b=(   0, 2.7952, 2.7648), c=(2.7600,     0, 2.7648)$\\
\begin{tabular}{@{}l@{}}Atom coordinates \\ (in basis vectors)\end{tabular} &
{\begin{tabular}{@{}l@{}}$U1=(0.5000, 0.5000, 0.5000)$ \\$O1=(0.2500, 0.2500, 0.2500)$, $O2=(0.7500, 0.7500, 0.7500)$\end{tabular}} \\ 
Strain & \begin{tabular}{@{}l@{}}($\epsilon_{xx},\epsilon_{yy},\epsilon_{zz},\epsilon_{xy},\epsilon_{xz},\epsilon_{yz}$)=(-0.0048, 0.0079, -0.0031,     0,     0,     0), \\($\epsilon_{A1g},\epsilon_{Eg.0},\epsilon_{Eg.1},\epsilon_{T2g.0},\epsilon_{T2g.1},\epsilon_{T2g.2}$)=(    0, -0.0090, -0.0038,     0,     0,     0).\end{tabular}\\
Spin moments ($\mu_B$)& (2.018, 0.000, 0.000)\\
Energy (eV) and lattice (\AA) of und & -29.095, 2.7733
\end{tabular}
\end{ruledtabular}
\end{table*}

$\occmat^\uparrow=
\setstackgap{L}{1.1\baselineskip}
\fixTABwidth{T}
\parenMatrixstack{
0.5179 &     0 & 0.4803 &     0 &     0 &     0 &     0 \\
    0 & 0.1346 &     0 &     0 &     0 &     0 &     0 \\
0.4803 &     0 & 0.5079 &     0 &     0 &     0 &     0 \\
    0 &     0 &     0 & 0.5192 &     0 & -0.4786 &     0 \\
    0 &     0 &     0 &     0 & 0.0348 &     0 & -0.0066 \\
    0 &     0 &     0 & -0.4786 &     0 & 0.5138 &     0 \\
    0 &     0 &     0 &     0 & -0.0066 &     0 & 0.0391
}$,

$\occmat^\downarrow=
\setstackgap{L}{1.1\baselineskip}
\fixTABwidth{T}
\parenMatrixstack{
0.0293 &     0 & 0.0033 &     0 &     0 &     0 &     0 \\
    0 & 0.1102 &     0 &     0 &     0 &     0 &     0 \\
0.0033 &     0 & 0.0269 &     0 &     0 &     0 &     0 \\
    0 &     0 &     0 & 0.0335 &     0 & 0.0022 &     0 \\
    0 &     0 &     0 &     0 & 0.0288 &     0 & -0.0054 \\
    0 &     0 &     0 & 0.0022 &     0 & 0.0229 &     0 \\
    0 &     0 &     0 &     0 & -0.0054 &     0 & 0.0319
}$.

\subsubsection{Properties of S$_2^3$}

\begin{table*}[h]%
\begin{ruledtabular}
\begin{tabular}{ll}
\textrm{Properties (unit)}&
\textrm{Values}\\
\colrule
Energy (eV) & -29.098 \\
Basis vectors (\AA)& $a=(2.7952, 2.7600,     0), b=(  0, 2.7600, 2.7648), c=(2.7952,     0, 2.7648)$\\
\begin{tabular}{@{}l@{}}Atom coordinates \\ (in basis vectors)\end{tabular} &
{\begin{tabular}{@{}l@{}}$U1=(0.5000, 0.5000, 0.5000)$ \\$O1=(0.2500, 0.2500, 0.2500)$, $O2=(0.7500, 0.7500, 0.7500)$\end{tabular}} \\ 
Strain & \begin{tabular}{@{}l@{}}($\epsilon_{xx},\epsilon_{yy},\epsilon_{zz},\epsilon_{xy},\epsilon_{xz},\epsilon_{yz}$)=(0.0079, -0.0048, -0.0031,     0,     0,     0), \\($\epsilon_{A1g},\epsilon_{Eg.0},\epsilon_{Eg.1},\epsilon_{T2g.0},\epsilon_{T2g.1},\epsilon_{T2g.2}$)=(     0, 0.0090, -0.0038,     0,     0,     0).\end{tabular}\\
Spin moments ($\mu_B$)& (2.018, 0.000, 0.000)\\
Energy (eV) and lattice (\AA) of und & -29.095, 2.7733
\end{tabular}
\end{ruledtabular}
\end{table*}

$\occmat^\uparrow=
\setstackgap{L}{1.1\baselineskip}
\fixTABwidth{T}
\parenMatrixstack{
0.0391 &     0 & 0.0066 &     0 &     0 &     0 &     0 \\
    0 & 0.1346 &     0 &     0 &     0 &     0 &     0 \\
0.0066 &     0 & 0.0348 &     0 &     0 &     0 &     0 \\
    0 &     0 &     0 & 0.5192 &     0 & 0.4786 &     0 \\
    0 &     0 &     0 &     0 & 0.5079 &     0 & -0.4803 \\
    0 &     0 &     0 & 0.4786 &     0 & 0.5138 &     0 \\
    0 &     0 &     0 &     0 & -0.4803 &     0 & 0.5179
}$,

$\occmat^\downarrow=
\setstackgap{L}{1.1\baselineskip}
\fixTABwidth{T}
\parenMatrixstack{
0.0319 &     0 & 0.0054 &     0 &     0 &     0 &     0 \\
    0 & 0.1102 &     0 &     0 &     0 &     0 &     0 \\
0.0054 &     0 & 0.0288 &     0 &     0 &     0 &     0 \\
    0 &     0 &     0 & 0.0335 &     0 & -0.0022 &     0 \\
    0 &     0 &     0 &     0 & 0.0269 &     0 & -0.0033 \\
    0 &     0 &     0 & -0.0022 &     0 & 0.0229 &     0 \\
    0 &     0 &     0 &     0 & -0.0033 &     0 & 0.0293
}$.

\subsubsection{Properties of S$_2^4$}

\begin{table*}[h]%
\begin{ruledtabular}
\begin{tabular}{ll}
\textrm{Properties (unit)}&
\textrm{Values}\\
\colrule
Energy (eV) & -29.098 \\
Basis vectors (\AA)& $a=(2.7952, 2.7648,     0), b=(  0, 2.7648, 2.7600), c=(2.7952,     0, 2.7600)$\\
\begin{tabular}{@{}l@{}}Atom coordinates \\ (in basis vectors)\end{tabular} &
{\begin{tabular}{@{}l@{}}$U1=(0.5000, 0.5000, 0.5000)$ \\$O1=(0.2500, 0.2500, 0.2500)$, $O2=(0.7500, 0.7500, 0.7500)$\end{tabular}} \\ 
Strain & \begin{tabular}{@{}l@{}}($\epsilon_{xx},\epsilon_{yy},\epsilon_{zz},\epsilon_{xy},\epsilon_{xz},\epsilon_{yz}$)=(0.0079, -0.0031, -0.0048,     0,     0,     0), \\($\epsilon_{A1g},\epsilon_{Eg.0},\epsilon_{Eg.1},\epsilon_{T2g.0},\epsilon_{T2g.1},\epsilon_{T2g.2}$)=(     0, 0.0078, -0.0059,     0,     0,     0).\end{tabular}\\
Spin moments ($\mu_B$)& (2.018, 0.000, 0.000)\\
Energy (eV) and lattice (\AA) of und & -29.095, 2.7733
\end{tabular}
\end{ruledtabular}
\end{table*}

$\occmat^\uparrow=
\setstackgap{L}{1.1\baselineskip}
\fixTABwidth{T}
\parenMatrixstack{
0.0538 &     0 & 0.1223 &     0 &     0 &     0 &     0 \\
    0 & 0.1346 &     0 &     0 &     0 &     0 &     0 \\
0.1223 &     0 & 0.9793 &     0 &     0 &     0 &     0 \\
    0 &     0 &     0 & 0.0439 &     0 & -0.0004 &     0 \\
    0 &     0 &     0 &     0 & 0.2848 &     0 & -0.4227 \\
    0 &     0 &     0 & -0.0004 &     0 & 0.0300 &     0 \\
    0 &     0 &     0 &     0 & -0.4227 &     0 & 0.7411
}$,

$\occmat^\downarrow=
\setstackgap{L}{1.1\baselineskip}
\fixTABwidth{T}
\parenMatrixstack{
0.0316 &     0 & 0.0046 &     0 &     0 &     0 &     0 \\
    0 & 0.1102 &     0 &     0 &     0 &     0 &     0 \\
0.0046 &     0 & 0.0248 &     0 &     0 &     0 &     0 \\
    0 &     0 &     0 & 0.0359 &     0 & -0.0001 &     0 \\
    0 &     0 &     0 &     0 & 0.0276 &     0 & -0.0034 \\
    0 &     0 &     0 & -0.0001 &     0 & 0.0248 &     0 \\
    0 &     0 &     0 &     0 & -0.0034 &     0 & 0.0286
}$.

\subsubsection{Properties of S$_2^5$}

\begin{table*}[h]%
\begin{ruledtabular}
\begin{tabular}{ll}
\textrm{Properties (unit)}&
\textrm{Values}\\
\colrule
Energy (eV) & -29.098 \\
Basis vectors (\AA)& $a=(2.7600, 2.7648,     0), b=(  0, 2.7648, 2.7952), c=(2.7600,     0, 2.7952)$\\
\begin{tabular}{@{}l@{}}Atom coordinates \\ (in basis vectors)\end{tabular} &
{\begin{tabular}{@{}l@{}}$U1=(0.5000, 0.5000, 0.5000)$ \\$O1=(0.2500, 0.2500, 0.2500)$, $O2=(0.7500, 0.7500, 0.7500)$\end{tabular}} \\ 
Strain & \begin{tabular}{@{}l@{}}($\epsilon_{xx},\epsilon_{yy},\epsilon_{zz},\epsilon_{xy},\epsilon_{xz},\epsilon_{yz}$)=(-0.0048, -0.0031, 0.0079,     0,     0,     0), \\($\epsilon_{A1g},\epsilon_{Eg.0},\epsilon_{Eg.1},\epsilon_{T2g.0},\epsilon_{T2g.1},\epsilon_{T2g.2}$)=(    0, -0.0012, 0.0097,     0,     0,     0).\end{tabular}\\
Spin moments ($\mu_B$)& (2.018, 0.000, 0.000)\\
Energy (eV) and lattice (\AA) of und & -29.095, 2.7733
\end{tabular}
\end{ruledtabular}
\end{table*}

$\occmat^\uparrow=
\setstackgap{L}{1.1\baselineskip}
\fixTABwidth{T}
\parenMatrixstack{
0.9806 &     0 & -0.1170 &     0 &     0 &     0 &     0 \\
    0 & 0.1346 &     0 &     0 &     0 &     0 &     0 \\
-0.1170 &     0 & 0.0525 &     0 &     0 &     0 &     0 \\
    0 &     0 &     0 & 0.9793 &     0 & 0.1152 &     0 \\
    0 &     0 &     0 &     0 & 0.0357 &     0 & -0.0068 \\
    0 &     0 &     0 & 0.1152 &     0 & 0.0466 &     0 \\
    0 &     0 &     0 &     0 & -0.0068 &     0 & 0.0383
}$,

$\occmat^\downarrow=
\setstackgap{L}{1.1\baselineskip}
\fixTABwidth{T}
\parenMatrixstack{
0.0274 &     0 & 0.0056 &     0 &     0 &     0 &     0 \\
    0 & 0.1102 &     0 &     0 &     0 &     0 &     0 \\
0.0056 &     0 & 0.0290 &     0 &     0 &     0 &     0 \\
    0 &     0 &     0 & 0.0316 &     0 & -0.0003 &     0 \\
    0 &     0 &     0 &     0 & 0.0291 &     0 & -0.0054 \\
    0 &     0 &     0 & -0.0003 &     0 & 0.0246 &     0 \\
    0 &     0 &     0 &     0 & -0.0054 &     0 & 0.0316
}$.
\clearpage

\subsubsection{Properties of S$_2^6$}

\begin{table*}[h]%
\begin{ruledtabular}
\begin{tabular}{ll}
\textrm{Properties (unit)}&
\textrm{Values}\\
\colrule
Energy (eV) & -29.098 \\
Basis vectors (\AA)& $a=(2.7648, 2.7952,     0), b=( 0, 2.7952, 2.7600), c=(2.7648,     0, 2.7600)$\\
\begin{tabular}{@{}l@{}}Atom coordinates \\ (in basis vectors)\end{tabular} &
{\begin{tabular}{@{}l@{}}$U1=(0.5000, 0.5000, 0.5000)$ \\$O1=(0.2500, 0.2500, 0.2500)$, $O2=(0.7500, 0.7500, 0.7500)$\end{tabular}} \\ 
Strain & \begin{tabular}{@{}l@{}}($\epsilon_{xx},\epsilon_{yy},\epsilon_{zz},\epsilon_{xy},\epsilon_{xz},\epsilon_{yz}$)=(-0.0031, 0.0079, -0.0048,     0,     0,     0), \\($\epsilon_{A1g},\epsilon_{Eg.0},\epsilon_{Eg.1},\epsilon_{T2g.0},\epsilon_{T2g.1},\epsilon_{T2g.2}$)=(       0, -0.0078, -0.0059,     0,     0,     0).\end{tabular}\\
Spin moments ($\mu_B$)& (2.018, 0.000, 0.000)\\
Energy (eV) and lattice (\AA) of und & -29.095, 2.7733
\end{tabular}
\end{ruledtabular}
\end{table*}

$\occmat^\uparrow=
\setstackgap{L}{1.1\baselineskip}
\fixTABwidth{T}
\parenMatrixstack{
0.7411 &     0 & 0.4227 &     0 &     0 &     0 &     0 \\
    0 & 0.1346 &     0 &     0 &     0 &     0 &     0 \\
0.4227 &     0 & 0.2848 &     0 &     0 &     0 &     0 \\
    0 &     0 &     0 & 0.0439 &     0 & 0.0004 &     0 \\
    0 &     0 &     0 &     0 & 0.9793 &     0 & -0.1223 \\
    0 &     0 &     0 & 0.0004 &     0 & 0.0300 &     0 \\
    0 &     0 &     0 &     0 & -0.1223 &     0 & 0.0538
}$,

$\occmat^\downarrow=
\setstackgap{L}{1.1\baselineskip}
\fixTABwidth{T}
\parenMatrixstack{
0.0286 &     0 & 0.0034 &     0 &     0 &     0 &     0 \\
    0 & 0.1102 &     0 &     0 &     0 &     0 &     0 \\
0.0034 &     0 & 0.0276 &     0 &     0 &     0 &     0 \\
    0 &     0 &     0 & 0.0359 &     0 & 0.0001 &     0 \\
    0 &     0 &     0 &     0 & 0.0248 &     0 & -0.0046 \\
    0 &     0 &     0 & 0.0001 &     0 & 0.0248 &     0 \\
    0 &     0 &     0 &     0 & -0.0046 &     0 & 0.0316
}$.

\subsubsection{Properties of S$_3^1$}

\begin{table*}[h]%
\begin{ruledtabular}
\begin{tabular}{ll}
\textrm{Properties (unit)}&
\textrm{Values}\\
\colrule
Energy (eV) & -29.097 \\
Basis vectors (\AA)& $a=(2.7690, 2.7691,     0), b=( 0.0066, 2.7625, 2.7952), c=(2.7625, 0.0066, 2.7952)$\\
\begin{tabular}{@{}l@{}}Atom coordinates \\ (in basis vectors)\end{tabular} &
{\begin{tabular}{@{}l@{}}$U1=(0.5000, 0.5000, 0.5000)$ \\$O1=(0.2505, 0.2495, 0.2495)$, $O2=(0.7495, 0.7505, 0.7505)$\end{tabular}} \\ 
Strain & \begin{tabular}{@{}l@{}}($\epsilon_{xx},\epsilon_{yy},\epsilon_{zz},\epsilon_{xy},\epsilon_{xz},\epsilon_{yz}$)=(-0.0039, -0.0039, 0.0079, 0.0024,     0,     0), \\($\epsilon_{A1g},\epsilon_{Eg.0},\epsilon_{Eg.1},\epsilon_{T2g.0},\epsilon_{T2g.1},\epsilon_{T2g.2}$)=(      0,     0, 0.0096, 0.0024,     0,     0).\end{tabular}\\
Spin moments ($\mu_B$)& (2.017, 0.000, 0.000)\\
Energy (eV) and lattice (\AA) of und & -29.093, 2.7734
\end{tabular}
\end{ruledtabular}
\end{table*}

$\occmat^\uparrow=
\setstackgap{L}{1.1\baselineskip}
\fixTABwidth{T}
\parenMatrixstack{
0.5248 &     0 & -0.0179 &     0 & 0.0207 & -0.0003 & 0.4775 \\
    0 & 0.1376 &     0 & -0.0403 &     0 & -0.0002 &     0 \\
-0.0179 &     0 & 0.0374 & 0.0006 & -0.0030 &     0 & -0.0235 \\
    0 & -0.0403 & 0.0006 & 0.9918 & 0.0006 & 0.0032 &     0 \\
0.0207 &     0 & -0.0030 & 0.0006 & 0.0371 &     0 & 0.0142 \\
-0.0003 & -0.0002 &     0 & 0.0032 &     0 & 0.0318 & -0.0002 \\
0.4775 &     0 & -0.0235 &     0 & 0.0142 & -0.0002 & 0.5056
}$,

$\occmat^\downarrow=
\setstackgap{L}{1.1\baselineskip}
\fixTABwidth{T}
\parenMatrixstack{
0.0292 &     0 & 0.0053 &     0 & -0.0001 &     0 & -0.0016 \\
    0 & 0.1103 &     0 & 0.0014 &     0 &     0 &     0 \\
0.0053 &     0 & 0.0290 &     0 & -0.0007 &     0 & 0.0001 \\
    0 & 0.0014 &     0 & 0.0315 &     0 &     0 &     0 \\
-0.0001 &     0 & -0.0007 &     0 & 0.0290 &     0 & -0.0053 \\
    0 &     0 &     0 &     0 &     0 & 0.0247 &     0 \\
-0.0016 &     0 & 0.0001 &     0 & -0.0053 &     0 & 0.0293
}$.

\subsubsection{Properties of S$_4^1$}

\begin{table*}[h]%
\begin{ruledtabular}
\begin{tabular}{ll}
\textrm{Properties (unit)}&
\textrm{Values}\\
\colrule
Energy (eV) & -29.088 \\
Basis vectors (\AA)& $a=(2.7851, 2.7506,     0), b=( 0, 2.7506, 2.7851), c=(2.7851,     0, 2.7851)$\\
\begin{tabular}{@{}l@{}}Atom coordinates \\ (in basis vectors)\end{tabular} &
{\begin{tabular}{@{}l@{}}$U1=(0.5000, 0.5000, 0.5000)$ \\$O1=(0.2500, 0.2500, 0.2500)$, $O2=(0.7500, 0.7500, 0.7500)$\end{tabular}} \\ 
Strain & \begin{tabular}{@{}l@{}}($\epsilon_{xx},\epsilon_{yy},\epsilon_{zz},\epsilon_{xy},\epsilon_{xz},\epsilon_{yz}$)=(0.0042, -0.0083, 0.0042,     0,     0,     0), \\($\epsilon_{A1g},\epsilon_{Eg.0},\epsilon_{Eg.1},\epsilon_{T2g.0},\epsilon_{T2g.1},\epsilon_{T2g.2}$)=(     0, 0.0088, 0.0051,     0,     0,     0).\end{tabular}\\
Spin moments ($\mu_B$)& (2.017, 0.000, 0.000)\\
Energy (eV) and lattice (\AA) of und & -29.084, 2.7736
\end{tabular}
\end{ruledtabular}
\end{table*}

$\occmat^\uparrow=
\setstackgap{L}{1.1\baselineskip}
\fixTABwidth{T}
\parenMatrixstack{
0.0380 &     0 & 0.0057 &     0 &     0 &     0 &     0 \\
    0 & 0.1378 &     0 &     0 &     0 &     0 &     0 \\
0.0057 &     0 & 0.0351 &     0 &     0 &     0 &     0 \\
    0 &     0 &     0 & 0.9126 &     0 & 0.2673 &     0 \\
    0 &     0 &     0 &     0 & 0.1548 &     0 & -0.3197 \\
    0 &     0 &     0 & 0.2673 &     0 & 0.1142 &     0 \\
    0 &     0 &     0 &     0 & -0.3197 &     0 & 0.8719
}$,

$\occmat^\downarrow=
\setstackgap{L}{1.1\baselineskip}
\fixTABwidth{T}
\parenMatrixstack{
0.0311 &     0 & 0.0052 &     0 &     0 &     0 &     0 \\
    0 & 0.1105 &     0 &     0 &     0 &     0 &     0 \\
0.0052 &     0 & 0.0284 &     0 &     0 &     0 &     0 \\
    0 &     0 &     0 & 0.0315 &     0 & -0.0014 &     0 \\
    0 &     0 &     0 &     0 & 0.0285 &     0 & -0.0038 \\
    0 &     0 &     0 & -0.0014 &     0 & 0.0245 &     0 \\
    0 &     0 &     0 &     0 & -0.0038 &     0 & 0.0275
}$.

\subsubsection{Properties of S$_4^2$}

\begin{table*}[h]%
\begin{ruledtabular}
\begin{tabular}{ll}
\textrm{Properties (unit)}&
\textrm{Values}\\
\colrule
Energy (eV) & -29.088 \\
Basis vectors (\AA)& $a=(2.7851, 2.7851,     0), b=(0, 2.7851, 2.7506), c=(2.7851,     0, 2.7506)$\\
\begin{tabular}{@{}l@{}}Atom coordinates \\ (in basis vectors)\end{tabular} &
{\begin{tabular}{@{}l@{}}$U1=(0.5000, 0.5000, 0.5000)$ \\$O1=(0.2500, 0.2500, 0.2500)$, $O2=(0.7500, 0.7500, 0.7500)$\end{tabular}} \\ 
Strain & \begin{tabular}{@{}l@{}}($\epsilon_{xx},\epsilon_{yy},\epsilon_{zz},\epsilon_{xy},\epsilon_{xz},\epsilon_{yz}$)=(0.0042, 0.0042, -0.0083,     0,     0,     0), \\($\epsilon_{A1g},\epsilon_{Eg.0},\epsilon_{Eg.1},\epsilon_{T2g.0},\epsilon_{T2g.1},\epsilon_{T2g.2}$)=(     0,     0, -0.0101,     0,     0,     0).\end{tabular}\\
Spin moments ($\mu_B$)& (2.017, 0.000, 0.000)\\
Energy (eV) and lattice (\AA) of und & -29.084, 2.7736
\end{tabular}
\end{ruledtabular}
\end{table*}

$\occmat^\uparrow=
\setstackgap{L}{1.1\baselineskip}
\fixTABwidth{T}
\parenMatrixstack{
0.3543 &     0 & 0.4533 & -0.0001 &     0 & -0.0001 &     0 \\
    0 & 0.1378 &     0 &     0 &     0 &     0 &     0 \\
0.4533 &     0 & 0.6724 & -0.0001 &     0 & -0.0001 &     0 \\
-0.0001 &     0 & -0.0001 & 0.0424 & 0.0001 &     0 & -0.0001 \\
    0 &     0 &     0 & 0.0001 & 0.6723 & -0.0002 & -0.4533 \\
-0.0001 &     0 & -0.0001 &     0 & -0.0002 & 0.0307 & 0.0001 \\
    0 &     0 &     0 & -0.0001 & -0.4533 & 0.0001 & 0.3545
}$,

$\occmat^\downarrow=
\setstackgap{L}{1.1\baselineskip}
\fixTABwidth{T}
\parenMatrixstack{
0.0303 &     0 & 0.0031 &     0 &     0 &     0 &     0 \\
    0 & 0.1105 &     0 &     0 &     0 &     0 &     0 \\
0.0031 &     0 & 0.0257 &     0 &     0 &     0 &     0 \\
    0 &     0 &     0 & 0.0351 &     0 &     0 &     0 \\
    0 &     0 &     0 &     0 & 0.0257 &     0 & -0.0031 \\
    0 &     0 &     0 &     0 &     0 & 0.0244 &     0 \\
    0 &     0 &     0 &     0 & -0.0031 &     0 & 0.0303
}$.

\subsubsection{Properties of S$_5^1$}

\begin{table*}[h]%
\begin{ruledtabular}
\begin{tabular}{ll}
\textrm{Properties (unit)}&
\textrm{Values}\\
\colrule
Energy (eV) & -29.038 \\
Basis vectors (\AA)& $a=(2.7640, 2.7762, -0.0001), b=(-0.0001, 2.7762, 2.7762), c=(2.7639,     0, 2.7761)$\\
\begin{tabular}{@{}l@{}}Atom coordinates \\ (in basis vectors)\end{tabular} &
{\begin{tabular}{@{}l@{}}$U1=(0.5000, 0.5000, 0.5000)$ \\$O1=(0.2500, 0.2500, 0.2500)$, $O2=(0.7500, 0.7500, 0.7500)$\end{tabular}} \\ 
Strain & \begin{tabular}{@{}l@{}}($\epsilon_{xx},\epsilon_{yy},\epsilon_{zz},\epsilon_{xy},\epsilon_{xz},\epsilon_{yz}$)=(-0.0029, 0.0015, 0.0015,     0,     0,     0), \\($\epsilon_{A1g},\epsilon_{Eg.0},\epsilon_{Eg.1},\epsilon_{T2g.0},\epsilon_{T2g.1},\epsilon_{T2g.2}$)=( 0.0001, -0.0031, 0.0018,     0,     0,     0).\end{tabular}\\
Spin moments ($\mu_B$)& (2.023, 0.000, 0.000)\\
Energy (eV) and lattice (\AA) of und & -29.037, 2.7721
\end{tabular}
\end{ruledtabular}
\end{table*}

$\occmat^\uparrow=
\setstackgap{L}{1.1\baselineskip}
\fixTABwidth{T}
\parenMatrixstack{
0.0361 & 0.0001 & 0.0104 &     0 &     0 &     0 & 0.0002 \\
0.0001 & 0.1309 &     0 &     0 &     0 &     0 &     0 \\
0.0104 &     0 & 0.0431 &     0 & 0.0001 &     0 &     0 \\
    0 &     0 &     0 & 0.0488 & -0.0045 & 0.0060 & 0.0008 \\
    0 &     0 & 0.0001 & -0.0045 & 0.9947 & 0.0025 & 0.0043 \\
    0 &     0 &     0 & 0.0060 & 0.0025 & 0.0304 & -0.0017 \\
0.0002 &     0 &     0 & 0.0008 & 0.0043 & -0.0017 & 0.9925
}$,

$\occmat^\downarrow=
\setstackgap{L}{1.1\baselineskip}
\fixTABwidth{T}
\parenMatrixstack{
0.0328 &     0 & 0.0066 &     0 &     0 &     0 &     0 \\
    0 & 0.1096 &     0 &     0 &     0 &     0 &     0 \\
0.0066 &     0 & 0.0306 &     0 &     0 &     0 &     0 \\
    0 &     0 &     0 & 0.0384 & 0.0001 & 0.0006 &     0 \\
    0 &     0 &     0 & 0.0001 & 0.0242 &     0 & -0.0040 \\
    0 &     0 &     0 & 0.0006 &     0 & 0.0250 &     0 \\
    0 &     0 &     0 &     0 & -0.0040 &     0 & 0.026
}$.

\subsubsection{Properties of S$_5^2$}

\begin{table*}[h]%
\begin{ruledtabular}
\begin{tabular}{ll}
\textrm{Properties (unit)}&
\textrm{Values}\\
\colrule
Energy (eV) & -29.038 \\
Basis vectors (\AA)& $a=(
2.7762, 2.7762,     0
), b=(
0, 2.7762, 2.7640
), c=(
2.7762,     0, 2.7640
)$\\
\begin{tabular}{@{}l@{}}Atom coordinates \\ (in basis vectors)\end{tabular} &
{\begin{tabular}{@{}l@{}}$U1=(0.5000, 0.5000, 0.5000)$ \\$O1=(0.2500, 0.2500, 0.2500)$, $O2=(0.7500, 0.7500, 0.7500)$\end{tabular}} \\ 
Strain & \begin{tabular}{@{}l@{}}($\epsilon_{xx},\epsilon_{yy},\epsilon_{zz},\epsilon_{xy},\epsilon_{xz},\epsilon_{yz}$)=(
0.0015, 0.0015, -0.0029,     0,     0,     0
), \\($\epsilon_{A1g},\epsilon_{Eg.0},\epsilon_{Eg.1},\epsilon_{T2g.0},\epsilon_{T2g.1},\epsilon_{T2g.2}$)=( 
 0,     0, -0.0036,     0,     0,     0
).\end{tabular}\\
Spin moments ($\mu_B$)& (
2.023, 0.000, 0.000
)\\
Energy (eV) and lattice (\AA) of und & -29.037, 2.7721
\end{tabular}
\end{ruledtabular}
\end{table*}

$\occmat^\uparrow=
\setstackgap{L}{1.1\baselineskip}
\fixTABwidth{T}
\parenMatrixstack{
0.0477 &     0 & 0.0074 &     0 &     0 &     0 &     0 \\
    0 & 0.1309 &     0 &     0 &     0 &     0 &     0 \\
0.0074 &     0 & 0.0315 &     0 &     0 &     0 &     0 \\
    0 &     0 &     0 & 0.9892 &     0 &     0 &     0 \\
    0 &     0 &     0 &     0 & 0.0315 &     0 & -0.0074 \\
    0 &     0 &     0 &     0 &     0 & 0.9980 &     0 \\
    0 &     0 &     0 &     0 & -0.0074 &     0 & 0.0477
}$,

$\occmat^\downarrow=
\setstackgap{L}{1.1\baselineskip}
\fixTABwidth{T}
\parenMatrixstack{
0.0340 &     0 & 0.0063 &     0 &     0 &     0 &     0 \\
    0 & 0.1096 &     0 &     0 &     0 &     0 &     0 \\
0.0063 &     0 & 0.0295 &     0 &     0 &     0 &     0 \\
    0 &     0 &     0 & 0.0293 &     0 &     0 &     0 \\
    0 &     0 &     0 &     0 & 0.0295 &     0 & -0.0063 \\
    0 &     0 &     0 &     0 &     0 & 0.0211 &     0 \\
    0 &     0 &     0 &     0 & -0.0063 &     0 & 0.0340
}$.

\subsubsection{Properties of S$_5^3$}

\begin{table*}[h]%
\begin{ruledtabular}
\begin{tabular}{ll}
\textrm{Properties (unit)}&
\textrm{Values}\\
\colrule
Energy (eV) & -29.038 \\
Basis vectors (\AA)& $a=(
2.7762, 2.7640, 0.0003
), b=(
0, 2.7643, 2.7765
), c=(
2.7762, 0.0003, 2.7762
)$\\
\begin{tabular}{@{}l@{}}Atom coordinates \\ (in basis vectors)\end{tabular} &
{\begin{tabular}{@{}l@{}}$U1=(0.5000, 0.5000, 0.5000)$ \\$O1=(0.2500, 0.2500, 0.2500)$, $O2=(0.7500, 0.7500, 0.7500)$\end{tabular}} \\ 
Strain & \begin{tabular}{@{}l@{}}($\epsilon_{xx},\epsilon_{yy},\epsilon_{zz},\epsilon_{xy},\epsilon_{xz},\epsilon_{yz}$)=(
0.0015, -0.0029, 0.0015,     0,     0, 0.0001
), \\($\epsilon_{A1g},\epsilon_{Eg.0},\epsilon_{Eg.1},\epsilon_{T2g.0},\epsilon_{T2g.1},\epsilon_{T2g.2}$)=( 
0.0001, 0.0031, 0.0018,     0, 0.0001,     0
).\end{tabular}\\
Spin moments ($\mu_B$)& (
2.023, 0.000, 0.000
)\\
Energy (eV) and lattice (\AA) of und & -29.037, 2.7721
\end{tabular}
\end{ruledtabular}
\end{table*}

$\occmat^\uparrow=
\setstackgap{L}{1.1\baselineskip}
\fixTABwidth{T}
\parenMatrixstack{
0.9925 &     0 & -0.0043 & -0.0005 &     0 & 0.0034 &     0 \\
    0 & 0.1309 &     0 &     0 & -0.0001 &     0 & 0.0001 \\
-0.0043 &     0 & 0.9946 & 0.0083 &     0 & 0.0036 &     0 \\
-0.0005 &     0 & 0.0083 & 0.0489 &     0 & -0.0060 &     0 \\
    0 & -0.0001 &     0 &     0 & 0.0431 &     0 & -0.0104 \\
0.0034 &     0 & 0.0036 & -0.0060 &     0 & 0.0304 &     0 \\
    0 & 0.0001 &     0 &     0 & -0.0104 &     0 & 0.0361
}$,

$\occmat^\downarrow=
\setstackgap{L}{1.1\baselineskip}
\fixTABwidth{T}
\parenMatrixstack{
0.0262 &     0 & 0.0040 &     0 &     0 &     0 &     0 \\
    0 & 0.1096 &     0 &     0 & -0.0001 &     0 & 0.0001 \\
0.0040 &     0 & 0.0242 & -0.0001 &     0 &     0 &     0 \\
    0 &     0 & -0.0001 & 0.0384 &     0 & -0.0006 &     0 \\
    0 & -0.0001 &     0 &     0 & 0.0306 &     0 & -0.0066 \\
    0 &     0 &     0 & -0.0006 &     0 & 0.0250 &     0 \\
    0 & 0.0001 &     0 &     0 & -0.0066 &     0 & 0.0328
}$.
\clearpage

\subsubsection{Properties of S$_6^1$}

\begin{table*}[h]%
\begin{ruledtabular}
\begin{tabular}{ll}
\textrm{Properties (unit)}&
\textrm{Values}\\
\colrule
Energy (eV) & -29.042 \\
Basis vectors (\AA)& $a=(
2.8291, 2.8291,     0
), b=(
0.0504, 2.7788, 2.7642
), c=(
2.7788, 0.0504, 2.7642
)$\\
\begin{tabular}{@{}l@{}}Atom coordinates \\ (in basis vectors)\end{tabular} &
{\begin{tabular}{@{}l@{}}$U1=(
0.5000, 0.5000, 0.5000
)$ \\$O1=(
0.2518, 0.2482, 0.2482
)$, $O2=(
0.7482, 0.7518, 0.7518
)$\end{tabular}} \\ 
Strain & \begin{tabular}{@{}l@{}}($\epsilon_{xx},\epsilon_{yy},\epsilon_{zz},\epsilon_{xy},\epsilon_{xz},\epsilon_{yz}$)=(
0.0024, 0.0024, -0.0029, 0.0182,     0,     0
), \\($\epsilon_{A1g},\epsilon_{Eg.0},\epsilon_{Eg.1},\epsilon_{T2g.0},\epsilon_{T2g.1},\epsilon_{T2g.2}$)=( 
0.0011,     0, -0.0043, 0.0182,     0,     0
).\end{tabular}\\
Spin moments ($\mu_B$)& (
2.025, 0.000, 0.000
)\\
Energy (eV) and lattice (\AA) of und & -29.023, 2.722
\end{tabular}
\end{ruledtabular}
\end{table*}

$\occmat^\uparrow=
\setstackgap{L}{1.1\baselineskip}
\fixTABwidth{T}
\parenMatrixstack{
0.4910 &     0 & 0.2619 &     0 & -0.0308 &     0 & 0.3972 \\
    0 & 0.1371 &     0 & 0.0201 &     0 &     0 &     0 \\
0.2619 &     0 & 0.5395 &     0 & 0.3951 &     0 & 0.0308 \\
    0 & 0.0201 &     0 & 0.0488 &     0 &     0 &     0 \\
-0.0308 &     0 & 0.3951 &     0 & 0.5397 &     0 & -0.2619 \\
    0 &     0 &     0 &     0 &     0 & 0.0303 &     0 \\
0.3972 &     0 & 0.0308 &     0 & -0.2619 &     0 & 0.4909
}$,

$\occmat^\downarrow=
\setstackgap{L}{1.1\baselineskip}
\fixTABwidth{T}
\parenMatrixstack{
0.0287 &     0 & 0.0036 &     0 &     0 &     0 & -0.0023 \\
    0 & 0.1102 &     0 & 0.0111 &     0 &     0 &     0 \\
0.0036 &     0 & 0.0272 &     0 & -0.0070 &     0 &     0 \\
    0 & 0.0111 &     0 & 0.0387 &     0 &     0 &     0 \\
    0 &     0 & -0.0070 &     0 & 0.0272 &     0 & -0.0036 \\
    0 &     0 &     0 &     0 &     0 & 0.0245 &     0 \\
-0.0023 &     0 &     0 &     0 & -0.0036 &     0 & 0.0287
}$.

\subsubsection{Properties of S$_7^1$}

\begin{table*}[h]%
\begin{ruledtabular}
\begin{tabular}{ll}
\textrm{Properties (unit)}&
\textrm{Values}\\
\colrule
Energy (eV) & -29.020 \\
Basis vectors (\AA)& $a=(
2.7842, 2.7839,     0
), b=(
0, 2.7839, 2.7487
), c=(
2.7842,     0, 2.7487
)$\\
\begin{tabular}{@{}l@{}}Atom coordinates \\ (in basis vectors)\end{tabular} &
{\begin{tabular}{@{}l@{}}$U1=(
0.5000, 0.5000, 0.5000
)$ \\$O1=(
0.2500, 0.2500, 0.2500
)$, $O2=(
0.7500, 0.7500, 0.7500
)$\end{tabular}} \\ 
Strain & \begin{tabular}{@{}l@{}}($\epsilon_{xx},\epsilon_{yy},\epsilon_{zz},\epsilon_{xy},\epsilon_{xz},\epsilon_{yz}$)=(
0.0043, 0.0042, -0.0085,     0,     0,     0
), \\($\epsilon_{A1g},\epsilon_{Eg.0},\epsilon_{Eg.1},\epsilon_{T2g.0},\epsilon_{T2g.1},\epsilon_{T2g.2}$)=( 
0.0001, 0.0001, -0.0104,     0,     0,     0
).\end{tabular}\\
Spin moments ($\mu_B$)& (
2.025, 0.000, 0.000
)\\
Energy (eV) and lattice (\AA) of und & -29.017, 2.7722
\end{tabular}
\end{ruledtabular}
\end{table*}

$\occmat^\uparrow=
\setstackgap{L}{1.1\baselineskip}
\fixTABwidth{T}
\parenMatrixstack{
0.0442 &     0 & 0.0087 & 0.0001 &     0 & 0.0002 &     0 \\
    0 & 0.1263 &     0 &     0 &     0 &     0 &     0 \\
0.0087 &     0 & 0.0343 &     0 &     0 & 0.0001 &     0 \\
0.0001 &     0 &     0 & 0.2029 &     0 & 0.3617 &     0 \\
    0 &     0 &     0 &     0 & 0.9917 &     0 & -0.0389 \\
0.0002 &     0 & 0.0001 & 0.3617 &     0 & 0.8304 &     0 \\
    0 &     0 &     0 &     0 & -0.0389 &     0 & 0.0513
}$,

$\occmat^\downarrow=
\setstackgap{L}{1.1\baselineskip}
\fixTABwidth{T}
\parenMatrixstack{
0.0343 &     0 & 0.0062 &     0 &     0 &     0 &     0 \\
    0 & 0.1089 &     0 &     0 &     0 &     0 &     0 \\
0.0062 &     0 & 0.0290 &     0 &     0 &     0 &     0 \\
    0 &     0 &     0 & 0.0332 &     0 & -0.0016 &     0 \\
    0 &     0 &     0 &     0 & 0.0262 &     0 & -0.0062 \\
    0 &     0 &     0 & -0.0016 &     0 & 0.0216 &     0 \\
    0 &     0 &     0 &     0 & -0.0062 &     0 & 0.0345
}$.

\subsubsection{Properties of S$_7^2$}

\begin{table*}[h]%
\begin{ruledtabular}
\begin{tabular}{ll}
\textrm{Properties (unit)}&
\textrm{Values}\\
\colrule
Energy (eV) & -29.020 \\
Basis vectors (\AA)& $a=(
2.7839, 2.7842,     0
), b=(
0, 2.7842, 2.7487
), c=(
2.7839,     0, 2.7487
)$\\
\begin{tabular}{@{}l@{}}Atom coordinates \\ (in basis vectors)\end{tabular} &
{\begin{tabular}{@{}l@{}}$U1=(
0.5000, 0.5000, 0.5000
)$ \\$O1=(
0.2500, 0.2500, 0.2500
)$, $O2=(
0.7500, 0.7500, 0.7500
)$\end{tabular}} \\ 
Strain & \begin{tabular}{@{}l@{}}($\epsilon_{xx},\epsilon_{yy},\epsilon_{zz},\epsilon_{xy},\epsilon_{xz},\epsilon_{yz}$)=(
0.0042, 0.0043, -0.0085,     0,     0,     0
), \\($\epsilon_{A1g},\epsilon_{Eg.0},\epsilon_{Eg.1},\epsilon_{T2g.0},\epsilon_{T2g.1},\epsilon_{T2g.2}$)=( 
0.0001, -0.0001, -0.0104,     0,     0,     0
).\end{tabular}\\
Spin moments ($\mu_B$)& (
2.025, 0.000, 0.000
)\\
Energy (eV) and lattice (\AA) of und & -29.017, 2.7722
\end{tabular}
\end{ruledtabular}
\end{table*}

$\occmat^\uparrow=
\setstackgap{L}{1.1\baselineskip}
\fixTABwidth{T}
\parenMatrixstack{
0.0513 &     0 & 0.0388 &     0 &     0 & 0.0001 &     0 \\
    0 & 0.1263 &     0 &     0 &     0 &     0 &     0 \\
0.0388 &     0 & 0.9918 &     0 &     0 &     0 &     0 \\
    0 &     0 &     0 & 0.2029 &     0 & -0.3617 &     0 \\
    0 &     0 &     0 &     0 & 0.0342 &     0 & -0.0087 \\
0.0001 &     0 &     0 & -0.3617 &     0 & 0.8305 & 0.0001 \\
    0 &     0 &     0 &     0 & -0.0087 & 0.0001 & 0.0442
}$,

$\occmat^\downarrow=
\setstackgap{L}{1.1\baselineskip}
\fixTABwidth{T}
\parenMatrixstack{
0.0345 &     0 & 0.0062 &     0 &     0 &     0 &     0 \\
    0 & 0.1089 &     0 &     0 &     0 &     0 &     0 \\
0.0062 &     0 & 0.0262 &     0 &     0 &     0 &     0 \\
    0 &     0 &     0 & 0.0332 &     0 & 0.0016 &     0 \\
    0 &     0 &     0 &     0 & 0.0290 &     0 & -0.0062 \\
    0 &     0 &     0 & 0.0016 &     0 & 0.0216 &     0 \\
    0 &     0 &     0 &     0 & -0.0062 &     0 & 0.0343
}$.

\subsubsection{Properties of S$_8^1$}

\begin{table*}[h]%
\begin{ruledtabular}
\begin{tabular}{ll}
\textrm{Properties (unit)}&
\textrm{Values}\\
\colrule
Energy (eV) & -29.028 \\
Basis vectors (\AA)& $a=(
2.7169, 2.7172,     0
), b=(
-0.0590, 2.7762, 2.7753
), c=(
2.7759, -0.0589, 2.7753
)$\\
\begin{tabular}{@{}l@{}}Atom coordinates \\ (in basis vectors)\end{tabular} &
{\begin{tabular}{@{}l@{}}$U1=(
0.5000, 0.5000, 0.5000
)$ \\$O1=(
0.2483, 0.2517, 0.2517
)$, $O2=(
0.7517, 0.7483, 0.7483
)$\end{tabular}} \\ 
Strain & \begin{tabular}{@{}l@{}}($\epsilon_{xx},\epsilon_{yy},\epsilon_{zz},\epsilon_{xy},\epsilon_{xz},\epsilon_{yz}$)=(
0.0010, 0.0011, 0.0008, -0.0213,     0,     0
), \\($\epsilon_{A1g},\epsilon_{Eg.0},\epsilon_{Eg.1},\epsilon_{T2g.0},\epsilon_{T2g.1},\epsilon_{T2g.2}$)=( 
0.0016, -0.0001, -0.0002, -0.0213,     0,     0
).\end{tabular}\\
Spin moments ($\mu_B$)& (
2.026, 0.000, 0.000
)\\
Energy (eV) and lattice (\AA) of und & -29.006, 2.7732
\end{tabular}
\end{ruledtabular}
\end{table*}

$\occmat^\uparrow=
\setstackgap{L}{1.1\baselineskip}
\fixTABwidth{T}
\parenMatrixstack{
0.5214 &     0 & 0.0030 &     0 & 0.0052 &     0 & 0.4794 \\
    0 & 0.1308 &     0 & -0.0193 &     0 &     0 &     0 \\
0.0030 &     0 & 0.5189 &     0 & -0.4745 &     0 & -0.0038 \\
    0 & -0.0193 &     0 & 0.0500 &     0 &     0 &     0 \\
0.0052 &     0 & -0.4745 &     0 & 0.5222 &     0 & -0.0027 \\
    0 &     0 &     0 &     0 &     0 & 0.0306 &     0 \\
0.4794 &     0 & -0.0038 &     0 & -0.0027 &     0 & 0.5079
}$,

$\occmat^\downarrow=
\setstackgap{L}{1.1\baselineskip}
\fixTABwidth{T}
\parenMatrixstack{
0.0289 &     0 & 0.0048 &     0 & 0.0004 &     0 & -0.0004 \\
    0 & 0.1099 &     0 & -0.0132 &     0 &     0 &     0 \\
0.0048 &     0 & 0.0280 &     0 & 0.0083 &     0 & -0.0004 \\
    0 & -0.0132 &     0 & 0.0394 &     0 &     0 &     0 \\
0.0004 &     0 & 0.0083 &     0 & 0.0279 &     0 & -0.0048 \\
    0 &     0 &     0 &     0 &     0 & 0.0246 &     0 \\
-0.0004 &     0 & -0.0004 &     0 & -0.0048 &     0 & 0.0290
}$.

\clearpage

\subsection{\label{sec:sm1kafmnosoc}1\textbf{k} AFM without SOC}

Here, we document our results for \label{sec:sm1kafmnosoc}1\textbf{k} AFM without SOC, which, though not presented
in our paper, is included here for the sake of completeness.
Initializing from the occupation matrices of the 21 FM states (from S$_0$ to S$_5$ in Section~\ref{sec:smfmnosoc}), we obtained 19 1\textbf{k} AFM states using GGA+$U$ ($U=4$ eV) without SOC. Table~\ref{tab:tables2} summarizes the energy and strain distortion of all the resulting states, as compared with FM states. Note that no oxygen cage distortion is observed for these resulting 1\textbf{k} AFM states. The 1\textbf{k} AFM magnetic ordering only slightly perturbs the system: only small changes in energy differences, strain distortions, and occupation matrices $\occmat$ are observed when comparing the initial FM state and the resulting 1\textbf{k} AFM state. 
Therefore, the $S_n^m$ of the initial FM state is a reasonable proxy for the resulting 1\textbf{k} AFM state. Due to the symmetry in the 1\textbf{k} AFM structure, the spin moments and $\occmat$ of the first two U atoms, U1 and U2, are equal to those of U3 and U4, respectively. Therefore, we only present the properties for U1 and U2 in this section. 

\begin{table*}[h]
\caption{\label{tab:tables2}%
Results for 1\textbf{k} AFM states, calculated using GGA+$U$ ($U=4$ eV) without SOC, and compared with the FM states. 
The results are sorted by increasing relative energy with
respect to the fully relaxed crystal structure having occupation matrix $S_0$ and energy $E_{min}$.}
\begin{ruledtabular}
\begin{tabular}{l|ccc|ccc}
\multirow{3}{*}{States}&\multicolumn{3}{c}{FM}&
\multicolumn{3}{c}{1\textbf{k} AFM}\\ \cline{2-7}
 & $E_{und}-E_{min}$ & $E_{dis}-E_{min}$ & Strain distortion& $E_{und}-E_{min}$ & $E_{dis}-E_{min}$ & Strain distortion\\
 & (meV/UO\textsubscript{2})&(meV/UO\textsubscript{2})& ($\epsilon_{xx},\epsilon_{yy},\epsilon_{zz},\epsilon_{xy},\epsilon_{xz},\epsilon_{yz}$)$\times10^3$ & (meV/UO\textsubscript{2})&(meV/UO\textsubscript{2})& ($\epsilon_{xx},\epsilon_{yy},\epsilon_{zz},\epsilon_{xy},\epsilon_{xz},\epsilon_{yz}$)$\times10^3$ \\
\colrule
S$_0^1$ & \multirow{6}{*}{9.9} & \multirow{6}{*}{0} & (0, 0, 0, 17, 0, 0) & 9.5 & 0 & (0, 0, 1, 16, 0, 0) \\
S$_0^2$ & & & (0, 0, 0, 0, 0, 17) & 9.5 & 0 & (1, 0, 0, 0, 0, 16) \\
S$_0^3$ & & & (0, 0, 0, 0, 17, 0) & 11.1 & 2.4 & (0, 0, 0, 0, 16, 0) \\
S$_0^4$ & & & (0, 0, 0, -17, 0, 0) & 9.5 & 0 & (0, 0, 1, -16, 0, 0) \\
S$_0^5$ & & & (0, 0, 0, 0, 0, -17) & 9.5 & 0 & (1, 0, 0, 0, 0, -16) \\
S$_0^6$ & & & (0, 0, 0, 0, -17, 0) & 11.1 & 2.4 & (0, 0, 0, 0, -16, 0) \\
\hline
S$_1^1$ & \multirow{3}{*}{13.0} & \multirow{3}{*}{12.9}  & (0, 1, 0, 0, 0, 0) & 12.2 & - & Converged to S$_0^6$  \\
S$_1^2$ & & & (0, 0, 1, 0, 0, 0) & 15.1 & 15.0 & (-1, -1, 1, 0, 0, 0) \\
S$_1^3$ & & & (1, 0, 0, 0, 0, 0) & 15.1 & 15.0 & (-1, -1, 1, 0, 0, 0) \\
\hline
S$_2^1$ & \multirow{6}{*}{42.1} & \multirow{6}{*}{38.9}  & (-3, -5, 8, 0, 0, 0) & 43.5 & 40.3 & (-3, -5, 8, 0, 0, 0) \\
S$_2^2$ & & & (-5, 8, -3, 0, 0, 0) & 38.4 & 35.3 & (-5, 8, -3, 0, 0, 0)\\
S$_2^3$ & & & (8, -5, -3, 0, 0, 0) & 43.5 & 40.3 & (8, -5, -3, 0, 0, 0)\\
S$_2^4$ & & & (8, -3, -5, 0, 0, 0) & 33.1 & 30.0 & (8, -2, -5, 0, 0, 0)\\
S$_2^5$ & & & (-5, -3, 8, 0, 0, 0) & 33.1 & 30.0 & (-5, -2, 8, 0, 0, 0)\\
S$_2^6$ & & & (-3, 8, -5, 0, 0, 0) & 38.4 & 35.3 & (-3, 8, -5, 0, 0, 0)\\
\hline
S$_3^1$ & 43.8 & 40.4 & (-4, -4, 8, 2, 0, 0) & - & - & Not converged\\
\hline
S$_4^1$ & \multirow{2}{*}{52.7} & \multirow{2}{*}{49.2} & (4, -8, 4, 0, 0, 0)& 54.5 & 51.4 & (4, -8, 4, 0, 0, 0) \\
S$_4^2$ & & & (4, 4, -8, 0, 0, 0) & 43.9 & 40.6 & (3, 5, -8, 0, 0, 0) \\
\hline
S$_5^1$ & \multirow{3}{*}{100} & \multirow{3}{*}{99.5} & (-3, 2, 2, 0, 0, 0) & 97.4 & 96.9 & (-4, 2, 2, 0, 0, 0)\\
S$_5^2$ & & & (2, 2, -3, 0, 0, 0) & 97.4 & 96.9 & (2, 2, -4, 0, 0, 0) \\
S$_5^3$ & & & (2, -3, 2, 0, 0, 0) & 104.9 & 104.5 & (2, -3, 2, 0, 0, 0) 
\\

\end{tabular}
\end{ruledtabular}
\end{table*}
\clearpage

\subsubsection{Properties of S$_0^1$ (1\textbf{k} AFM without SOC)}

\begin{table*}[!h]%
\begin{ruledtabular}
\begin{tabular}{ll}
\textrm{Properties (unit)}&
\textrm{Values}\\

\colrule
Energy (eV) &  -116.561 \\
Basis vectors (\AA)& $a=(
5.5456,  0.0872,  0
)$, $b=(
0.0872,  5.5432,  0
)$, $c=(
0,  0,  5.5515
)$\\
\begin{tabular}{@{}l@{}}Atom coordinates \\ (in basis vectors)\end{tabular} & {\begin{tabular}{@{}l@{}}U1:$(
0.5000,  0,  0 )$, U2:$(
0.5000,  0.5000,  0.5000 )$, \\U3:$(
0,  0,  0.5000 )$, U4:$(
0,  0.5000,  0 )$, \\O1:$(
0.2500,  0.7500,  0.7498 )$, O2:$(
0.2500,  0.2500,  0.7502 )$, \\O3:$(
0.2500,  0.2500,  0.2498 )$, O4:$(
0.2500,  0.7500,  0.2502 )$, \\O5:$(
0.7500,  0.7500,  0.2498 )$, O6:$(
0.7500,  0.2500,  0.2502 )$, \\O7:$(
0.7500,  0.2500,  0.7498 )$, O8:$(
0.7500,  0.7500,  0.7502 )$\end{tabular}} \\ 
Strain & \begin{tabular}{@{}l@{}}$(\epsilon_{xx},\epsilon_{yy},\epsilon_{zz},\epsilon_{xy},\epsilon_{xz},\epsilon_{yz})=(
0.0001,  -0.0004,  0.0011,  0.0157,  0,  0 )$, \\$(\epsilon_{A1g},\epsilon_{Eg.0},\epsilon_{Eg.1},\epsilon_{T2g.0},\epsilon_{T2g.1},\epsilon_{T2g.2})=(
0.0005,  0.0003,  0.0010,  0.0157,  0,  0 )$.\end{tabular}\\
Spin moments ($\mu_B$)& U1:$(
2.020,  0,  0
)$, U2:$(
-2.020,  0,  0
)$\\
Energy (eV) and lattice (\AA) of und & 
-116.523, 5.5453
\end{tabular}
\end{ruledtabular}
\end{table*}

$\occmat(\textrm{U}1)$:

$\occmat^{\uparrow}=
\setstackgap{L}{1.1\baselineskip}
\fixTABwidth{T}
\parenMatrixstack{
0.31&  0&  -0.25&  0&  0.10&  0&  0.34 \\
0&  0.13&  0&  0.01&  0&  0&  0 \\
-0.25&  0&  0.60&  0&  0.34&  0&  -0.20 \\
0&  0.01&  0&  0.04&  0&  0&  0 \\
0.10&  0&  0.34&  0&  0.67&  0&  0.28 \\
0&  0&  0&  0&  0&  0.03&  0 \\
0.34&  0&  -0.20&  0&  0.28&  0&  0.50
}$,
$\occmat^{\downarrow}=
\setstackgap{L}{1.1\baselineskip}
\fixTABwidth{T}
\parenMatrixstack{
0.03&  0&  0.01&  0&  0&  0&  0 \\
0&  0.11&  0&  0.01&  0&  0&  0 \\
0.01&  0&  0.03&  0&  0&  0&  0 \\
0&  0.01&  0&  0.04&  0&  0&  0 \\
0&  0&  0&  0&  0.03&  0&  -0.01 \\
0&  0&  0&  0&  0&  0.02&  0 \\
0&  0&  0&  0&  -0.01&  0&  0.03
}$

$\occmat(\textrm{U}2)$:

$\occmat^{\uparrow}=
\setstackgap{L}{1.1\baselineskip}
\fixTABwidth{T}
\parenMatrixstack{
0.03&  0&  0.01&  0&  0&  0&  0 \\
0&  0.11&  0&  0.01&  0&  0&  0 \\
0.01&  0&  0.03&  0&  0&  0&  0 \\
0&  0.01&  0&  0.04&  0&  0&  0 \\
0&  0&  0&  0&  0.03&  0&  -0.01 \\
0&  0&  0&  0&  0&  0.02&  0 \\
0&  0&  0&  0&  -0.01&  0&  0.03
}$,
$\occmat^{\downarrow}=
\setstackgap{L}{1.1\baselineskip}
\fixTABwidth{T}
\parenMatrixstack{
0.31&  0&  -0.25&  0&  0.10&  0&  0.34 \\
0&  0.13&  0&  0.01&  0&  0&  0 \\
-0.25&  0&  0.60&  0&  0.34&  0&  -0.20 \\
0&  0.01&  0&  0.04&  0&  0&  0 \\
0.10&  0&  0.34&  0&  0.67&  0&  0.28 \\
0&  0&  0&  0&  0&  0.03&  0 \\
0.34&  0&  -0.20&  0&  0.28&  0&  0.50
}$

\subsubsection{Properties of S$_0^2$ (1\textbf{k} AFM without SOC)}

\begin{table*}[!h]%
\begin{ruledtabular}
\begin{tabular}{ll}
\textrm{Properties (unit)}&
\textrm{Values}\\

\colrule
Energy (eV) &  -116.561 \\
Basis vectors (\AA)& $a=(
5.5515,  0,  0
)$, $b=(
0,  5.5432,  0.0872
)$, $c=(
0,  0.0872,  5.5456
)$\\
\begin{tabular}{@{}l@{}}Atom coordinates \\ (in basis vectors)\end{tabular} & {\begin{tabular}{@{}l@{}}U1:$(
0.5000,  0,  0 )$, U2:$(
0.5000,  0.5000,  0.5000 )$, \\U3:$(
0,  0,  0.5000 )$, U4:$(
0,  0.5000,  0 )$, \\O1:$(
0.2498,  0.7500,  0.7500 )$, O2:$(
0.2502,  0.2500,  0.7500 )$, \\O3:$(
0.2498,  0.2500,  0.2500 )$, O4:$(
0.2502,  0.7500,  0.2500 )$, \\O5:$(
0.7498,  0.7500,  0.2500 )$, O6:$(
0.7502,  0.2500,  0.2500 )$, \\O7:$(
0.7498,  0.2500,  0.7500 )$, O8:$(
0.7502,  0.7500,  0.7500 )$\end{tabular}} \\ 
Strain & \begin{tabular}{@{}l@{}}$(\epsilon_{xx},\epsilon_{yy},\epsilon_{zz},\epsilon_{xy},\epsilon_{xz},\epsilon_{yz})=(
0.0011,  -0.0004,  0.0001,  0,  0,  0.0157 )$, \\$(\epsilon_{A1g},\epsilon_{Eg.0},\epsilon_{Eg.1},\epsilon_{T2g.0},\epsilon_{T2g.1},\epsilon_{T2g.2})=(
0.0005,  0.0011,  -0.0003,  0,  0.0157,  0 )$.\end{tabular}\\
Spin moments ($\mu_B$)& U1:$(
2.020,  0,  0
)$, U2:$(
-2.020,  0,  0
)$\\
Energy (eV) and lattice (\AA) of und & 
-116.523, 5.5453
\end{tabular}
\end{ruledtabular}
\end{table*}

$\occmat(\textrm{U}1)$:

$\occmat^{\uparrow}=
\setstackgap{L}{1.1\baselineskip}
\fixTABwidth{T}
\parenMatrixstack{
0.47&  0&  -0.29&  -0.30&  0&  0.21&  0 \\
0&  0.13&  0&  0&  -0.01&  0&  0.01 \\
-0.29&  0&  0.44&  0.29&  0&  0.23&  0 \\
-0.30&  0&  0.29&  0.30&  0&  -0.01&  0 \\
0&  -0.01&  0&  0&  0.04&  0&  -0.01 \\
0.21&  0&  0.23&  -0.01&  0&  0.87&  0 \\
0&  0.01&  0&  0&  -0.01&  0&  0.04
}$,
$\occmat^{\downarrow}=
\setstackgap{L}{1.1\baselineskip}
\fixTABwidth{T}
\parenMatrixstack{
0.03&  0&  0.01&  0&  0&  0&  0 \\
0&  0.11&  0&  0&  0&  0&  0.01 \\
0.01&  0&  0.03&  0&  0&  0&  0 \\
0&  0&  0&  0.04&  0&  0&  0 \\
0&  0&  0&  0&  0.03&  0&  -0.01 \\
0&  0&  0&  0&  0&  0.02&  0 \\
0&  0.01&  0&  0&  -0.01&  0&  0.03
}$

$\occmat(\textrm{U}2)$:

$\occmat^{\uparrow}=
\setstackgap{L}{1.1\baselineskip}
\fixTABwidth{T}
\parenMatrixstack{
0.03&  0&  0.01&  0&  0&  0&  0 \\
0&  0.11&  0&  0&  0&  0&  0.01 \\
0.01&  0&  0.03&  0&  0&  0&  0 \\
0&  0&  0&  0.04&  0&  0&  0 \\
0&  0&  0&  0&  0.03&  0&  -0.01 \\
0&  0&  0&  0&  0&  0.02&  0 \\
0&  0.01&  0&  0&  -0.01&  0&  0.03
}$,
$\occmat^{\downarrow}=
\setstackgap{L}{1.1\baselineskip}
\fixTABwidth{T}
\parenMatrixstack{
0.47&  0&  -0.29&  -0.30&  0&  0.21&  0 \\
0&  0.13&  0&  0&  -0.01&  0&  0.01 \\
-0.29&  0&  0.44&  0.29&  0&  0.23&  0 \\
-0.30&  0&  0.29&  0.30&  0&  -0.01&  0 \\
0&  -0.01&  0&  0&  0.04&  0&  -0.01 \\
0.21&  0&  0.23&  -0.01&  0&  0.87&  0 \\
0&  0.01&  0&  0&  -0.01&  0&  0.04
}$

\subsubsection{Properties of S$_0^3$ (1\textbf{k} AFM without SOC)}

\begin{table*}[!h]%
\begin{ruledtabular}
\begin{tabular}{ll}
\textrm{Properties (unit)}&
\textrm{Values}\\

\colrule
Energy (eV) &  -116.552 \\
Basis vectors (\AA)& $a=(
5.5473,  0,  0.0905
)$, $b=(
0,  5.5476,  0
)$, $c=(
0.0905,  0,  5.5473
)$\\
\begin{tabular}{@{}l@{}}Atom coordinates \\ (in basis vectors)\end{tabular} & {\begin{tabular}{@{}l@{}}U1:$(
0.5000,  0,  0 )$, U2:$(
0.5000,  0.5000,  0.5000 )$, \\U3:$(
0,  0,  0.5000 )$, U4:$(
0,  0.5000,  0 )$, \\O1:$(
0.2500,  0.7501,  0.7500 )$, O2:$(
0.2500,  0.2499,  0.7500 )$, \\O3:$(
0.2500,  0.2501,  0.2500 )$, O4:$(
0.2500,  0.7499,  0.2500 )$, \\O5:$(
0.7500,  0.7501,  0.2500 )$, O6:$(
0.7500,  0.2499,  0.2500 )$, \\O7:$(
0.7500,  0.2501,  0.7500 )$, O8:$(
0.7500,  0.7499,  0.7500 )$\end{tabular}} \\ 
Strain & \begin{tabular}{@{}l@{}}$(\epsilon_{xx},\epsilon_{yy},\epsilon_{zz},\epsilon_{xy},\epsilon_{xz},\epsilon_{yz})=(
0.0002,  0.0003,  0.0002,  0,  0.0163,  0 )$, \\$(\epsilon_{A1g},\epsilon_{Eg.0},\epsilon_{Eg.1},\epsilon_{T2g.0},\epsilon_{T2g.1},\epsilon_{T2g.2})=(
0.0005,  0,  0,  0,  0,  0.0163 )$.\end{tabular}\\
Spin moments ($\mu_B$)& U1:$(
2.022,  0,  0
)$, U2:$(
-2.022,  0,  0
)$\\
Energy (eV) and lattice (\AA) of und & 
-116.517, 5.5459
\end{tabular}
\end{ruledtabular}
\end{table*}

$\occmat(\textrm{U}1)$:

$\occmat^{\uparrow}=
\setstackgap{L}{1.1\baselineskip}
\fixTABwidth{T}
\parenMatrixstack{
0.04&  -0.01&  0.01&  0&  0&  0&  0 \\
-0.01&  0.12&  -0.01&  0&  0&  0&  0 \\
0.01&  -0.01&  0.03&  0&  0&  0&  0 \\
0&  0&  0&  0.21&  0.25&  0.05&  0.25 \\
0&  0&  0&  0.25&  0.55&  -0.26&  0.32 \\
0&  0&  0&  0.05&  -0.26&  0.83&  0.25 \\
0&  0&  0&  0.25&  0.32&  0.25&  0.49
}$,
$\occmat^{\downarrow}=
\setstackgap{L}{1.1\baselineskip}
\fixTABwidth{T}
\parenMatrixstack{
0.03&  -0.01&  0.01&  0&  0&  0&  0 \\
-0.01&  0.11&  0&  0&  0&  0&  0 \\
0.01&  0&  0.03&  0&  0&  0&  0 \\
0&  0&  0&  0.04&  0&  0&  0 \\
0&  0&  0&  0&  0.03&  0&  -0.01 \\
0&  0&  0&  0&  0&  0.02&  0 \\
0&  0&  0&  0&  -0.01&  0&  0.03
}$

$\occmat(\textrm{U}2)$:

$\occmat^{\uparrow}=
\setstackgap{L}{1.1\baselineskip}
\fixTABwidth{T}
\parenMatrixstack{
0.03&  -0.01&  0.01&  0&  0&  0&  0 \\
-0.01&  0.11&  0&  0&  0&  0&  0 \\
0.01&  0&  0.03&  0&  0&  0&  0 \\
0&  0&  0&  0.04&  0&  0&  0 \\
0&  0&  0&  0&  0.03&  0&  -0.01 \\
0&  0&  0&  0&  0&  0.02&  0 \\
0&  0&  0&  0&  -0.01&  0&  0.03
}$,
$\occmat^{\downarrow}=
\setstackgap{L}{1.1\baselineskip}
\fixTABwidth{T}
\parenMatrixstack{
0.04&  -0.01&  0.01&  0&  0&  0&  0 \\
-0.01&  0.12&  -0.01&  0&  0&  0&  0 \\
0.01&  -0.01&  0.03&  0&  0&  0&  0 \\
0&  0&  0&  0.21&  0.25&  0.05&  0.25 \\
0&  0&  0&  0.25&  0.55&  -0.26&  0.32 \\
0&  0&  0&  0.05&  -0.26&  0.83&  0.25 \\
0&  0&  0&  0.25&  0.32&  0.25&  0.49
}$
\clearpage
\subsubsection{Properties of S$_0^4$ (1\textbf{k} AFM without SOC)}

\begin{table*}[!h]%
\begin{ruledtabular}
\begin{tabular}{ll}
\textrm{Properties (unit)}&
\textrm{Values}\\

\colrule
Energy (eV) &  -116.561 \\
Basis vectors (\AA)& $a=(
5.5456,  -0.0872,  0
)$, $b=(
-0.0872,  5.5432,  0
)$, $c=(
0,  0,  5.5515
)$\\
\begin{tabular}{@{}l@{}}Atom coordinates \\ (in basis vectors)\end{tabular} & {\begin{tabular}{@{}l@{}}U1:$(
0.5000,  0,  0 )$, U2:$(
0.5000,  0.5000,  0.5000 )$, \\U3:$(
0,  0,  0.5000 )$, U4:$(
0,  0.5000,  0 )$, \\O1:$(
0.2500,  0.7500,  0.7502 )$, O2:$(
0.2500,  0.2500,  0.7498 )$, \\O3:$(
0.2500,  0.2500,  0.2502 )$, O4:$(
0.2500,  0.7500,  0.2498 )$, \\O5:$(
0.7500,  0.7500,  0.2502 )$, O6:$(
0.7500,  0.2500,  0.2498 )$, \\O7:$(
0.7500,  0.2500,  0.7502 )$, O8:$(
0.7500,  0.7500,  0.7498 )$\end{tabular}} \\ 
Strain & \begin{tabular}{@{}l@{}}$(\epsilon_{xx},\epsilon_{yy},\epsilon_{zz},\epsilon_{xy},\epsilon_{xz},\epsilon_{yz})=(
0.0001,  -0.0004,  0.0011,  -0.0157,  0,  0 )$, \\$(\epsilon_{A1g},\epsilon_{Eg.0},\epsilon_{Eg.1},\epsilon_{T2g.0},\epsilon_{T2g.1},\epsilon_{T2g.2})=(
0.0005,  0.0003,  0.0010,  -0.0157,  0,  0 )$.\end{tabular}\\
Spin moments ($\mu_B$)& U1:$(
2.020,  0,  0
)$, U2:$(
-2.020,  0,  0
)$\\
Energy (eV) and lattice (\AA) of und & 
-116.523, 5.5453
\end{tabular}
\end{ruledtabular}
\end{table*}

$\occmat(\textrm{U}1)$:

$\occmat^{\uparrow}=
\setstackgap{L}{1.1\baselineskip}
\fixTABwidth{T}
\parenMatrixstack{
0.30&  0&  -0.25&  0&  -0.10&  0&  -0.33 \\
0&  0.13&  0&  -0.01&  0&  0&  0 \\
-0.25&  0&  0.60&  0&  -0.34&  0&  0.20 \\
0&  -0.01&  0&  0.04&  0&  0&  0 \\
-0.10&  0&  -0.34&  0&  0.67&  0&  0.28 \\
0&  0&  0&  0&  0&  0.03&  0 \\
-0.33&  0&  0.20&  0&  0.28&  0&  0.50
}$,
$\occmat^{\downarrow}=
\setstackgap{L}{1.1\baselineskip}
\fixTABwidth{T}
\parenMatrixstack{
0.03&  0&  0.01&  0&  0&  0&  0 \\
0&  0.11&  0&  -0.01&  0&  0&  0 \\
0.01&  0&  0.03&  0&  0&  0&  0 \\
0&  -0.01&  0&  0.04&  0&  0&  0 \\
0&  0&  0&  0&  0.03&  0&  -0.01 \\
0&  0&  0&  0&  0&  0.02&  0 \\
0&  0&  0&  0&  -0.01&  0&  0.03
}$

$\occmat(\textrm{U}2)$:

$\occmat^{\uparrow}=
\setstackgap{L}{1.1\baselineskip}
\fixTABwidth{T}
\parenMatrixstack{
0.03&  0&  0.01&  0&  0&  0&  0 \\
0&  0.11&  0&  -0.01&  0&  0&  0 \\
0.01&  0&  0.03&  0&  0&  0&  0 \\
0&  -0.01&  0&  0.04&  0&  0&  0 \\
0&  0&  0&  0&  0.03&  0&  -0.01 \\
0&  0&  0&  0&  0&  0.02&  0 \\
0&  0&  0&  0&  -0.01&  0&  0.03
}$,
$\occmat^{\downarrow}=
\setstackgap{L}{1.1\baselineskip}
\fixTABwidth{T}
\parenMatrixstack{
0.30&  0&  -0.25&  0&  -0.10&  0&  -0.33 \\
0&  0.13&  0&  -0.01&  0&  0&  0 \\
-0.25&  0&  0.60&  0&  -0.34&  0&  0.20 \\
0&  -0.01&  0&  0.04&  0&  0&  0 \\
-0.10&  0&  -0.34&  0&  0.67&  0&  0.28 \\
0&  0&  0&  0&  0&  0.03&  0 \\
-0.33&  0&  0.20&  0&  0.28&  0&  0.50
}$

\subsubsection{Properties of S$_0^5$ (1\textbf{k} AFM without SOC)}

\begin{table*}[!h]%
\begin{ruledtabular}
\begin{tabular}{ll}
\textrm{Properties (unit)}&
\textrm{Values}\\

\colrule
Energy (eV) &  -116.561 \\
Basis vectors (\AA)& $a=(
5.5515,  0,  0
)$, $b=(
0,  5.5432,  -0.0872
)$, $c=(
0,  -0.0873,  5.5456
)$\\
\begin{tabular}{@{}l@{}}Atom coordinates \\ (in basis vectors)\end{tabular} & {\begin{tabular}{@{}l@{}}U1:$(
0.5000,  0,  0 )$, U2:$(
0.5000,  0.5000,  0.5000 )$, \\U3:$(
0,  0,  0.5000 )$, U4:$(
0,  0.5000,  0 )$, \\O1:$(
0.2502,  0.7500,  0.7500 )$, O2:$(
0.2498,  0.2500,  0.7500 )$, \\O3:$(
0.2502,  0.2500,  0.2500 )$, O4:$(
0.2498,  0.7500,  0.2500 )$, \\O5:$(
0.7502,  0.7500,  0.2500 )$, O6:$(
0.7498,  0.2500,  0.2500 )$, \\O7:$(
0.7502,  0.2500,  0.7500 )$, O8:$(
0.7498,  0.7500,  0.7500 )$\end{tabular}} \\ 
Strain & \begin{tabular}{@{}l@{}}$(\epsilon_{xx},\epsilon_{yy},\epsilon_{zz},\epsilon_{xy},\epsilon_{xz},\epsilon_{yz})=(
0.0011,  -0.0004,  0,  0,  0,  -0.0157 )$, \\$(\epsilon_{A1g},\epsilon_{Eg.0},\epsilon_{Eg.1},\epsilon_{T2g.0},\epsilon_{T2g.1},\epsilon_{T2g.2})=(
0.0005,  0.0011,  -0.0003,  0,  -0.0157,  0 )$.\end{tabular}\\
Spin moments ($\mu_B$)& U1:$(
2.020,  0,  0
)$, U2:$(
-2.020,  0,  0
)$\\
Energy (eV) and lattice (\AA) of und & 
-116.523, 5.5453
\end{tabular}
\end{ruledtabular}
\end{table*}

$\occmat(\textrm{U}1)$:

$\occmat^{\uparrow}=
\setstackgap{L}{1.1\baselineskip}
\fixTABwidth{T}
\parenMatrixstack{
0.47&  0&  -0.29&  0.30&  0&  -0.21&  0 \\
0&  0.13&  0&  0&  0.01&  0&  -0.01 \\
-0.29&  0&  0.45&  -0.29&  0&  -0.24&  0 \\
0.30&  0&  -0.29&  0.30&  0&  -0.01&  0 \\
0&  0.01&  0&  0&  0.04&  0&  -0.01 \\
-0.21&  0&  -0.24&  -0.01&  0&  0.87&  0 \\
0&  -0.01&  0&  0&  -0.01&  0&  0.04
}$,
$\occmat^{\downarrow}=
\setstackgap{L}{1.1\baselineskip}
\fixTABwidth{T}
\parenMatrixstack{
0.03&  0&  0.01&  0&  0&  0&  0 \\
0&  0.11&  0&  0&  0&  0&  -0.01 \\
0.01&  0&  0.03&  0&  0&  0&  0 \\
0&  0&  0&  0.04&  0&  0&  0 \\
0&  0&  0&  0&  0.03&  0&  -0.01 \\
0&  0&  0&  0&  0&  0.02&  0 \\
0&  -0.01&  0&  0&  -0.01&  0&  0.03
}$

$\occmat(\textrm{U}2)$:

$\occmat^{\uparrow}=
\setstackgap{L}{1.1\baselineskip}
\fixTABwidth{T}
\parenMatrixstack{
0.03&  0&  0.01&  0&  0&  0&  0 \\
0&  0.11&  0&  0&  0&  0&  -0.01 \\
0.01&  0&  0.03&  0&  0&  0&  0 \\
0&  0&  0&  0.04&  0&  0&  0 \\
0&  0&  0&  0&  0.03&  0&  -0.01 \\
0&  0&  0&  0&  0&  0.02&  0 \\
0&  -0.01&  0&  0&  -0.01&  0&  0.03
}$,
$\occmat^{\downarrow}=
\setstackgap{L}{1.1\baselineskip}
\fixTABwidth{T}
\parenMatrixstack{
0.47&  0&  -0.29&  0.30&  0&  -0.21&  0 \\
0&  0.13&  0&  0&  0.01&  0&  -0.01 \\
-0.29&  0&  0.45&  -0.29&  0&  -0.24&  0 \\
0.30&  0&  -0.29&  0.30&  0&  -0.01&  0 \\
0&  0.01&  0&  0&  0.04&  0&  -0.01 \\
-0.21&  0&  -0.24&  -0.01&  0&  0.87&  0 \\
0&  -0.01&  0&  0&  -0.01&  0&  0.04
}$

\subsubsection{Properties of S$_0^6$ (1\textbf{k} AFM without SOC)}

\begin{table*}[!h]%
\begin{ruledtabular}
\begin{tabular}{ll}
\textrm{Properties (unit)}&
\textrm{Values}\\

\colrule
Energy (eV) &  -116.552 \\
Basis vectors (\AA)& $a=(
5.5473,  0,  -0.0905
)$, $b=(
0,  5.5476,  0
)$, $c=(
-0.0905,  0,  5.5473
)$\\
\begin{tabular}{@{}l@{}}Atom coordinates \\ (in basis vectors)\end{tabular} & {\begin{tabular}{@{}l@{}}U1:$(
0.5000,  0,  0 )$, U2:$(
0.5000,  0.5000,  0.5000 )$, \\U3:$(
0,  0,  0.5000 )$, U4:$(
0,  0.5000,  0 )$, \\O1:$(
0.2500,  0.7499,  0.7500 )$, O2:$(
0.2500,  0.2501,  0.7500 )$, \\O3:$(
0.2500,  0.2499,  0.2500 )$, O4:$(
0.2500,  0.7501,  0.2500 )$, \\O5:$(
0.7500,  0.7499,  0.2500 )$, O6:$(
0.7500,  0.2501,  0.2500 )$, \\O7:$(
0.7500,  0.2499,  0.7500 )$, O8:$(
0.7500,  0.7501,  0.7500 )$\end{tabular}} \\ 
Strain & \begin{tabular}{@{}l@{}}$(\epsilon_{xx},\epsilon_{yy},\epsilon_{zz},\epsilon_{xy},\epsilon_{xz},\epsilon_{yz})=(
0.0002,  0.0003,  0.0002,  0,  -0.0163,  0 )$, \\$(\epsilon_{A1g},\epsilon_{Eg.0},\epsilon_{Eg.1},\epsilon_{T2g.0},\epsilon_{T2g.1},\epsilon_{T2g.2})=(
0.0005,  0,  0,  0,  0,  -0.0163 )$.\end{tabular}\\
Spin moments ($\mu_B$)& U1:$(
2.022,  0,  0
)$, U2:$(
-2.022,  0,  0
)$\\
Energy (eV) and lattice (\AA) of und & 
-116.517, 5.5459
\end{tabular}
\end{ruledtabular}
\end{table*}

$\occmat(\textrm{U}1)$:

$\occmat^{\uparrow}=
\setstackgap{L}{1.1\baselineskip}
\fixTABwidth{T}
\parenMatrixstack{
0.04&  0.01&  0.01&  0&  0&  0&  0 \\
0.01&  0.12&  0.01&  0&  0&  0&  0 \\
0.01&  0.01&  0.03&  0&  0&  0&  0 \\
0&  0&  0&  0.21&  -0.25&  0.05&  -0.25 \\
0&  0&  0&  -0.25&  0.55&  0.26&  0.32 \\
0&  0&  0&  0.05&  0.26&  0.83&  -0.25 \\
0&  0&  0&  -0.25&  0.32&  -0.25&  0.49
}$,
$\occmat^{\downarrow}=
\setstackgap{L}{1.1\baselineskip}
\fixTABwidth{T}
\parenMatrixstack{
0.03&  0.01&  0.01&  0&  0&  0&  0 \\
0.01&  0.11&  0&  0&  0&  0&  0 \\
0.01&  0&  0.03&  0&  0&  0&  0 \\
0&  0&  0&  0.04&  0&  0&  0 \\
0&  0&  0&  0&  0.03&  0&  -0.01 \\
0&  0&  0&  0&  0&  0.02&  0 \\
0&  0&  0&  0&  -0.01&  0&  0.03
}$

$\occmat(\textrm{U}2)$:

$\occmat^{\uparrow}=
\setstackgap{L}{1.1\baselineskip}
\fixTABwidth{T}
\parenMatrixstack{
0.03&  0.01&  0.01&  0&  0&  0&  0 \\
0.01&  0.11&  0&  0&  0&  0&  0 \\
0.01&  0&  0.03&  0&  0&  0&  0 \\
0&  0&  0&  0.04&  0&  0&  0 \\
0&  0&  0&  0&  0.03&  0&  -0.01 \\
0&  0&  0&  0&  0&  0.02&  0 \\
0&  0&  0&  0&  -0.01&  0&  0.03
}$,
$\occmat^{\downarrow}=
\setstackgap{L}{1.1\baselineskip}
\fixTABwidth{T}
\parenMatrixstack{
0.04&  0.01&  0.01&  0&  0&  0&  0 \\
0.01&  0.12&  0.01&  0&  0&  0&  0 \\
0.01&  0.01&  0.03&  0&  0&  0&  0 \\
0&  0&  0&  0.21&  -0.25&  0.05&  -0.25 \\
0&  0&  0&  -0.25&  0.55&  0.26&  0.32 \\
0&  0&  0&  0.05&  0.26&  0.83&  -0.25 \\
0&  0&  0&  -0.25&  0.32&  -0.25&  0.49
}$
\clearpage
\subsubsection{Properties of S$_1^2$ (1\textbf{k} AFM without SOC)}

\begin{table*}[!h]%
\begin{ruledtabular}
\begin{tabular}{ll}
\textrm{Properties (unit)}&
\textrm{Values}\\

\colrule
Energy (eV) &  -116.501 \\
Basis vectors (\AA)& $a=(
5.5428,  0,  0
)$, $b=(
0,  5.5420,  0
)$, $c=(
0,  0,  5.5535
)$\\
\begin{tabular}{@{}l@{}}Atom coordinates \\ (in basis vectors)\end{tabular} & {\begin{tabular}{@{}l@{}}U1:$(
0.5000,  0,  0 )$, U2:$(
0.5000,  0.5000,  0.5000 )$, \\U3:$(
0,  0,  0.5000 )$, U4:$(
0,  0.5000,  0 )$, \\O1:$(
0.2500,  0.7500,  0.7500 )$, O2:$(
0.2500,  0.2500,  0.7500 )$, \\O3:$(
0.2500,  0.2500,  0.2500 )$, O4:$(
0.2500,  0.7500,  0.2500 )$, \\O5:$(
0.7500,  0.7500,  0.2500 )$, O6:$(
0.7500,  0.2500,  0.2500 )$, \\O7:$(
0.7500,  0.2500,  0.7500 )$, O8:$(
0.7500,  0.7500,  0.7500 )$\end{tabular}} \\ 
Strain & \begin{tabular}{@{}l@{}}$(\epsilon_{xx},\epsilon_{yy},\epsilon_{zz},\epsilon_{xy},\epsilon_{xz},\epsilon_{yz})=(
-0.0006,  -0.0008,  0.0013,  0,  0,  0 )$, \\$(\epsilon_{A1g},\epsilon_{Eg.0},\epsilon_{Eg.1},\epsilon_{T2g.0},\epsilon_{T2g.1},\epsilon_{T2g.2})=(
0,  0.0001,  0.0016,  0,  0,  0 )$.\end{tabular}\\
Spin moments ($\mu_B$)& U1:$(
2.017,  0,  0
)$, U2:$(
-2.017,  0,  0
)$\\
Energy (eV) and lattice (\AA) of und & 
-116.501, 5.5462
\end{tabular}
\end{ruledtabular}
\end{table*}

$\occmat(\textrm{U}1)$:

$\occmat^{\uparrow}=
\setstackgap{L}{1.1\baselineskip}
\fixTABwidth{T}
\parenMatrixstack{
0.35&  0&  -0.44&  0&  0&  0&  0 \\
0&  0.12&  0&  0&  0&  0&  0 \\
-0.44&  0&  0.69&  0&  0&  0&  0 \\
0&  0&  0&  0.04&  0&  0&  0 \\
0&  0&  0&  0&  0.69&  0&  0.44 \\
0&  0&  0&  0&  0&  0.03&  0 \\
0&  0&  0&  0&  0.44&  0&  0.35
}$,
$\occmat^{\downarrow}=
\setstackgap{L}{1.1\baselineskip}
\fixTABwidth{T}
\parenMatrixstack{
0.04&  0&  0.01&  0&  0&  0&  0 \\
0&  0.11&  0&  0&  0&  0&  0 \\
0.01&  0&  0.03&  0&  0&  0&  0 \\
0&  0&  0&  0.04&  0&  0&  0 \\
0&  0&  0&  0&  0.03&  0&  -0.01 \\
0&  0&  0&  0&  0&  0.02&  0 \\
0&  0&  0&  0&  -0.01&  0&  0.04
}$

$\occmat(\textrm{U}2)$:

$\occmat^{\uparrow}=
\setstackgap{L}{1.1\baselineskip}
\fixTABwidth{T}
\parenMatrixstack{
0.04&  0&  0.01&  0&  0&  0&  0 \\
0&  0.11&  0&  0&  0&  0&  0 \\
0.01&  0&  0.03&  0&  0&  0&  0 \\
0&  0&  0&  0.04&  0&  0&  0 \\
0&  0&  0&  0&  0.03&  0&  -0.01 \\
0&  0&  0&  0&  0&  0.02&  0 \\
0&  0&  0&  0&  -0.01&  0&  0.04
}$,
$\occmat^{\downarrow}=
\setstackgap{L}{1.1\baselineskip}
\fixTABwidth{T}
\parenMatrixstack{
0.35&  0&  -0.44&  0&  0&  0&  0 \\
0&  0.12&  0&  0&  0&  0&  0 \\
-0.44&  0&  0.69&  0&  0&  0&  0 \\
0&  0&  0&  0.04&  0&  0&  0 \\
0&  0&  0&  0&  0.69&  0&  0.44 \\
0&  0&  0&  0&  0&  0.03&  0 \\
0&  0&  0&  0&  0.44&  0&  0.35
}$

\subsubsection{Properties of S$_1^3$ (1\textbf{k} AFM without SOC)}

\begin{table*}[!h]%
\begin{ruledtabular}
\begin{tabular}{ll}
\textrm{Properties (unit)}&
\textrm{Values}\\

\colrule
Energy (eV) &  -116.501 \\
Basis vectors (\AA)& $a=(
5.5535,  0,  0
)$, $b=(
0,  5.5420,  0
)$, $c=(
0,  0,  5.5428
)$\\
\begin{tabular}{@{}l@{}}Atom coordinates \\ (in basis vectors)\end{tabular} & {\begin{tabular}{@{}l@{}}U1:$(
0.5000,  0,  0 )$, U2:$(
0.5000,  0.5000,  0.5000 )$, \\U3:$(
0,  0,  0.5000 )$, U4:$(
0,  0.5000,  0 )$, \\O1:$(
0.2500,  0.7500,  0.7500 )$, O2:$(
0.2500,  0.2500,  0.7500 )$, \\O3:$(
0.2500,  0.2500,  0.2500 )$, O4:$(
0.2500,  0.7500,  0.2500 )$, \\O5:$(
0.7500,  0.7500,  0.2500 )$, O6:$(
0.7500,  0.2500,  0.2500 )$, \\O7:$(
0.7500,  0.2500,  0.7500 )$, O8:$(
0.7500,  0.7500,  0.7500 )$\end{tabular}} \\ 
Strain & \begin{tabular}{@{}l@{}}$(\epsilon_{xx},\epsilon_{yy},\epsilon_{zz},\epsilon_{xy},\epsilon_{xz},\epsilon_{yz})=(
0.0013,  -0.0008,  -0.0006,  0,  0,  0 )$, \\$(\epsilon_{A1g},\epsilon_{Eg.0},\epsilon_{Eg.1},\epsilon_{T2g.0},\epsilon_{T2g.1},\epsilon_{T2g.2})=(
0,  0.0015,  -0.0007,  0,  0,  0 )$.\end{tabular}\\
Spin moments ($\mu_B$)& U1:$(
2.017,  0,  0
)$, U2:$(
-2.017,  0,  0
)$\\
Energy (eV) and lattice (\AA) of und & 
-116.501, 5.5462
\end{tabular}
\end{ruledtabular}
\end{table*}

$\occmat(\textrm{U}1)$:

$\occmat^{\uparrow}=
\setstackgap{L}{1.1\baselineskip}
\fixTABwidth{T}
\parenMatrixstack{
0.46&  0&  -0.47&  0&  0&  0&  0 \\
0&  0.12&  0&  0&  0&  0&  0 \\
-0.47&  0&  0.58&  0&  0&  0&  0 \\
0&  0&  0&  0.05&  0&  -0.06&  0 \\
0&  0&  0&  0&  0.03&  0&  -0.01 \\
0&  0&  0&  -0.06&  0&  0.99&  0 \\
0&  0&  0&  0&  -0.01&  0&  0.04
}$,
$\occmat^{\downarrow}=
\setstackgap{L}{1.1\baselineskip}
\fixTABwidth{T}
\parenMatrixstack{
0.03&  0&  0.01&  0&  0&  0&  0 \\
0&  0.11&  0&  0&  0&  0&  0 \\
0.01&  0&  0.03&  0&  0&  0&  0 \\
0&  0&  0&  0.04&  0&  0&  0 \\
0&  0&  0&  0&  0.03&  0&  -0.01 \\
0&  0&  0&  0&  0&  0.02&  0 \\
0&  0&  0&  0&  -0.01&  0&  0.03
}$

$\occmat(\textrm{U}2)$:

$\occmat^{\uparrow}=
\setstackgap{L}{1.1\baselineskip}
\fixTABwidth{T}
\parenMatrixstack{
0.03&  0&  0.01&  0&  0&  0&  0 \\
0&  0.11&  0&  0&  0&  0&  0 \\
0.01&  0&  0.03&  0&  0&  0&  0 \\
0&  0&  0&  0.04&  0&  0&  0 \\
0&  0&  0&  0&  0.03&  0&  -0.01 \\
0&  0&  0&  0&  0&  0.02&  0 \\
0&  0&  0&  0&  -0.01&  0&  0.03
}$,
$\occmat^{\downarrow}=
\setstackgap{L}{1.1\baselineskip}
\fixTABwidth{T}
\parenMatrixstack{
0.46&  0&  -0.47&  0&  0&  0&  0 \\
0&  0.12&  0&  0&  0&  0&  0 \\
-0.47&  0&  0.58&  0&  0&  0&  0 \\
0&  0&  0&  0.05&  0&  -0.06&  0 \\
0&  0&  0&  0&  0.03&  0&  -0.01 \\
0&  0&  0&  -0.06&  0&  0.99&  0 \\
0&  0&  0&  0&  -0.01&  0&  0.04
}$

\subsubsection{Properties of S$_2^1$ (1\textbf{k} AFM without SOC)}

\begin{table*}[!h]%
\begin{ruledtabular}
\begin{tabular}{ll}
\textrm{Properties (unit)}&
\textrm{Values}\\

\colrule
Energy (eV) &  -116.400 \\
Basis vectors (\AA)& $a=(
5.5277,  0,  0
)$, $b=(
0,  5.5209,  0
)$, $c=(
0,  0,  5.5902
)$\\
\begin{tabular}{@{}l@{}}Atom coordinates \\ (in basis vectors)\end{tabular} & {\begin{tabular}{@{}l@{}}U1:$(
0.5000,  0,  0 )$, U2:$(
0.5000,  0.5000,  0.5000 )$, \\U3:$(
0,  0,  0.5000 )$, U4:$(
0,  0.5000,  0 )$, \\O1:$(
0.2500,  0.7500,  0.7500 )$, O2:$(
0.2500,  0.2500,  0.7500 )$, \\O3:$(
0.2500,  0.2500,  0.2500 )$, O4:$(
0.2500,  0.7500,  0.2500 )$, \\O5:$(
0.7500,  0.7500,  0.2500 )$, O6:$(
0.7500,  0.2500,  0.2500 )$, \\O7:$(
0.7500,  0.2500,  0.7500 )$, O8:$(
0.7500,  0.7500,  0.7500 )$\end{tabular}} \\ 
Strain & \begin{tabular}{@{}l@{}}$(\epsilon_{xx},\epsilon_{yy},\epsilon_{zz},\epsilon_{xy},\epsilon_{xz},\epsilon_{yz})=(
-0.0033,  -0.0046,  0.0079,  0,  0,  0 )$, \\$(\epsilon_{A1g},\epsilon_{Eg.0},\epsilon_{Eg.1},\epsilon_{T2g.0},\epsilon_{T2g.1},\epsilon_{T2g.2})=(
0,  0.0009,  0.0097,  0,  0,  0 )$.\end{tabular}\\
Spin moments ($\mu_B$)& U1:$(
2.006,  0,  0
)$, U2:$(
-2.006,  0,  0
)$\\
Energy (eV) and lattice (\AA) of und & 
-116.387, 5.5462
\end{tabular}
\end{ruledtabular}
\end{table*}

$\occmat(\textrm{U}1)$:

$\occmat^{\uparrow}=
\setstackgap{L}{1.1\baselineskip}
\fixTABwidth{T}
\parenMatrixstack{
0.04&  0&  0.01&  0&  0&  0&  0 \\
0&  0.13&  0&  0&  0&  0&  0 \\
0.01&  0&  0.04&  0&  0&  0&  0 \\
0&  0&  0&  0.98&  0&  -0.12&  0 \\
0&  0&  0&  0&  0.05&  0&  0.13 \\
0&  0&  0&  -0.12&  0&  0.05&  0 \\
0&  0&  0&  0&  0.13&  0&  0.98
}$,
$\occmat^{\downarrow}=
\setstackgap{L}{1.1\baselineskip}
\fixTABwidth{T}
\parenMatrixstack{
0.03&  0&  0&  0&  0&  0&  0 \\
0&  0.11&  0&  0&  0&  0&  0 \\
0&  0&  0.03&  0&  0&  0&  0 \\
0&  0&  0&  0.03&  0&  0&  0 \\
0&  0&  0&  0&  0.03&  0&  -0.01 \\
0&  0&  0&  0&  0&  0.02&  0 \\
0&  0&  0&  0&  -0.01&  0&  0.03
}$

$\occmat(\textrm{U}2)$:

$\occmat^{\uparrow}=
\setstackgap{L}{1.1\baselineskip}
\fixTABwidth{T}
\parenMatrixstack{
0.03&  0&  0&  0&  0&  0&  0 \\
0&  0.11&  0&  0&  0&  0&  0 \\
0&  0&  0.03&  0&  0&  0&  0 \\
0&  0&  0&  0.03&  0&  0&  0 \\
0&  0&  0&  0&  0.03&  0&  -0.01 \\
0&  0&  0&  0&  0&  0.02&  0 \\
0&  0&  0&  0&  -0.01&  0&  0.03
}$,
$\occmat^{\downarrow}=
\setstackgap{L}{1.1\baselineskip}
\fixTABwidth{T}
\parenMatrixstack{
0.04&  0&  0.01&  0&  0&  0&  0 \\
0&  0.13&  0&  0&  0&  0&  0 \\
0.01&  0&  0.04&  0&  0&  0&  0 \\
0&  0&  0&  0.98&  0&  -0.12&  0 \\
0&  0&  0&  0&  0.05&  0&  0.13 \\
0&  0&  0&  -0.12&  0&  0.05&  0 \\
0&  0&  0&  0&  0.13&  0&  0.98
}$
\clearpage
\subsubsection{Properties of S$_2^2$ (1\textbf{k} AFM without SOC)}

\begin{table*}[!h]%
\begin{ruledtabular}
\begin{tabular}{ll}
\textrm{Properties (unit)}&
\textrm{Values}\\

\colrule
Energy (eV) &  -116.421 \\
Basis vectors (\AA)& $a=(
5.5186,  0,  0
)$, $b=(
0,  5.5891,  0
)$, $c=(
0,  0,  5.5284
)$\\
\begin{tabular}{@{}l@{}}Atom coordinates \\ (in basis vectors)\end{tabular} & {\begin{tabular}{@{}l@{}}U1:$(
0.5000,  0,  0 )$, U2:$(
0.5000,  0.5000,  0.5000 )$, \\U3:$(
0,  0,  0.5000 )$, U4:$(
0,  0.5000,  0 )$, \\O1:$(
0.2500,  0.7500,  0.7500 )$, O2:$(
0.2500,  0.2500,  0.7500 )$, \\O3:$(
0.2500,  0.2500,  0.2500 )$, O4:$(
0.2500,  0.7500,  0.2500 )$, \\O5:$(
0.7500,  0.7500,  0.2500 )$, O6:$(
0.7500,  0.2500,  0.2500 )$, \\O7:$(
0.7500,  0.2500,  0.7500 )$, O8:$(
0.7500,  0.7500,  0.7500 )$\end{tabular}} \\ 
Strain & \begin{tabular}{@{}l@{}}$(\epsilon_{xx},\epsilon_{yy},\epsilon_{zz},\epsilon_{xy},\epsilon_{xz},\epsilon_{yz})=(
-0.0048,  0.0079,  -0.0031,  0,  0,  0 )$, \\$(\epsilon_{A1g},\epsilon_{Eg.0},\epsilon_{Eg.1},\epsilon_{T2g.0},\epsilon_{T2g.1},\epsilon_{T2g.2})=(
0,  -0.0090,  -0.0038,  0,  0,  0 )$.\end{tabular}\\
Spin moments ($\mu_B$)& U1:$(
1.996,  0,  0
)$, U2:$(
-1.996,  0,  0
)$\\
Energy (eV) and lattice (\AA) of und & 
-116.408, 5.5454
\end{tabular}
\end{ruledtabular}
\end{table*}

$\occmat(\textrm{U}1)$:

$\occmat^{\uparrow}=
\setstackgap{L}{1.1\baselineskip}
\fixTABwidth{T}
\parenMatrixstack{
0.54&  0&  0.48&  0&  0&  0&  0 \\
0&  0.14&  0&  0&  0&  0&  0 \\
0.48&  0&  0.48&  0&  0&  0&  0 \\
0&  0&  0&  0.54&  0&  -0.48&  0 \\
0&  0&  0&  0&  0.03&  0&  -0.01 \\
0&  0&  0&  -0.48&  0&  0.49&  0 \\
0&  0&  0&  0&  -0.01&  0&  0.04
}$,
$\occmat^{\downarrow}=
\setstackgap{L}{1.1\baselineskip}
\fixTABwidth{T}
\parenMatrixstack{
0.03&  0&  0&  0&  0&  0&  0 \\
0&  0.11&  0&  0&  0&  0&  0 \\
0&  0&  0.03&  0&  0&  0&  0 \\
0&  0&  0&  0.04&  0&  0&  0 \\
0&  0&  0&  0&  0.03&  0&  -0.01 \\
0&  0&  0&  0&  0&  0.02&  0 \\
0&  0&  0&  0&  -0.01&  0&  0.03
}$

$\occmat(\textrm{U}2)$:

$\occmat^{\uparrow}=
\setstackgap{L}{1.1\baselineskip}
\fixTABwidth{T}
\parenMatrixstack{
0.03&  0&  0&  0&  0&  0&  0 \\
0&  0.11&  0&  0&  0&  0&  0 \\
0&  0&  0.03&  0&  0&  0&  0 \\
0&  0&  0&  0.04&  0&  0&  0 \\
0&  0&  0&  0&  0.03&  0&  -0.01 \\
0&  0&  0&  0&  0&  0.02&  0 \\
0&  0&  0&  0&  -0.01&  0&  0.03
}$,
$\occmat^{\downarrow}=
\setstackgap{L}{1.1\baselineskip}
\fixTABwidth{T}
\parenMatrixstack{
0.54&  0&  0.48&  0&  0&  0&  0 \\
0&  0.14&  0&  0&  0&  0&  0 \\
0.48&  0&  0.48&  0&  0&  0&  0 \\
0&  0&  0&  0.54&  0&  -0.48&  0 \\
0&  0&  0&  0&  0.03&  0&  -0.01 \\
0&  0&  0&  -0.48&  0&  0.49&  0 \\
0&  0&  0&  0&  -0.01&  0&  0.04
}$

\subsubsection{Properties of S$_2^3$ (1\textbf{k} AFM without SOC)}

\begin{table*}[!h]%
\begin{ruledtabular}
\begin{tabular}{ll}
\textrm{Properties (unit)}&
\textrm{Values}\\

\colrule
Energy (eV) &  -116.400 \\
Basis vectors (\AA)& $a=(
5.5902,  0,  0
)$, $b=(
0,  5.5209,  0
)$, $c=(
0,  0,  5.5277
)$\\
\begin{tabular}{@{}l@{}}Atom coordinates \\ (in basis vectors)\end{tabular} & {\begin{tabular}{@{}l@{}}U1:$(
0.5000,  0,  0 )$, U2:$(
0.5000,  0.5000,  0.5000 )$, \\U3:$(
0,  0,  0.5000 )$, U4:$(
0,  0.5000,  0 )$, \\O1:$(
0.2500,  0.7500,  0.7500 )$, O2:$(
0.2500,  0.2500,  0.7500 )$, \\O3:$(
0.2500,  0.2500,  0.2500 )$, O4:$(
0.2500,  0.7500,  0.2500 )$, \\O5:$(
0.7500,  0.7500,  0.2500 )$, O6:$(
0.7500,  0.2500,  0.2500 )$, \\O7:$(
0.7500,  0.2500,  0.7500 )$, O8:$(
0.7500,  0.7500,  0.7500 )$\end{tabular}} \\ 
Strain & \begin{tabular}{@{}l@{}}$(\epsilon_{xx},\epsilon_{yy},\epsilon_{zz},\epsilon_{xy},\epsilon_{xz},\epsilon_{yz})=(
0.0079,  -0.0046,  -0.0033,  0,  0,  0 )$, \\$(\epsilon_{A1g},\epsilon_{Eg.0},\epsilon_{Eg.1},\epsilon_{T2g.0},\epsilon_{T2g.1},\epsilon_{T2g.2})=(
0,  0.0088,  -0.0041,  0,  0,  0 )$.\end{tabular}\\
Spin moments ($\mu_B$)& U1:$(
2.006,  0,  0
)$, U2:$(
-2.006,  0,  0
)$\\
Energy (eV) and lattice (\AA) of und & 
-116.387, 5.5462
\end{tabular}
\end{ruledtabular}
\end{table*}

$\occmat(\textrm{U}1)$:

$\occmat^{\uparrow}=
\setstackgap{L}{1.1\baselineskip}
\fixTABwidth{T}
\parenMatrixstack{
0.04&  0&  0.01&  0&  0&  0&  0 \\
0&  0.13&  0&  0&  0&  0&  0 \\
0.01&  0&  0.03&  0&  0&  0&  0 \\
0&  0&  0&  0.51&  0&  0.48&  0 \\
0&  0&  0&  0&  0.51&  0&  -0.48 \\
0&  0&  0&  0.48&  0&  0.52&  0 \\
0&  0&  0&  0&  -0.48&  0&  0.51
}$,
$\occmat^{\downarrow}=
\setstackgap{L}{1.1\baselineskip}
\fixTABwidth{T}
\setstacktabbedgap{9pt}
\parenMatrixstack{
0.03&  0&  0.01&  0&  0&  0&  0 \\
0&  0.11&  0&  0&  0&  0&  0 \\
0.01&  0&  0.03&  0&  0&  0&  0 \\
0&  0&  0&  0.03&  0&  0&  0 \\
0&  0&  0&  0&  0.03&  0&  0 \\
0&  0&  0&  0&  0&  0.02&  0 \\
0&  0&  0&  0&  0&  0&  0.03
}$

$\occmat(\textrm{U}2)$:

$\occmat^{\uparrow}=
\setstackgap{L}{1.1\baselineskip}
\fixTABwidth{T}
\setstacktabbedgap{9pt}
\parenMatrixstack{
0.03&  0&  0.01&  0&  0&  0&  0 \\
0&  0.11&  0&  0&  0&  0&  0 \\
0.01&  0&  0.03&  0&  0&  0&  0 \\
0&  0&  0&  0.03&  0&  0&  0 \\
0&  0&  0&  0&  0.03&  0&  0 \\
0&  0&  0&  0&  0&  0.02&  0 \\
0&  0&  0&  0&  0&  0&  0.03
}$,
$\occmat^{\downarrow}=
\setstackgap{L}{1.1\baselineskip}
\fixTABwidth{T}
\parenMatrixstack{
0.04&  0&  0.01&  0&  0&  0&  0 \\
0&  0.13&  0&  0&  0&  0&  0 \\
0.01&  0&  0.03&  0&  0&  0&  0 \\
0&  0&  0&  0.51&  0&  0.48&  0 \\
0&  0&  0&  0&  0.51&  0&  -0.48 \\
0&  0&  0&  0.48&  0&  0.52&  0 \\
0&  0&  0&  0&  -0.48&  0&  0.51
}$

\subsubsection{Properties of S$_2^4$ (1\textbf{k} AFM without SOC)}

\begin{table*}[!h]%
\begin{ruledtabular}
\begin{tabular}{ll}
\textrm{Properties (unit)}&
\textrm{Values}\\

\colrule
Energy (eV) &  -116.441 \\
Basis vectors (\AA)& $a=(
5.5867,  0,  0
)$, $b=(
0,  5.5310,  0
)$, $c=(
0,  0,  5.5152
)$\\
\begin{tabular}{@{}l@{}}Atom coordinates \\ (in basis vectors)\end{tabular} & {\begin{tabular}{@{}l@{}}U1:$(
0.5000,  0,  0 )$, U2:$(
0.5000,  0.5000,  0.5000 )$, \\U3:$(
0,  0,  0.5000 )$, U4:$(
0,  0.5000,  0 )$, \\O1:$(
0.2500,  0.7500,  0.7500 )$, O2:$(
0.2500,  0.2500,  0.7500 )$, \\O3:$(
0.2500,  0.2500,  0.2500 )$, O4:$(
0.2500,  0.7500,  0.2500 )$, \\O5:$(
0.7500,  0.7500,  0.2500 )$, O6:$(
0.7500,  0.2500,  0.2500 )$, \\O7:$(
0.7500,  0.2500,  0.7500 )$, O8:$(
0.7500,  0.7500,  0.7500 )$\end{tabular}} \\ 
Strain & \begin{tabular}{@{}l@{}}$(\epsilon_{xx},\epsilon_{yy},\epsilon_{zz},\epsilon_{xy},\epsilon_{xz},\epsilon_{yz})=(
0.0076,  -0.0024,  -0.0052,  0,  0,  0 )$, \\$(\epsilon_{A1g},\epsilon_{Eg.0},\epsilon_{Eg.1},\epsilon_{T2g.0},\epsilon_{T2g.1},\epsilon_{T2g.2})=(
0,  0.0071,  -0.0064,  0,  0,  0 )$.\end{tabular}\\
Spin moments ($\mu_B$)& U1:$(
2.002,  0,  0
)$, U2:$(
-2.002,  0,  0
)$\\
Energy (eV) and lattice (\AA) of und & 
-116.429, 5.5443
\end{tabular}
\end{ruledtabular}
\end{table*}

$\occmat(\textrm{U}1)$:

$\occmat^{\uparrow}=
\setstackgap{L}{1.1\baselineskip}
\fixTABwidth{T}
\parenMatrixstack{
0.06&  0&  0.15&  0&  0&  0&  0 \\
0&  0.13&  0&  0&  0&  0&  0 \\
0.15&  0&  0.97&  0&  0&  0&  0 \\
0&  0&  0&  0.04&  0&  0&  0 \\
0&  0&  0&  0&  0.31&  0&  -0.44 \\
0&  0&  0&  0&  0&  0.03&  0 \\
0&  0&  0&  0&  -0.44&  0&  0.71
}$,
$\occmat^{\downarrow}=
\setstackgap{L}{1.1\baselineskip}
\fixTABwidth{T}
\parenMatrixstack{
0.03&  0&  0&  0&  0&  0&  0 \\
0&  0.11&  0&  0&  0&  0&  0 \\
0&  0&  0.03&  0&  0&  0&  0 \\
0&  0&  0&  0.04&  0&  0&  0 \\
0&  0&  0&  0&  0.03&  0&  -0.01 \\
0&  0&  0&  0&  0&  0.03&  0 \\
0&  0&  0&  0&  -0.01&  0&  0.03
}$

$\occmat(\textrm{U}2)$:

$\occmat^{\uparrow}=
\setstackgap{L}{1.1\baselineskip}
\fixTABwidth{T}
\parenMatrixstack{
0.03&  0&  0&  0&  0&  0&  0 \\
0&  0.11&  0&  0&  0&  0&  0 \\
0&  0&  0.03&  0&  0&  0&  0 \\
0&  0&  0&  0.04&  0&  0&  0 \\
0&  0&  0&  0&  0.03&  0&  -0.01 \\
0&  0&  0&  0&  0&  0.03&  0 \\
0&  0&  0&  0&  -0.01&  0&  0.03
}$,
$\occmat^{\downarrow}=
\setstackgap{L}{1.1\baselineskip}
\fixTABwidth{T}
\parenMatrixstack{
0.06&  0&  0.15&  0&  0&  0&  0 \\
0&  0.13&  0&  0&  0&  0&  0 \\
0.15&  0&  0.97&  0&  0&  0&  0 \\
0&  0&  0&  0.04&  0&  0&  0 \\
0&  0&  0&  0&  0.31&  0&  -0.44 \\
0&  0&  0&  0&  0&  0.03&  0 \\
0&  0&  0&  0&  -0.44&  0&  0.71
}$
\clearpage
\subsubsection{Properties of S$_2^5$ (1\textbf{k} AFM without SOC)}

\begin{table*}[!h]%
\begin{ruledtabular}
\begin{tabular}{ll}
\textrm{Properties (unit)}&
\textrm{Values}\\

\colrule
Energy (eV) &  -116.441 \\
Basis vectors (\AA)& $a=(
5.5152,  0,  0
)$, $b=(
0,  5.5310,  0
)$, $c=(
0,  0,  5.5867
)$\\
\begin{tabular}{@{}l@{}}Atom coordinates \\ (in basis vectors)\end{tabular} & {\begin{tabular}{@{}l@{}}U1:$(
0.5000,  0,  0 )$, U2:$(
0.5000,  0.5000,  0.5000 )$, \\U3:$(
0,  0,  0.5000 )$, U4:$(
0,  0.5000,  0 )$, \\O1:$(
0.2500,  0.7500,  0.7500 )$, O2:$(
0.2500,  0.2500,  0.7500 )$, \\O3:$(
0.2500,  0.2500,  0.2500 )$, O4:$(
0.2500,  0.7500,  0.2500 )$, \\O5:$(
0.7500,  0.7500,  0.2500 )$, O6:$(
0.7500,  0.2500,  0.2500 )$, \\O7:$(
0.7500,  0.2500,  0.7500 )$, O8:$(
0.7500,  0.7500,  0.7500 )$\end{tabular}} \\ 
Strain & \begin{tabular}{@{}l@{}}$(\epsilon_{xx},\epsilon_{yy},\epsilon_{zz},\epsilon_{xy},\epsilon_{xz},\epsilon_{yz})=(
-0.0052,  -0.0024,  0.0076,  0,  0,  0 )$, \\$(\epsilon_{A1g},\epsilon_{Eg.0},\epsilon_{Eg.1},\epsilon_{T2g.0},\epsilon_{T2g.1},\epsilon_{T2g.2})=(
0,  -0.0020,  0.0094,  0,  0,  0 )$.\end{tabular}\\
Spin moments ($\mu_B$)& U1:$(
2.002,  0,  0
)$, U2:$(
-2.002,  0,  0
)$\\
Energy (eV) and lattice (\AA) of und & 
-116.429, 5.5443
\end{tabular}
\end{ruledtabular}
\end{table*}

$\occmat(\textrm{U}1)$:

$\occmat^{\uparrow}=
\setstackgap{L}{1.1\baselineskip}
\fixTABwidth{T}
\parenMatrixstack{
0.99&  0&  -0.08&  0&  0&  0&  0 \\
0&  0.13&  0&  0&  0&  0&  0 \\
-0.08&  0&  0.05&  0&  0&  0&  0 \\
0&  0&  0&  0.98&  0&  0.08&  0 \\
0&  0&  0&  0&  0.04&  0&  -0.01 \\
0&  0&  0&  0.08&  0&  0.04&  0 \\
0&  0&  0&  0&  -0.01&  0&  0.04
}$,
$\occmat^{\downarrow}=
\setstackgap{L}{1.1\baselineskip}
\fixTABwidth{T}
\setstacktabbedgap{9pt}
\parenMatrixstack{
0.03&  0&  0.01&  0&  0&  0&  0 \\
0&  0.11&  0&  0&  0&  0&  0 \\
0.01&  0&  0.03&  0&  0&  0&  0 \\
0&  0&  0&  0.04&  0&  0&  0 \\
0&  0&  0&  0&  0.03&  0&  0 \\
0&  0&  0&  0&  0&  0.03&  0 \\
0&  0&  0&  0&  0&  0&  0.03
}$

$\occmat(\textrm{U}2)$:

$\occmat^{\uparrow}=
\setstackgap{L}{1.1\baselineskip}
\fixTABwidth{T}
\setstacktabbedgap{9pt}
\parenMatrixstack{
0.03&  0&  0.01&  0&  0&  0&  0 \\
0&  0.11&  0&  0&  0&  0&  0 \\
0.01&  0&  0.03&  0&  0&  0&  0 \\
0&  0&  0&  0.04&  0&  0&  0 \\
0&  0&  0&  0&  0.03&  0&  0 \\
0&  0&  0&  0&  0&  0.03&  0 \\
0&  0&  0&  0&  0&  0&  0.03
}$,
$\occmat^{\downarrow}=
\setstackgap{L}{1.1\baselineskip}
\fixTABwidth{T}
\parenMatrixstack{
0.99&  0&  -0.08&  0&  0&  0&  0 \\
0&  0.13&  0&  0&  0&  0&  0 \\
-0.08&  0&  0.05&  0&  0&  0&  0 \\
0&  0&  0&  0.98&  0&  0.08&  0 \\
0&  0&  0&  0&  0.04&  0&  -0.01 \\
0&  0&  0&  0.08&  0&  0.04&  0 \\
0&  0&  0&  0&  -0.01&  0&  0.04
}$

\subsubsection{Properties of S$_2^6$ (1\textbf{k} AFM without SOC)}

\begin{table*}[!h]%
\begin{ruledtabular}
\begin{tabular}{ll}
\textrm{Properties (unit)}&
\textrm{Values}\\

\colrule
Energy (eV) &  -116.421 \\
Basis vectors (\AA)& $a=(
5.5284,  0,  0
)$, $b=(
0,  5.5891,  0
)$, $c=(
0,  0,  5.5186
)$\\
\begin{tabular}{@{}l@{}}Atom coordinates \\ (in basis vectors)\end{tabular} & {\begin{tabular}{@{}l@{}}U1:$(
0.5000,  0,  0 )$, U2:$(
0.5000,  0.5000,  0.5000 )$, \\U3:$(
0,  0,  0.5000 )$, U4:$(
0,  0.5000,  0 )$, \\O1:$(
0.2500,  0.7500,  0.7500 )$, O2:$(
0.2500,  0.2500,  0.7500 )$, \\O3:$(
0.2500,  0.2500,  0.2500 )$, O4:$(
0.2500,  0.7500,  0.2500 )$, \\O5:$(
0.7500,  0.7500,  0.2500 )$, O6:$(
0.7500,  0.2500,  0.2500 )$, \\O7:$(
0.7500,  0.2500,  0.7500 )$, O8:$(
0.7500,  0.7500,  0.7500 )$\end{tabular}} \\ 
Strain & \begin{tabular}{@{}l@{}}$(\epsilon_{xx},\epsilon_{yy},\epsilon_{zz},\epsilon_{xy},\epsilon_{xz},\epsilon_{yz})=(
-0.0031,  0.0079,  -0.0048,  0,  0,  0 )$, \\$(\epsilon_{A1g},\epsilon_{Eg.0},\epsilon_{Eg.1},\epsilon_{T2g.0},\epsilon_{T2g.1},\epsilon_{T2g.2})=(
0,  -0.0077,  -0.0059,  0,  0,  0 )$.\end{tabular}\\
Spin moments ($\mu_B$)& U1:$(
1.996,  0,  0
)$, U2:$(
-1.996,  0,  0
)$\\
Energy (eV) and lattice (\AA) of und & 
-116.408, 5.5454
\end{tabular}
\end{ruledtabular}
\end{table*}

$\occmat(\textrm{U}1)$:

$\occmat^{\uparrow}=
\setstackgap{L}{1.1\baselineskip}
\fixTABwidth{T}
\parenMatrixstack{
0.72&  0&  0.43&  0&  0&  0&  0 \\
0&  0.14&  0&  0&  0&  0&  0 \\
0.43&  0&  0.30&  0&  0&  0&  0 \\
0&  0&  0&  0.04&  0&  0&  0 \\
0&  0&  0&  0&  0.97&  0&  -0.15 \\
0&  0&  0&  0&  0&  0.03&  0 \\
0&  0&  0&  0&  -0.15&  0&  0.06
}$,
$\occmat^{\downarrow}=
\setstackgap{L}{1.1\baselineskip}
\fixTABwidth{T}
\setstacktabbedgap{9pt}
\parenMatrixstack{
0.03&  0&  0&  0&  0&  0&  0 \\
0&  0.11&  0&  0&  0&  0&  0 \\
0&  0&  0.03&  0&  0&  0&  0 \\
0&  0&  0&  0.04&  0&  0&  0 \\
0&  0&  0&  0&  0.03&  0&  0 \\
0&  0&  0&  0&  0&  0.03&  0 \\
0&  0&  0&  0&  0&  0&  0.03
}$

$\occmat(\textrm{U}2)$:

$\occmat^{\uparrow}=
\setstackgap{L}{1.1\baselineskip}
\fixTABwidth{T}
\setstacktabbedgap{9pt}
\parenMatrixstack{
0.03&  0&  0&  0&  0&  0&  0 \\
0&  0.11&  0&  0&  0&  0&  0 \\
0&  0&  0.03&  0&  0&  0&  0 \\
0&  0&  0&  0.04&  0&  0&  0 \\
0&  0&  0&  0&  0.03&  0&  0 \\
0&  0&  0&  0&  0&  0.03&  0 \\
0&  0&  0&  0&  0&  0&  0.03
}$,
$\occmat^{\downarrow}=
\setstackgap{L}{1.1\baselineskip}
\fixTABwidth{T}
\parenMatrixstack{
0.72&  0&  0.43&  0&  0&  0&  0 \\
0&  0.14&  0&  0&  0&  0&  0 \\
0.43&  0&  0.30&  0&  0&  0&  0 \\
0&  0&  0&  0.04&  0&  0&  0 \\
0&  0&  0&  0&  0.97&  0&  -0.15 \\
0&  0&  0&  0&  0&  0.03&  0 \\
0&  0&  0&  0&  -0.15&  0&  0.06
}$

\subsubsection{Properties of S$_4^1$ (1\textbf{k} AFM without SOC)}

\begin{table*}[!h]%
\begin{ruledtabular}
\begin{tabular}{ll}
\textrm{Properties (unit)}&
\textrm{Values}\\

\colrule
Energy (eV) &  -116.356 \\
Basis vectors (\AA)& $a=(
5.5683,  0,  0
)$, $b=(
0,  5.5036,  0
)$, $c=(
0,  0,  5.5683
)$\\
\begin{tabular}{@{}l@{}}Atom coordinates \\ (in basis vectors)\end{tabular} & {\begin{tabular}{@{}l@{}}U1:$(
0.5000,  0,  0 )$, U2:$(
0.5000,  0.5000,  0.5000 )$, \\U3:$(
0,  0,  0.5000 )$, U4:$(
0,  0.5000,  0 )$, \\O1:$(
0.2500,  0.7500,  0.7500 )$, O2:$(
0.2500,  0.2500,  0.7500 )$, \\O3:$(
0.2500,  0.2500,  0.2500 )$, O4:$(
0.2500,  0.7500,  0.2500 )$, \\O5:$(
0.7500,  0.7500,  0.2500 )$, O6:$(
0.7500,  0.2500,  0.2500 )$, \\O7:$(
0.7500,  0.2500,  0.7500 )$, O8:$(
0.7500,  0.7500,  0.7500 )$\end{tabular}} \\ 
Strain & \begin{tabular}{@{}l@{}}$(\epsilon_{xx},\epsilon_{yy},\epsilon_{zz},\epsilon_{xy},\epsilon_{xz},\epsilon_{yz})=(
0.0039,  -0.0078,  0.0039,  0,  0,  0 )$, \\$(\epsilon_{A1g},\epsilon_{Eg.0},\epsilon_{Eg.1},\epsilon_{T2g.0},\epsilon_{T2g.1},\epsilon_{T2g.2})=(
0,  0.0083,  0.0048,  0,  0,  0 )$.\end{tabular}\\
Spin moments ($\mu_B$)& U1:$(
2.001,  0,  0
)$, U2:$(
-2.001,  0,  0
)$\\
Energy (eV) and lattice (\AA) of und & 
-116.343, 5.5466
\end{tabular}
\end{ruledtabular}
\end{table*}

$\occmat(\textrm{U}1)$:

$\occmat^{\uparrow}=
\setstackgap{L}{1.1\baselineskip}
\fixTABwidth{T}
\parenMatrixstack{
0.04&  0&  0.01&  0&  0&  0&  0 \\
0&  0.14&  0&  0&  0&  0&  0 \\
0.01&  0&  0.03&  0&  0&  0&  0 \\
0&  0&  0&  0.91&  0&  0.27&  0 \\
0&  0&  0&  0&  0.16&  0&  -0.32 \\
0&  0&  0&  0.27&  0&  0.11&  0 \\
0&  0&  0&  0&  -0.32&  0&  0.87
}$,
$\occmat^{\downarrow}=
\setstackgap{L}{1.1\baselineskip}
\fixTABwidth{T}
\setstacktabbedgap{9pt}
\parenMatrixstack{
0.03&  0&  0&  0&  0&  0&  0 \\
0&  0.11&  0&  0&  0&  0&  0 \\
0&  0&  0.03&  0&  0&  0&  0 \\
0&  0&  0&  0.03&  0&  0&  0 \\
0&  0&  0&  0&  0.03&  0&  0 \\
0&  0&  0&  0&  0&  0.02&  0 \\
0&  0&  0&  0&  0&  0&  0.03
}$

$\occmat(\textrm{U}2)$:

$\occmat^{\uparrow}=
\setstackgap{L}{1.1\baselineskip}
\fixTABwidth{T}
\setstacktabbedgap{9pt}
\parenMatrixstack{
0.03&  0&  0&  0&  0&  0&  0 \\
0&  0.11&  0&  0&  0&  0&  0 \\
0&  0&  0.03&  0&  0&  0&  0 \\
0&  0&  0&  0.03&  0&  0&  0 \\
0&  0&  0&  0&  0.03&  0&  0 \\
0&  0&  0&  0&  0&  0.02&  0 \\
0&  0&  0&  0&  0&  0&  0.03
}$,
$\occmat^{\downarrow}=
\setstackgap{L}{1.1\baselineskip}
\fixTABwidth{T}
\parenMatrixstack{
0.04&  0&  0.01&  0&  0&  0&  0 \\
0&  0.14&  0&  0&  0&  0&  0 \\
0.01&  0&  0.03&  0&  0&  0&  0 \\
0&  0&  0&  0.91&  0&  0.27&  0 \\
0&  0&  0&  0&  0.16&  0&  -0.32 \\
0&  0&  0&  0.27&  0&  0.11&  0 \\
0&  0&  0&  0&  -0.32&  0&  0.87
}$
\clearpage
\subsubsection{Properties of S$_4^2$ (1\textbf{k} AFM without SOC)}

\begin{table*}[!h]%
\begin{ruledtabular}
\begin{tabular}{ll}
\textrm{Properties (unit)}&
\textrm{Values}\\

\colrule
Energy (eV) &  -116.399 \\
Basis vectors (\AA)& $a=(
5.5611,  0,  0
)$, $b=(
0,  5.5727,  0
)$, $c=(
0,  0,  5.5011
)$\\
\begin{tabular}{@{}l@{}}Atom coordinates \\ (in basis vectors)\end{tabular} & {\begin{tabular}{@{}l@{}}U1:$(
0.5000,  0,  0 )$, U2:$(
0.5000,  0.5000,  0.5000 )$, \\U3:$(
0,  0,  0.5000 )$, U4:$(
0,  0.5000,  0 )$, \\O1:$(
0.2500,  0.7500,  0.7500 )$, O2:$(
0.2500,  0.2500,  0.7500 )$, \\O3:$(
0.2500,  0.2500,  0.2500 )$, O4:$(
0.2500,  0.7500,  0.2500 )$, \\O5:$(
0.7500,  0.7500,  0.2500 )$, O6:$(
0.7500,  0.2500,  0.2500 )$, \\O7:$(
0.7500,  0.2500,  0.7500 )$, O8:$(
0.7500,  0.7500,  0.7500 )$\end{tabular}} \\ 
Strain & \begin{tabular}{@{}l@{}}$(\epsilon_{xx},\epsilon_{yy},\epsilon_{zz},\epsilon_{xy},\epsilon_{xz},\epsilon_{yz})=(
0.0029,  0.0050,  -0.0079,  0,  0,  0 )$, \\$(\epsilon_{A1g},\epsilon_{Eg.0},\epsilon_{Eg.1},\epsilon_{T2g.0},\epsilon_{T2g.1},\epsilon_{T2g.2})=(
0,  -0.0015,  -0.0097,  0,  0,  0 )$.\end{tabular}\\
Spin moments ($\mu_B$)& U1:$(
1.995,  0,  0
)$, U2:$(
-1.995,  0,  0
)$\\
Energy (eV) and lattice (\AA) of und & 
-116.386, 5.5450
\end{tabular}
\end{ruledtabular}
\end{table*}

$\occmat(\textrm{U}1)$:

$\occmat^{\uparrow}=
\setstackgap{L}{1.1\baselineskip}
\fixTABwidth{T}
\parenMatrixstack{
0.41&  0&  0.47&  0&  0&  0&  0 \\
0&  0.14&  0&  0&  0&  0&  0 \\
0.47&  0&  0.62&  0&  0&  0&  0 \\
0&  0&  0&  0.04&  0&  0&  0 \\
0&  0&  0&  0&  0.73&  0&  -0.43 \\
0&  0&  0&  0&  0&  0.03&  0 \\
0&  0&  0&  0&  -0.43&  0&  0.30
}$,
$\occmat^{\downarrow}=
\setstackgap{L}{1.1\baselineskip}
\fixTABwidth{T}
\setstacktabbedgap{9pt}
\parenMatrixstack{
0.03&  0&  0&  0&  0&  0&  0 \\
0&  0.11&  0&  0&  0&  0&  0 \\
0&  0&  0.03&  0&  0&  0&  0 \\
0&  0&  0&  0.04&  0&  0&  0 \\
0&  0&  0&  0&  0.03&  0&  0 \\
0&  0&  0&  0&  0&  0.03&  0 \\
0&  0&  0&  0&  0&  0&  0.03
}$

$\occmat(\textrm{U}2)$:

$\occmat^{\uparrow}=
\setstackgap{L}{1.1\baselineskip}
\fixTABwidth{T}
\setstacktabbedgap{9pt}
\parenMatrixstack{
0.03&  0&  0&  0&  0&  0&  0 \\
0&  0.11&  0&  0&  0&  0&  0 \\
0&  0&  0.03&  0&  0&  0&  0 \\
0&  0&  0&  0.04&  0&  0&  0 \\
0&  0&  0&  0&  0.03&  0&  0 \\
0&  0&  0&  0&  0&  0.03&  0 \\
0&  0&  0&  0&  0&  0&  0.03
}$,
$\occmat^{\downarrow}=
\setstackgap{L}{1.1\baselineskip}
\fixTABwidth{T}
\parenMatrixstack{
0.41&  0&  0.47&  0&  0&  0&  0 \\
0&  0.14&  0&  0&  0&  0&  0 \\
0.47&  0&  0.62&  0&  0&  0&  0 \\
0&  0&  0&  0.04&  0&  0&  0 \\
0&  0&  0&  0&  0.73&  0&  -0.43 \\
0&  0&  0&  0&  0&  0.03&  0 \\
0&  0&  0&  0&  -0.43&  0&  0.30
}$

\subsubsection{Properties of S$_5^1$ (1\textbf{k} AFM without SOC)}

\begin{table*}[!h]%
\begin{ruledtabular}
\begin{tabular}{ll}
\textrm{Properties (unit)}&
\textrm{Values}\\

\colrule
Energy (eV) &  -116.174 \\
Basis vectors (\AA)& $a=(
5.5237,  0,  0.0037
)$, $b=(
0,  5.5514,  0
)$, $c=(
0.0037,  0,  5.5534
)$\\
\begin{tabular}{@{}l@{}}Atom coordinates \\ (in basis vectors)\end{tabular} & {\begin{tabular}{@{}l@{}}U1:$(
0.5000,  0,  0 )$, U2:$(
0.5000,  0.5000,  0.5000 )$, \\U3:$(
0,  0,  0.5000 )$, U4:$(
0,  0.5000,  0 )$, \\O1:$(
0.2500,  0.7499,  0.7500 )$, O2:$(
0.2500,  0.2501,  0.7500 )$, \\O3:$(
0.2500,  0.2499,  0.2500 )$, O4:$(
0.2500,  0.7501,  0.2500 )$, \\O5:$(
0.7500,  0.7499,  0.2500 )$, O6:$(
0.7500,  0.2501,  0.2500 )$, \\O7:$(
0.7500,  0.2499,  0.7500 )$, O8:$(
0.7500,  0.7501,  0.7500 )$\end{tabular}} \\ 
Strain & \begin{tabular}{@{}l@{}}$(\epsilon_{xx},\epsilon_{yy},\epsilon_{zz},\epsilon_{xy},\epsilon_{xz},\epsilon_{yz})=(
-0.0035,  0.0015,  0.0019,  0,  0.0007,  0 )$, \\$(\epsilon_{A1g},\epsilon_{Eg.0},\epsilon_{Eg.1},\epsilon_{T2g.0},\epsilon_{T2g.1},\epsilon_{T2g.2})=(
0,  -0.0035,  0.0023,  0,  0,  0.0007 )$.\end{tabular}\\
Spin moments ($\mu_B$)& U1:$(
2.011,  0,  0
)$, U2:$(
-2.011,  0,  0
)$\\
Energy (eV) and lattice (\AA) of und & 
-116.172, 5.5429
\end{tabular}
\end{ruledtabular}
\end{table*}

$\occmat(\textrm{U}1)$:

$\occmat^{\uparrow}=
\setstackgap{L}{1.1\baselineskip}
\fixTABwidth{T}
\parenMatrixstack{
0.04&  0&  0.01&  0&  0&  0&  0 \\
0&  0.13&  0&  0&  0&  0&  0 \\
0.01&  0&  0.04&  0&  0&  0&  0 \\
0&  0&  0&  0.05&  0.03&  0.01&  0 \\
0&  0&  0&  0.03&  0.99&  -0.01&  0 \\
0&  0&  0&  0.01&  -0.01&  0.03&  0.01 \\
0&  0&  0&  0&  0&  0.01&  0.99
}$,
$\occmat^{\downarrow}=
\setstackgap{L}{1.1\baselineskip}
\fixTABwidth{T}
\setstacktabbedgap{9pt}
\parenMatrixstack{
0.03&  0&  0.01&  0&  0&  0&  0 \\
0&  0.11&  0&  0&  0&  0&  0 \\
0.01&  0&  0.03&  0&  0&  0&  0 \\
0&  0&  0&  0.04&  0&  0&  0 \\
0&  0&  0&  0&  0.03&  0&  0 \\
0&  0&  0&  0&  0&  0.02&  0 \\
0&  0&  0&  0&  0&  0&  0.03
}$

$\occmat(\textrm{U}2)$:

$\occmat^{\uparrow}=
\setstackgap{L}{1.1\baselineskip}
\fixTABwidth{T}
\setstacktabbedgap{9pt}
\parenMatrixstack{
0.03&  0&  0.01&  0&  0&  0&  0 \\
0&  0.11&  0&  0&  0&  0&  0 \\
0.01&  0&  0.03&  0&  0&  0&  0 \\
0&  0&  0&  0.04&  0&  0&  0 \\
0&  0&  0&  0&  0.03&  0&  0 \\
0&  0&  0&  0&  0&  0.02&  0 \\
0&  0&  0&  0&  0&  0&  0.03
}$,
$\occmat^{\downarrow}=
\setstackgap{L}{1.1\baselineskip}
\fixTABwidth{T}
\parenMatrixstack{
0.04&  0&  0.01&  0&  0&  0&  0 \\
0&  0.13&  0&  0&  0&  0&  0 \\
0.01&  0&  0.04&  0&  0&  0&  0 \\
0&  0&  0&  0.05&  0.03&  0.01&  0 \\
0&  0&  0&  0.03&  0.99&  -0.01&  0 \\
0&  0&  0&  0.01&  -0.01&  0.03&  0.01 \\
0&  0&  0&  0&  0&  0.01&  0.99
}$

\subsubsection{Properties of S$_5^2$ (1\textbf{k} AFM without SOC)}

\begin{table*}[!h]%
\begin{ruledtabular}
\begin{tabular}{ll}
\textrm{Properties (unit)}&
\textrm{Values}\\

\colrule
Energy (eV) &  -116.174 \\
Basis vectors (\AA)& $a=(
5.5534,  0,  0
)$, $b=(
0,  5.5514,  0
)$, $c=(
0,  0,  5.5237
)$\\
\begin{tabular}{@{}l@{}}Atom coordinates \\ (in basis vectors)\end{tabular} & {\begin{tabular}{@{}l@{}}U1:$(
0.5000,  0,  0 )$, U2:$(
0.5000,  0.5000,  0.5000 )$, \\U3:$(
0,  0,  0.5000 )$, U4:$(
0,  0.5000,  0 )$, \\O1:$(
0.2500,  0.7500,  0.7500 )$, O2:$(
0.2500,  0.2500,  0.7500 )$, \\O3:$(
0.2500,  0.2500,  0.2500 )$, O4:$(
0.2500,  0.7500,  0.2500 )$, \\O5:$(
0.7500,  0.7500,  0.2500 )$, O6:$(
0.7500,  0.2500,  0.2500 )$, \\O7:$(
0.7500,  0.2500,  0.7500 )$, O8:$(
0.7500,  0.7500,  0.7500 )$\end{tabular}} \\ 
Strain & \begin{tabular}{@{}l@{}}$(\epsilon_{xx},\epsilon_{yy},\epsilon_{zz},\epsilon_{xy},\epsilon_{xz},\epsilon_{yz})=(
0.0019,  0.0015,  -0.0035,  0,  0,  0 )$, \\$(\epsilon_{A1g},\epsilon_{Eg.0},\epsilon_{Eg.1},\epsilon_{T2g.0},\epsilon_{T2g.1},\epsilon_{T2g.2})=(
0,  0.0002,  -0.0042,  0,  0,  0 )$.\end{tabular}\\
Spin moments ($\mu_B$)& U1:$(
2.011,  0,  0
)$, U2:$(
-2.011,  0,  0
)$\\
Energy (eV) and lattice (\AA) of und & 
-116.172, 5.5429
\end{tabular}
\end{ruledtabular}
\end{table*}

$\occmat(\textrm{U}1)$:

$\occmat^{\uparrow}=
\setstackgap{L}{1.1\baselineskip}
\fixTABwidth{T}
\parenMatrixstack{
0.04&  0&  0.01&  0&  0&  0&  0 \\
0&  0.13&  0&  0&  0&  0&  0 \\
0.01&  0&  0.03&  0&  0&  0&  0 \\
0&  0&  0&  0.99&  0&  0&  0 \\
0&  0&  0&  0&  0.03&  0&  -0.01 \\
0&  0&  0&  0&  0&  1.00&  0 \\
0&  0&  0&  0&  -0.01&  0&  0.05
}$,
$\occmat^{\downarrow}=
\setstackgap{L}{1.1\baselineskip}
\fixTABwidth{T}
\parenMatrixstack{
0.04&  0&  0.01&  0&  0&  0&  0 \\
0&  0.11&  0&  0&  0&  0&  0 \\
0.01&  0&  0.03&  0&  0&  0&  0 \\
0&  0&  0&  0.03&  0&  0&  0 \\
0&  0&  0&  0&  0.03&  0&  -0.01 \\
0&  0&  0&  0&  0&  0.02&  0 \\
0&  0&  0&  0&  -0.01&  0&  0.03
}$

$\occmat(\textrm{U}2)$:

$\occmat^{\uparrow}=
\setstackgap{L}{1.1\baselineskip}
\fixTABwidth{T}
\parenMatrixstack{
0.04&  0&  0.01&  0&  0&  0&  0 \\
0&  0.11&  0&  0&  0&  0&  0 \\
0.01&  0&  0.03&  0&  0&  0&  0 \\
0&  0&  0&  0.03&  0&  0&  0 \\
0&  0&  0&  0&  0.03&  0&  -0.01 \\
0&  0&  0&  0&  0&  0.02&  0 \\
0&  0&  0&  0&  -0.01&  0&  0.03
}$,
$\occmat^{\downarrow}=
\setstackgap{L}{1.1\baselineskip}
\fixTABwidth{T}
\parenMatrixstack{
0.04&  0&  0.01&  0&  0&  0&  0 \\
0&  0.13&  0&  0&  0&  0&  0 \\
0.01&  0&  0.03&  0&  0&  0&  0 \\
0&  0&  0&  0.99&  0&  0&  0 \\
0&  0&  0&  0&  0.03&  0&  -0.01 \\
0&  0&  0&  0&  0&  1.00&  0 \\
0&  0&  0&  0&  -0.01&  0&  0.05
}$
\clearpage
\subsubsection{Properties of S$_5^3$ (1\textbf{k} AFM without SOC)}

\begin{table*}[!h]%
\begin{ruledtabular}
\begin{tabular}{ll}
\textrm{Properties (unit)}&
\textrm{Values}\\

\colrule
Energy (eV) &  -116.143 \\
Basis vectors (\AA)& $a=(
5.5526,  0,  0
)$, $b=(
0,  5.5268,  0.0010
)$, $c=(
0,  0.0010,  5.5526
)$\\
\begin{tabular}{@{}l@{}}Atom coordinates \\ (in basis vectors)\end{tabular} & {\begin{tabular}{@{}l@{}}U1:$(
0.5000,  0,  0 )$, U2:$(
0.5000,  0.5000,  0.5000 )$, \\U3:$(
0,  0,  0.5000 )$, U4:$(
0,  0.5000,  0 )$, \\O1:$(
0.2500,  0.7500,  0.7500 )$, O2:$(
0.2500,  0.2500,  0.7500 )$, \\O3:$(
0.2500,  0.2500,  0.2500 )$, O4:$(
0.2500,  0.7500,  0.2500 )$, \\O5:$(
0.7500,  0.7500,  0.2500 )$, O6:$(
0.7500,  0.2500,  0.2500 )$, \\O7:$(
0.7500,  0.2500,  0.7500 )$, O8:$(
0.7500,  0.7500,  0.7500 )$\end{tabular}} \\ 
Strain & \begin{tabular}{@{}l@{}}$(\epsilon_{xx},\epsilon_{yy},\epsilon_{zz},\epsilon_{xy},\epsilon_{xz},\epsilon_{yz})=(
0.0015,  -0.0031,  0.0015,  0,  0,  0.0002 )$, \\$(\epsilon_{A1g},\epsilon_{Eg.0},\epsilon_{Eg.1},\epsilon_{T2g.0},\epsilon_{T2g.1},\epsilon_{T2g.2})=(
-0.0001,  0.0033,  0.0019,  0,  0.0002,  0 )$.\end{tabular}\\
Spin moments ($\mu_B$)& U1:$(
2.002,  0,  0
)$, U2:$(
-2.002,  0,  0
)$\\
Energy (eV) and lattice (\AA) of und & 
-116.142, 5.5441
\end{tabular}
\end{ruledtabular}
\end{table*}

$\occmat(\textrm{U}1)$:

$\occmat^{\uparrow}=
\setstackgap{L}{1.1\baselineskip}
\fixTABwidth{T}
\parenMatrixstack{
0.99&  0&  0&  0&  0&  0.01&  0 \\
0&  0.13&  0&  0&  0&  0&  0 \\
0&  0&  1.00&  0.01&  0&  0&  0 \\
0&  0&  0.01&  0.05&  0&  0&  0 \\
0&  0&  0&  0&  0.04&  0&  -0.01 \\
0.01&  0&  0&  0&  0&  0.03&  0 \\
0&  0&  0&  0&  -0.01&  0&  0.04
}$,
$\occmat^{\downarrow}=
\setstackgap{L}{1.1\baselineskip}
\fixTABwidth{T}
\parenMatrixstack{
0.03&  0&  0&  0&  0&  0&  0 \\
0&  0.11&  0&  0&  0&  0&  0 \\
0&  0&  0.02&  0&  0&  0&  0 \\
0&  0&  0&  0.04&  0&  0&  0 \\
0&  0&  0&  0&  0.03&  0&  -0.01 \\
0&  0&  0&  0&  0&  0.03&  0 \\
0&  0&  0&  0&  -0.01&  0&  0.03
}$

$\occmat(\textrm{U}2)$:

$\occmat^{\uparrow}=
\setstackgap{L}{1.1\baselineskip}
\fixTABwidth{T}
\parenMatrixstack{
0.03&  0&  0&  0&  0&  0&  0 \\
0&  0.11&  0&  0&  0&  0&  0 \\
0&  0&  0.02&  0&  0&  0&  0 \\
0&  0&  0&  0.04&  0&  0&  0 \\
0&  0&  0&  0&  0.03&  0&  -0.01 \\
0&  0&  0&  0&  0&  0.03&  0 \\
0&  0&  0&  0&  -0.01&  0&  0.03
}$,
$\occmat^{\downarrow}=
\setstackgap{L}{1.1\baselineskip}
\fixTABwidth{T}
\parenMatrixstack{
0.99&  0&  0&  0&  0&  0.01&  0 \\
0&  0.13&  0&  0&  0&  0&  0 \\
0&  0&  1.00&  0.01&  0&  0&  0 \\
0&  0&  0.01&  0.05&  0&  0&  0 \\
0&  0&  0&  0&  0.04&  0&  -0.01 \\
0.01&  0&  0&  0&  0&  0.03&  0 \\
0&  0&  0&  0&  -0.01&  0&  0.04
}$

\clearpage

\subsection{\label{sec:sm3kafmnosoc}3\textbf{k} AFM without SOC}
Here, we document the results for 3\textbf{k} AFM without SOC, which, though not presented
in our paper, is included here for the sake of completeness. It should be noted that these
calculations are extremely challenging to converge,
and we were only successful in obtaining one converged state using GGA+$U$ ($U=4$ eV).

\begin{table*}[!h]%
\begin{ruledtabular}
\begin{tabular}{ll}
\textrm{Properties (unit)}&
\textrm{Values}\\

\colrule
Energy (eV) &  -116.257 \\
Basis vectors (\AA)& $a=(
5.5445,  -0.0001,  -0.0001
)$, $b=(
-0.0001,  5.5447,  0
)$, $c=(
-0.0001,  0,  5.5445
)$\\
\begin{tabular}{@{}l@{}}Atom coordinates \\ (in basis vectors)\end{tabular} & {\begin{tabular}{@{}l@{}}U1:$(
0.5000,  0,  0 )$, U2:$(
0.5000,  0.5000,  0.5000 )$, \\U3:$(
0,  0,  0.5000 )$, U4:$(
0,  0.5000,  0 )$, \\O1:$(
0.2526,  0.7525,  0.7475 )$, O2:$(
0.2525,  0.2525,  0.7525 )$, \\O3:$(
0.2475,  0.2525,  0.2525 )$, O4:$(
0.2475,  0.7525,  0.2475 )$, \\O5:$(
0.7475,  0.7475,  0.2475 )$, O6:$(
0.7474,  0.2475,  0.2525 )$, \\O7:$(
0.7525,  0.2475,  0.7525 )$, O8:$(
0.7525,  0.7475,  0.7475 )$\end{tabular}} \\ 
Strain & \begin{tabular}{@{}l@{}}$(\epsilon_{xx},\epsilon_{yy},\epsilon_{zz},\epsilon_{xy},\epsilon_{xz},\epsilon_{yz})=(
-0.0002,  -0.0001,  -0.0002,  0,  0,  0 )$, \\$(\epsilon_{A1g},\epsilon_{Eg.0},\epsilon_{Eg.1},\epsilon_{T2g.0},\epsilon_{T2g.1},\epsilon_{T2g.2})=(
-0.0003,  0,  0,  0,  0,  0 )$.\end{tabular}\\
Spin moments ($\mu_B$)& \begin{tabular}{@{}l@{}} U1:$(
1.031,  -1.309,  -1.127
)$, U2:$(
-1.119,  1.167,  -1.196
)$,\\ U3:$(
1.089,  1.193,  1.198
)$, U4:$(
-1.102,  -1.209,  1.170
)$\end{tabular}\\
Energy (eV) and lattice (\AA) of und & -116.248, 5.5455
\end{tabular}
\end{ruledtabular}
\end{table*}

$\occmat(\textrm{U}1)$:

$\occmat^{\uparrow\uparrow}=
\setstackgap{L}{1.1\baselineskip}
\fixTABwidth{T}
\parenMatrixstack{
0.11 &    0 & 0.02 & -0.04 & -0.01 &    0 & 0.07 \\
   0 & 0.11 &    0 & -0.01 &    0 &    0 &    0 \\
0.02 &    0 & 0.09 & -0.05 & -0.02 & -0.04 & 0.01 \\
-0.04 & -0.01 & -0.05 & 0.12 & 0.06 & 0.01 & -0.03 \\
-0.01 &    0 & -0.02 & 0.06 & 0.10 & -0.04 & -0.02 \\
   0 &    0 & -0.04 & 0.01 & -0.04 & 0.09 &    0 \\
0.07 &    0 & 0.01 & -0.03 & -0.02 &    0 & 0.10
}+
\setstackgap{L}{1.1\baselineskip}
\fixTABwidth{T}
\parenMatrixstack{
   0 &    0 & 0.02 &    0 & 0.03 & -0.04 &    0 \\
   0 &    0 &    0 &    0 &    0 &    0 &    0 \\
-0.02 &    0 &    0 & -0.01 & -0.04 & 0.03 & -0.02 \\
   0 &    0 & 0.01 &    0 & 0.02 & -0.03 &    0 \\
-0.03 &    0 & 0.04 & -0.02 &    0 & -0.04 & -0.02 \\
0.04 &    0 & -0.03 & 0.03 & 0.04 &    0 & 0.04 \\
   0 &    0 & 0.02 &    0 & 0.02 & -0.04 &    0 
}i$,

$\occmat^{\uparrow\downarrow}=
\setstackgap{L}{1.1\baselineskip}
\fixTABwidth{T}
\parenMatrixstack{
0.09 &    0 & -0.03 & -0.04 & -0.06 & 0.06 & 0.08 \\
   0 & 0.01 &    0 &    0 &    0 &    0 &    0 \\
0.05 &    0 & 0.07 & -0.04 & 0.04 & -0.09 & 0.05 \\
-0.05 & -0.01 & -0.08 & 0.10 & 0.04 & 0.04 & -0.05 \\
0.03 &    0 & -0.08 & 0.09 & 0.09 & 0.01 & 0.02 \\
-0.07 &    0 &    0 & -0.03 & -0.10 & 0.07 & -0.06 \\
0.09 &    0 & -0.02 & -0.04 & -0.05 & 0.06 & 0.08 
}+
\setstackgap{L}{1.1\baselineskip}
\fixTABwidth{T}
\parenMatrixstack{
0.12 &    0 & 0.04 & -0.06 & 0.02 & -0.05 & 0.11 \\
0.01 & 0.01 &    0 & -0.01 &    0 &    0 & 0.01 \\
-0.02 &    0 & 0.09 & -0.10 & -0.08 & -0.02 & -0.01 \\
-0.06 & -0.01 & -0.07 & 0.12 & 0.11 & -0.02 & -0.05 \\
-0.05 &    0 & 0.02 & 0.07 & 0.11 & -0.10 & -0.04 \\
0.05 &    0 & -0.09 & 0.04 & -0.01 & 0.09 & 0.04 \\
0.11 &    0 & 0.04 & -0.06 & 0.01 & -0.05 & 0.10
}i$,

$\occmat^{\downarrow\uparrow}=
\setstackgap{L}{1.1\baselineskip}
\fixTABwidth{T}
\parenMatrixstack{
0.09 &    0 & 0.05 & -0.05 & 0.03 & -0.07 & 0.09 \\
   0 & 0.01 &    0 & -0.01 &    0 &    0 &    0 \\
-0.03 &    0 & 0.07 & -0.08 & -0.08 &    0 & -0.02 \\
-0.04 &    0 & -0.04 & 0.10 & 0.09 & -0.03 & -0.04 \\
-0.06 &    0 & 0.04 & 0.04 & 0.09 & -0.10 & -0.05 \\
0.06 &    0 & -0.09 & 0.04 & 0.01 & 0.07 & 0.06 \\
0.08 &    0 & 0.05 & -0.05 & 0.02 & -0.06 & 0.08
}+
\setstackgap{L}{1.1\baselineskip}
\fixTABwidth{T}
\parenMatrixstack{
-0.12 & -0.01 & 0.02 & 0.06 & 0.05 & -0.05 & -0.11 \\
   0 & -0.01 &    0 & 0.01 &    0 &    0 &    0 \\
-0.04 &    0 & -0.09 & 0.07 & -0.02 & 0.09 & -0.04 \\
0.06 & 0.01 & 0.10 & -0.12 & -0.07 & -0.04 & 0.06 \\
-0.02 &    0 & 0.08 & -0.11 & -0.11 & 0.01 & -0.01 \\
0.05 &    0 & 0.02 & 0.02 & 0.10 & -0.09 & 0.05 \\
-0.11 & -0.01 & 0.01 & 0.05 & 0.04 & -0.04 & -0.10
}i$,

$\occmat^{\downarrow\downarrow}=
\setstackgap{L}{1.1\baselineskip}
\fixTABwidth{T}
\parenMatrixstack{
0.32 & 0.01 & 0.04 & -0.13 & -0.04 &    0 & 0.26 \\
0.01 & 0.13 & 0.01 & -0.02 & -0.01 &    0 & 0.01 \\
0.04 & 0.01 & 0.25 & -0.19 & -0.07 & -0.13 & 0.03 \\
-0.13 & -0.02 & -0.19 & 0.33 & 0.20 & 0.02 & -0.13 \\
-0.04 & -0.01 & -0.07 & 0.20 & 0.30 & -0.13 & -0.04 \\
   0 &    0 & -0.13 & 0.02 & -0.13 & 0.24 & -0.01 \\
0.26 & 0.01 & 0.03 & -0.13 & -0.04 & -0.01 & 0.27
}+
\setstackgap{L}{1.1\baselineskip}
\fixTABwidth{T}
\parenMatrixstack{
   0 &    0 & 0.09 &    0 & 0.10 & -0.16 & 0.01 \\
   0 &    0 &    0 &    0 &    0 &    0 &    0 \\
-0.09 &    0 &    0 & -0.05 & -0.15 & 0.11 & -0.08 \\
   0 &    0 & 0.05 &    0 & 0.06 & -0.09 & 0.01 \\
-0.10 &    0 & 0.15 & -0.06 &    0 & -0.13 & -0.09 \\
0.16 &    0 & -0.11 & 0.09 & 0.13 &    0 & 0.14 \\
-0.01 &    0 & 0.08 & -0.01 & 0.09 & -0.14 &    0
}i$.

$\occmat(\textrm{U}2)$:

$\occmat^{\uparrow\uparrow}=
\setstackgap{L}{1.1\baselineskip}
\fixTABwidth{T}
\parenMatrixstack{
0.11 &    0 & 0.01 & 0.03 & -0.01 &    0 & 0.07 \\
   0 & 0.11 &    0 & -0.01 &    0 &    0 &    0 \\
0.01 &    0 & 0.09 & 0.05 & -0.02 & 0.03 & 0.01 \\
0.03 & -0.01 & 0.05 & 0.11 & -0.05 & 0.01 & 0.03 \\
-0.01 &    0 & -0.02 & -0.05 & 0.10 & 0.03 & -0.02 \\
   0 &    0 & 0.03 & 0.01 & 0.03 & 0.08 &    0 \\
0.07 &    0 & 0.01 & 0.03 & -0.02 &    0 & 0.10 
}+
\setstackgap{L}{1.1\baselineskip}
\fixTABwidth{T}
\parenMatrixstack{
   0 &    0 & 0.02 &    0 & 0.03 & 0.04 &    0 \\
   0 &    0 &    0 &    0 &    0 &    0 &    0 \\
-0.02 &    0 &    0 & 0.01 & -0.04 & -0.03 & -0.02 \\
   0 &    0 & -0.01 &    0 & -0.02 & -0.02 &    0 \\
-0.03 &    0 & 0.04 & 0.02 &    0 & 0.03 & -0.02 \\
-0.04 &    0 & 0.03 & 0.02 & -0.03 &    0 & -0.04 \\
   0 &    0 & 0.02 &    0 & 0.02 & 0.04 &    0 
}i$,

$\occmat^{\uparrow\downarrow}=
\setstackgap{L}{1.1\baselineskip}
\fixTABwidth{T}
\parenMatrixstack{
-0.10 &    0 & 0.02 & -0.05 & 0.05 & 0.06 & -0.09 \\
   0 & -0.01 &    0 & 0.01 &    0 &    0 &    0 \\
-0.04 &    0 & -0.08 & -0.05 & -0.03 & -0.09 & -0.04 \\
-0.05 & 0.01 & -0.09 & -0.10 & 0.05 & -0.04 & -0.05 \\
-0.02 &    0 & 0.08 & 0.10 & -0.10 &    0 & -0.02 \\
-0.06 &    0 & -0.01 & 0.02 & -0.09 & -0.08 & -0.06 \\
-0.09 &    0 & 0.02 & -0.04 & 0.05 & 0.05 & -0.09
}+
\setstackgap{L}{1.1\baselineskip}
\fixTABwidth{T}
\parenMatrixstack{
-0.11 &    0 & -0.04 & -0.05 & -0.02 & -0.06 & -0.10 \\
   0 & -0.01 &    0 & 0.01 &    0 &    0 &    0 \\
0.02 &    0 & -0.08 & -0.09 & 0.08 & -0.01 & 0.02 \\
-0.05 & 0.01 & -0.06 & -0.11 & 0.10 & 0.02 & -0.05 \\
0.05 &    0 & -0.02 & 0.05 & -0.10 & -0.09 & 0.05 \\
0.06 &    0 & -0.09 & -0.04 &    0 & -0.08 & 0.05 \\
-0.09 &    0 & -0.04 & -0.05 & -0.02 & -0.05 & -0.09
}i$,

$\occmat^{\downarrow\uparrow}=
\setstackgap{L}{1.1\baselineskip}
\fixTABwidth{T}
\parenMatrixstack{
-0.10 &    0 & -0.04 & -0.05 & -0.02 & -0.06 & -0.09 \\
   0 & -0.01 &    0 & 0.01 &    0 &    0 &    0 \\
0.02 &    0 & -0.08 & -0.09 & 0.08 & -0.01 & 0.02 \\
-0.05 & 0.01 & -0.05 & -0.10 & 0.10 & 0.02 & -0.04 \\
0.05 &    0 & -0.03 & 0.05 & -0.10 & -0.09 & 0.05 \\
0.06 &    0 & -0.09 & -0.04 &    0 & -0.08 & 0.05 \\
-0.09 &    0 & -0.04 & -0.05 & -0.02 & -0.06 & -0.09
}+
\setstackgap{L}{1.1\baselineskip}
\fixTABwidth{T}
\parenMatrixstack{
0.11 &    0 & -0.02 & 0.05 & -0.05 & -0.06 & 0.09 \\
   0 & 0.01 &    0 & -0.01 &    0 &    0 &    0 \\
0.04 &    0 & 0.08 & 0.06 & 0.02 & 0.09 & 0.04 \\
0.05 & -0.01 & 0.09 & 0.11 & -0.05 & 0.04 & 0.05 \\
0.02 &    0 & -0.08 & -0.10 & 0.10 &    0 & 0.02 \\
0.06 &    0 & 0.01 & -0.02 & 0.09 & 0.08 & 0.05 \\
0.10 &    0 & -0.02 & 0.05 & -0.05 & -0.05 & 0.09
}i$,

$\occmat^{\downarrow\downarrow}=
\setstackgap{L}{1.1\baselineskip}
\fixTABwidth{T}
\parenMatrixstack{
0.32 & -0.01 & 0.04 & 0.14 & -0.05 &    0 & 0.26 \\
-0.01 & 0.13 & -0.01 & -0.02 & 0.01 &    0 & -0.01 \\
0.04 & -0.01 & 0.26 & 0.20 & -0.08 & 0.14 & 0.03 \\
0.14 & -0.02 & 0.20 & 0.33 & -0.21 & 0.02 & 0.13 \\
-0.05 & 0.01 & -0.08 & -0.21 & 0.30 & 0.13 & -0.05 \\
   0 &    0 & 0.14 & 0.02 & 0.13 & 0.25 & 0.01 \\
0.26 & -0.01 & 0.03 & 0.13 & -0.05 & 0.01 & 0.28 
}+
\setstackgap{L}{1.1\baselineskip}
\fixTABwidth{T}
\parenMatrixstack{
   0 &    0 & 0.09 & 0.01 & 0.11 & 0.16 & 0.01 \\
   0 &    0 &    0 &    0 &    0 &    0 &    0 \\
-0.09 &    0 &    0 & 0.05 & -0.15 & -0.11 & -0.08 \\
-0.01 &    0 & -0.05 &    0 & -0.06 & -0.09 & -0.01 \\
-0.11 &    0 & 0.15 & 0.06 &    0 & 0.13 & -0.09 \\
-0.16 &    0 & 0.11 & 0.09 & -0.13 &    0 & -0.14 \\
-0.01 &    0 & 0.08 & 0.01 & 0.09 & 0.14 &    0
}i$.

$\occmat(\textrm{U}3)$:

$\occmat^{\uparrow\uparrow}=
\setstackgap{L}{1.1\baselineskip}
\fixTABwidth{T}
\parenMatrixstack{
0.32 & -0.01 & 0.04 & -0.14 & 0.04 &    0 & -0.26 \\
-0.01 & 0.13 & -0.01 & 0.02 & -0.01 &    0 & 0.01 \\
0.04 & -0.01 & 0.26 & -0.20 & 0.08 & -0.14 & -0.03 \\
-0.14 & 0.02 & -0.20 & 0.33 & -0.21 & 0.02 & 0.13 \\
0.04 & -0.01 & 0.08 & -0.21 & 0.30 & 0.13 & -0.05 \\
   0 &    0 & -0.14 & 0.02 & 0.13 & 0.25 & 0.01 \\
-0.26 & 0.01 & -0.03 & 0.13 & -0.05 & 0.01 & 0.28
}+
\setstackgap{L}{1.1\baselineskip}
\fixTABwidth{T}
\parenMatrixstack{
   0 &    0 & 0.09 &    0 & -0.10 & -0.16 & -0.01 \\
   0 &    0 &    0 &    0 &    0 &    0 &    0 \\
-0.09 &    0 &    0 & -0.05 & 0.15 & 0.11 & 0.08 \\
   0 &    0 & 0.05 &    0 & -0.06 & -0.09 & -0.01 \\
0.10 &    0 & -0.15 & 0.06 &    0 & 0.13 & -0.09 \\
0.16 &    0 & -0.11 & 0.09 & -0.13 &    0 & -0.14 \\
0.01 &    0 & -0.08 & 0.01 & 0.09 & 0.14 &    0 
}i$,

$\occmat^{\uparrow\downarrow}=
\setstackgap{L}{1.1\baselineskip}
\fixTABwidth{T}
\parenMatrixstack{
0.10 &    0 & 0.04 & -0.05 & -0.02 & -0.06 & -0.09 \\
   0 & 0.01 &    0 & 0.01 &    0 &    0 &    0 \\
-0.02 &    0 & 0.08 & -0.09 & 0.08 &    0 & 0.02 \\
-0.04 & 0.01 & -0.05 & 0.10 & -0.10 & -0.03 & 0.04 \\
0.05 &    0 & -0.03 & -0.05 & 0.09 & 0.09 & -0.05 \\
0.06 &    0 & -0.09 & 0.04 &    0 & 0.08 & -0.05 \\
-0.09 &    0 & -0.04 & 0.05 & 0.02 & 0.06 & 0.08
}+
\setstackgap{L}{1.1\baselineskip}
\fixTABwidth{T}
\parenMatrixstack{
-0.11 &    0 & 0.02 & 0.05 & -0.05 & -0.05 & 0.10 \\
   0 & -0.01 &    0 & -0.01 &    0 &    0 &    0 \\
-0.04 &    0 & -0.09 & 0.06 & 0.02 & 0.09 & 0.04 \\
0.05 & -0.01 & 0.09 & -0.11 & 0.06 & -0.04 & -0.05 \\
0.02 &    0 & -0.08 & 0.10 & -0.10 & -0.01 & -0.02 \\
0.05 &    0 & 0.01 & 0.02 & -0.09 & -0.08 & -0.05 \\
0.10 &    0 & -0.02 & -0.05 & 0.05 & 0.05 & -0.09
}i$,

$\occmat^{\downarrow\uparrow}=
\setstackgap{L}{1.1\baselineskip}
\fixTABwidth{T}
\parenMatrixstack{
0.10 &    0 & -0.02 & -0.04 & 0.05 & 0.06 & -0.09 \\
   0 & 0.01 &    0 & 0.01 &    0 &    0 &    0 \\
0.04 &    0 & 0.08 & -0.05 & -0.03 & -0.09 & -0.04 \\
-0.05 & 0.01 & -0.09 & 0.10 & -0.05 & 0.04 & 0.05 \\
-0.02 &    0 & 0.08 & -0.10 & 0.09 &    0 & 0.02 \\
-0.06 &    0 &    0 & -0.03 & 0.09 & 0.08 & 0.06 \\
-0.09 &    0 & 0.02 & 0.04 & -0.05 & -0.05 & 0.08
}+
\setstackgap{L}{1.1\baselineskip}
\fixTABwidth{T}
\parenMatrixstack{
0.11 &    0 & 0.04 & -0.05 & -0.02 & -0.05 & -0.10 \\
   0 & 0.01 &    0 & 0.01 &    0 &    0 &    0 \\
-0.02 &    0 & 0.09 & -0.09 & 0.08 & -0.01 & 0.02 \\
-0.05 & 0.01 & -0.06 & 0.11 & -0.10 & -0.02 & 0.05 \\
0.05 &    0 & -0.02 & -0.06 & 0.10 & 0.09 & -0.05 \\
0.05 &    0 & -0.09 & 0.04 & 0.01 & 0.08 & -0.05 \\
-0.10 &    0 & -0.04 & 0.05 & 0.02 & 0.05 & 0.09
}i$,

$\occmat^{\downarrow\downarrow}=
\setstackgap{L}{1.1\baselineskip}
\fixTABwidth{T}
\parenMatrixstack{
0.10 &    0 & 0.01 & -0.03 & 0.01 &    0 & -0.07 \\
   0 & 0.11 &    0 & 0.01 &    0 &    0 &    0 \\
0.01 &    0 & 0.09 & -0.05 & 0.02 & -0.03 & -0.01 \\
-0.03 & 0.01 & -0.05 & 0.11 & -0.05 & 0.01 & 0.03 \\
0.01 &    0 & 0.02 & -0.05 & 0.10 & 0.03 & -0.02 \\
   0 &    0 & -0.03 & 0.01 & 0.03 & 0.08 &    0 \\
-0.07 &    0 & -0.01 & 0.03 & -0.02 &    0 & 0.10
}+
\setstackgap{L}{1.1\baselineskip}
\fixTABwidth{T}
\parenMatrixstack{
   0 &    0 & 0.02 &    0 & -0.03 & -0.04 &    0 \\
   0 &    0 &    0 &    0 &    0 &    0 &    0 \\
-0.02 &    0 &    0 & -0.01 & 0.04 & 0.03 & 0.02 \\
   0 &    0 & 0.01 &    0 & -0.02 & -0.02 &    0 \\
0.03 &    0 & -0.04 & 0.02 &    0 & 0.03 & -0.02 \\
0.04 &    0 & -0.03 & 0.02 & -0.03 &    0 & -0.04 \\
   0 &    0 & -0.02 &    0 & 0.02 & 0.04 &    0
}i$.

$\occmat(\textrm{U}4)$:

$\occmat^{\uparrow\uparrow}=
\setstackgap{L}{1.1\baselineskip}
\fixTABwidth{T}
\parenMatrixstack{
0.32 & 0.01 & 0.04 & 0.14 & 0.04 &    0 & -0.26 \\
0.01 & 0.13 & 0.01 & 0.02 & 0.01 &    0 & -0.01 \\
0.04 & 0.01 & 0.25 & 0.20 & 0.08 & 0.14 & -0.03 \\
0.14 & 0.02 & 0.20 & 0.33 & 0.21 & 0.02 & -0.13 \\
0.04 & 0.01 & 0.08 & 0.21 & 0.30 & -0.13 & -0.05 \\
   0 &    0 & 0.14 & 0.02 & -0.13 & 0.25 & -0.01 \\
-0.26 & -0.01 & -0.03 & -0.13 & -0.05 & -0.01 & 0.28
}+
\setstackgap{L}{1.1\baselineskip}
\fixTABwidth{T}
\parenMatrixstack{
   0 &    0 & 0.09 & 0.01 & -0.10 & 0.16 & -0.01 \\
   0 &    0 &    0 &    0 &    0 &    0 &    0 \\
-0.09 &    0 &    0 & 0.05 & 0.15 & -0.11 & 0.08 \\
-0.01 &    0 & -0.05 &    0 & 0.06 & -0.09 & 0.01 \\
0.10 &    0 & -0.15 & -0.06 &    0 & -0.13 & -0.09 \\
-0.16 &    0 & 0.11 & 0.09 & 0.13 &    0 & 0.14 \\
0.01 &    0 & -0.08 & -0.01 & 0.09 & -0.14 &    0
}i$,

$\occmat^{\uparrow\downarrow}=
\setstackgap{L}{1.1\baselineskip}
\fixTABwidth{T}
\parenMatrixstack{
-0.10 &    0 & -0.04 & -0.05 & 0.03 & -0.06 & 0.09 \\
   0 & -0.01 &    0 & -0.01 &    0 &    0 &    0 \\
0.02 &    0 & -0.08 & -0.09 & -0.08 &    0 & -0.02 \\
-0.05 & -0.01 & -0.05 & -0.10 & -0.10 & 0.03 & 0.04 \\
-0.06 &    0 & 0.03 & -0.05 & -0.10 & 0.09 & 0.05 \\
0.06 &    0 & -0.09 & -0.04 &    0 & -0.08 & -0.05 \\
0.09 &    0 & 0.04 & 0.05 & -0.02 & 0.06 & -0.09
}+
\setstackgap{L}{1.1\baselineskip}
\fixTABwidth{T}
\parenMatrixstack{
0.11 &    0 & -0.02 & 0.05 & 0.05 & -0.05 & -0.10 \\
   0 & 0.01 &    0 & 0.01 &    0 &    0 &    0 \\
0.04 &    0 & 0.09 & 0.06 & -0.02 & 0.09 & -0.04 \\
0.05 & 0.01 & 0.09 & 0.11 & 0.06 & 0.04 & -0.05 \\
-0.02 &    0 & 0.08 & 0.10 & 0.10 & -0.01 & 0.02 \\
0.06 &    0 & 0.01 & -0.02 & -0.09 & 0.08 & -0.05 \\
-0.10 &    0 & 0.02 & -0.05 & -0.05 & 0.04 & 0.09
}i$,

$\occmat^{\downarrow\uparrow}=
\setstackgap{L}{1.1\baselineskip}
\fixTABwidth{T}
\parenMatrixstack{
-0.10 &    0 & 0.02 & -0.05 & -0.06 & 0.06 & 0.09 \\
   0 & -0.01 &    0 & -0.01 &    0 &    0 &    0 \\
-0.04 &    0 & -0.08 & -0.05 & 0.03 & -0.09 & 0.04 \\
-0.05 & -0.01 & -0.09 & -0.10 & -0.05 & -0.04 & 0.05 \\
0.03 &    0 & -0.08 & -0.10 & -0.10 &    0 & -0.02 \\
-0.06 &    0 &    0 & 0.03 & 0.09 & -0.08 & 0.06 \\
0.09 &    0 & -0.02 & 0.04 & 0.05 & -0.05 & -0.09 
}+
\setstackgap{L}{1.1\baselineskip}
\fixTABwidth{T}
\parenMatrixstack{
-0.11 &    0 & -0.04 & -0.05 & 0.02 & -0.06 & 0.10 \\
   0 & -0.01 &    0 & -0.01 &    0 &    0 &    0 \\
0.02 &    0 & -0.09 & -0.09 & -0.08 & -0.01 & -0.02 \\
-0.05 & -0.01 & -0.06 & -0.11 & -0.10 & 0.02 & 0.05 \\
-0.05 &    0 & 0.02 & -0.06 & -0.10 & 0.09 & 0.05 \\
0.05 &    0 & -0.09 & -0.04 & 0.01 & -0.08 & -0.04 \\
0.10 &    0 & 0.04 & 0.05 & -0.02 & 0.05 & -0.09
}i$,

$\occmat^{\downarrow\downarrow}=
\setstackgap{L}{1.1\baselineskip}
\fixTABwidth{T}
\parenMatrixstack{
0.11 &    0 & 0.01 & 0.03 & 0.01 &    0 & -0.07 \\
   0 & 0.11 &    0 & 0.01 &    0 &    0 &    0 \\
0.01 &    0 & 0.09 & 0.05 & 0.02 & 0.04 & -0.01 \\
0.03 & 0.01 & 0.05 & 0.11 & 0.05 & 0.01 & -0.03 \\
0.01 &    0 & 0.02 & 0.05 & 0.10 & -0.03 & -0.02 \\
   0 &    0 & 0.04 & 0.01 & -0.03 & 0.08 &    0 \\
-0.07 &    0 & -0.01 & -0.03 & -0.02 &    0 & 0.10
}+
\setstackgap{L}{1.1\baselineskip}
\fixTABwidth{T}
\parenMatrixstack{
   0 &    0 & 0.02 &    0 & -0.03 & 0.04 &    0 \\
   0 &    0 &    0 &    0 &    0 &    0 &    0 \\
-0.02 &    0 &    0 & 0.01 & 0.04 & -0.03 & 0.02 \\
   0 &    0 & -0.01 &    0 & 0.02 & -0.02 &    0 \\
0.03 &    0 & -0.04 & -0.02 &    0 & -0.03 & -0.02 \\
-0.04 &    0 & 0.03 & 0.02 & 0.03 &    0 & 0.04 \\
   0 &    0 & -0.02 &    0 & 0.02 & -0.04 &    0
}i$.

\clearpage

\subsection{\label{sec:smfmsoc}FM with SOC (the stage 2)}

In the stage 2 of the ground state search, which initializes from all 25 FM states in Section~\ref{sec:smfmnosoc}, we obtained eight states in four energy classes by using GGA+$U$ (U=4eV), denoted as $\mathbb{S}_0$-$\mathbb{S}_3$. The state $\mathbb{S}_n^{m}$ represents the $m$-th state in the energy class $\mathbb{S}_n$, and the lowest-energy class is represented by $\mathbb{S}_0$. 

\begin{table*}[h]%
\caption{\label{tab:tables3}%
Results for FM states found in the stage 2 of our ground state search, using GGA+$U$ ($U=4$ eV) with SOC. 
The results are sorted by increasing relative energy with
respect to the fully relaxed crystal having occupation matrix $\mathbb{S}_0$ and energy $E_{min}$.}
\begin{ruledtabular}
\begin{tabular}{lccccr}
\textrm{State}&
\textrm{Initialized from states}&
\textrm{\begin{tabular}{@{}c@{}}$E_{und}-E_{min}$ \\ (meV/UO\textsubscript{2})\end{tabular} }&
\textrm{\begin{tabular}{@{}c@{}}$E_{dis}-E_{min}$ \\ (meV/UO\textsubscript{2})\end{tabular} }&
\textrm{{\begin{tabular}{@{}c@{}}External distortion \\ ($\epsilon_{xx},\epsilon_{yy},\epsilon_{zz},\epsilon_{xy},\epsilon_{xz},\epsilon_{yz}$,)$\times10^3$\end{tabular} } }&
\textrm{\begin{tabular}{@{}c@{}}Internal distortion \\ oxygen displacement\end{tabular} }\\
\colrule
$\mathbb{S}_0^{1}$ & S$_0^{1-6}$, S$_1^2$, S$_2^{4,6}$, S$_4^2$, S$_5^{1-3}$, S$_6$, S$_8$ & 0.3 & 0 & (-1, -1, 2, 0, 0, 0) & - \\
$\mathbb{S}_1^{1}$ & S$_2^1$, S$_7^1$ & 68.9 & 67.7 & (-5, 2, 3, 0, 0, 0)& - \\
$\mathbb{S}_1^{2}$ & S$_2^5$, S$_4^1$ & 68.9 & 67.7 & (2, -5, 3, 0, 0, 0)& - \\
$\mathbb{S}_2^{1}$ & S$_1^1$ & 72.3 & 70.8 & (-4, 5, 0, 0, 0, 0)& - \\
$\mathbb{S}_2^{2}$ & S$_1^3$ & 72.3 & 70.8 & (5, -4, 0, 0, 0, 0)& - \\
$\mathbb{S}_2^{3}$ & S$_2^2$ & 72.3 & 70.8 & (-4, 5, 0, 0, 0, 0)& - \\
$\mathbb{S}_2^{4}$ & S$_2^3$ & 72.3 & 70.8 & (5, -4, 0, 0, 0, 0)& - \\
$\mathbb{S}_3^{1}$ & S$_3^1$ & 72.9 & 71.5 & (-3, -3, 5, 0, 0, 0)& - \\
\end{tabular}
\end{ruledtabular}
\end{table*}

\subsubsection{Properties of $\mathbb{S}_0^{1}$}

\begin{table*}[!h]%
\begin{ruledtabular}
\begin{tabular}{ll}
\textrm{Properties (unit)}&
\textrm{Values}\\
\colrule
Initiated from states & S$_0^{1-6}$, S$_1^2$, S$_2^{4,6}$, S$_4^2$, S$_5^{1-3}$, S$_6^1$, S$_8^1$ \\
Energy (eV) & -32.058 \\
Basis vectors (\AA)& $a=(
2.7711, 2.7711,     0
), b=(
0, 2.7711, 2.7801
), c=(
2.7711,     0, 2.7801
)$\\
\begin{tabular}{@{}l@{}}Atom coordinates \\ (in basis vectors)\end{tabular} &
{\begin{tabular}{@{}l@{}}$U1=(
0.5000, 0.5000, 0.5000
)$ \\$O1=(
0.2500, 0.2500, 0.2500
)$, $O2=(
0.7500, 0.7500, 0.7500
)$\end{tabular}} \\ 
Strain & \begin{tabular}{@{}l@{}}($\epsilon_{xx},\epsilon_{yy},\epsilon_{zz},\epsilon_{xy},\epsilon_{xz},\epsilon_{yz}$)=(
-0.0011, -0.0011, 0.0022,     0,     0,     0
), \\($\epsilon_{A1g},\epsilon_{Eg.0},\epsilon_{Eg.1},\epsilon_{T2g.0},\epsilon_{T2g.1},\epsilon_{T2g.2}$)=(
 0,     0, 0.0026,     0,     0,     0
).\end{tabular}\\
Spin moments ($\mu_B$)& (-0.001, -0.001, 1.526)\\
Orbit moments ($\mu_B$)& (0.003, 0.003, -3.545)\\
Total magnetic moments ($\mu_B$)& (0.002, 0.002, -2.019)\\
Energy (eV) and lattice (\AA) of und & -32.058, 2.7731
\end{tabular}
\end{ruledtabular}
\end{table*}

$\occmat^{\uparrow\uparrow}=
\setstackgap{L}{1.1\baselineskip}
\setstacktabbedgap{9pt}
\fixTABwidth{T}
\parenMatrixstack{
0.48 &    0 &    0 &    0 &    0 &    0 &    0 \\
   0 & 0.13 &    0 &    0 &    0 &    0 &    0 \\
   0 &    0 & 0.43 &    0 &    0 &    0 &    0 \\
   0 &    0 &    0 & 0.05 &    0 &    0 &    0 \\
   0 &    0 &    0 &    0 & 0.43 &    0 &    0 \\
   0 &    0 &    0 &    0 &    0 & 0.03 &    0 \\
   0 &    0 &    0 &    0 &    0 &    0 & 0.48
}+$
$\setstackgap{L}{1.1\baselineskip}
\fixTABwidth{T}
\parenMatrixstack{
   0 &    0 &    0 &    0 &    0 &    0 & 0.44 \\
   0 &    0 &    0 &    0 &    0 &    0 &    0 \\
   0 &    0 &    0 &    0 & 0.39 &    0 &    0 \\
   0 &    0 &    0 &    0 &    0 &    0 &    0 \\
   0 &    0 & -0.39 &    0 &    0 &    0 &    0 \\
   0 &    0 &    0 &    0 &    0 &    0 &    0 \\
-0.44 &    0 &    0 &    0 &    0 &    0 &    0
}i$,

$\occmat^{\uparrow\downarrow}=
\setstackgap{L}{1.1\baselineskip}
\fixTABwidth{T}
\parenMatrixstack{
   0 &    0 &    0 &    0 &    0 &    0 &    0 \\
-0.09 &    0 &    0 &    0 &    0 &    0 &    0 \\
   0 &    0 &    0 &    0 &    0 &    0 &    0 \\
   0 &    0 &    0 &    0 & -0.26 &    0 &    0 \\
   0 &    0 &    0 &    0 &    0 &    0 &    0 \\
   0 &    0 &    0 &    0 &    0 &    0 & -0.15 \\
   0 &    0 &    0 &    0 &    0 &    0 &    0
}+
\setstackgap{L}{1.1\baselineskip}
\fixTABwidth{T}
\parenMatrixstack{
   0 &    0 &    0 &    0 &    0 &    0 &    0 \\
   0 &    0 &    0 &    0 &    0 &    0 & -0.09 \\
   0 &    0 &    0 &    0 &    0 &    0 &    0 \\
   0 &    0 & 0.26 &    0 &    0 &    0 &    0 \\
   0 &    0 &    0 &    0 &    0 &    0 &    0 \\
0.15 &    0 &    0 &    0 &    0 &    0 &    0 \\
   0 &    0 &    0 &    0 &    0 &    0 &    0
}i$,

$\occmat^{\downarrow\uparrow}=
\setstackgap{L}{1.1\baselineskip}
\fixTABwidth{T}
\parenMatrixstack{
   0 & -0.09 &    0 &    0 &    0 &    0 &    0 \\
   0 &    0 &    0 &    0 &    0 &    0 &    0 \\
   0 &    0 &    0 &    0 &    0 &    0 &    0 \\
   0 &    0 &    0 &    0 &    0 &    0 &    0 \\
   0 &    0 &    0 & -0.26 &    0 &    0 &    0 \\
   0 &    0 &    0 &    0 &    0 &    0 &    0 \\
   0 &    0 &    0 &    0 &    0 & -0.15 &    0
}+
\setstackgap{L}{1.1\baselineskip}
\fixTABwidth{T}
\parenMatrixstack{
   0 &    0 &    0 &    0 &    0 & -0.15 &    0 \\
   0 &    0 &    0 &    0 &    0 &    0 &    0 \\
   0 &    0 &    0 & -0.26 &    0 &    0 &    0 \\
   0 &    0 &    0 &    0 &    0 &    0 &    0 \\
   0 &    0 &    0 &    0 &    0 &    0 &    0 \\
   0 &    0 &    0 &    0 &    0 &    0 &    0 \\
   0 & 0.09 &    0 &    0 &    0 &    0 &    0
}i$,

$\occmat^{\downarrow\downarrow}=
\setstackgap{L}{1.1\baselineskip}
\fixTABwidth{T}
\parenMatrixstack{
0.03 &    0 & 0.01 &    0 &    0 &    0 &    0 \\
   0 & 0.13 &    0 &    0 &    0 &    0 &    0 \\
0.01 &    0 & 0.03 &    0 &    0 &    0 &    0 \\
   0 &    0 &    0 & 0.21 &    0 &    0 &    0 \\
   0 &    0 &    0 &    0 & 0.03 &    0 & -0.01 \\
   0 &    0 &    0 &    0 &    0 & 0.08 &    0 \\
   0 &    0 &    0 &    0 & -0.01 &    0 & 0.03
}+
\setstackgap{L}{1.1\baselineskip}
\fixTABwidth{T}
\parenMatrixstack{
   0 &    0 &    0 &    0 &    0 &    0 &    0 \\
   0 &    0 &    0 &    0 &    0 & 0.03 &    0 \\
   0 &    0 &    0 &    0 &    0 &    0 &    0 \\
   0 &    0 &    0 &    0 &    0 &    0 &    0 \\
   0 &    0 &    0 &    0 &    0 &    0 &    0 \\
   0 & -0.03 &    0 &    0 &    0 &    0 &    0 \\
   0 &    0 &    0 &    0 &    0 &    0 &    0
}i$.

\bigskip
\noindent Relaxed crystal structure in the format of POSCAR:\\
U1 O2          \\  
   1.00000000000000     \\
     2.7711040498960133    2.7711040049104594    0.0000144396275418\\
     0.0000066600828028    2.7711136130644425    2.7801061695202218\\
     2.7711139512681968    0.0000063740393353    2.7801064520672041\\
   U    O \\
     1     2\\
Direct\\
  0.4999999960908179  0.4999999955506112  0.5000000032544837\\
  0.2499987579137400  0.2500000727852644  0.2500001107265577\\
  0.7500012459954422  0.7499999316641245  0.7499998860189514
  
\subsubsection{Properties of $\mathbb{S}_1^{1}$}

\begin{table*}[!h]%
\begin{ruledtabular}
\begin{tabular}{ll}
\textrm{Properties (unit)}&
\textrm{Values}\\
\colrule
Initiated from states & 
S$_2^1$, S$_7^1$ \\
Energy (eV) & -31.991 \\
Basis vectors (\AA)& $a=(
2.7603, 2.7800, 0.0003
), b=(
0.0002, 2.7802, 2.7812
), c=(
2.7605, 0.0001, 2.7813
)$\\
\begin{tabular}{@{}l@{}}Atom coordinates \\ (in basis vectors)\end{tabular} &
{\begin{tabular}{@{}l@{}}$U1=(
0.5000, 0.5000, 0.5000
)$ \\$O1=(
0.2500, 0.2500, 0.2500
)$, $O2=(
0.7500, 0.7500, 0.7500
)$\end{tabular}} \\ 
Strain & \begin{tabular}{@{}l@{}}($\epsilon_{xx},\epsilon_{yy},\epsilon_{zz},\epsilon_{xy},\epsilon_{xz},\epsilon_{yz}$)=(
-0.0049, 0.0022, 0.0026,     0, 0.0001,     0
), \\($\epsilon_{A1g},\epsilon_{Eg.0},\epsilon_{Eg.1},\epsilon_{T2g.0},\epsilon_{T2g.1},\epsilon_{T2g.2}$)=(
 0, -0.0050, 0.0032,     0,     0, 0.0001
).\end{tabular}\\
Spin moments ($\mu_B$)& (-0.004, -0.008, 1.387)\\
Orbit moments ($\mu_B$)& (0.006, 0.0020, -2.267)\\
Total magnetic moments ($\mu_B$)& (0.002, 0.012, -0.880)\\
Energy (eV) and lattice (\AA) of und & -31.990, 2.7738
\end{tabular}
\end{ruledtabular}
\end{table*}

$\occmat^{\uparrow\uparrow}=
\setstackgap{L}{1.1\baselineskip}
\fixTABwidth{T}
\parenMatrixstack{
0.39 &    0 & 0.20 &    0 &    0 &    0 &    0 \\
   0 & 0.14 &    0 &    0 &    0 &    0 &    0 \\
0.20 &    0 & 0.14 &    0 &    0 &    0 &    0 \\
   0 &    0 &    0 & 0.66 &    0 & -0.27 &    0 \\
   0 &    0 &    0 &    0 & 0.06 &    0 & 0.09 \\
   0 &    0 &    0 & -0.27 &    0 & 0.15 &    0 \\
   0 &    0 &    0 &    0 & 0.09 &    0 & 0.41
}+
\setstackgap{L}{1.1\baselineskip}
\fixTABwidth{T}
\parenMatrixstack{
  0 &    0 &    0 &    0 & 0.09 &    0 & 0.36 \\
   0 &    0 &    0 & -0.07 &    0 & 0.04 &    0 \\
   0 &    0 &    0 &    0 & 0.05 &    0 & 0.20 \\
   0 & 0.07 &    0 &    0 &    0 &    0 &    0 \\
-0.09 &    0 & -0.05 &    0 &    0 &    0 &    0 \\
   0 & -0.04 &    0 &    0 &    0 &    0 &    0 \\
-0.36 &    0 & -0.20 &    0 &    0 &    0 &    0
}i$,

$\occmat^{\uparrow\downarrow}=
\setstackgap{L}{1.1\baselineskip}
\fixTABwidth{T}
\parenMatrixstack{
   0 & -0.01 &    0 &    0 &    0 &    0 &    0 \\
-0.04 &    0 & -0.02 &    0 &    0 &    0 &    0 \\
   0 & 0.02 &    0 &    0 &    0 &    0 &    0 \\
   0 &    0 &    0 &    0 & -0.03 &    0 & -0.09 \\
   0 &    0 &    0 & 0.29 &    0 & -0.13 &    0 \\
   0 &    0 &    0 &    0 & -0.04 &    0 & -0.16 \\
   0 &    0 &    0 & -0.04 &    0 & 0.02 &    0
}+
\setstackgap{L}{1.1\baselineskip}
\fixTABwidth{T}
\parenMatrixstack{
  0 &    0 &    0 & 0.06 &    0 & -0.03 &    0 \\
   0 &    0 &    0 &    0 & -0.01 &    0 & -0.04 \\
   0 &    0 &    0 & -0.20 &    0 & 0.09 &    0 \\
0.09 &    0 & 0.05 &    0 &    0 &    0 &    0 \\
   0 & 0.04 &    0 &    0 &    0 &    0 &    0 \\
0.16 &    0 & 0.09 &    0 &    0 &    0 &    0 \\
   0 &    0 &    0 &    0 &    0 &    0 &    0
}i$,

$\occmat^{\downarrow\uparrow}=
\setstackgap{L}{1.1\baselineskip}
\fixTABwidth{T}
\parenMatrixstack{
   0 & -0.04 &    0 &    0 &    0 &    0 &    0 \\
-0.01 &    0 & 0.02 &    0 &    0 &    0 &    0 \\
   0 & -0.02 &    0 &    0 &    0 &    0 &    0 \\
   0 &    0 &    0 &    0 & 0.29 &    0 & -0.04 \\
   0 &    0 &    0 & -0.03 &    0 & -0.04 &    0 \\
   0 &    0 &    0 &    0 & -0.13 &    0 & 0.02 \\
   0 &    0 &    0 & -0.09 &    0 & -0.16 &    0
}+
\setstackgap{L}{1.1\baselineskip}
\fixTABwidth{T}
\parenMatrixstack{
    0 &    0 &    0 & -0.09 &    0 & -0.16 &    0 \\
   0 &    0 &    0 &    0 & -0.04 &    0 &    0 \\
   0 &    0 &    0 & -0.05 &    0 & -0.09 &    0 \\
-0.06 &    0 & 0.20 &    0 &    0 &    0 &    0 \\
   0 & 0.01 &    0 &    0 &    0 &    0 &    0 \\
0.03 &    0 & -0.09 &    0 &    0 &    0 &    0 \\
   0 & 0.04 &    0 &    0 &    0 &    0 &    0
}i$,

$\occmat^{\downarrow\downarrow}=
\setstackgap{L}{1.1\baselineskip}
\fixTABwidth{T}
\parenMatrixstack{
0.04 &    0 & -0.02 &    0 &    0 &    0 &    0 \\
   0 & 0.12 &    0 &    0 &    0 &    0 &    0 \\
-0.02 &    0 & 0.09 &    0 &    0 &    0 &    0 \\
   0 &    0 &    0 & 0.06 &    0 & 0.04 &    0 \\
   0 &    0 &    0 &    0 & 0.16 &    0 & -0.02 \\
   0 &    0 &    0 & 0.04 &    0 & 0.10 &    0 \\
   0 &    0 &    0 &    0 & -0.02 &    0 & 0.03
}+
\setstackgap{L}{1.1\baselineskip}
\fixTABwidth{T}
\parenMatrixstack{
   0 &    0 &    0 &    0 & 0.03 &    0 & -0.01 \\
   0 &    0 &    0 & 0.01 &    0 & 0.02 &    0 \\
   0 &    0 &    0 &    0 & -0.09 &    0 & 0.01 \\
   0 & -0.01 &    0 &    0 &    0 &    0 &    0 \\
-0.03 &    0 & 0.09 &    0 &    0 &    0 &    0 \\
   0 & -0.02 &    0 &    0 &    0 &    0 &    0 \\
0.01 &    0 & -0.01 &    0 &    0 &    0 &    0
}i$.

\subsubsection{Properties of $\mathbb{S}_1^{2}$}

\begin{table*}[!h]%
\begin{ruledtabular}
\begin{tabular}{ll}
\textrm{Properties (unit)}&
\textrm{Values}\\
\colrule
Initiated from states & 
S$_2^5$, S$_4^1$ \\
Energy (eV) & -31.991 \\
Basis vectors (\AA)& $a=(
2.7800, 2.7603,     0
), b=(
0, 2.7603, 2.7811
), c=(
2.7800,     0, 2.7811
)$\\
\begin{tabular}{@{}l@{}}Atom coordinates \\ (in basis vectors)\end{tabular} &
{\begin{tabular}{@{}l@{}}$U1=(
0.5000, 0.5000, 0.5000
)$ \\$O1=(
0.2500, 0.2500, 0.2500
)$, $O2=(
0.7500, 0.7500, 0.7500
)$\end{tabular}} \\ 
Strain & \begin{tabular}{@{}l@{}}($\epsilon_{xx},\epsilon_{yy},\epsilon_{zz},\epsilon_{xy},\epsilon_{xz},\epsilon_{yz}$)=(
0.0022, -0.0049, 0.0026,     0,     0,     0
), \\($\epsilon_{A1g},\epsilon_{Eg.0},\epsilon_{Eg.1},\epsilon_{T2g.0},\epsilon_{T2g.1},\epsilon_{T2g.2}$)=(
  0, 0.0050, 0.0032,     0,     0,     0
).\end{tabular}\\
Spin moments ($\mu_B$)& (0.002, -0.003, 1.387)\\
Orbit moments ($\mu_B$)& (-0.004, 0.004, -2.268)\\
Total magnetic moments ($\mu_B$)& (-0.002, 0.001, -0.881)\\
Energy (eV) and lattice (\AA) of und & -31.990, 2.7738
\end{tabular}
\end{ruledtabular}
\end{table*}

$\occmat^{\uparrow\uparrow}=
\setstackgap{L}{1.1\baselineskip}
\fixTABwidth{T}
\parenMatrixstack{
0.41 &    0 & -0.09 &    0 &    0 &    0 &    0 \\
   0 & 0.14 &    0 &    0 &    0 &    0 &    0 \\
-0.09 &    0 & 0.06 &    0 &    0 &    0 &    0 \\
   0 &    0 &    0 & 0.66 &    0 & 0.27 &    0 \\
   0 &    0 &    0 &    0 & 0.14 &    0 & -0.20 \\
   0 &    0 &    0 & 0.27 &    0 & 0.15 &    0 \\
   0 &    0 &    0 &    0 & -0.20 &    0 & 0.39
}+
\setstackgap{L}{1.1\baselineskip}
\fixTABwidth{T}
\parenMatrixstack{
   0 &    0 &    0 &    0 & -0.19 &    0 & 0.36 \\
   0 &    0 &    0 & 0.07 &    0 & 0.04 &    0 \\
   0 &    0 &    0 &    0 & 0.05 &    0 & -0.09 \\
   0 & -0.07 &    0 &    0 &    0 &    0 &    0 \\
0.19 &    0 & -0.05 &    0 &    0 &    0 &    0 \\
   0 & -0.04 &    0 &    0 &    0 &    0 &    0 \\
-0.36 &    0 & 0.09 &    0 &    0 &    0 &    0
}i$,

$\occmat^{\uparrow\downarrow}=
\setstackgap{L}{1.1\baselineskip}
\fixTABwidth{T}
\parenMatrixstack{
   0 &    0 &    0 &    0 &    0 &    0 &    0 \\
-0.05 &    0 & 0.01 &    0 &    0 &    0 &    0 \\
   0 & -0.04 &    0 &    0 &    0 &    0 &    0 \\
   0 &    0 &    0 &    0 & -0.05 &    0 & 0.09 \\
   0 &    0 &    0 & 0.20 &    0 & 0.09 &    0 \\
   0 &    0 &    0 &    0 & 0.09 &    0 & -0.16 \\
   0 &    0 &    0 & 0.06 &    0 & 0.03 &    0
}+
\setstackgap{L}{1.1\baselineskip}
\fixTABwidth{T}
\parenMatrixstack{
  0 &    0 &    0 & -0.04 &    0 & -0.02 &    0 \\
   0 &    0 &    0 &    0 & 0.02 &    0 & -0.04 \\
   0 &    0 &    0 & -0.29 &    0 & -0.13 &    0 \\
-0.09 &    0 & 0.03 &    0 &    0 &    0 &    0 \\
   0 & -0.02 &    0 &    0 &    0 &    0 &    0 \\
0.16 &    0 & -0.04 &    0 &    0 &    0 &    0 \\
   0 & -0.01 &    0 &    0 &    0 &    0 &    0
}i$,

$\occmat^{\downarrow\uparrow}=
\setstackgap{L}{1.1\baselineskip}
\fixTABwidth{T}
\parenMatrixstack{
  0 & -0.05 &    0 &    0 &    0 &    0 &    0 \\
   0 &    0 & -0.04 &    0 &    0 &    0 &    0 \\
   0 & 0.01 &    0 &    0 &    0 &    0 &    0 \\
   0 &    0 &    0 &    0 & 0.20 &    0 & 0.06 \\
   0 &    0 &    0 & -0.05 &    0 & 0.09 &    0 \\
   0 &    0 &    0 &    0 & 0.09 &    0 & 0.03 \\
   0 &    0 &    0 & 0.09 &    0 & -0.16 &    0
}+
\setstackgap{L}{1.1\baselineskip}
\fixTABwidth{T}
\parenMatrixstack{
  0 &    0 &    0 & 0.09 &    0 & -0.16 &    0 \\
   0 &    0 &    0 &    0 & 0.02 &    0 & 0.01 \\
   0 &    0 &    0 & -0.03 &    0 & 0.04 &    0 \\
0.04 &    0 & 0.29 &    0 &    0 &    0 &    0 \\
   0 & -0.02 &    0 &    0 &    0 &    0 &    0 \\
0.02 &    0 & 0.13 &    0 &    0 &    0 &    0 \\
   0 & 0.04 &    0 &    0 &    0 &    0 &    0
}i$,

$\occmat^{\downarrow\downarrow}=
\setstackgap{L}{1.1\baselineskip}
\fixTABwidth{T}
\parenMatrixstack{
0.03 &    0 & 0.02 &    0 &    0 &    0 &    0 \\
   0 & 0.12 &    0 &    0 &    0 &    0 &    0 \\
0.02 &    0 & 0.16 &    0 &    0 &    0 &    0 \\
   0 &    0 &    0 & 0.06 &    0 & -0.04 &    0 \\
   0 &    0 &    0 &    0 & 0.09 &    0 & 0.02 \\
   0 &    0 &    0 & -0.04 &    0 & 0.10 &    0 \\
   0 &    0 &    0 &    0 & 0.02 &    0 & 0.04
}+
\setstackgap{L}{1.1\baselineskip}
\fixTABwidth{T}
\parenMatrixstack{
   0 &    0 &    0 &    0 & -0.01 &    0 & -0.01 \\
   0 &    0 &    0 & -0.01 &    0 & 0.02 &    0 \\
   0 &    0 &    0 &    0 & -0.09 &    0 & -0.03 \\
   0 & 0.01 &    0 &    0 &    0 &    0 &    0 \\
0.01 &    0 & 0.09 &    0 &    0 &    0 &    0 \\
   0 & -0.02 &    0 &    0 &    0 &    0 &    0 \\
0.01 &    0 & 0.03 &    0 &    0 &    0 &    0
}i$.

\subsubsection{Properties of $\mathbb{S}_2^{1}$}

\begin{table*}[!h]%
\begin{ruledtabular}
\begin{tabular}{ll}
\textrm{Properties (unit)}&
\textrm{Values}\\
\colrule
Initiated from states & 
S$_1^1$ \\
Energy (eV) & -31.988 \\
Basis vectors (\AA)& $a=(
2.7623, 2.7873, -0.0005
), b=(
-0.0004, 2.7873, 2.7728
), c=(
2.7619,     0, 2.7723
)$\\
\begin{tabular}{@{}l@{}}Atom coordinates \\ (in basis vectors)\end{tabular} &
{\begin{tabular}{@{}l@{}}$U1=(
0.5000, 0.5000, 0.5000
)$ \\$O1=(
0.2500, 0.2500, 0.2500
)$, $O2=(
0.7500, 0.7500, 0.7500
)$\end{tabular}} \\ 
Strain & \begin{tabular}{@{}l@{}}($\epsilon_{xx},\epsilon_{yy},\epsilon_{zz},\epsilon_{xy},\epsilon_{xz},\epsilon_{yz}$)=(
-0.0042, 0.0047, -0.0005,     0, -0.0001,     0
), \\($\epsilon_{A1g},\epsilon_{Eg.0},\epsilon_{Eg.1},\epsilon_{T2g.0},\epsilon_{T2g.1},\epsilon_{T2g.2}$)=(
  0, -0.0063, -0.0006,     0,     0, -0.0001
).\end{tabular}\\
Spin moments ($\mu_B$)& (0.018, 0.003, 1.411)\\
Orbit moments ($\mu_B$)& (-0.034, -0.004, -2.310)\\
Total magnetic moments ($\mu_B$)& (-0.016, -0.001, -0.899)\\
Energy (eV) and lattice (\AA) of und & -31.986, 2.7739
\end{tabular}
\end{ruledtabular}
\end{table*}

$\occmat^{\uparrow\uparrow}=
\setstackgap{L}{1.1\baselineskip}
\fixTABwidth{T}
\parenMatrixstack{
0.30 &    0 & 0.24 &    0 &    0 &    0 &    0 \\
   0 & 0.16 &    0 &    0 &    0 &    0 &    0 \\
0.24 &    0 & 0.25 &    0 &    0 &    0 &    0 \\
   0 &    0 &    0 & 0.37 &    0 & -0.37 &    0 \\
   0 &    0 &    0 &    0 & 0.08 & -0.01 & 0.12 \\
   0 &    0 &    0 & -0.37 & -0.01 & 0.45 &    0 \\
   0 &    0 &    0 &    0 & 0.12 &    0 & 0.34 
}+
\setstackgap{L}{1.1\baselineskip}
\fixTABwidth{T}
\parenMatrixstack{
  0 &    0 &    0 &    0 & 0.11 &    0 & 0.28 \\
   0 &    0 &    0 & -0.10 &    0 & 0.12 &    0 \\
   0 &    0 &    0 &    0 & 0.10 &    0 & 0.25 \\
   0 & 0.10 &    0 &    0 &    0 &    0 &    0 \\
-0.11 &    0 & -0.10 &    0 &    0 &    0 &    0 \\
   0 & -0.12 &    0 &    0 &    0 &    0 &    0 \\
-0.28 &    0 & -0.25 &    0 &    0 &    0 &    0
}i$,

$\occmat^{\uparrow\downarrow}=
\setstackgap{L}{1.1\baselineskip}
\fixTABwidth{T}
\parenMatrixstack{
   0 & -0.02 &    0 &    0 &    0 &    0 &    0 \\
-0.03 &    0 & -0.02 &    0 &    0 &    0 &    0 \\
   0 & 0.01 &    0 &    0 &    0 &    0 &    0 \\
   0 &    0 &    0 &    0 & -0.06 &    0 & -0.13 \\
   0 &    0 &    0 & 0.22 &    0 & -0.25 &    0 \\
   0 &    0 &    0 &    0 & -0.06 &    0 & -0.14 \\
   0 &    0 &    0 & -0.04 &    0 & 0.04 &    0
}+
\setstackgap{L}{1.1\baselineskip}
\fixTABwidth{T}
\parenMatrixstack{
  0 &    0 &    0 & 0.06 &    0 & -0.08 &    0 \\
   0 &    0 &    0 &    0 & -0.01 &    0 & -0.03 \\
   0 &    0 &    0 & -0.06 &    0 & 0.06 &    0 \\
0.13 &    0 & 0.12 &    0 &    0 &    0 &    0 \\
   0 & 0.07 &    0 &    0 &    0 &    0 &    0 \\
0.13 &    0 & 0.12 &    0 &    0 &    0 &    0 \\
   0 & -0.01 &    0 &    0 &    0 &    0 &    0
}i$,

$\occmat^{\downarrow\uparrow}=
\setstackgap{L}{1.1\baselineskip}
\fixTABwidth{T}
\parenMatrixstack{
  0 & -0.03 &    0 &    0 &    0 &    0 &    0 \\
-0.02 &    0 & 0.01 &    0 &    0 &    0 &    0 \\
   0 & -0.02 &    0 &    0 &    0 &    0 &    0 \\
   0 &    0 &    0 &    0 & 0.22 &    0 & -0.04 \\
   0 &    0 &    0 & -0.06 &    0 & -0.06 &    0 \\
   0 &    0 &    0 &    0 & -0.25 &    0 & 0.04 \\
   0 &    0 &    0 & -0.13 &    0 & -0.14 &    0
}+
\setstackgap{L}{1.1\baselineskip}
\fixTABwidth{T}
\parenMatrixstack{
  0 &    0 &    0 & -0.13 &    0 & -0.13 &    0 \\
   0 &    0 &    0 &    0 & -0.07 &    0 & 0.01 \\
   0 &    0 &    0 & -0.12 &    0 & -0.12 &    0 \\
-0.06 &    0 & 0.06 &    0 &    0 &    0 &    0 \\
   0 & 0.01 &    0 &    0 &    0 &    0 &    0 \\
0.08 &    0 & -0.06 &    0 &    0 &    0 &    0 \\
   0 & 0.03 &    0 &    0 &    0 &    0 &    0
}i$,

$\occmat^{\downarrow\downarrow}=
\setstackgap{L}{1.1\baselineskip}
\fixTABwidth{T}
\parenMatrixstack{
0.04 &    0 & -0.01 &    0 &    0 &    0 &    0 \\
   0 & 0.11 &    0 &    0 &    0 &    0 &    0 \\
-0.01 &    0 & 0.04 &    0 &    0 &    0 &    0 \\
   0 &    0 &    0 & 0.10 &    0 & 0.06 &    0 \\
   0 &    0 &    0 &    0 & 0.17 &    0 & -0.03 \\
   0 &    0 &    0 & 0.06 &    0 & 0.09 &    0 \\
   0 &    0 &    0 &    0 & -0.03 &    0 & 0.04
}+
\setstackgap{L}{1.1\baselineskip}
\fixTABwidth{T}
\parenMatrixstack{
   0 &    0 &    0 &    0 & 0.04 &    0 & -0.01 \\
   0 &    0 &    0 & 0.01 &    0 & 0.01 &    0 \\
   0 &    0 &    0 &    0 & -0.03 &    0 & 0.01 \\
   0 & -0.01 &    0 &    0 &    0 &    0 &    0 \\
-0.04 &    0 & 0.03 &    0 &    0 &    0 &    0 \\
   0 & -0.01 &    0 &    0 &    0 &    0 &    0 \\
0.01 &    0 & -0.01 &    0 &    0 &    0 &    0 
}i$.

\subsubsection{Properties of $\mathbb{S}_2^{2}$}

\begin{table*}[!h]%
\begin{ruledtabular}
\begin{tabular}{ll}
\textrm{Properties (unit)}&
\textrm{Values}\\
\colrule
Initiated from states & 
S$_1^3$ \\
Energy (eV) & -31.988 \\
Basis vectors (\AA)& $a=(
2.7872, 2.7625, -0.0001
), b=(
 0, 2.7624, 2.7726
), c=(
2.7872, -0.0001, 2.7727
)$\\
\begin{tabular}{@{}l@{}}Atom coordinates \\ (in basis vectors)\end{tabular} &
{\begin{tabular}{@{}l@{}}$U1=(
0.5000, 0.5000, 0.5000
)$ \\$O1=(
0.2500, 0.2500, 0.2500
)$, $O2=(
0.7500, 0.7500, 0.7500
)$\end{tabular}} \\ 
Strain & \begin{tabular}{@{}l@{}}($\epsilon_{xx},\epsilon_{yy},\epsilon_{zz},\epsilon_{xy},\epsilon_{xz},\epsilon_{yz}$)=(
0.0047, -0.0042, -0.0005,     0,     0,       0
), \\($\epsilon_{A1g},\epsilon_{Eg.0},\epsilon_{Eg.1},\epsilon_{T2g.0},\epsilon_{T2g.1},\epsilon_{T2g.2}$)=(
0, 0.0063, -0.0006,     0,     0,     0
).\end{tabular}\\
Spin moments ($\mu_B$)& (0.003, 0.007, 1.414)\\
Orbit moments ($\mu_B$)& (-0.006, -0.014, -2.322)\\
Total magnetic moments ($\mu_B$)& (-0.003, -0.007, -0.908)\\
Energy (eV) and lattice (\AA) of und & -31.986, 2.7741
\end{tabular}
\end{ruledtabular}
\end{table*}

$\occmat^{\uparrow\uparrow}=
\setstackgap{L}{1.1\baselineskip}
\fixTABwidth{T}
\parenMatrixstack{
0.34 &    0 & -0.12 &    0 &    0 &    0 &    0 \\
   0 & 0.16 &    0 &    0 &    0 &    0 &    0 \\
-0.12 &    0 & 0.09 &    0 &    0 &    0 &    0 \\
   0 &    0 &    0 & 0.36 &    0 & 0.37 &    0 \\
   0 &    0 &    0 &    0 & 0.25 &    0 & -0.24 \\
   0 &    0 &    0 & 0.37 &    0 & 0.47 &    0 \\
   0 &    0 &    0 &    0 & -0.24 &    0 & 0.30 
}+
\setstackgap{L}{1.1\baselineskip}
\fixTABwidth{T}
\parenMatrixstack{
   0 &    0 &    0 &    0 & -0.25 &    0 & 0.28 \\
   0 &    0 &    0 & 0.10 &    0 & 0.13 &    0 \\
   0 &    0 &    0 &    0 & 0.11 &    0 & -0.11 \\
   0 & -0.10 &    0 &    0 &    0 &    0 &    0 \\
0.25 &    0 & -0.11 &    0 &    0 &    0 &    0 \\
   0 & -0.13 &    0 &    0 &    0 &    0 &    0 \\
-0.28 &    0 & 0.11 &    0 &    0 &    0 &    0
}i$,

$\occmat^{\uparrow\downarrow}=
\setstackgap{L}{1.1\baselineskip}
\fixTABwidth{T}
\parenMatrixstack{
  0 & -0.01 &    0 &    0 &    0 &    0 &    0 \\
-0.03 &    0 & 0.01 &    0 &    0 &    0 &    0 \\
   0 & -0.07 &    0 &    0 &    0 &    0 &    0 \\
   0 &    0 &    0 &    0 & -0.12 &    0 & 0.13 \\
   0 &    0 &    0 & 0.05 &    0 & 0.05 &    0 \\
   0 &    0 &    0 &    0 & 0.12 &    0 & -0.13 \\
   0 &    0 &    0 & 0.06 &    0 & 0.08 &    0
}+
\setstackgap{L}{1.1\baselineskip}
\fixTABwidth{T}
\parenMatrixstack{
   0 &    0 &    0 & -0.04 &    0 & -0.05 &    0 \\
   0 &    0 &    0 &    0 & 0.02 &    0 & -0.03 \\
   0 &    0 &    0 & -0.21 &    0 & -0.25 &    0 \\
-0.13 &    0 & 0.06 &    0 &    0 &    0 &    0 \\
   0 & -0.01 &    0 &    0 &    0 &    0 &    0 \\
0.14 &    0 & -0.06 &    0 &    0 &    0 &    0 \\
   0 & -0.02 &    0 &    0 &    0 &    0 &    0
}i$,

$\occmat^{\downarrow\uparrow}=
\setstackgap{L}{1.1\baselineskip}
\fixTABwidth{T}
\parenMatrixstack{
  0 & -0.03 &    0 &    0 &    0 &    0 &    0 \\
-0.01 &    0 & -0.07 &    0 &    0 &    0 &    0 \\
   0 & 0.01 &    0 &    0 &    0 &    0 &    0 \\
   0 &    0 &    0 &    0 & 0.05 &    0 & 0.06 \\
   0 &    0 &    0 & -0.12 &    0 & 0.12 &    0 \\
   0 &    0 &    0 &    0 & 0.05 &    0 & 0.08 \\
   0 &    0 &    0 & 0.13 &    0 & -0.13 &    0
}+
\setstackgap{L}{1.1\baselineskip}
\fixTABwidth{T}
\parenMatrixstack{
 0 &    0 &    0 & 0.13 &    0 & -0.14 &    0 \\
   0 &    0 &    0 &    0 & 0.01 &    0 & 0.02 \\
   0 &    0 &    0 & -0.06 &    0 & 0.06 &    0 \\
0.04 &    0 & 0.21 &    0 &    0 &    0 &    0 \\
   0 & -0.02 &    0 &    0 &    0 &    0 &    0 \\
0.05 &    0 & 0.25 &    0 &    0 &    0 &    0 \\
   0 & 0.03 &    0 &    0 &    0 &    0 &    0
}i$,

$\occmat^{\downarrow\downarrow}=
\setstackgap{L}{1.1\baselineskip}
\fixTABwidth{T}
\parenMatrixstack{
0.04 &    0 & 0.03 &    0 &    0 &    0 &    0 \\
   0 & 0.11 &    0 &    0 &    0 &    0 &    0 \\
0.03 &    0 & 0.17 &    0 &    0 &    0 &    0 \\
   0 &    0 &    0 & 0.10 &    0 & -0.06 &    0 \\
   0 &    0 &    0 &    0 & 0.04 &    0 &    0 \\
   0 &    0 &    0 & -0.06 &    0 & 0.09 &    0 \\
   0 &    0 &    0 &    0 &    0 &    0 & 0.04 
}+
\setstackgap{L}{1.1\baselineskip}
\fixTABwidth{T}
\parenMatrixstack{
  0 &    0 &    0 &    0 & -0.01 &    0 & -0.01 \\
   0 &    0 &    0 & -0.01 &    0 & 0.01 &    0 \\
   0 &    0 &    0 &    0 & -0.03 &    0 & -0.04 \\
   0 & 0.01 &    0 &    0 &    0 &    0 &    0 \\
0.01 &    0 & 0.03 &    0 &    0 &    0 &    0 \\
   0 & -0.01 &    0 &    0 &    0 &    0 &    0 \\
0.01 &    0 & 0.04 &    0 &    0 &    0 &    0
}i$.

\subsubsection{Properties of $\mathbb{S}_2^{3}$}

\begin{table*}[!h]%
\begin{ruledtabular}
\begin{tabular}{ll}
\textrm{Properties (unit)}&
\textrm{Values}\\
\colrule
Initiated from states & 
S$_2^2$ \\
Energy (eV) & -31.988 \\
Basis vectors (\AA)& $a=(
2.7618, 2.7871,     0
), b=(
  0, 2.7871, 2.7732
), c=(
2.7617,     0, 2.7731
)$\\
\begin{tabular}{@{}l@{}}Atom coordinates \\ (in basis vectors)\end{tabular} &
{\begin{tabular}{@{}l@{}}$U1=(
0.5000, 0.5000, 0.5000
)$ \\$O1=(
0.2500, 0.2500, 0.2500
)$, $O2=(
0.7500, 0.7500, 0.7500
)$\end{tabular}} \\ 
Strain & \begin{tabular}{@{}l@{}}($\epsilon_{xx},\epsilon_{yy},\epsilon_{zz},\epsilon_{xy},\epsilon_{xz},\epsilon_{yz}$)=(
-0.0044, 0.0047, -0.0003,     0,     0,     0
), \\($\epsilon_{A1g},\epsilon_{Eg.0},\epsilon_{Eg.1},\epsilon_{T2g.0},\epsilon_{T2g.1},\epsilon_{T2g.2}$)=(
0, -0.0065, -0.0004,     0,     0,     0
).\end{tabular}\\
Spin moments ($\mu_B$)& (0.002, 0.000, 1.406)\\
Orbit moments ($\mu_B$)& (-0.004, 0.000, -2.286)\\
Total magnetic moments ($\mu_B$)& (-0.002, 0.000, -0.880)\\
Energy (eV) and lattice (\AA) of und & -31.986, 2.7740
\end{tabular}
\end{ruledtabular}
\end{table*}

$\occmat^{\uparrow\uparrow}=
\setstackgap{L}{1.1\baselineskip}
\fixTABwidth{T}
\parenMatrixstack{
0.31 &    0 & 0.24 &    0 &    0 &    0 &    0 \\
   0 & 0.16 &    0 &    0 &    0 &    0 &    0 \\
0.24 &    0 & 0.25 &    0 &    0 &    0 &    0 \\
   0 &    0 &    0 & 0.40 &    0 & -0.37 &    0 \\
   0 &    0 &    0 &    0 & 0.08 &    0 & 0.11 \\
   0 &    0 &    0 & -0.37 &    0 & 0.43 &    0 \\
   0 &    0 &    0 &    0 & 0.11 &    0 & 0.34
}+
\setstackgap{L}{1.1\baselineskip}
\fixTABwidth{T}
\parenMatrixstack{
   0 &    0 &    0 &    0 & 0.11 &    0 & 0.28 \\
   0 &    0 &    0 & -0.10 &    0 & 0.12 &    0 \\
   0 &    0 &    0 &    0 & 0.10 &    0 & 0.25 \\
   0 & 0.10 &    0 &    0 &    0 &    0 &    0 \\
-0.11 &    0 & -0.10 &    0 &    0 &    0 &    0 \\
   0 & -0.12 &    0 &    0 &    0 &    0 &    0 \\
-0.28 &    0 & -0.25 &    0 &    0 &    0 &    0 
}i$,

$\occmat^{\uparrow\downarrow}=
\setstackgap{L}{1.1\baselineskip}
\fixTABwidth{T}
\parenMatrixstack{
   0 & -0.02 &    0 &    0 &    0 &    0 &    0 \\
-0.03 &    0 & -0.02 &    0 &    0 &    0 &    0 \\
   0 & 0.02 &    0 &    0 &    0 &    0 &    0 \\
   0 &    0 &    0 &    0 & -0.06 &    0 & -0.13 \\
   0 &    0 &    0 & 0.22 &    0 & -0.24 &    0 \\
   0 &    0 &    0 &    0 & -0.06 &    0 & -0.14 \\
   0 &    0 &    0 & -0.04 &    0 & 0.04 &    0
}+
\setstackgap{L}{1.1\baselineskip}
\fixTABwidth{T}
\parenMatrixstack{
   0 &    0 &    0 & 0.07 &    0 & -0.07 &    0 \\
   0 &    0 &    0 &    0 & -0.01 &    0 & -0.03 \\
   0 &    0 &    0 & -0.07 &    0 & 0.07 &    0 \\
0.12 &    0 & 0.11 &    0 &    0 &    0 &    0 \\
   0 & 0.07 &    0 &    0 &    0 &    0 &    0 \\
0.13 &    0 & 0.12 &    0 &    0 &    0 &    0 \\
   0 & -0.01 &    0 &    0 &    0 &    0 &    0
}i$,

$\occmat^{\downarrow\uparrow}=
\setstackgap{L}{1.1\baselineskip}
\fixTABwidth{T}
\parenMatrixstack{
  0 & -0.03 &    0 &    0 &    0 &    0 &    0 \\
-0.02 &    0 & 0.02 &    0 &    0 &    0 &    0 \\
   0 & -0.02 &    0 &    0 &    0 &    0 &    0 \\
   0 &    0 &    0 &    0 & 0.22 &    0 & -0.04 \\
   0 &    0 &    0 & -0.06 &    0 & -0.06 &    0 \\
   0 &    0 &    0 &    0 & -0.24 &    0 & 0.04 \\
   0 &    0 &    0 & -0.13 &    0 & -0.14 &    0
}+
\setstackgap{L}{1.1\baselineskip}
\fixTABwidth{T}
\parenMatrixstack{
   0 &    0 &    0 & -0.12 &    0 & -0.13 &    0 \\
   0 &    0 &    0 &    0 & -0.07 &    0 & 0.01 \\
   0 &    0 &    0 & -0.11 &    0 & -0.12 &    0 \\
-0.07 &    0 & 0.07 &    0 &    0 &    0 &    0 \\
   0 & 0.01 &    0 &    0 &    0 &    0 &    0 \\
0.07 &    0 & -0.07 &    0 &    0 &    0 &    0 \\
   0 & 0.03 &    0 &    0 &    0 &    0 &    0
}i$,

$\occmat^{\downarrow\downarrow}=
\setstackgap{L}{1.1\baselineskip}
\fixTABwidth{T}
\parenMatrixstack{
0.04 &    0 & -0.01 &    0 &    0 &    0 &    0 \\
   0 & 0.11 &    0 &    0 &    0 &    0 &    0 \\
-0.01 &    0 & 0.04 &    0 &    0 &    0 &    0 \\
   0 &    0 &    0 & 0.09 &    0 & 0.06 &    0 \\
   0 &    0 &    0 &    0 & 0.17 &    0 & -0.03 \\
   0 &    0 &    0 & 0.06 &    0 & 0.09 &    0 \\
   0 &    0 &    0 &    0 & -0.03 &    0 & 0.04 
}+
\setstackgap{L}{1.1\baselineskip}
\fixTABwidth{T}
\parenMatrixstack{
   0 &    0 &    0 &    0 & 0.04 &    0 & -0.01 \\
   0 &    0 &    0 & 0.01 &    0 & 0.01 &    0 \\
   0 &    0 &    0 &    0 & -0.04 &    0 & 0.01 \\
   0 & -0.01 &    0 &    0 &    0 &    0 &    0 \\
-0.04 &    0 & 0.04 &    0 &    0 &    0 &    0 \\
   0 & -0.01 &    0 &    0 &    0 &    0 &    0 \\
0.01 &    0 & -0.01 &    0 &    0 &    0 &    0
}i$.

\subsubsection{Properties of $\mathbb{S}_2^{4}$}

\begin{table*}[!h]%
\begin{ruledtabular}
\begin{tabular}{ll}
\textrm{Properties (unit)}&
\textrm{Values}\\
\colrule
Initiated from states & 
S$_2^3$ \\
Energy (eV) & -31.988 \\
Basis vectors (\AA)& $a=(
2.7871, 2.7616,     0
), b=(
  0, 2.7616, 2.7733
), c=(
2.7871,     0, 2.7733
)$\\
\begin{tabular}{@{}l@{}}Atom coordinates \\ (in basis vectors)\end{tabular} &
{\begin{tabular}{@{}l@{}}$U1=(
0.5000, 0.5000, 0.5000
)$ \\$O1=(
0.2500, 0.2500, 0.2500
)$, $O2=(
0.7500, 0.7500, 0.7500
)$\end{tabular}} \\ 
Strain & \begin{tabular}{@{}l@{}}($\epsilon_{xx},\epsilon_{yy},\epsilon_{zz},\epsilon_{xy},\epsilon_{xz},\epsilon_{yz}$)=(
0.0047, -0.0045, -0.0003,     0,     0,     0
), \\($\epsilon_{A1g},\epsilon_{Eg.0},\epsilon_{Eg.1},\epsilon_{T2g.0},\epsilon_{T2g.1},\epsilon_{T2g.2}$)=(
0, 0.0065, -0.0003,     0,     0,     0
).\end{tabular}\\
Spin moments ($\mu_B$)& (0.000, -0.001, 1.405)\\
Orbit moments ($\mu_B$)& (0.001, 0.002, -2.281)\\
Total magnetic moments ($\mu_B$)& (0.001, 0.001, -0.876)\\
Energy (eV) and lattice (\AA) of und & -31.986, 2.7740
\end{tabular}
\end{ruledtabular}
\end{table*}

$\occmat^{\uparrow\uparrow}=
\setstackgap{L}{1.1\baselineskip}
\fixTABwidth{T}
\parenMatrixstack{
0.34 &    0 & -0.11 &    0 &    0 &    0 &    0 \\
   0 & 0.16 &    0 &    0 &    0 &    0 &    0 \\
-0.11 &    0 & 0.08 &    0 &    0 &    0 &    0 \\
   0 &    0 &    0 & 0.40 &    0 & 0.37 &    0 \\
   0 &    0 &    0 &    0 & 0.24 &    0 & -0.24 \\
   0 &    0 &    0 & 0.37 &    0 & 0.42 &    0 \\
   0 &    0 &    0 &    0 & -0.24 &    0 & 0.31
}+
\setstackgap{L}{1.1\baselineskip}
\fixTABwidth{T}
\parenMatrixstack{
   0 &    0 &    0 &    0 & -0.25 &    0 & 0.29 \\
   0 &    0 &    0 & 0.10 &    0 & 0.12 &    0 \\
   0 &    0 &    0 &    0 & 0.10 &    0 & -0.11 \\
   0 & -0.10 &    0 &    0 &    0 &    0 &    0 \\
0.25 &    0 & -0.10 &    0 &    0 &    0 &    0 \\
   0 & -0.12 &    0 &    0 &    0 &    0 &    0 \\
-0.29 &    0 & 0.11 &    0 &    0 &    0 &    0
}i$,

$\occmat^{\uparrow\downarrow}=
\setstackgap{L}{1.1\baselineskip}
\fixTABwidth{T}
\parenMatrixstack{
  0 & -0.01 &    0 &    0 &    0 &    0 &    0 \\
-0.03 &    0 & 0.01 &    0 &    0 &    0 &    0 \\
   0 & -0.07 &    0 &    0 &    0 &    0 &    0 \\
   0 &    0 &    0 &    0 & -0.11 &    0 & 0.12 \\
   0 &    0 &    0 & 0.07 &    0 & 0.07 &    0 \\
   0 &    0 &    0 &    0 & 0.12 &    0 & -0.13 \\
   0 &    0 &    0 & 0.07 &    0 & 0.07 &    0
}+
\setstackgap{L}{1.1\baselineskip}
\fixTABwidth{T}
\parenMatrixstack{
   0 &    0 &    0 & -0.04 &    0 & -0.04 &    0 \\
   0 &    0 &    0 &    0 & 0.02 &    0 & -0.03 \\
   0 &    0 &    0 & -0.23 &    0 & -0.24 &    0 \\
-0.13 &    0 & 0.05 &    0 &    0 &    0 &    0 \\
   0 & -0.02 &    0 &    0 &    0 &    0 &    0 \\
0.14 &    0 & -0.06 &    0 &    0 &    0 &    0 \\
   0 & -0.02 &    0 &    0 &    0 &    0 &    0
}i$,

$\occmat^{\downarrow\uparrow}=
\setstackgap{L}{1.1\baselineskip}
\fixTABwidth{T}
\parenMatrixstack{
   0 & -0.03 &    0 &    0 &    0 &    0 &    0 \\
-0.01 &    0 & -0.07 &    0 &    0 &    0 &    0 \\
   0 & 0.01 &    0 &    0 &    0 &    0 &    0 \\
   0 &    0 &    0 &    0 & 0.07 &    0 & 0.07 \\
   0 &    0 &    0 & -0.11 &    0 & 0.12 &    0 \\
   0 &    0 &    0 &    0 & 0.07 &    0 & 0.07 \\
   0 &    0 &    0 & 0.12 &    0 & -0.13 &    0  
}+
\setstackgap{L}{1.1\baselineskip}
\fixTABwidth{T}
\parenMatrixstack{
   0 &    0 &    0 & 0.13 &    0 & -0.14 &    0 \\
   0 &    0 &    0 &    0 & 0.02 &    0 & 0.02 \\
   0 &    0 &    0 & -0.05 &    0 & 0.06 &    0 \\
0.04 &    0 & 0.23 &    0 &    0 &    0 &    0 \\
   0 & -0.02 &    0 &    0 &    0 &    0 &    0 \\
0.04 &    0 & 0.24 &    0 &    0 &    0 &    0 \\
   0 & 0.03 &    0 &    0 &    0 &    0 &    0
}i$,

$\occmat^{\downarrow\downarrow}=
\setstackgap{L}{1.1\baselineskip}
\fixTABwidth{T}
\parenMatrixstack{
0.04 &    0 & 0.03 &    0 &    0 &    0 &    0 \\
   0 & 0.11 &    0 &    0 &    0 &    0 &    0 \\
0.03 &    0 & 0.17 &    0 &    0 &    0 &    0 \\
   0 &    0 &    0 & 0.09 &    0 & -0.06 &    0 \\
   0 &    0 &    0 &    0 & 0.04 &    0 & 0.01 \\
   0 &    0 &    0 & -0.06 &    0 & 0.09 &    0 \\
   0 &    0 &    0 &    0 & 0.01 &    0 & 0.04
}+
\setstackgap{L}{1.1\baselineskip}
\fixTABwidth{T}
\parenMatrixstack{
  0 &    0 &    0 &    0 & -0.01 &    0 & -0.01 \\
   0 &    0 &    0 & -0.01 &    0 & 0.01 &    0 \\
   0 &    0 &    0 &    0 & -0.04 &    0 & -0.04 \\
   0 & 0.01 &    0 &    0 &    0 &    0 &    0 \\
0.01 &    0 & 0.04 &    0 &    0 &    0 &    0 \\
   0 & -0.01 &    0 &    0 &    0 &    0 &    0 \\
0.01 &    0 & 0.04 &    0 &    0 &    0 &    0
}i$.

\subsubsection{Properties of $\mathbb{S}_3^{1}$}

\begin{table*}[!h]%
\begin{ruledtabular}
\begin{tabular}{ll}
\textrm{Properties (unit)}&
\textrm{Values}\\
\colrule
Initiated from states & 
S$_3^1$ \\
Energy (eV) & -31.987 \\
Basis vectors (\AA)& $a=(
2.7669, 2.7667,     0
), b=(
  0, 2.7667, 2.7883
), c=(
2.7669,     0, 2.7883
)$\\
\begin{tabular}{@{}l@{}}Atom coordinates \\ (in basis vectors)\end{tabular} &
{\begin{tabular}{@{}l@{}}$U1=(
0.5000, 0.5000, 0.5000
)$ \\$O1=(
0.2500, 0.2500, 0.2500
)$, $O2=(
0.7500, 0.7500, 0.7500
)$\end{tabular}} \\ 
Strain & \begin{tabular}{@{}l@{}}($\epsilon_{xx},\epsilon_{yy},\epsilon_{zz},\epsilon_{xy},\epsilon_{xz},\epsilon_{yz}$)=(
-0.0025, -0.0026, 0.0052,     0,     0,     0
), \\($\epsilon_{A1g},\epsilon_{Eg.0},\epsilon_{Eg.1},\epsilon_{T2g.0},\epsilon_{T2g.1},\epsilon_{T2g.2}$)=(
 0, 0.0001, 0.0063,     0,     0,     0
).\end{tabular}\\
Spin moments ($\mu_B$)& (-0.002, -0.002, 1.419)\\
Orbit moments ($\mu_B$)& (0.003, 0.003, -2.508)\\
Total magnetic moments ($\mu_B$)& (0.001, 0.001, -1.089)\\
Energy (eV) and lattice (\AA) of und & -31.986, 2.7740
\end{tabular}
\end{ruledtabular}
\end{table*}

$\occmat^{\uparrow\uparrow}=
\setstackgap{L}{1.1\baselineskip}
\fixTABwidth{T}
\parenMatrixstack{
0.47 &    0 & 0.03 &    0 &    0 &    0 &    0 \\
   0 & 0.13 &    0 &    0 &    0 &    0 &    0 \\
0.03 &    0 & 0.04 &    0 &    0 &    0 &    0 \\
   0 &    0 &    0 & 0.78 &    0 &    0 &    0 \\
   0 &    0 &    0 &    0 & 0.04 &    0 & -0.03 \\
   0 &    0 &    0 &    0 &    0 & 0.03 &    0 \\
   0 &    0 &    0 &    0 & -0.03 &    0 & 0.47
}+
\setstackgap{L}{1.1\baselineskip}
\fixTABwidth{T}
\parenMatrixstack{
   0 &    0 &    0 &    0 & -0.02 &    0 & 0.44 \\
   0 &    0 &    0 &    0 &    0 &    0 &    0 \\
   0 &    0 &    0 &    0 &    0 &    0 & 0.02 \\
   0 &    0 &    0 &    0 &    0 &    0 &    0 \\
0.02 &    0 &    0 &    0 &    0 &    0 &    0 \\
   0 &    0 &    0 &    0 &    0 &    0 &    0 \\
-0.44 &    0 & -0.02 &    0 &    0 &    0 &    0
}i$,

$\occmat^{\uparrow\downarrow}=
\setstackgap{L}{1.1\baselineskip}
\fixTABwidth{T}
\parenMatrixstack{
   0 &    0 &    0 &    0 &    0 &    0 &    0 \\
-0.07 &    0 &    0 &    0 &    0 &    0 &    0 \\
   0 &    0 &    0 &    0 &    0 &    0 &    0 \\
   0 &    0 &    0 &    0 &    0 &    0 &    0 \\
   0 &    0 &    0 & 0.28 &    0 &    0 &    0 \\
   0 &    0 &    0 &    0 & 0.01 &    0 & -0.17 \\
   0 &    0 &    0 &    0 &    0 &    0 &    0 
}+
\setstackgap{L}{1.1\baselineskip}
\fixTABwidth{T}
\parenMatrixstack{
   0 &    0 &    0 &    0 &    0 &    0 &    0 \\
   0 &    0 &    0 &    0 &    0 &    0 & -0.07 \\
   0 &    0 &    0 & -0.28 &    0 &    0 &    0 \\
   0 &    0 &    0 &    0 &    0 &    0 &    0 \\
   0 &    0 &    0 &    0 &    0 &    0 &    0 \\
0.17 &    0 & 0.01 &    0 &    0 &    0 &    0 \\
   0 &    0 &    0 &    0 &    0 &    0 &    0
}i$,

$\occmat^{\downarrow\uparrow}=
\setstackgap{L}{1.1\baselineskip}
\fixTABwidth{T}
\parenMatrixstack{
  0 & -0.07 &    0 &    0 &    0 &    0 &    0 \\
   0 &    0 &    0 &    0 &    0 &    0 &    0 \\
   0 &    0 &    0 &    0 &    0 &    0 &    0 \\
   0 &    0 &    0 &    0 & 0.28 &    0 &    0 \\
   0 &    0 &    0 &    0 &    0 & 0.01 &    0 \\
   0 &    0 &    0 &    0 &    0 &    0 &    0 \\
   0 &    0 &    0 &    0 &    0 & -0.17 &    0
}+$
$\setstackgap{L}{1.1\baselineskip}
\fixTABwidth{T}
\parenMatrixstack{
   0 &    0 &    0 &    0 &    0 & -0.17 &    0 \\
   0 &    0 &    0 &    0 &    0 &    0 &    0 \\
   0 &    0 &    0 &    0 &    0 & -0.01 &    0 \\
   0 &    0 & 0.28 &    0 &    0 &    0 &    0 \\
   0 &    0 &    0 &    0 &    0 &    0 &    0 \\
   0 &    0 &    0 &    0 &    0 &    0 &    0 \\
   0 & 0.07 &    0 &    0 &    0 &    0 &    0
}i$,

$\occmat^{\downarrow\downarrow}=
\setstackgap{L}{1.1\baselineskip}
\setstacktabbedgap{9pt}
\fixTABwidth{T}
\parenMatrixstack{
0.03 &    0 &    0 &    0 &    0 &    0 &    0 \\
   0 & 0.12 &    0 &    0 &    0 &    0 &    0 \\
   0 &    0 & 0.14 &    0 &    0 &    0 &    0 \\
   0 &    0 &    0 & 0.03 &    0 &    0 &    0 \\
   0 &    0 &    0 &    0 & 0.13 &    0 &    0 \\
   0 &    0 &    0 &    0 &    0 & 0.09 &    0 \\
   0 &    0 &    0 &    0 &    0 &    0 & 0.03
}+$
$\setstackgap{L}{1.1\baselineskip}
\fixTABwidth{T}
\parenMatrixstack{
  0 &    0 &    0 &    0 &    0 &    0 &    0 \\
   0 &    0 &    0 &    0 &    0 & 0.02 &    0 \\
   0 &    0 &    0 &    0 & -0.11 &    0 &    0 \\
   0 &    0 &    0 &    0 &    0 &    0 &    0 \\
   0 &    0 & 0.11 &    0 &    0 &    0 &    0 \\
   0 & -0.02 &    0 &    0 &    0 &    0 &    0 \\
   0 &    0 &    0 &    0 &    0 &    0 &    0
}i$.

\clearpage

\subsection{\label{sec:sm1kafmsoc}1\textbf{k} AFM with SOC (the stage 3)}

Initialized by the occupation matrices of $\mathbb{S}_0$ in Section~\ref{sec:smfmsoc}, we obtained 1\textbf{k} AFM state using GGA+$U$ ($U=4$ eV). The resulting $\occmat$ was further used to initialize the calculations of GGA+$U$ and LDA+$U$ using different values of $U$. Due to the symmetry of the 1\textbf{k} AFM structure, the spin moments, orbit moments, total magnetic moments, and $\occmat$ of the first two U atoms, U1 and U2, are equal to those of U3 and U4, respectively. Therefore, these properties are only reported for U1 and U2. Also, only the $\occmat$ of $U=4$ eV are presented, as the $\occmat$ of other $U$ values are approximately identical to $U=4$ eV.

\subsubsection{Properties of $\mathbb{S}_0$ (1\textbf{k} AFM, GGA+U, U=2eV)}

\begin{table*}[!h]%
\begin{ruledtabular}
\begin{tabular}{ll}
\textrm{Properties (unit)}&
\textrm{Values}\\
\colrule
Energy (eV) & -130.700 \\
Basis vectors (\AA)& $a=(
5.5107,  0 ,  0)$, $b=(0 , 5.5072 , 0)$, $c=(0 ,     0 , 5.5240)$\\
\begin{tabular}{@{}l@{}}Atom coordinates \\ (in basis vectors)\end{tabular} &
{\begin{tabular}{@{}l@{}}U1:$(0.5000 ,     0 ,     0)$, U2:$(0.5000 , 0.5000 , 0.5000)$ \\U3:$(   0 ,     0 , 0.5000)$, U4:$(  0 , 0.5000 ,     0 )$ \\O1:$(0.2500 , 0.7500 , 0.7500)$, O2:$(0.2500 , 0.2500 , 0.7500)$\\O3:$(0.2500 , 0.2500 , 0.2500)$, O4:$(0.2500 , 0.7500 , 0.2500)$\\O5:$(0.7500 , 0.7500 , 0.2500)$, O6:$(0.7500 , 0.2500 , 0.2500)$\\O7:$(0.7500 , 0.2500 , 0.7500)$, O8:$(0.7500 , 0.7500 , 0.7500)$\end{tabular}} \\ 
Strain & \begin{tabular}{@{}l@{}}$(\epsilon_{xx},\epsilon_{yy},\epsilon_{zz},\epsilon_{xy},\epsilon_{xz},\epsilon_{yz})=(
-0.0006 , -0.0012 , 0.0018 ,     0 ,     0 ,     0
)$, \\$(\epsilon_{A1g},\epsilon_{Eg.0},\epsilon_{Eg.1},\epsilon_{T2g.0},\epsilon_{T2g.1},\epsilon_{T2g.2})=(
  0 , 0.0005 , 0.0022 ,     0 ,     0 ,     0
)$.\end{tabular}\\
Spin moments ($\mu_B$)& U1:$(0.000, 0.000, 1.467)$, U2:$(0.000, 0.000, -1.467)$\\
Orbit moments ($\mu_B$)& U1:$(0.000, 0.000, -3.363)$, U2:$(0.000, 0.000, 3.363)$\\
Total magnetic moments ($\mu_B$)& U1:$(0.000, 0.000, -1.896)$, U2:$(0.000, 0.000, 1.896)$\\
Energy (eV) and lattice (\AA) of und & -130.703, 5.5138
\end{tabular}
\end{ruledtabular}
\end{table*}

\subsubsection{Properties of $\mathbb{S}_0$ (1\textbf{k} AFM, GGA+U, U=3eV)}

\begin{table*}[!h]%
\begin{ruledtabular}
\begin{tabular}{ll}
\textrm{Properties (unit)}&
\textrm{Values}\\
\colrule
Energy (eV) & -129.395 \\
Basis vectors (\AA)& $a=(
5.5286,  0 ,  0
)$, $b=(
0 , 5.5263 , 0
)$, $c=(
0 ,     0 , 5.5415
)$\\
\begin{tabular}{@{}l@{}}Atom coordinates \\ (in basis vectors)\end{tabular} &
{\begin{tabular}{@{}l@{}}U1:$(0.5000 ,     0 ,     0)$, U2:$(0.5000 , 0.5000 , 0.5000)$ \\U3:$(   0 ,     0 , 0.5000)$, U4:$(  0 , 0.5000 ,     0 )$ \\O1:$(0.2500 , 0.7500 , 0.7500)$, O2:$(0.2500 , 0.2500 , 0.7500)$\\O3:$(0.2500 , 0.2500 , 0.2500)$, O4:$(0.2500 , 0.7500 , 0.2500)$\\O5:$(0.7500 , 0.7500 , 0.2500)$, O6:$(0.7500 , 0.2500 , 0.2500)$\\O7:$(0.7500 , 0.2500 , 0.7500)$, O8:$(0.7500 , 0.7500 , 0.7500)$\end{tabular}} \\ 
Strain & \begin{tabular}{@{}l@{}}$(\epsilon_{xx},\epsilon_{yy},\epsilon_{zz},\epsilon_{xy},\epsilon_{xz},\epsilon_{yz})=(
-0.0006, -0.0010, 0.0017,     0 ,     0 ,     0
)$, \\$(\epsilon_{A1g},\epsilon_{Eg.0},\epsilon_{Eg.1},\epsilon_{T2g.0},\epsilon_{T2g.1},\epsilon_{T2g.2})=(
  0, 0.0003, 0.0021,     0,     0,     0
)$.\end{tabular}\\
Spin moments ($\mu_B$)& U1:$(0.000, 0.000, 1.487)$, U2:$(0.000, 0.000, -1.487)$\\
Orbit moments ($\mu_B$)& U1:$(0.000, 0.000, -3.448)$, U2:$(0.000, 0.000, 3.448)$\\
Total magnetic moments ($\mu_B$)& U1:$(0.000, 0.000, -1.961)$, U2:$(0.000, 0.000, 1.961)$\\
Energy (eV) and lattice (\AA) of und & -129.396, 5.5321
\end{tabular}
\end{ruledtabular}
\end{table*}
\clearpage

\subsubsection{Properties of $\mathbb{S}_0$ (1\textbf{k} AFM, GGA+U, U=4eV)}

\begin{table*}[!h]%
\begin{ruledtabular}
\begin{tabular}{ll}
\textrm{Properties (unit)}&
\textrm{Values}\\
\colrule
Energy (eV) & -128.250 \\
Basis vectors (\AA)& $a=(
5.5435,  0 ,  0)$, $b=(0 , 5.5420 , 0)$, $c=(0 ,     0 , 5.5561)$\\
\begin{tabular}{@{}l@{}}Atom coordinates \\ (in basis vectors)\end{tabular} &
{\begin{tabular}{@{}l@{}}U1:$(0.5000 ,     0 ,     0)$, U2:$(0.5000 , 0.5000 , 0.5000)$ \\U3:$(   0 ,     0 , 0.5000)$, U4:$(  0 , 0.5000 ,     0 )$ \\O1:$(0.2500 , 0.7500 , 0.7500)$, O2:$(0.2500 , 0.2500 , 0.7500)$\\O3:$(0.2500 , 0.2500 , 0.2500)$, O4:$(0.2500 , 0.7500 , 0.2500)$\\O5:$(0.7500 , 0.7500 , 0.2500)$, O6:$(0.7500 , 0.2500 , 0.2500)$\\O7:$(0.7500 , 0.2500 , 0.7500)$, O8:$(0.7500 , 0.7500 , 0.7500)$\end{tabular}} \\ 
Strain & \begin{tabular}{@{}l@{}}$(\epsilon_{xx},\epsilon_{yy},\epsilon_{zz},\epsilon_{xy},\epsilon_{xz},\epsilon_{yz})=(
-0.0005, -0.0007, 0.0018,     0 ,     0 ,     0
)$, \\$(\epsilon_{A1g},\epsilon_{Eg.0},\epsilon_{Eg.1},\epsilon_{T2g.0},\epsilon_{T2g.1},\epsilon_{T2g.2})=(
0.0003, 0.0002, 0.0020,     0 ,     0 ,     0
)$.\end{tabular}\\
Spin moments ($\mu_B$)& U1:$(0.000, 0.000, 1.495)$, U2:$(0.000, 0.000, -1.495)$\\
Orbit moments ($\mu_B$)& U1:$(0.000, 0.000, -3.498)$, U2:$(0.000, 0.000, 3.498)$\\
Total magnetic moments ($\mu_B$)& U1:$(0.000, 0.000, -2.003)$, U2:$(0.000, 0.000, 2.003)$\\
Energy (eV) and lattice (\AA) of und & -128.249, 5.5461
\end{tabular}
\end{ruledtabular}
\end{table*}
$\occmat(\textrm{U}1)$:

$\occmat^{\uparrow\uparrow}=
\setstackgap{L}{1.1\baselineskip}
\fixTABwidth{T}
\parenMatrixstack{
0.46 &    0 & 0.04 &    0 &    0 &    0 &    0 \\
   0 & 0.13 &    0 &    0 &    0 &    0 &    0 \\
0.04 &    0 & 0.43 &    0 &    0 &    0 &    0 \\
   0 &    0 &    0 & 0.05 &    0 &    0 &    0 \\
   0 &    0 &    0 &    0 & 0.43 &    0 & -0.04 \\
   0 &    0 &    0 &    0 &    0 & 0.03 &    0 \\
   0 &    0 &    0 &    0 & -0.04 &    0 & 0.48
}+
\setstackgap{L}{1.1\baselineskip}
\fixTABwidth{T}
\parenMatrixstack{
   0 &    0 &    0 &    0 & -0.03 &    0 & 0.44 \\
   0 &    0 &    0 &    0 &    0 &    0 &    0 \\
   0 &    0 &    0 &    0 & 0.39 &    0 & 0.03 \\
   0 &    0 &    0 &    0 &    0 &    0 &    0 \\
0.03 &    0 & -0.39 &    0 &    0 &    0 &    0 \\
   0 &    0 &    0 &    0 &    0 &    0 &    0 \\
-0.44 &    0 & -0.03 &    0 &    0 &    0 &    0
}i$,

$\occmat^{\uparrow\downarrow}=
\setstackgap{L}{1.1\baselineskip}
\fixTABwidth{T}
\parenMatrixstack{
   0 &    0 &    0 &    0 &    0 &    0 &    0 \\
-0.07 &    0 &    0 &    0 &    0 &    0 &    0 \\
   0 &    0 &    0 &    0 &    0 &    0 &    0 \\
   0 &    0 &    0 &    0 & -0.26 &    0 &    0 \\
   0 &    0 &    0 &    0 &    0 &    0 &    0 \\
   0 &    0 &    0 &    0 & 0.01 &    0 & -0.17 \\
   0 &    0 &    0 &    0 &    0 &    0 &    0
}+
\setstackgap{L}{1.1\baselineskip}
\fixTABwidth{T}
\parenMatrixstack{
  0 &    0 &    0 &    0 &    0 &    0 &    0 \\
   0 &    0 &    0 &    0 &    0 &    0 & -0.08 \\
   0 &    0 &    0 &    0 &    0 &    0 &    0 \\
   0 &    0 & 0.26 &    0 &    0 &    0 &    0 \\
   0 &    0 &    0 &    0 &    0 &    0 &    0 \\
0.16 &    0 & 0.01 &    0 &    0 &    0 &    0 \\
   0 &    0 &    0 &    0 &    0 &    0 &    0
}i$,

$\occmat^{\downarrow\uparrow}=
\setstackgap{L}{1.1\baselineskip}
\fixTABwidth{T}
\parenMatrixstack{
   0 & -0.07 &    0 &    0 &    0 &    0 &    0 \\
   0 &    0 &    0 &    0 &    0 &    0 &    0 \\
   0 &    0 &    0 &    0 &    0 &    0 &    0 \\
   0 &    0 &    0 &    0 &    0 &    0 &    0 \\
   0 &    0 &    0 & -0.26 &    0 & 0.01 &    0 \\
   0 &    0 &    0 &    0 &    0 &    0 &    0 \\
   0 &    0 &    0 &    0 &    0 & -0.17 &    0
}+
\setstackgap{L}{1.1\baselineskip}
\fixTABwidth{T}
\parenMatrixstack{
  0 &    0 &    0 &    0 &    0 & -0.16 &    0 \\
   0 &    0 &    0 &    0 &    0 &    0 &    0 \\
   0 &    0 &    0 & -0.26 &    0 & -0.01 &    0 \\
   0 &    0 &    0 &    0 &    0 &    0 &    0 \\
   0 &    0 &    0 &    0 &    0 &    0 &    0 \\
   0 &    0 &    0 &    0 &    0 &    0 &    0 \\
   0 & 0.08 &    0 &    0 &    0 &    0 &    0
}i$,

$\occmat^{\downarrow\downarrow}=
\setstackgap{L}{1.1\baselineskip}
\fixTABwidth{T}
\parenMatrixstack{
0.03 &    0 & 0.01 &    0 &    0 &    0 &    0 \\
   0 & 0.12 &    0 &    0 &    0 &    0 &    0 \\
0.01 &    0 & 0.03 &    0 &    0 &    0 &    0 \\
   0 &    0 &    0 & 0.21 &    0 &    0 &    0 \\
   0 &    0 &    0 &    0 & 0.03 &    0 & -0.01 \\
   0 &    0 &    0 &    0 &    0 & 0.09 &    0 \\
   0 &    0 &    0 &    0 & -0.01 &    0 & 0.03
}+
\setstackgap{L}{1.1\baselineskip}
\fixTABwidth{T}
\parenMatrixstack{
   0 &    0 &    0 &    0 &    0 &    0 &    0 \\
   0 &    0 &    0 &    0 &    0 & 0.02 &    0 \\
   0 &    0 &    0 &    0 &    0 &    0 &    0 \\
   0 &    0 &    0 &    0 &    0 &    0 &    0 \\
   0 &    0 &    0 &    0 &    0 &    0 &    0 \\
   0 & -0.02 &    0 &    0 &    0 &    0 &    0 \\
   0 &    0 &    0 &    0 &    0 &    0 &    0
}i$.

$\occmat(\textrm{U}2)$:

$\occmat^{\uparrow\uparrow}=
\setstackgap{L}{1.1\baselineskip}
\fixTABwidth{T}
\parenMatrixstack{
0.03 &    0 & 0.01 &    0 &    0 &    0 &    0 \\
   0 & 0.12 &    0 &    0 &    0 &    0 &    0 \\
0.01 &    0 & 0.03 &    0 &    0 &    0 &    0 \\
   0 &    0 &    0 & 0.21 &    0 &    0 &    0 \\
   0 &    0 &    0 &    0 & 0.03 &    0 & -0.01 \\
   0 &    0 &    0 &    0 &    0 & 0.09 &    0 \\
   0 &    0 &    0 &    0 & -0.01 &    0 & 0.03 
}+
\setstackgap{L}{1.1\baselineskip}
\fixTABwidth{T}
\parenMatrixstack{
  0 &    0 &    0 &    0 &    0 &    0 &    0 \\
   0 &    0 &    0 &    0 &    0 & -0.02 &    0 \\
   0 &    0 &    0 &    0 &    0 &    0 &    0 \\
   0 &    0 &    0 &    0 &    0 &    0 &    0 \\
   0 &    0 &    0 &    0 &    0 &    0 &    0 \\
   0 & 0.02 &    0 &    0 &    0 &    0 &    0 \\
   0 &    0 &    0 &    0 &    0 &    0 &    0
}i$,

$\occmat^{\uparrow\downarrow}=
\setstackgap{L}{1.1\baselineskip}
\fixTABwidth{T}
\parenMatrixstack{
   0 & 0.07 &    0 &    0 &    0 &    0 &    0 \\
   0 &    0 &    0 &    0 &    0 &    0 &    0 \\
   0 &    0 &    0 &    0 &    0 &    0 &    0 \\
   0 &    0 &    0 &    0 &    0 &    0 &    0 \\
   0 &    0 &    0 & 0.26 &    0 & -0.01 &    0 \\
   0 &    0 &    0 &    0 &    0 &    0 &    0 \\
   0 &    0 &    0 &    0 &    0 & 0.17 &    0
}+
\setstackgap{L}{1.1\baselineskip}
\fixTABwidth{T}
\parenMatrixstack{
  0 &    0 &    0 &    0 &    0 & -0.16 &    0 \\
   0 &    0 &    0 &    0 &    0 &    0 &    0 \\
   0 &    0 &    0 & -0.26 &    0 & -0.01 &    0 \\
   0 &    0 &    0 &    0 &    0 &    0 &    0 \\
   0 &    0 &    0 &    0 &    0 &    0 &    0 \\
   0 &    0 &    0 &    0 &    0 &    0 &    0 \\
   0 & 0.08 &    0 &    0 &    0 &    0 &    0 
}i$,

$\occmat^{\downarrow\uparrow}=
\setstackgap{L}{1.1\baselineskip}
\fixTABwidth{T}
\parenMatrixstack{
   0 &    0 &    0 &    0 &    0 &    0 &    0 \\
0.07 &    0 &    0 &    0 &    0 &    0 &    0 \\
   0 &    0 &    0 &    0 &    0 &    0 &    0 \\
   0 &    0 &    0 &    0 & 0.26 &    0 &    0 \\
   0 &    0 &    0 &    0 &    0 &    0 &    0 \\
   0 &    0 &    0 &    0 & -0.01 &    0 & 0.17 \\
   0 &    0 &    0 &    0 &    0 &    0 &    0
}+
\setstackgap{L}{1.1\baselineskip}
\fixTABwidth{T}
\parenMatrixstack{
   0 &    0 &    0 &    0 &    0 &    0 &    0 \\
   0 &    0 &    0 &    0 &    0 &    0 & -0.08 \\
   0 &    0 &    0 &    0 &    0 &    0 &    0 \\
   0 &    0 & 0.26 &    0 &    0 &    0 &    0 \\
   0 &    0 &    0 &    0 &    0 &    0 &    0 \\
0.16 &    0 & 0.01 &    0 &    0 &    0 &    0 \\
   0 &    0 &    0 &    0 &    0 &    0 &    0
}i$,

$\occmat^{\downarrow\downarrow}=
\setstackgap{L}{1.1\baselineskip}
\fixTABwidth{T}
\parenMatrixstack{
0.46 &    0 & 0.04 &    0 &    0 &    0 &    0 \\
   0 & 0.13 &    0 &    0 &    0 &    0 &    0 \\
0.04 &    0 & 0.43 &    0 &    0 &    0 &    0 \\
   0 &    0 &    0 & 0.05 &    0 &    0 &    0 \\
   0 &    0 &    0 &    0 & 0.43 &    0 & -0.04 \\
   0 &    0 &    0 &    0 &    0 & 0.03 &    0 \\
   0 &    0 &    0 &    0 & -0.04 &    0 & 0.48
}+
\setstackgap{L}{1.1\baselineskip}
\fixTABwidth{T}
\parenMatrixstack{
   0 &    0 &    0 &    0 & 0.03 &    0 & -0.44 \\
   0 &    0 &    0 &    0 &    0 &    0 &    0 \\
   0 &    0 &    0 &    0 & -0.39 &    0 & -0.03 \\
   0 &    0 &    0 &    0 &    0 &    0 &    0 \\
-0.03 &    0 & 0.39 &    0 &    0 &    0 &    0 \\
   0 &    0 &    0 &    0 &    0 &    0 &    0 \\
0.44 &    0 & 0.03 &    0 &    0 &    0 &    0
}i$.

\bigskip
\noindent Relaxed crystal structure in the format of POSCAR:\\
U4 O8       \\                            
   1.00000000000000    \\ 
     5.5434830476892865    0.0000000272409901    0.0000160363414482\\
     0.0000000853804913    5.5420213397748066   -0.0000125008649321\\
     0.0000165190435380   -0.0000127030963634    5.5561139766636076\\
   U    O \\
     4     8\\
Direct\\
  0.5000000003016624  0.0000000006865173  0.0000000003244056\\
  0.5000000009495792  0.5000000001614202  0.5000000012394971\\
  0.0000000018931981  0.9999999984055028  0.5000000020256418\\
  0.0000000009185781  0.5000000000994982  0.0000000013592359\\
  0.2500006838439653  0.7499993836991272  0.7499997375344126\\
  0.2499992752355948  0.2500006462170571  0.7499990163392894\\
  0.2500007374953186  0.2499993349276320  0.2499990313260836\\
  0.2499992904683926  0.7500006160008827  0.2499997137984222\\
  0.7500007235291865  0.7499993538252078  0.2500009864284182\\
  0.7499993136005475  0.2500006179527046  0.2500002592009659\\
  0.7500007068553984  0.2499993841300096  0.7500002806192926\\
  0.7499992649085787  0.7500006638944333  0.7500009698043353
\clearpage

\subsubsection{Properties of $\mathbb{S}_0$ (1\textbf{k} AFM, GGA+U, U=5eV)}

\begin{table*}[!h]%
\begin{ruledtabular}
\begin{tabular}{ll}
\textrm{Properties (unit)}&
\textrm{Values}\\
\colrule
Energy (eV) & -127.236 \\
Basis vectors (\AA)& $a=(
5.5566,  0 ,  0)$, $b=(0 , 5.5557 , 0)$, $c=(0 ,     0 , 5.5688)$\\
\begin{tabular}{@{}l@{}}Atom coordinates \\ (in basis vectors)\end{tabular} &
{\begin{tabular}{@{}l@{}}U1:$(0.5000 ,     0 ,     0)$, U2:$(0.5000 , 0.5000 , 0.5000)$ \\U3:$(   0 ,     0 , 0.5000)$, U4:$(  0 , 0.5000 ,     0 )$ \\O1:$(0.2500 , 0.7500 , 0.7500)$, O2:$(0.2500 , 0.2500 , 0.7500)$\\O3:$(0.2500 , 0.2500 , 0.2500)$, O4:$(0.2500 , 0.7500 , 0.2500)$\\O5:$(0.7500 , 0.7500 , 0.2500)$, O6:$(0.7500 , 0.2500 , 0.2500)$\\O7:$(0.7500 , 0.2500 , 0.7500)$, O8:$(0.7500 , 0.7500 , 0.7500)$\end{tabular}} \\ 
Strain & \begin{tabular}{@{}l@{}}$(\epsilon_{xx},\epsilon_{yy},\epsilon_{zz},\epsilon_{xy},\epsilon_{xz},\epsilon_{yz})=(
-0.0007, -0.0009, 0.0015,     0 ,     0 ,     0
)$, \\$(\epsilon_{A1g},\epsilon_{Eg.0},\epsilon_{Eg.1},\epsilon_{T2g.0},\epsilon_{T2g.1},\epsilon_{T2g.2})=(
    0, 0.0001, 0.0019,     0 ,     0 ,     0
)$.\end{tabular}\\
Spin moments ($\mu_B$)& U1:$(0.000, 0.000, 1.501)$, U2:$(0.000, 0.000, -1.501)$\\
Orbit moments ($\mu_B$)& U1:$(0.000, 0.000, -3.534)$, U2:$(0.000, 0.000, 3.534)$\\
Total magnetic moments ($\mu_B$)& U1:$(0.000, 0.000, -2.033)$, U2:$(0.000, 0.000, 2.033)$\\
Energy (eV) and lattice (\AA) of und & -127.234, 5.5605
\end{tabular}
\end{ruledtabular}
\end{table*}

\subsubsection{Properties of $\mathbb{S}_0$ (1\textbf{k} AFM, LDA+U, U=2eV)}

\begin{table*}[!h]%
\begin{ruledtabular}
\begin{tabular}{ll}
\textrm{Properties (unit)}&
\textrm{Values}\\
\colrule
Energy (eV) & -142.037 \\
Basis vectors (\AA)& $a=(
5.4115,  0 ,  0)$, $b=(0 , 5.4058 , 0)$, $c=(0 ,     0 , 5.4201)$\\
\begin{tabular}{@{}l@{}}Atom coordinates \\ (in basis vectors)\end{tabular} &
{\begin{tabular}{@{}l@{}}U1:$(0.5000 ,     0 ,     0)$, U2:$(0.5000 , 0.5000 , 0.5000)$ \\U3:$(   0 ,     0 , 0.5000)$, U4:$(  0 , 0.5000 ,     0 )$ \\O1:$(0.2500 , 0.7500 , 0.7500)$, O2:$(0.2500 , 0.2500 , 0.7500)$\\O3:$(0.2500 , 0.2500 , 0.2500)$, O4:$(0.2500 , 0.7500 , 0.2500)$\\O5:$(0.7500 , 0.7500 , 0.2500)$, O6:$(0.7500 , 0.2500 , 0.2500)$\\O7:$(0.7500 , 0.2500 , 0.7500)$, O8:$(0.7500 , 0.7500 , 0.7500)$\end{tabular}} \\ 
Strain & \begin{tabular}{@{}l@{}}$(\epsilon_{xx},\epsilon_{yy},\epsilon_{zz},\epsilon_{xy},\epsilon_{xz},\epsilon_{yz})=(
    0, -0.0011, 0.0016,     0 ,     0 ,     0
)$, \\$(\epsilon_{A1g},\epsilon_{Eg.0},\epsilon_{Eg.1},\epsilon_{T2g.0},\epsilon_{T2g.1},\epsilon_{T2g.2})=(
0.0002, 0.0008, 0.0017,     0 ,     0 ,     0
)$.\end{tabular}\\
Spin moments ($\mu_B$)& U1:$(0.001, -0.002, -1.194)$, U2:$(0.002, -0.001, 1.194)$\\
Orbit moments ($\mu_B$)& U1:$(-0.002, 0.005, 3.063)$, U2:$(-0.005, 0.002, -3.063)$\\
Total magnetic moments ($\mu_B$)& U1:$(-0.001, 0.003, 1.869)$, U2:$(-0.003, 0.001, -1.869)$\\
Energy (eV) and lattice (\AA) of und & -142.042, 5.4117
\end{tabular}
\end{ruledtabular}
\end{table*}

\clearpage
\subsubsection{Properties of $\mathbb{S}_0$ (1\textbf{k} AFM, LDA+U, U=3eV)}

\begin{table*}[!h]%
\begin{ruledtabular}
\begin{tabular}{ll}
\textrm{Properties (unit)}&
\textrm{Values}\\
\colrule
Energy (eV) & -140.579 \\
Basis vectors (\AA)& $a=(
5.4320,  0 ,  0)$, $b=(0 , 5.4282 , 0)$, $c=(0 ,     0 , 5.4409)$\\
\begin{tabular}{@{}l@{}}Atom coordinates \\ (in basis vectors)\end{tabular} &
{\begin{tabular}{@{}l@{}}U1:$(0.5000 ,     0 ,     0)$, U2:$(0.5000 , 0.5000 , 0.5000)$ \\U3:$(   0 ,     0 , 0.5000)$, U4:$(  0 , 0.5000 ,     0 )$ \\O1:$(0.2500 , 0.7500 , 0.7500)$, O2:$(0.2500 , 0.2500 , 0.7500)$\\O3:$(0.2500 , 0.2500 , 0.2500)$, O4:$(0.2500 , 0.7500 , 0.2500)$\\O5:$(0.7500 , 0.7500 , 0.2500)$, O6:$(0.7500 , 0.2500 , 0.2500)$\\O7:$(0.7500 , 0.2500 , 0.7500)$, O8:$(0.7500 , 0.7500 , 0.7500)$\end{tabular}} \\ 
Strain & \begin{tabular}{@{}l@{}}$(\epsilon_{xx},\epsilon_{yy},\epsilon_{zz},\epsilon_{xy},\epsilon_{xz},\epsilon_{yz})=(
-0.0002, -0.0009, 0.0014,     0 ,     0 ,     0
)$, \\$(\epsilon_{A1g},\epsilon_{Eg.0},\epsilon_{Eg.1},\epsilon_{T2g.0},\epsilon_{T2g.1},\epsilon_{T2g.2})=(
0.0002, 0.0005, 0.0016,     0 ,     0 ,     0
)$.\end{tabular}\\
Spin moments ($\mu_B$)& U1:$(0.003, -0.007, -1.234)$, U2:$(0.007, 0.000, 1.234)$\\
Orbit moments ($\mu_B$)& U1:$(-0.007, 0.020, 3.194)$, U2:$(-0.021, 0.001, -3.194)$\\
Total magnetic moments ($\mu_B$)& U1:$(-0.004, 0.013, 1.960)$, U2:$(-0.014, 0.001, 1.960)$\\
Energy (eV) and lattice (\AA) of und & -140.580, 5.4331
\end{tabular}
\end{ruledtabular}
\end{table*}

\subsubsection{Properties of $\mathbb{S}_0$ (1\textbf{k} AFM, LDA+U, U=4eV)}

\begin{table*}[!h]%
\begin{ruledtabular}
\begin{tabular}{ll}
\textrm{Properties (unit)}&
\textrm{Values}\\
\colrule
Energy (eV) & -139.310 \\
Basis vectors (\AA)& $a=(
5.4483,  0 ,  0)$, $b=(0 , 5.4457 , 0)$, $c=(0 ,     0 , 5.4576)$\\
\begin{tabular}{@{}l@{}}Atom coordinates \\ (in basis vectors)\end{tabular} &
{\begin{tabular}{@{}l@{}}U1:$(0.5000 ,     0 ,     0)$, U2:$(0.5000 , 0.5000 , 0.5000)$ \\U3:$(   0 ,     0 , 0.5000)$, U4:$(  0 , 0.5000 ,     0 )$ \\O1:$(0.2500 , 0.7500 , 0.7500)$, O2:$(0.2500 , 0.2500 , 0.7500)$\\O3:$(0.2500 , 0.2500 , 0.2500)$, O4:$(0.2500 , 0.7500 , 0.2500)$\\O5:$(0.7500 , 0.7500 , 0.2500)$, O6:$(0.7500 , 0.2500 , 0.2500)$\\O7:$(0.7500 , 0.2500 , 0.7500)$, O8:$(0.7500 , 0.7500 , 0.7500)$\end{tabular}} \\ 
Strain & \begin{tabular}{@{}l@{}}$(\epsilon_{xx},\epsilon_{yy},\epsilon_{zz},\epsilon_{xy},\epsilon_{xz},\epsilon_{yz})=(
-0.0003, -0.0008, 0.0014,     0 ,     0 ,     0
)$, \\$(\epsilon_{A1g},\epsilon_{Eg.0},\epsilon_{Eg.1},\epsilon_{T2g.0},\epsilon_{T2g.1},\epsilon_{T2g.2})=(
0.0002, 0.0003, 0.0016,     0 ,     0 ,     0
)$.\end{tabular}\\
Spin moments ($\mu_B$)& U1:$(0.001, -0.006, -1.253)$, U2:$(0.006, -0.001, 1.253)$\\
Orbit moments ($\mu_B$)& U1:$(-0.004, 0.018, 3.271)$, U2:$(-0.017, 0.004, -3.271)$\\
Total magnetic moments ($\mu_B$)& U1:$(-0.003, 0.012, -2.018)$, U1:$(-0.011, 0.003, 2.018)$\\
Energy (eV) and lattice (\AA) of und & -139.309, 5.4500
\end{tabular}
\end{ruledtabular}
\end{table*}

$\occmat(\textrm{U}1)$:

$\occmat^{\uparrow\uparrow}=
\setstackgap{L}{1.1\baselineskip}
\fixTABwidth{T}
\parenMatrixstack{
0.04 &    0 & 0.01 &    0 &    0 &    0 &    0 \\
   0 & 0.13 &    0 &    0 &    0 &    0 &    0 \\
0.01 &    0 & 0.03 &    0 &    0 &    0 &    0 \\
   0 &    0 &    0 & 0.27 &    0 &    0 &    0 \\
   0 &    0 &    0 &    0 & 0.03 &    0 & -0.01 \\
   0 &    0 &    0 &    0 &    0 & 0.14 &    0 \\
   0 &    0 &    0 &    0 & -0.01 &    0 & 0.04 
}+
\setstackgap{L}{1.1\baselineskip}
\fixTABwidth{T}
\parenMatrixstack{
   0 &    0 &    0 &    0 &    0 &    0 &    0 \\
   0 &    0 &    0 &    0 &    0 & -0.03 &    0 \\
   0 &    0 &    0 &    0 &    0 &    0 &    0 \\
   0 &    0 &    0 &    0 &    0 &    0 &    0 \\
   0 &    0 &    0 &    0 &    0 &    0 &    0 \\
   0 & 0.03 &    0 &    0 &    0 &    0 &    0 \\
   0 &    0 &    0 &    0 &    0 &    0 &    0 
}i$,

$\occmat^{\uparrow\downarrow}=
\setstackgap{L}{1.1\baselineskip}
\fixTABwidth{T}
\parenMatrixstack{
   0 & 0.07 &    0 &    0 &    0 &    0 &    0 \\
   0 &    0 &    0 &    0 &    0 &    0 &    0 \\
   0 &    0 &    0 &    0 &    0 &    0 &    0 \\
   0 &    0 &    0 &    0 &    0 &    0 &    0 \\
   0 &    0 &    0 & 0.29 &    0 & -0.02 &    0 \\
   0 &    0 &    0 &    0 &    0 &    0 &    0 \\
   0 &    0 &    0 &    0 &    0 & 0.21 &    0
}+
\setstackgap{L}{1.1\baselineskip}
\fixTABwidth{T}
\parenMatrixstack{
   0 &    0 &    0 &    0 &    0 & -0.21 &    0 \\
   0 &    0 &    0 &    0 &    0 &    0 &    0 \\
   0 &    0 &    0 & -0.29 &    0 & -0.03 &    0 \\
   0 &    0 &    0 &    0 &    0 &    0 &    0 \\
   0 & -0.01 &    0 &    0 &    0 &    0 &    0 \\
   0 &    0 &    0 &    0 &    0 &    0 &    0 \\
   0 & 0.07 &    0 &    0 &    0 &    0 &    0
}i$,

$\occmat^{\downarrow\uparrow}=
\setstackgap{L}{1.1\baselineskip}
\fixTABwidth{T}
\parenMatrixstack{
   0 &    0 &    0 &    0 &    0 &    0 &    0 \\
0.07 &    0 &    0 &    0 &    0 &    0 &    0 \\
   0 &    0 &    0 &    0 &    0 &    0 &    0 \\
   0 &    0 &    0 &    0 & 0.29 &    0 &    0 \\
   0 &    0 &    0 &    0 &    0 &    0 &    0 \\
   0 &    0 &    0 &    0 & -0.02 &    0 & 0.21 \\
   0 &    0 &    0 &    0 &    0 &    0 &    0
}+
\setstackgap{L}{1.1\baselineskip}
\fixTABwidth{T}
\parenMatrixstack{
   0 &    0 &    0 &    0 &    0 &    0 &    0 \\
   0 &    0 &    0 &    0 & 0.01 &    0 & -0.07 \\
   0 &    0 &    0 &    0 &    0 &    0 &    0 \\
   0 &    0 & 0.29 &    0 &    0 &    0 &    0 \\
   0 &    0 &    0 &    0 &    0 &    0 &    0 \\
0.21 &    0 & 0.03 &    0 &    0 &    0 &    0 \\
   0 &    0 &    0 &    0 &    0 &    0 &    0 
}i$,

$\occmat^{\downarrow\downarrow}=
\setstackgap{L}{1.1\baselineskip}
\fixTABwidth{T}
\parenMatrixstack{
0.44 &    0 & 0.04 &    0 &    0 &    0 &    0 \\
   0 & 0.13 &    0 &    0 &    0 &    0 &    0 \\
0.04 &    0 & 0.40 &    0 &    0 &    0 &    0 \\
   0 &    0 &    0 & 0.05 &    0 &    0 &    0 \\
   0 &    0 &    0 &    0 & 0.40 &    0 & -0.04 \\
   0 &    0 &    0 &    0 &    0 & 0.03 &    0 \\
   0 &    0 &    0 &    0 & -0.04 &    0 & 0.45 
}+
\setstackgap{L}{1.1\baselineskip}
\fixTABwidth{T}
\parenMatrixstack{
   0 &    0 &    0 &    0 & 0.05 &    0 & -0.41 \\
   0 &    0 &    0 &    0 &    0 &    0 &    0 \\
   0 &    0 &    0 &    0 & -0.36 &    0 & -0.05 \\
   0 &    0 &    0 &    0 &    0 &    0 &    0 \\
-0.05 &    0 & 0.36 &    0 &    0 &    0 &    0 \\
   0 &    0 &    0 &    0 &    0 &    0 &    0 \\
0.41 &    0 & 0.05 &    0 &    0 &    0 &    0
}i$.

$\occmat(\textrm{U}2)$:

$\occmat^{\uparrow\uparrow}=
\setstackgap{L}{1.1\baselineskip}
\fixTABwidth{T}
\parenMatrixstack{
0.44 &    0 & 0.04 &    0 &    0 &    0 &    0 \\
   0 & 0.13 &    0 &    0 &    0 &    0 &    0 \\
0.04 &    0 & 0.40 &    0 &    0 &    0 &    0 \\
   0 &    0 &    0 & 0.05 &    0 &    0 &    0 \\
   0 &    0 &    0 &    0 & 0.40 &    0 & -0.04 \\
   0 &    0 &    0 &    0 &    0 & 0.03 &    0 \\
   0 &    0 &    0 &    0 & -0.04 &    0 & 0.45  
}+
\setstackgap{L}{1.1\baselineskip}
\fixTABwidth{T}
\parenMatrixstack{
   0 &    0 &    0 &    0 & -0.05 &    0 & 0.41 \\
   0 &    0 &    0 &    0 &    0 &    0 &    0 \\
   0 &    0 &    0 &    0 & 0.36 &    0 & 0.05 \\
   0 &    0 &    0 &    0 &    0 &    0 &    0 \\
0.05 &    0 & -0.36 &    0 &    0 &    0 &    0 \\
   0 &    0 &    0 &    0 &    0 &    0 &    0 \\
-0.41 &    0 & -0.05 &    0 &    0 &    0 &    0
}i$,

$\occmat^{\uparrow\downarrow}=
\setstackgap{L}{1.1\baselineskip}
\fixTABwidth{T}
\parenMatrixstack{
   0 &    0 &    0 &    0 &    0 &    0 &    0 \\
-0.07 &    0 &    0 &    0 &    0 &    0 &    0 \\
   0 &    0 &    0 &    0 &    0 &    0 &    0 \\
   0 &    0 &    0 &    0 & -0.29 &    0 &    0 \\
   0 &    0 &    0 &    0 &    0 &    0 &    0 \\
   0 &    0 &    0 &    0 & 0.02 &    0 & -0.21 \\
   0 &    0 &    0 &    0 &    0 &    0 &    0
}+
\setstackgap{L}{1.1\baselineskip}
\fixTABwidth{T}
\parenMatrixstack{
   0 &    0 &    0 &    0 &    0 &    0 &    0 \\
   0 &    0 &    0 &    0 & 0.01 &    0 & -0.07 \\
   0 &    0 &    0 &    0 &    0 &    0 &    0 \\
   0 &    0 & 0.29 &    0 &    0 &    0 &    0 \\
   0 &    0 &    0 &    0 &    0 &    0 &    0 \\
0.21 &    0 & 0.03 &    0 &    0 &    0 &    0 \\
   0 &    0 &    0 &    0 &    0 &    0 &    0
}i$,

$\occmat^{\downarrow\uparrow}=
\setstackgap{L}{1.1\baselineskip}
\fixTABwidth{T}
\parenMatrixstack{
   0 & -0.07 &    0 &    0 &    0 &    0 &    0 \\
   0 &    0 &    0 &    0 &    0 &    0 &    0 \\
   0 &    0 &    0 &    0 &    0 &    0 &    0 \\
   0 &    0 &    0 &    0 &    0 &    0 &    0 \\
   0 &    0 &    0 & -0.29 &    0 & 0.02 &    0 \\
   0 &    0 &    0 &    0 &    0 &    0 &    0 \\
   0 &    0 &    0 &    0 &    0 & -0.21 &    0
}+
\setstackgap{L}{1.1\baselineskip}
\fixTABwidth{T}
\parenMatrixstack{
   0 &    0 &    0 &    0 &    0 & -0.21 &    0 \\
   0 &    0 &    0 &    0 &    0 &    0 &    0 \\
   0 &    0 &    0 & -0.29 &    0 & -0.03 &    0 \\
   0 &    0 &    0 &    0 &    0 &    0 &    0 \\
   0 & -0.01 &    0 &    0 &    0 &    0 &    0 \\
   0 &    0 &    0 &    0 &    0 &    0 &    0 \\
   0 & 0.07 &    0 &    0 &    0 &    0 &    0
}i$,

$\occmat^{\downarrow\downarrow}=
\setstackgap{L}{1.1\baselineskip}
\fixTABwidth{T}
\parenMatrixstack{
0.04 &    0 & 0.01 &    0 &    0 &    0 &    0 \\
   0 & 0.13 &    0 &    0 &    0 &    0 &    0 \\
0.01 &    0 & 0.03 &    0 &    0 &    0 &    0 \\
   0 &    0 &    0 & 0.27 &    0 &    0 &    0 \\
   0 &    0 &    0 &    0 & 0.03 &    0 & -0.01 \\
   0 &    0 &    0 &    0 &    0 & 0.14 &    0 \\
   0 &    0 &    0 &    0 & -0.01 &    0 & 0.04
}+
\setstackgap{L}{1.1\baselineskip}
\fixTABwidth{T}
\parenMatrixstack{
   0 &    0 &    0 &    0 &    0 &    0 &    0 \\
   0 &    0 &    0 &    0 &    0 & 0.03 &    0 \\
   0 &    0 &    0 &    0 &    0 &    0 &    0 \\
   0 &    0 &    0 &    0 &    0 &    0 &    0 \\
   0 &    0 &    0 &    0 &    0 &    0 &    0 \\
   0 & -0.03 &    0 &    0 &    0 &    0 &    0 \\
   0 &    0 &    0 &    0 &    0 &    0 &    0 
}i$.

\bigskip
\noindent Relaxed crystal structure in the format of POSCAR:\\
U4 O8          \\                         
   1.00000000000000     \\
     5.4482864073235309   -0.0000004510084203    0.0000236285273868\\
     0.0000001029388739    5.4457184715294256    0.0000341828404409\\
     0.0000280639735352    0.0000457713739367    5.4575894734366708\\
   U    O \\
     4     8\\
Direct\\
  0.4999999998533022 -0.0000000002314476 -0.0000000003322726\\
  0.4999999999736697  0.5000000002061282  0.5000000002980081\\
  0.0000000000850247 -0.0000000001843882  0.5000000004565766\\
  0.0000000001751999  0.5000000001858915 -0.0000000003890123\\
  0.2499911724798545  0.7500010808300769  0.7500130591179383\\
  0.2499998566789327  0.2500033197509033  0.7500028040315000\\
  0.2500001300957470  0.2499964411744289  0.2500028533756473\\
  0.2500089557706671  0.7499987288876079  0.2500130166086011\\
  0.7500001433224749  0.7499966801965211  0.2499971959686774\\
  0.7500088275140546  0.2499989191876259  0.2499869409706972\\
  0.7499910441922029  0.2500012711363568  0.7499869832922026\\
  0.7499998698588626  0.7500035588602955  0.7499971466014579

\subsubsection{Properties of $\mathbb{S}_0$ (1\textbf{k} AFM, LDA+U, U=5eV)}

\begin{table*}[!h]%
\begin{ruledtabular}
\begin{tabular}{ll}
\textrm{Properties (unit)}&
\textrm{Values}\\
\colrule
Energy (eV) & -138.176 \\
Basis vectors (\AA)& $a=(
5.4625,     0, -0.0006)$, $b=(    0, 5.4606, 0.0006)$, $c=(-0.0007, 0.0007, 5.4730)$\\
\begin{tabular}{@{}l@{}}Atom coordinates \\ (in basis vectors)\end{tabular} &
{\begin{tabular}{@{}l@{}}U1:$(0.5000 ,     0 ,     0)$, U2:$(0.5000 , 0.5000 , 0.5000)$ \\U3:$(   0 ,     0 , 0.5000)$, U4:$(  0 , 0.5000 ,     0 )$ \\O1:$(0.2500 , 0.7500 , 0.7500)$, O2:$(0.2500 , 0.2500 , 0.7500)$\\O3:$(0.2500 , 0.2500 , 0.2500)$, O4:$(0.2500 , 0.7500 , 0.2500)$\\O5:$(0.7500 , 0.7500 , 0.2500)$, O6:$(0.7500 , 0.2500 , 0.2500)$\\O7:$(0.7500 , 0.2500 , 0.7500)$, O8:$(0.7500 , 0.7500 , 0.7500)$\end{tabular}} \\ 
Strain & \begin{tabular}{@{}l@{}}$(\epsilon_{xx},\epsilon_{yy},\epsilon_{zz},\epsilon_{xy},\epsilon_{xz},\epsilon_{yz})=(
-0.0004, -0.0007, 0.0015,     0, -0.0001, 0.0001
)$, \\$(\epsilon_{A1g},\epsilon_{Eg.0},\epsilon_{Eg.1},\epsilon_{T2g.0},\epsilon_{T2g.1},\epsilon_{T2g.2})=(
0.0002, 0.0002, 0.0017,     0, 0.0001, -0.0001
)$.\end{tabular}\\
Spin moments ($\mu_B$)& U1:$(-0.116, 0.066, -1.256)$, U2:$(0.204, -0.204, 1.221)$\\
Orbit moments ($\mu_B$)& U1:$(0.358, -0.205, 3.314)$, U2:$(-0.626, 0.631, -3.249)$\\
Total magnetic moments ($\mu_B$)& U1:$(0.242, -0.139, 2.058)$, U1:$(-0.422, 0.427, -2.028)$\\
Energy (eV) and lattice (\AA) of und & -138.175, 5.4646
\end{tabular}
\end{ruledtabular}
\end{table*}
\clearpage

\subsection{\label{sec:sm3kafmsoc}3\textbf{k} AFM with SOC (the stage 3)}

Initialized by the occupation matrices of $\mathbb{S}_0$ in Section~\ref{sec:smfmsoc}, we obtained 3\textbf{k} AFM states by using GGA+$U$ ($U=4$ eV). The resulting $\occmat$ was further used to initialize the calculations of GGA+$U$ and LDA+$U$ using different values of $U$. Only the $\occmat$ of $U=4$ eV are reported, as the $\occmat$ of other $U$ are approximately identical to $U=4$ eV.

\subsubsection{Properties of $\mathbb{S}_0$ (3\textbf{k} AFM, GGA+U, U=2eV)}

\begin{table*}[!h]%
\begin{ruledtabular}
\begin{tabular}{ll}
\textrm{Properties (unit)}&
\textrm{Values}\\

\colrule
Energy (eV) &  -130.689 \\
Basis vectors (\AA)& $a=(
5.5136,  0.0001,  0.0001
)$, $b=(
0.0001,  5.5133,  0
)$, $c=(
0.0001,  0,  5.5137
)$\\
\begin{tabular}{@{}l@{}}Atom coordinates \\ (in basis vectors)\end{tabular} & {\begin{tabular}{@{}l@{}}U1:$(
0.5,  0,  0 )$, U2:$(
0.5,  0.5,  0.5 )$, \\U3:$(
0,  0,  0.5 )$, U4:$(
0,  0.5,  0 )$, \\O1:$(
0.2473,  0.7474,  0.7526 )$, O2:$(
0.2473,  0.2474,  0.7474 )$, \\O3:$(
0.2527,  0.2474,  0.2474 )$, O4:$(
0.2527,  0.7474,  0.2526 )$, \\O5:$(
0.7527,  0.7526,  0.2526 )$, O6:$(
0.7527,  0.2526,  0.2474 )$, \\O7:$(
0.7473,  0.2526,  0.7474 )$, O8:$(
0.7473,  0.7526,  0.7526 )$\end{tabular}} \\
Strain & \begin{tabular}{@{}l@{}}$(\epsilon_{xx},\epsilon_{yy},\epsilon_{zz},\epsilon_{xy},\epsilon_{xz},\epsilon_{yz})=(
0.0001,     0, 0.0001,     0 ,     0 ,     0
)$, \\$(\epsilon_{A1g},\epsilon_{Eg.0},\epsilon_{Eg.1},\epsilon_{T2g.0},\epsilon_{T2g.1},\epsilon_{T2g.2})=(
0.0002,     0,     0,     0 ,     0 ,     0
)$.\end{tabular}\\
Spin moments ($\mu_B$)& \begin{tabular}{@{}l@{}}U1:$(
 0.843, -0.776, -0.809
)$, U2:$(
-0.830,  0.795, -0.805
)$ \\ U3:$(
 0.803,  0.804,  0.825
)$, U4:$(
-0.822, -0.775,  0.833
)$\end{tabular}\\
Orbit moments ($\mu_B$)& \begin{tabular}{@{}l@{}}U1:$(
-2.030,  1.879,  1.955
)$, U2:$(
 2.002, -1.919,  1.944
)$ \\ U3:$(
-1.940, -1.939, -1.988
)$, U4:$(
 1.983,  1.877, -2.006
)$\end{tabular}\\
Total magnetic moments ($\mu_B$)& \begin{tabular}{@{}l@{}}U1:$(
-1.187, 1.103, 1.146
)$, U2:$(
1.172, -1.124, 1.139
)$ \\ U3:$(
-1.137, -1.135, -1.163
)$, U4:$(
1.161, 1.102, -1.173
)$\end{tabular}\\
Energy (eV) and lattice (\AA) of und & -130.686, 5.5130
\end{tabular}
\end{ruledtabular}
\end{table*}

\subsubsection{Properties of $\mathbb{S}_0$ (3\textbf{k} AFM, GGA+U, U=3eV)}

\begin{table*}[!h]%
\begin{ruledtabular}
\begin{tabular}{ll}
\textrm{Properties (unit)}&
\textrm{Values}\\
\colrule
Energy (eV) & -129.381 \\
Basis vectors (\AA)& $a=(
5.5320, 0.0001, 0.0001
)$, $b=(
0.0001, 5.5318,     0
)$, $c=(
    0,     0, 5.5320
)$\\
\begin{tabular}{@{}l@{}}Atom coordinates \\ (in basis vectors)\end{tabular} &
{\begin{tabular}{@{}l@{}}U1:$(0.5000 ,     0 ,     0)$, U2:$(0.5000 , 0.5000 , 0.5000)$ \\U3:$(   0 ,     0 , 0.5000)$, U4:$(  0 , 0.5000 ,     0 )$ \\O1:$(
0.2479, 0.7479, 0.7521
)$, O2:$(
0.2479, 0.2479, 0.7479
)$\\O3:$(
0.2521, 0.2479, 0.2479
)$, O4:$(
0.2521, 0.7479, 0.2521
)$\\O5:$(
0.7521, 0.7521, 0.2521
)$, O6:$(
0.7521, 0.2521, 0.2479
)$\\O7:$(
0.7479, 0.2521, 0.7479
)$, O8:$(
0.7479, 0.7521, 0.7521
)$\end{tabular}} \\ 
Strain & \begin{tabular}{@{}l@{}}$(\epsilon_{xx},\epsilon_{yy},\epsilon_{zz},\epsilon_{xy},\epsilon_{xz},\epsilon_{yz})=(
0.0001,     0, 0.0001,     0 ,     0 ,     0
)$, \\$(\epsilon_{A1g},\epsilon_{Eg.0},\epsilon_{Eg.1},\epsilon_{T2g.0},\epsilon_{T2g.1},\epsilon_{T2g.2})=(
0.0001,     0,     0,     0 ,     0 ,     0
)$.\end{tabular}\\
Spin moments ($\mu_B$)& \begin{tabular}{@{}l@{}}U1:$(
 0.836, -0.807, -0.819
)$, U2:$(
-0.831,  0.812, -0.820
)$ \\ U3:$(
 0.818,  0.818,  0.826
)$, U4:$(
-0.826, -0.809,  0.828
)$\end{tabular}\\
Orbit moments ($\mu_B$)& \begin{tabular}{@{}l@{}}U1:$(
-2.020,  1.953,  1.981
)$, U2:$(
 2.008, -1.963,  1.983
)$ \\ U3:$(
-1.980, -1.979, -1.996
)$, U4:$(
 1.996,  1.958, -2.001
)$\end{tabular}\\
Total magnetic moments ($\mu_B$)& \begin{tabular}{@{}l@{}}U1:$(
-1.184, 1.146, 1.162
)$, U2:$(
1.177, -1.151, 1.163
)$ \\ U3:$(
-1.162, -1.161, -1.170
)$, U4:$(
1.170, 1.149, -1.173
)$\end{tabular}\\
Energy (eV) and lattice (\AA) of und & -129.377, 5.5316
\end{tabular}
\end{ruledtabular}
\end{table*}
\clearpage
\subsubsection{Properties of $\mathbb{S}_0$ (3\textbf{k} AFM, GGA+U, U=4eV)}

\begin{table*}[!h]%
\begin{ruledtabular}
\begin{tabular}{ll}
\textrm{Properties (unit)}&
\textrm{Values}\\
\colrule
Energy (eV) & -128.234 \\
Basis vectors (\AA)& $a=(
5.5471,     0,     0
)$, $b=(
    0, 5.5471,     0
)$, $c=(
    0,     0, 5.5471\
)$\\
\begin{tabular}{@{}l@{}}Atom coordinates \\ (in basis vectors)\end{tabular} &
{\begin{tabular}{@{}l@{}}U1:$(0.5000 ,     0 ,     0)$, U2:$(0.5000 , 0.5000 , 0.5000)$ \\U3:$(   0 ,     0 , 0.5000)$, U4:$(  0 , 0.5000 ,     0 )$ \\O1:$(
0.2483, 0.7483, 0.7517
)$, O2:$(
0.2483, 0.2483, 0.7483
)$\\O3:$(
0.2517, 0.2483, 0.2483
)$, O4:$(
0.2517, 0.7484, 0.2517
)$\\O5:$(
0.7517, 0.7517, 0.2517
)$, O6:$(
0.7517, 0.2517, 0.2483
)$\\O7:$(
0.7483, 0.2516, 0.7483
)$, O8:$(
0.7483, 0.7517, 0.7517
)$\end{tabular}} \\ 
Strain & \begin{tabular}{@{}l@{}}$(\epsilon_{xx},\epsilon_{yy},\epsilon_{zz},\epsilon_{xy},\epsilon_{xz},\epsilon_{yz})=(
0.0002, 0.0002, 0.0002,     0 ,     0 ,     0
)$, \\$(\epsilon_{A1g},\epsilon_{Eg.0},\epsilon_{Eg.1},\epsilon_{T2g.0},\epsilon_{T2g.1},\epsilon_{T2g.2})=(
0.0003,     0,     0,     0 ,     0 ,     0
)$.\end{tabular}\\
Spin moments ($\mu_B$)& \begin{tabular}{@{}l@{}}U1:$(
 0.838, -0.820, -0.826
)$, U2:$(
-0.835,  0.822, -0.828
)$ \\ U3:$(
 0.826,  0.827,  0.831
)$, U4:$(
-0.830, -0.821,  0.832
)$\end{tabular}\\
Orbit moments ($\mu_B$)& \begin{tabular}{@{}l@{}}U1:$(
-2.024,  1.982,  1.997
)$, U2:$(
 2.017, -1.986,  2.001
)$ \\ U3:$(
-1.996, -2.000, -2.008
)$, U4:$(
 2.007,  1.985, -2.011
)$\end{tabular}\\
Total magnetic moments ($\mu_B$)& \begin{tabular}{@{}l@{}}U1:$(
-1.186, 1.162, 1.171
)$, U2:$(
1.182, -1.164, 1.173
)$ \\ U3:$(
-1.170, -1.173, -1.177
)$, U4:$(
1.177, 1.164, -1.179
)$\end{tabular}\\
Energy (eV) and lattice (\AA) of und & -128.230, 5.5461
\end{tabular}
\end{ruledtabular}
\end{table*}

$\occmat(\textrm{U}1)$:

$\occmat^{\uparrow\uparrow}=
\setstackgap{L}{1.1\baselineskip}
\fixTABwidth{T}
\parenMatrixstack{
0.06 &    0 & 0.02 & -0.02 & -0.01 & 0.02 & 0.02 \\
   0 & 0.12 &    0 &    0 &    0 &    0 &    0 \\
0.02 &    0 & 0.11 & 0.03 & 0.04 & -0.02 & 0.01 \\
-0.02 &    0 & 0.03 & 0.25 & -0.03 &    0 & -0.02 \\
-0.01 &    0 & 0.04 & -0.03 & 0.10 & -0.02 & -0.02 \\
0.02 &    0 & -0.02 &    0 & -0.02 & 0.17 & -0.02 \\
0.02 &    0 & 0.01 & -0.02 & -0.02 & -0.02 & 0.06
}+
\setstackgap{L}{1.1\baselineskip}
\fixTABwidth{T}
\parenMatrixstack{
   0 &    0 &    0 & -0.01 & -0.01 & -0.06 & 0.02 \\
   0 &    0 &    0 &    0 &    0 &    0 &    0 \\
   0 &    0 &    0 & -0.11 & 0.04 & -0.04 & 0.01 \\
0.01 &    0 & 0.11 &    0 & 0.11 & -0.01 & -0.01 \\
0.01 &    0 & -0.04 & -0.11 &    0 & 0.04 &    0 \\
0.06 &    0 & 0.04 & 0.01 & -0.04 &    0 & 0.06 \\
-0.02 &    0 & -0.01 & 0.01 &    0 & -0.06 &    0 
}i$,

$\occmat^{\uparrow\downarrow}=
\setstackgap{L}{1.1\baselineskip}
\fixTABwidth{T}
\parenMatrixstack{
0.09 &    0 & 0.06 & -0.02 & -0.06 & 0.03 & 0.09 \\
   0 &    0 &    0 &    0 &    0 &    0 &    0 \\
0.03 &    0 & 0.13 & 0.01 & 0.09 & -0.02 & 0.01 \\
-0.01 &    0 & -0.05 & 0.15 & -0.10 & 0.02 &    0 \\
-0.03 &    0 & 0.04 & 0.21 & -0.02 & -0.05 & -0.01 \\
-0.02 &    0 & -0.01 & 0.03 & 0.02 & 0.07 & -0.04 \\
0.04 &    0 &    0 & -0.01 & -0.02 & 0.21 & -0.01
}+
\setstackgap{L}{1.1\baselineskip}
\fixTABwidth{T}
\parenMatrixstack{
-0.01 &    0 & 0.01 & -0.01 &    0 & -0.21 & 0.04 \\
   0 &    0 &    0 &    0 &    0 &    0 &    0 \\
0.02 &    0 & -0.02 & -0.21 & 0.04 & -0.05 & 0.02 \\
   0 &    0 & 0.11 & 0.14 & 0.05 & -0.03 & -0.01 \\
-0.01 &    0 & 0.09 & -0.01 & 0.12 & -0.02 & -0.03 \\
0.04 &    0 & 0.02 & -0.03 & -0.01 & 0.07 & 0.02 \\
0.08 &    0 & 0.05 & -0.02 & -0.06 & -0.03 & 0.09
}i$,

$\occmat^{\downarrow\uparrow}=
\setstackgap{L}{1.1\baselineskip}
\fixTABwidth{T}
\parenMatrixstack{
0.09 &    0 & 0.03 & -0.01 & -0.03 & -0.02 & 0.04 \\
   0 &    0 &    0 &    0 &    0 &    0 &    0 \\
0.06 &    0 & 0.13 & -0.05 & 0.04 & -0.01 &    0 \\
-0.02 &    0 & 0.01 & 0.15 & 0.21 & 0.03 & -0.01 \\
-0.06 &    0 & 0.09 & -0.10 & -0.02 & 0.02 & -0.02 \\
0.03 &    0 & -0.02 & 0.02 & -0.05 & 0.07 & 0.21 \\
0.09 &    0 & 0.01 &    0 & -0.01 & -0.04 & -0.01
}+
\setstackgap{L}{1.1\baselineskip}
\fixTABwidth{T}
\parenMatrixstack{
0.01 &    0 & -0.02 &    0 & 0.01 & -0.04 & -0.08 \\
   0 &    0 &    0 &    0 &    0 &    0 &    0 \\
-0.01 &    0 & 0.02 & -0.11 & -0.09 & -0.02 & -0.05 \\
0.01 &    0 & 0.21 & -0.14 & 0.01 & 0.03 & 0.02 \\
   0 &    0 & -0.04 & -0.05 & -0.12 & 0.01 & 0.06 \\
0.21 &    0 & 0.05 & 0.03 & 0.02 & -0.07 & 0.03 \\
-0.04 &    0 & -0.02 & 0.01 & 0.03 & -0.02 & -0.09
}i$,

$\occmat^{\downarrow\downarrow}=
\setstackgap{L}{1.1\baselineskip}
\fixTABwidth{T}
\parenMatrixstack{
0.35 &    0 & 0.09 & 0.01 & -0.02 & -0.09 & 0.08 \\
   0 & 0.12 &    0 &    0 &    0 &    0 &    0 \\
0.09 &    0 & 0.25 & -0.12 & 0.02 &    0 & 0.02 \\
0.01 &    0 & -0.12 & 0.24 & 0.13 &    0 & 0.01 \\
-0.02 &    0 & 0.02 & 0.13 & 0.25 &    0 & -0.08 \\
-0.09 &    0 &    0 &    0 &    0 & 0.11 & 0.09 \\
0.08 &    0 & 0.02 & 0.01 & -0.08 & 0.09 & 0.35
}+
\setstackgap{L}{1.1\baselineskip}
\fixTABwidth{T}
\parenMatrixstack{
   0 &    0 &    0 & -0.01 & 0.07 & -0.12 & -0.30 \\
   0 &    0 &    0 &    0 &    0 &    0 &    0 \\
   0 &    0 &    0 & -0.16 & -0.18 & -0.05 & -0.07 \\
0.01 &    0 & 0.16 &    0 & 0.15 & 0.02 & -0.01 \\
-0.07 &    0 & 0.18 & -0.15 &    0 & 0.05 &    0 \\
0.12 &    0 & 0.05 & -0.02 & -0.05 &    0 & 0.12 \\
0.30 &    0 & 0.07 & 0.01 &    0 & -0.12 &    0 
}i$.

$\occmat(\textrm{U}2)$:

$\occmat^{\uparrow\uparrow}=
\setstackgap{L}{1.1\baselineskip}
\fixTABwidth{T}
\parenMatrixstack{
0.06 &    0 & 0.02 & 0.02 & -0.01 & -0.02 & 0.02 \\
   0 & 0.12 &    0 &    0 &    0 &    0 &    0 \\
0.02 &    0 & 0.11 & -0.03 & 0.04 & 0.02 & 0.01 \\
0.02 &    0 & -0.03 & 0.25 & 0.03 &    0 & 0.02 \\
-0.01 &    0 & 0.04 & 0.03 & 0.10 & 0.02 & -0.02 \\
-0.02 &    0 & 0.02 &    0 & 0.02 & 0.17 & 0.02 \\
0.02 &    0 & 0.01 & 0.02 & -0.02 & 0.02 & 0.06
}+
\setstackgap{L}{1.1\baselineskip}
\fixTABwidth{T}
\parenMatrixstack{
   0 &    0 &    0 & 0.01 & -0.01 & 0.06 & 0.02 \\
   0 &    0 &    0 &    0 &    0 &    0 &    0 \\
   0 &    0 &    0 & 0.11 & 0.04 & 0.04 & 0.01 \\
-0.01 &    0 & -0.11 &    0 & -0.11 & -0.01 & 0.01 \\
0.01 &    0 & -0.04 & 0.11 &    0 & -0.04 &    0 \\
-0.06 &    0 & -0.04 & 0.01 & 0.04 &    0 & -0.06 \\
-0.02 &    0 & -0.01 & -0.01 &    0 & 0.06 &    0
}i$,

$\occmat^{\uparrow\downarrow}=
\setstackgap{L}{1.1\baselineskip}
\fixTABwidth{T}
\parenMatrixstack{
-0.09 &    0 & -0.06 & -0.02 & 0.06 & 0.03 & -0.09 \\
   0 &    0 &    0 &    0 &    0 &    0 &    0 \\
-0.03 &    0 & -0.13 & 0.01 & -0.09 & -0.02 & -0.01 \\
-0.01 &    0 & -0.05 & -0.15 & -0.10 & -0.02 &    0 \\
0.03 &    0 & -0.04 & 0.21 & 0.02 & -0.05 & 0.01 \\
-0.02 &    0 & -0.01 & -0.03 & 0.02 & -0.07 & -0.04 \\
-0.04 &    0 &    0 & -0.01 & 0.02 & 0.21 & 0.01
}+
\setstackgap{L}{1.1\baselineskip}
\fixTABwidth{T}
\parenMatrixstack{
0.01 &    0 & -0.01 & -0.01 &    0 & -0.21 & -0.04 \\
   0 &    0 &    0 &    0 &    0 &    0 &    0 \\
-0.02 &    0 & 0.02 & -0.21 & -0.04 & -0.05 & -0.02 \\
   0 &    0 & 0.11 & -0.14 & 0.05 & 0.03 & -0.01 \\
0.01 &    0 & -0.09 & -0.01 & -0.12 & -0.02 & 0.03 \\
0.04 &    0 & 0.02 & 0.02 & -0.01 & -0.07 & 0.02 \\
-0.08 &    0 & -0.05 & -0.02 & 0.06 & -0.02 & -0.09
}i$,

$\occmat^{\downarrow\uparrow}=
\setstackgap{L}{1.1\baselineskip}
\fixTABwidth{T}
\parenMatrixstack{
-0.09 &    0 & -0.03 & -0.01 & 0.03 & -0.02 & -0.04 \\
   0 &    0 &    0 &    0 &    0 &    0 &    0 \\
-0.06 &    0 & -0.13 & -0.05 & -0.04 & -0.01 &    0 \\
-0.02 &    0 & 0.01 & -0.15 & 0.21 & -0.03 & -0.01 \\
0.06 &    0 & -0.09 & -0.10 & 0.02 & 0.02 & 0.02 \\
0.03 &    0 & -0.02 & -0.02 & -0.05 & -0.07 & 0.21 \\
-0.09 &    0 & -0.01 &    0 & 0.01 & -0.04 & 0.01
}+
\setstackgap{L}{1.1\baselineskip}
\fixTABwidth{T}
\parenMatrixstack{
-0.01 &    0 & 0.02 &    0 & -0.01 & -0.04 & 0.08 \\
   0 &    0 &    0 &    0 &    0 &    0 &    0 \\
0.01 &    0 & -0.02 & -0.11 & 0.09 & -0.02 & 0.05 \\
0.01 &    0 & 0.21 & 0.14 & 0.01 & -0.02 & 0.02 \\
   0 &    0 & 0.04 & -0.05 & 0.12 & 0.01 & -0.06 \\
0.21 &    0 & 0.05 & -0.03 & 0.02 & 0.07 & 0.02 \\
0.04 &    0 & 0.02 & 0.01 & -0.03 & -0.02 & 0.09 
}i$,

$\occmat^{\downarrow\downarrow}=
\setstackgap{L}{1.1\baselineskip}
\fixTABwidth{T}
\parenMatrixstack{
0.35 &    0 & 0.09 & -0.01 & -0.02 & 0.09 & 0.08 \\
   0 & 0.12 &    0 &    0 &    0 &    0 &    0 \\
0.09 &    0 & 0.25 & 0.13 & 0.02 &    0 & 0.02 \\
-0.01 &    0 & 0.13 & 0.24 & -0.13 &    0 & -0.01 \\
-0.02 &    0 & 0.02 & -0.13 & 0.25 &    0 & -0.08 \\
0.09 &    0 &    0 &    0 &    0 & 0.11 & -0.09 \\
0.08 &    0 & 0.02 & -0.01 & -0.08 & -0.09 & 0.35
}+
\setstackgap{L}{1.1\baselineskip}
\fixTABwidth{T}
\parenMatrixstack{
   0 &    0 &    0 & 0.01 & 0.07 & 0.12 & -0.30 \\
   0 &    0 &    0 &    0 &    0 &    0 &    0 \\
   0 &    0 &    0 & 0.16 & -0.18 & 0.05 & -0.07 \\
-0.01 &    0 & -0.16 &    0 & -0.15 & 0.02 & 0.01 \\
-0.07 &    0 & 0.18 & 0.15 &    0 & -0.05 &    0 \\
-0.12 &    0 & -0.05 & -0.02 & 0.05 &    0 & -0.12 \\
0.30 &    0 & 0.07 & -0.01 &    0 & 0.12 &    0
}i$.

$\occmat(\textrm{U}3)$:

$\occmat^{\uparrow\uparrow}=
\setstackgap{L}{1.1\baselineskip}
\fixTABwidth{T}
\parenMatrixstack{
0.35 &    0 & 0.09 & 0.01 & 0.02 & -0.09 & -0.08 \\
   0 & 0.12 &    0 &    0 &    0 &    0 &    0 \\
0.09 &    0 & 0.25 & -0.13 & -0.02 &    0 & -0.02 \\
0.01 &    0 & -0.13 & 0.24 & -0.13 &    0 & -0.01 \\
0.02 &    0 & -0.02 & -0.13 & 0.25 &    0 & -0.08 \\
-0.09 &    0 &    0 &    0 &    0 & 0.11 & -0.09 \\
-0.08 &    0 & -0.02 & -0.01 & -0.08 & -0.09 & 0.35
}+
\setstackgap{L}{1.1\baselineskip}
\fixTABwidth{T}
\parenMatrixstack{
   0 &    0 &    0 & -0.01 & -0.07 & -0.12 & 0.31 \\
   0 &    0 &    0 &    0 &    0 &    0 &    0 \\
   0 &    0 &    0 & -0.16 & 0.18 & -0.04 & 0.07 \\
0.01 &    0 & 0.16 &    0 & -0.15 & 0.02 & 0.01 \\
0.07 &    0 & -0.18 & 0.15 &    0 & -0.05 &    0 \\
0.12 &    0 & 0.04 & -0.02 & 0.05 &    0 & -0.12 \\
-0.31 &    0 & -0.07 & -0.01 &    0 & 0.12 &    0 
}i$,

$\occmat^{\uparrow\downarrow}=
\setstackgap{L}{1.1\baselineskip}
\fixTABwidth{T}
\parenMatrixstack{
0.09 &    0 & 0.03 & -0.01 & 0.03 & -0.02 & -0.04 \\
   0 &    0 &    0 &    0 &    0 &    0 &    0 \\
0.05 &    0 & 0.13 & -0.05 & -0.04 & -0.01 &    0 \\
-0.02 &    0 & 0.01 & 0.15 & -0.21 & 0.03 & 0.01 \\
0.06 &    0 & -0.09 & 0.10 & -0.02 & -0.03 & -0.01 \\
0.03 &    0 & -0.02 & 0.02 & 0.05 & 0.07 & -0.21 \\
-0.09 &    0 & -0.01 &    0 & -0.01 & 0.04 & -0.01 
}+
\setstackgap{L}{1.1\baselineskip}
\fixTABwidth{T}
\parenMatrixstack{
0.01 &    0 & -0.01 &    0 & -0.01 & -0.04 & 0.08 \\
   0 &    0 &    0 &    0 &    0 &    0 &    0 \\
-0.01 &    0 & 0.02 & -0.11 & 0.09 & -0.02 & 0.05 \\
0.01 &    0 & 0.21 & -0.14 & -0.01 & 0.02 & -0.02 \\
   0 &    0 & 0.04 & 0.05 & -0.12 & -0.01 & 0.06 \\
0.21 &    0 & 0.05 & 0.03 & -0.02 & -0.07 & -0.02 \\
0.04 &    0 & 0.02 & -0.01 & 0.03 & 0.02 & -0.09
}i$,

$\occmat^{\downarrow\uparrow}=
\setstackgap{L}{1.1\baselineskip}
\fixTABwidth{T}
\parenMatrixstack{
0.09 &    0 & 0.05 & -0.02 & 0.06 & 0.03 & -0.09 \\
   0 &    0 &    0 &    0 &    0 &    0 &    0 \\
0.03 &    0 & 0.13 & 0.01 & -0.09 & -0.02 & -0.01 \\
-0.01 &    0 & -0.05 & 0.15 & 0.10 & 0.02 &    0 \\
0.03 &    0 & -0.04 & -0.21 & -0.02 & 0.05 & -0.01 \\
-0.02 &    0 & -0.01 & 0.03 & -0.03 & 0.07 & 0.04 \\
-0.04 &    0 &    0 & 0.01 & -0.01 & -0.21 & -0.01
}+
\setstackgap{L}{1.1\baselineskip}
\fixTABwidth{T}
\parenMatrixstack{
-0.01 &    0 & 0.01 & -0.01 &    0 & -0.21 & -0.04 \\
   0 &    0 &    0 &    0 &    0 &    0 &    0 \\
0.01 &    0 & -0.02 & -0.21 & -0.04 & -0.05 & -0.02 \\
   0 &    0 & 0.11 & 0.14 & -0.05 & -0.03 & 0.01 \\
0.01 &    0 & -0.09 & 0.01 & 0.12 & 0.02 & -0.03 \\
0.04 &    0 & 0.02 & -0.02 & 0.01 & 0.07 & -0.02 \\
-0.08 &    0 & -0.05 & 0.02 & -0.06 & 0.02 & 0.09
}i$,

$\occmat^{\downarrow\downarrow}=
\setstackgap{L}{1.1\baselineskip}
\fixTABwidth{T}
\parenMatrixstack{
0.06 &    0 & 0.02 & -0.02 & 0.01 & 0.02 & -0.02 \\
   0 & 0.12 &    0 &    0 &    0 &    0 &    0 \\
0.02 &    0 & 0.10 & 0.03 & -0.04 & -0.02 & -0.01 \\
-0.02 &    0 & 0.03 & 0.25 & 0.03 &    0 & 0.02 \\
0.01 &    0 & -0.04 & 0.03 & 0.10 & 0.02 & -0.02 \\
0.02 &    0 & -0.02 &    0 & 0.02 & 0.17 & 0.02 \\
-0.02 &    0 & -0.01 & 0.02 & -0.02 & 0.02 & 0.06 
}+
\setstackgap{L}{1.1\baselineskip}
\fixTABwidth{T}
\parenMatrixstack{
   0 &    0 &    0 & -0.01 & 0.01 & -0.06 & -0.02 \\
   0 &    0 &    0 &    0 &    0 &    0 &    0 \\
   0 &    0 &    0 & -0.11 & -0.04 & -0.04 & -0.01 \\
0.01 &    0 & 0.11 &    0 & -0.11 & -0.01 & 0.01 \\
-0.01 &    0 & 0.04 & 0.11 &    0 & -0.04 &    0 \\
0.06 &    0 & 0.04 & 0.01 & 0.04 &    0 & -0.06 \\
0.02 &    0 & 0.01 & -0.01 &    0 & 0.06 &    0
}i$.

$\occmat(\textrm{U}4)$:

$\occmat^{\uparrow\uparrow}=
\setstackgap{L}{1.1\baselineskip}
\fixTABwidth{T}
\parenMatrixstack{
0.35 &    0 & 0.09 & -0.01 & 0.02 & 0.09 & -0.08 \\
   0 & 0.12 &    0 &    0 &    0 &    0 &    0 \\
0.09 &    0 & 0.25 & 0.13 & -0.02 &    0 & -0.02 \\
-0.01 &    0 & 0.13 & 0.24 & 0.13 &    0 & 0.01 \\
0.02 &    0 & -0.02 & 0.13 & 0.25 &    0 & -0.08 \\
0.09 &    0 &    0 &    0 &    0 & 0.10 & 0.09 \\
-0.08 &    0 & -0.02 & 0.01 & -0.08 & 0.09 & 0.35
}+
\setstackgap{L}{1.1\baselineskip}
\fixTABwidth{T}
\parenMatrixstack{
   0 &    0 &    0 & 0.01 & -0.07 & 0.12 & 0.31 \\
   0 &    0 &    0 &    0 &    0 &    0 &    0 \\
   0 &    0 &    0 & 0.16 & 0.18 & 0.04 & 0.07 \\
-0.01 &    0 & -0.16 &    0 & 0.15 & 0.02 & -0.01 \\
0.07 &    0 & -0.18 & -0.15 &    0 & 0.05 &    0 \\
-0.12 &    0 & -0.04 & -0.02 & -0.05 &    0 & 0.12 \\
-0.31 &    0 & -0.07 & 0.01 &    0 & -0.12 &    0
}i$,

$\occmat^{\uparrow\downarrow}=
\setstackgap{L}{1.1\baselineskip}
\fixTABwidth{T}
\parenMatrixstack{
-0.09 &    0 & -0.03 & -0.01 & -0.03 & -0.02 & 0.04 \\
   0 &    0 &    0 &    0 &    0 &    0 &    0 \\
-0.05 &    0 & -0.13 & -0.05 & 0.04 & -0.01 &    0 \\
-0.02 &    0 & 0.01 & -0.15 & -0.21 & -0.03 & 0.01 \\
-0.06 &    0 & 0.09 & 0.10 & 0.02 & -0.02 & 0.01 \\
0.03 &    0 & -0.02 & -0.02 & 0.05 & -0.07 & -0.21 \\
0.09 &    0 & 0.01 &    0 & 0.01 & 0.04 & 0.01
}+
\setstackgap{L}{1.1\baselineskip}
\fixTABwidth{T}
\parenMatrixstack{
-0.01 &    0 & 0.02 &    0 & 0.01 & -0.04 & -0.08 \\
   0 &    0 &    0 &    0 &    0 &    0 &    0 \\
0.01 &    0 & -0.02 & -0.11 & -0.09 & -0.02 & -0.05 \\
0.01 &    0 & 0.21 & 0.14 & -0.01 & -0.02 & -0.02 \\
   0 &    0 & -0.04 & 0.05 & 0.12 & -0.01 & -0.06 \\
0.21 &    0 & 0.05 & -0.03 & -0.02 & 0.07 & -0.02 \\
-0.04 &    0 & -0.02 & -0.01 & -0.03 & 0.02 & 0.09
}i$,

$\occmat^{\downarrow\uparrow}=
\setstackgap{L}{1.1\baselineskip}
\fixTABwidth{T}
\parenMatrixstack{
-0.09 &    0 & -0.05 & -0.02 & -0.06 & 0.03 & 0.09 \\
   0 &    0 &    0 &    0 &    0 &    0 &    0 \\
-0.03 &    0 & -0.13 & 0.01 & 0.09 & -0.02 & 0.01 \\
-0.01 &    0 & -0.05 & -0.15 & 0.10 & -0.02 &    0 \\
-0.03 &    0 & 0.04 & -0.21 & 0.02 & 0.05 & 0.01 \\
-0.02 &    0 & -0.01 & -0.03 & -0.02 & -0.07 & 0.04 \\
0.04 &    0 &    0 & 0.01 & 0.01 & -0.21 & 0.01
}+
\setstackgap{L}{1.1\baselineskip}
\fixTABwidth{T}
\parenMatrixstack{
0.01 &    0 & -0.01 & -0.01 &    0 & -0.21 & 0.04 \\
   0 &    0 &    0 &    0 &    0 &    0 &    0 \\
-0.02 &    0 & 0.02 & -0.21 & 0.04 & -0.05 & 0.02 \\
   0 &    0 & 0.11 & -0.14 & -0.05 & 0.03 & 0.01 \\
-0.01 &    0 & 0.09 & 0.01 & -0.12 & 0.02 & 0.03 \\
0.04 &    0 & 0.02 & 0.02 & 0.01 & -0.07 & -0.02 \\
0.08 &    0 & 0.05 & 0.02 & 0.06 & 0.02 & -0.09 
}i$,

$\occmat^{\downarrow\downarrow}=
\setstackgap{L}{1.1\baselineskip}
\fixTABwidth{T}
\parenMatrixstack{
0.06 &    0 & 0.02 & 0.02 & 0.01 & -0.02 & -0.02 \\
   0 & 0.12 &    0 &    0 &    0 &    0 &    0 \\
0.02 &    0 & 0.10 & -0.03 & -0.04 & 0.02 & -0.01 \\
0.02 &    0 & -0.03 & 0.25 & -0.03 &    0 & -0.02 \\
0.01 &    0 & -0.04 & -0.03 & 0.10 & -0.02 & -0.02 \\
-0.02 &    0 & 0.02 &    0 & -0.02 & 0.17 & -0.02 \\
-0.02 &    0 & -0.01 & -0.02 & -0.02 & -0.02 & 0.06
}+
\setstackgap{L}{1.1\baselineskip}
\fixTABwidth{T}
\parenMatrixstack{
   0 &    0 &    0 & 0.01 & 0.01 & 0.06 & -0.02 \\
   0 &    0 &    0 &    0 &    0 &    0 &    0 \\
   0 &    0 &    0 & 0.11 & -0.04 & 0.04 & -0.01 \\
-0.01 &    0 & -0.11 &    0 & 0.11 & -0.01 & -0.01 \\
-0.01 &    0 & 0.04 & -0.11 &    0 & 0.04 &    0 \\
-0.06 &    0 & -0.04 & 0.01 & -0.04 &    0 & 0.06 \\
0.02 &    0 & 0.01 & 0.01 &    0 & -0.06 &    0 
}i$.

\bigskip
\noindent Relaxed crystal structure in the format of POSCAR:\\
U4 O8        \\                           
   1.00000000000000     \\
     5.5471006157960288 -0.0000195440696708   -0.0000226247298930\\
    -0.0000380456513832    5.5470514412708383    0.0000085892324893\\
    -0.0000241444917248    0.0000083201018715    5.5470879387012255\\
   U    O \\
     4     8\\
Direct\\
  0.5000000000486770  0.0000000002015143 -0.0000000000554230\\
  0.4999999998946900  0.5000000001391827  0.4999999997516726\\
  0.9999999994784251  0.0000000000631235  0.4999999998174839\\
  0.0000000003374192  0.4999999997836156  0.9999999997513933\\
  0.2483258018748809  0.7483296225922117  0.7516584033698774\\
  0.2483232963540089  0.2483324080283958  0.7483499571158573\\
  0.2516707776885484  0.2483482938673824  0.2483480735644173\\
  0.2516732557247472  0.7483502615945291  0.2516598548648705\\
  0.7516767039580829  0.7516675919644482  0.2516500439413922\\
  0.7516741978692347  0.2516703773285294  0.2483415961427774\\
  0.7483267446842400  0.2516497383693724  0.7483401445212374\\
  0.7483292220870458  0.7516517060677094  0.7516519272144510

\subsubsection{Properties of $\mathbb{S}_0$ (3\textbf{k} AFM, GGA+U, U=5eV)}

\begin{table*}[!h]%
\begin{ruledtabular}
\begin{tabular}{ll}
\textrm{Properties (unit)}&
\textrm{Values}\\
\colrule
Energy (eV) & -127.220 \\
Basis vectors (\AA)& $a=(
5.5603,     0,     0
)$, $b=(
    0, 5.5604,     0
)$, $c=(
    0,     0, 5.5603
)$\\
\begin{tabular}{@{}l@{}}Atom coordinates \\ (in basis vectors)\end{tabular} &
{\begin{tabular}{@{}l@{}}U1:$(0.5000 ,     0 ,     0)$, U2:$(0.5000 , 0.5000 , 0.5000)$ \\U3:$(   0 ,     0 , 0.5000)$, U4:$(  0 , 0.5000 ,     0 )$ \\O1:$(
0.2487, 0.7487, 0.7514
)$, O2:$(
0.2487, 0.2487, 0.7486
)$\\O3:$(
0.2513, 0.2486, 0.2486
)$, O4:$(
0.2513, 0.7486, 0.2514
)$\\O5:$(
0.7513, 0.7513, 0.2514
)$, O6:$(
0.7513, 0.2513, 0.2486
)$\\O7:$(
0.7487, 0.2514, 0.7486
)$, O8:$(
0.7487, 0.7514, 0.7514
)$\end{tabular}} \\ 
Strain & \begin{tabular}{@{}l@{}}$(\epsilon_{xx},\epsilon_{yy},\epsilon_{zz},\epsilon_{xy},\epsilon_{xz},\epsilon_{yz})=(
    0,     0,     0,     0,     0,     0
)$, \\$(\epsilon_{A1g},\epsilon_{Eg.0},\epsilon_{Eg.1},\epsilon_{T2g.0},\epsilon_{T2g.1},\epsilon_{T2g.2})=(
    0,     0,     0,     0,     0,     0
)$.\end{tabular}\\
Spin moments ($\mu_B$)& \begin{tabular}{@{}l@{}}U1:$(
 0.828, -0.841, -0.837
)$, U2:$(
-0.830,  0.840, -0.835
)$ \\ U3:$(
 0.835,  0.837,  0.833
)$, U4:$(
-0.831, -0.841,  0.833
)$\end{tabular}\\
Orbit moments ($\mu_B$)& \begin{tabular}{@{}l@{}}U1:$(
-1.996,  2.025,  2.016
)$, U2:$(
 2.001, -2.024,  2.012
)$ \\ U3:$(
-2.012, -2.017, -2.008
)$, U4:$(
 2.003,  2.026, -2.007
)$\end{tabular}\\
Total magnetic moments ($\mu_B$)& \begin{tabular}{@{}l@{}}U1:$(
-1.168, 1.184, 1.179
)$, U2:$(
1.171, -1.184, 1.177
)$ \\ U3:$(
-1.177, -1.180, -1.175
)$, U4:$(
1.172, 1.185, -1.174
)$\end{tabular}\\
Energy (eV) and lattice (\AA) of und & -127.216, 5.5603
\end{tabular}
\end{ruledtabular}
\end{table*}

\subsubsection{Properties of $\mathbb{S}_0$ (3\textbf{k} AFM, LDA+U, U=2eV)}

\begin{table*}[!h]%
\begin{ruledtabular}
\begin{tabular}{ll}
\textrm{Properties (unit)}&
\textrm{Values}\\
\colrule
Energy (eV) & -142.079 \\
Basis vectors (\AA)& $a=(
5.4120,     0,     0
)$, $b=(
    0, 5.4120,     0
)$, $c=(
    0,     0, 5.4120
)$\\
\begin{tabular}{@{}l@{}}Atom coordinates \\ (in basis vectors)\end{tabular} &
{\begin{tabular}{@{}l@{}}U1:$(0.5000 ,     0 ,     0)$, U2:$(0.5000 , 0.5000 , 0.5000)$ \\U3:$(   0 ,     0 , 0.5000)$, U4:$(  0 , 0.5000 ,     0 )$ \\O1:$(
0.2477, 0.7477, 0.7523
)$, O2:$(
0.2477, 0.2477, 0.7477
)$\\O3:$(
0.2523, 0.2477, 0.2477
)$, O4:$(
0.2523, 0.7477, 0.2523
)$\\O5:$(
0.7523, 0.7523, 0.2523
)$, O6:$(
0.7523, 0.2523, 0.2477
)$\\O7:$(
0.7477, 0.2523, 0.7477
)$, O8:$(
0.7477, 0.7523, 0.7523
)$\end{tabular}} \\ 
Strain & \begin{tabular}{@{}l@{}}$(\epsilon_{xx},\epsilon_{yy},\epsilon_{zz},\epsilon_{xy},\epsilon_{xz},\epsilon_{yz})=(
0.0002, 0.0002, 0.0002,     0,     0,     0
)$, \\$(\epsilon_{A1g},\epsilon_{Eg.0},\epsilon_{Eg.1},\epsilon_{T2g.0},\epsilon_{T2g.1},\epsilon_{T2g.2})=(
0.0004,     0,     0,     0,     0,     0
)$.\end{tabular}\\
Spin moments ($\mu_B$)& \begin{tabular}{@{}l@{}}U1:$(
 0.659, -0.659, -0.659
)$, U2:$(
-0.659,  0.659, -0.659
)$ \\ U3:$(
 0.659,  0.659,  0.659
)$, U4:$(
-0.659, -0.659,  0.659
)$\end{tabular}\\
Orbit moments ($\mu_B$)& \begin{tabular}{@{}l@{}}U1:$(
-1.891,  1.891,  1.891
)$, U2:$(
 1.890, -1.891,  1.891
)$ \\ U3:$(
-1.891, -1.891, -1.890
)$, U4:$(
 1.891,  1.891, -1.890
)$\end{tabular}\\
Total magnetic moments ($\mu_B$)& \begin{tabular}{@{}l@{}}U1:$(
-1.232, 1.232, 1.232
)$, U2:$(
1.231, -1.232, 1.232
)$ \\ U3:$(
-1.232, -1.232, -1.231
)$, U4:$(
1.232, 1.232, -1.231
)$\end{tabular}\\
Energy (eV) and lattice (\AA) of und & -142.075, 5.4109
\end{tabular}
\end{ruledtabular}
\end{table*}
\clearpage

\subsubsection{Properties of $\mathbb{S}_0$ (3\textbf{k} AFM, LDA+U, U=3eV)}

\begin{table*}[!h]%
\begin{ruledtabular}
\begin{tabular}{ll}
\textrm{Properties (unit)}&
\textrm{Values}\\
\colrule
Energy (eV) & -140.615 \\
Basis vectors (\AA)& $a=(
5.4334,     0,     0
)$, $b=(
    0, 5.4333,     0
)$, $c=(
    0,     0, 5.4336
)$\\
\begin{tabular}{@{}l@{}}Atom coordinates \\ (in basis vectors)\end{tabular} &
{\begin{tabular}{@{}l@{}}U1:$(0.5000 ,     0 ,     0)$, U2:$(0.5000 , 0.5000 , 0.5000)$ \\U3:$(   0 ,     0 , 0.5000)$, U4:$(  0 , 0.5000 ,     0 )$ \\O1:$(
0.2480, 0.7480, 0.7520
)$, O2:$(
0.2480, 0.2480, 0.7480
)$\\O3:$(
0.2520, 0.2480, 0.2480
)$, O4:$(
0.2520, 0.7480, 0.2520
)$\\O5:$(
0.7520, 0.7520, 0.2520
)$, O6:$(
0.7520, 0.2520, 0.2480
)$\\O7:$(
0.7480, 0.2520, 0.7480
)$, O8:$(
0.7480, 0.7520, 0.7520
)$\end{tabular}} \\ 
Strain & \begin{tabular}{@{}l@{}}$(\epsilon_{xx},\epsilon_{yy},\epsilon_{zz},\epsilon_{xy},\epsilon_{xz},\epsilon_{yz})=(
0.0001, 0.0001, 0.0002,     0,     0,     0
)$, \\$(\epsilon_{A1g},\epsilon_{Eg.0},\epsilon_{Eg.1},\epsilon_{T2g.0},\epsilon_{T2g.1},\epsilon_{T2g.2})=(
0.0003,     0,     0,     0,     0,     0
)$.\end{tabular}\\
Spin moments ($\mu_B$)& \begin{tabular}{@{}l@{}}U1:$(
 0.674, -0.662, -0.672
)$, U2:$(
-0.669,  0.670, -0.669
)$ \\ U3:$(
 0.669,  0.666,  0.674
)$, U4:$(
-0.664, -0.666,  0.678
)$\end{tabular}\\
Orbit moments ($\mu_B$)& \begin{tabular}{@{}l@{}}U1:$(
-1.968,  1.935,  1.962
)$, U2:$(
 1.954, -1.957,  1.954
)$ \\ U3:$(
-1.954, -1.945, -1.967
)$, U4:$(
 1.941,  1.947, -1.977
)$\end{tabular}\\
Total magnetic moments ($\mu_B$)& \begin{tabular}{@{}l@{}}U1:$(
-1.294, 1.273, 1.290
)$, U2:$(
1.285, -1.287, 1.285
)$ \\ U3:$(
-1.285, -1.279, -1.293
)$, U4:$(
1.277, 1.281, -1.299
)$\end{tabular}\\
Energy (eV) and lattice (\AA) of und & -140.610, 5.4326
\end{tabular}
\end{ruledtabular}
\end{table*}

\subsubsection{Properties of $\mathbb{S}_0$ (3\textbf{k} AFM, LDA+U, U=4eV)}

\begin{table*}[!h]%
\begin{ruledtabular}
\begin{tabular}{ll}
\textrm{Properties (unit)}&
\textrm{Values}\\
\colrule
Energy (eV) & -139.344 \\
Basis vectors (\AA)& $a=(
5.4503,     0,     0
)$, $b=(
    0, 5.4501,     0
)$, $c=(
    0,     0, 5.4505
)$\\
\begin{tabular}{@{}l@{}}Atom coordinates \\ (in basis vectors)\end{tabular} &
{\begin{tabular}{@{}l@{}}U1:$(0.5000 ,     0 ,     0)$, U2:$(0.5000 , 0.5000 , 0.5000)$ \\U3:$(   0 ,     0 , 0.5000)$, U4:$(  0 , 0.5000 ,     0 )$ \\O1:$(
0.2483, 0.7483, 0.7517
)$, O2:$(
0.2483, 0.2483, 0.7483
)$\\O3:$(
0.2517, 0.2483, 0.2483
)$, O4:$(
0.2517, 0.7483, 0.2517
)$\\O5:$(
0.7517, 0.7517, 0.2517
)$, O6:$(
0.7517, 0.2517, 0.2483
)$\\O7:$(
0.7483, 0.2517, 0.7483
)$, O8:$(
0.7483, 0.7517, 0.7517
)$\end{tabular}} \\ 
Strain & \begin{tabular}{@{}l@{}}$(\epsilon_{xx},\epsilon_{yy},\epsilon_{zz},\epsilon_{xy},\epsilon_{xz},\epsilon_{yz})=(
    0,     0, 0.0001,     0,     0,     0
)$, \\$(\epsilon_{A1g},\epsilon_{Eg.0},\epsilon_{Eg.1},\epsilon_{T2g.0},\epsilon_{T2g.1},\epsilon_{T2g.2})=(
0.0001,     0,     0,     0,     0,     0
)$.\end{tabular}\\
Spin moments ($\mu_B$)& \begin{tabular}{@{}l@{}}U1:$(
 0.684, -0.656, -0.680
)$, U2:$(
-0.673,  0.673, -0.675
)$ \\ U3:$(
 0.673,  0.663,  0.684
)$, U4:$(
-0.664, -0.665,  0.692
)$\end{tabular}\\
Orbit moments ($\mu_B$)& \begin{tabular}{@{}l@{}}U1:$(
-2.018,  1.942,  2.008
)$, U2:$(
 1.988, -1.987,  1.993
)$ \\ U3:$(
-1.989, -1.962, -2.018
)$, U4:$(
 1.963,  1.965, -2.039
)$\end{tabular}\\
Total magnetic moments ($\mu_B$)& \begin{tabular}{@{}l@{}}U1:$(
-1.334, 1.286, 1.328
)$, U2:$(
1.315, -1.314, 1.318
)$ \\ U3:$(
-1.316, -1.299, -1.334
)$, U4:$(
1.299, 1.300, -1.347
)$\end{tabular}\\
Energy (eV) and lattice (\AA) of und & -139.339, 5.4501
\end{tabular}
\end{ruledtabular}
\end{table*}

$\occmat(\textrm{U}1)$:

$\occmat^{\uparrow\uparrow}=
\setstackgap{L}{1.1\baselineskip}
\fixTABwidth{T}
\parenMatrixstack{
0.06 &    0 & 0.02 & -0.03 &    0 & 0.03 & 0.01 \\
   0 & 0.13 &    0 &    0 &    0 &    0 &    0 \\
0.02 &    0 & 0.13 & 0.05 & 0.04 & -0.02 &    0 \\
-0.03 &    0 & 0.05 & 0.26 & -0.06 &    0 & -0.02 \\
   0 &    0 & 0.04 & -0.06 & 0.13 & -0.02 & -0.02 \\
0.03 &    0 & -0.02 &    0 & -0.02 & 0.20 & -0.02 \\
0.01 &    0 &    0 & -0.02 & -0.02 & -0.02 & 0.06 
}+
\setstackgap{L}{1.1\baselineskip}
\fixTABwidth{T}
\parenMatrixstack{
   0 &    0 & -0.01 & -0.01 & -0.02 & -0.04 & 0.02 \\
   0 &    0 &    0 &    0 &    0 &    0 &    0 \\
0.01 &    0 &    0 & -0.13 & 0.06 & -0.05 & 0.01 \\
0.01 &    0 & 0.13 &    0 & 0.11 & -0.01 & -0.02 \\
0.02 &    0 & -0.06 & -0.11 &    0 & 0.05 &    0 \\
0.04 &    0 & 0.05 & 0.01 & -0.05 &    0 & 0.05 \\
-0.02 &    0 & -0.01 & 0.02 &    0 & -0.05 &    0
}i$,

$\occmat^{\uparrow\downarrow}=
\setstackgap{L}{1.1\baselineskip}
\fixTABwidth{T}
\parenMatrixstack{
0.07 & 0.01 & 0.05 & -0.02 & -0.07 & 0.01 & 0.07 \\
   0 &    0 & 0.01 &    0 & 0.01 & 0.01 &    0 \\
0.02 &    0 & 0.14 & 0.01 & 0.09 & -0.01 &    0 \\
-0.01 &    0 & -0.03 & 0.14 & -0.12 & 0.03 & 0.01 \\
-0.03 &    0 & 0.06 & 0.20 & -0.04 & -0.05 & -0.01 \\
-0.01 &    0 & 0.01 & 0.04 & 0.03 & 0.06 & -0.04 \\
0.04 &    0 &    0 &    0 & -0.01 & 0.23 & -0.03
}+
\setstackgap{L}{1.1\baselineskip}
\fixTABwidth{T}
\parenMatrixstack{
-0.04 &    0 & 0.01 &    0 &    0 & -0.23 & 0.04 \\
   0 &    0 &    0 & -0.01 &    0 &    0 &    0 \\
0.02 &    0 & -0.04 & -0.21 & 0.06 & -0.05 & 0.02 \\
-0.01 &    0 & 0.12 & 0.12 & 0.03 & -0.04 & -0.01 \\
   0 &    0 & 0.10 & -0.01 & 0.12 & -0.01 & -0.03 \\
0.04 &    0 & 0.03 & -0.04 & 0.01 & 0.07 & 0.01 \\
0.06 & 0.01 & 0.06 & -0.02 & -0.06 & -0.01 & 0.07
}i$,

$\occmat^{\downarrow\uparrow}=
\setstackgap{L}{1.1\baselineskip}
\fixTABwidth{T}
\parenMatrixstack{
0.07 &    0 & 0.02 & -0.01 & -0.03 & -0.01 & 0.04 \\
0.01 &    0 &    0 &    0 &    0 &    0 &    0 \\
0.05 & 0.01 & 0.14 & -0.03 & 0.06 & 0.01 &    0 \\
-0.02 &    0 & 0.01 & 0.14 & 0.20 & 0.04 &    0 \\
-0.07 & 0.01 & 0.09 & -0.12 & -0.04 & 0.03 & -0.01 \\
0.01 & 0.01 & -0.01 & 0.03 & -0.05 & 0.06 & 0.23 \\
0.07 &    0 &    0 & 0.01 & -0.01 & -0.04 & -0.03
}+
\setstackgap{L}{1.1\baselineskip}
\fixTABwidth{T}
\parenMatrixstack{
0.04 &    0 & -0.02 & 0.01 &    0 & -0.04 & -0.06 \\
   0 &    0 &    0 &    0 &    0 &    0 & -0.01 \\
-0.01 &    0 & 0.04 & -0.12 & -0.10 & -0.03 & -0.06 \\
   0 & 0.01 & 0.21 & -0.12 & 0.01 & 0.04 & 0.02 \\
   0 &    0 & -0.06 & -0.03 & -0.12 & -0.01 & 0.06 \\
0.23 &    0 & 0.05 & 0.04 & 0.01 & -0.07 & 0.01 \\
-0.04 &    0 & -0.02 & 0.01 & 0.03 & -0.01 & -0.07
}i$,

$\occmat^{\downarrow\downarrow}=
\setstackgap{L}{1.1\baselineskip}
\fixTABwidth{T}
\parenMatrixstack{
0.35 &    0 & 0.07 & 0.03 & -0.02 & -0.09 & 0.03 \\
   0 & 0.13 & 0.01 &    0 &    0 &    0 & 0.01 \\
0.07 & 0.01 & 0.24 & -0.10 & 0.02 & 0.02 & 0.02 \\
0.03 &    0 & -0.10 & 0.21 & 0.11 &    0 & 0.03 \\
-0.02 &    0 & 0.02 & 0.11 & 0.23 & 0.02 & -0.06 \\
-0.09 &    0 & 0.02 &    0 & 0.02 & 0.10 & 0.09 \\
0.03 & 0.01 & 0.02 & 0.03 & -0.06 & 0.09 & 0.35
}+
\setstackgap{L}{1.1\baselineskip}
\fixTABwidth{T}
\parenMatrixstack{
   0 & -0.01 & -0.01 & -0.02 & 0.06 & -0.10 & -0.31 \\
0.01 &    0 &    0 &    0 & -0.01 & -0.01 &    0 \\
0.01 &    0 &    0 & -0.14 & -0.17 & -0.05 & -0.07 \\
0.02 &    0 & 0.14 &    0 & 0.13 & 0.03 & -0.03 \\
-0.06 & 0.01 & 0.17 & -0.13 &    0 & 0.05 & 0.02 \\
0.10 & 0.01 & 0.05 & -0.03 & -0.05 &    0 & 0.10 \\
0.31 &    0 & 0.07 & 0.03 & -0.02 & -0.10 &    0
}i$.

$\occmat(\textrm{U}2)$:

$\occmat^{\uparrow\uparrow}=
\setstackgap{L}{1.1\baselineskip}
\fixTABwidth{T}
\parenMatrixstack{
0.06 &    0 & 0.02 & 0.03 &    0 & -0.03 & 0.01 \\
   0 & 0.13 &    0 &    0 &    0 &    0 &    0 \\
0.02 &    0 & 0.13 & -0.05 & 0.04 & 0.02 &    0 \\
0.03 &    0 & -0.05 & 0.26 & 0.06 &    0 & 0.02 \\
   0 &    0 & 0.04 & 0.06 & 0.13 & 0.02 & -0.03 \\
-0.03 &    0 & 0.02 &    0 & 0.02 & 0.20 & 0.02 \\
0.01 &    0 &    0 & 0.02 & -0.03 & 0.02 & 0.06
}+
\setstackgap{L}{1.1\baselineskip}
\fixTABwidth{T}
\parenMatrixstack{
   0 &    0 & -0.01 & 0.01 & -0.02 & 0.04 & 0.02 \\
   0 &    0 &    0 &    0 &    0 &    0 &    0 \\
0.01 &    0 &    0 & 0.12 & 0.07 & 0.05 & 0.01 \\
-0.01 &    0 & -0.12 &    0 & -0.12 & -0.01 & 0.02 \\
0.02 &    0 & -0.07 & 0.12 &    0 & -0.05 &    0 \\
-0.04 &    0 & -0.05 & 0.01 & 0.05 &    0 & -0.05 \\
-0.02 &    0 & -0.01 & -0.02 &    0 & 0.05 &    0
}i$,

$\occmat^{\uparrow\downarrow}=
\setstackgap{L}{1.1\baselineskip}
\fixTABwidth{T}
\parenMatrixstack{
-0.07 & 0.01 & -0.05 & -0.02 & 0.07 & 0.01 & -0.07 \\
   0 &    0 & 0.01 &    0 & 0.01 & -0.01 &    0 \\
-0.02 &    0 & -0.13 & 0.01 & -0.09 & -0.01 &    0 \\
-0.01 &    0 & -0.03 & -0.14 & -0.12 & -0.03 & 0.01 \\
0.03 &    0 & -0.06 & 0.20 & 0.04 & -0.05 & 0.01 \\
-0.01 &    0 & 0.01 & -0.04 & 0.03 & -0.06 & -0.04 \\
-0.04 &    0 &    0 & 0.01 & 0.01 & 0.23 & 0.03
}+
\setstackgap{L}{1.1\baselineskip}
\fixTABwidth{T}
\parenMatrixstack{
0.04 &    0 & -0.01 &    0 &    0 & -0.23 & -0.04 \\
   0 &    0 &    0 & 0.01 &    0 &    0 &    0 \\
-0.02 &    0 & 0.04 & -0.21 & -0.06 & -0.05 & -0.02 \\
-0.01 &    0 & 0.12 & -0.13 & 0.03 & 0.04 & -0.01 \\
   0 &    0 & -0.10 & -0.01 & -0.13 & -0.01 & 0.03 \\
0.04 &    0 & 0.03 & 0.04 & 0.01 & -0.07 & 0.01 \\
-0.06 & 0.01 & -0.06 & -0.02 & 0.06 & -0.01 & -0.07
}i$,

$\occmat^{\downarrow\uparrow}=
\setstackgap{L}{1.1\baselineskip}
\fixTABwidth{T}
\parenMatrixstack{
-0.07 &    0 & -0.02 & -0.01 & 0.03 & -0.01 & -0.04 \\
0.01 &    0 &    0 &    0 &    0 &    0 &    0 \\
-0.05 & 0.01 & -0.13 & -0.03 & -0.06 & 0.01 &    0 \\
-0.02 &    0 & 0.01 & -0.14 & 0.20 & -0.04 & 0.01 \\
0.07 & 0.01 & -0.09 & -0.12 & 0.04 & 0.03 & 0.01 \\
0.01 & -0.01 & -0.01 & -0.03 & -0.05 & -0.06 & 0.23 \\
-0.07 &    0 &    0 & 0.01 & 0.01 & -0.04 & 0.03 
}+
\setstackgap{L}{1.1\baselineskip}
\fixTABwidth{T}
\parenMatrixstack{
-0.04 &    0 & 0.02 & 0.01 &    0 & -0.04 & 0.06 \\
   0 &    0 &    0 &    0 &    0 &    0 & -0.01 \\
0.01 &    0 & -0.04 & -0.12 & 0.10 & -0.03 & 0.06 \\
   0 & -0.01 & 0.21 & 0.13 & 0.01 & -0.04 & 0.02 \\
   0 &    0 & 0.06 & -0.03 & 0.13 & -0.01 & -0.06 \\
0.23 &    0 & 0.05 & -0.04 & 0.01 & 0.07 & 0.01 \\
0.04 &    0 & 0.02 & 0.01 & -0.03 & -0.01 & 0.07 
}i$,

$\occmat^{\downarrow\downarrow}=
\setstackgap{L}{1.1\baselineskip}
\fixTABwidth{T}
\parenMatrixstack{
0.35 &    0 & 0.07 & -0.03 & -0.02 & 0.09 & 0.03 \\
   0 & 0.13 & -0.01 &    0 &    0 &    0 & -0.01 \\
0.07 & -0.01 & 0.24 & 0.10 & 0.02 & -0.02 & 0.02 \\
-0.03 &    0 & 0.10 & 0.21 & -0.11 &    0 & -0.03 \\
-0.02 &    0 & 0.02 & -0.11 & 0.23 & -0.02 & -0.06 \\
0.09 &    0 & -0.02 &    0 & -0.02 & 0.10 & -0.09 \\
0.03 & -0.01 & 0.02 & -0.03 & -0.06 & -0.09 & 0.35 
}+
\setstackgap{L}{1.1\baselineskip}
\fixTABwidth{T}
\parenMatrixstack{
   0 & 0.01 & -0.01 & 0.02 & 0.06 & 0.10 & -0.31 \\
-0.01 &    0 &    0 &    0 & 0.01 & -0.01 &    0 \\
0.01 &    0 &    0 & 0.14 & -0.17 & 0.05 & -0.07 \\
-0.02 &    0 & -0.14 &    0 & -0.13 & 0.03 & 0.03 \\
-0.06 & -0.01 & 0.17 & 0.13 &    0 & -0.06 & 0.02 \\
-0.10 & 0.01 & -0.05 & -0.03 & 0.06 &    0 & -0.10 \\
0.31 &    0 & 0.07 & -0.03 & -0.02 & 0.10 &    0 
}i$.

$\occmat(\textrm{U}3)$:

$\occmat^{\uparrow\uparrow}=
\setstackgap{L}{1.1\baselineskip}
\fixTABwidth{T}
\parenMatrixstack{
0.35 &    0 & 0.07 & 0.03 & 0.02 & -0.09 & -0.03 \\
   0 & 0.13 & -0.01 &    0 &    0 &    0 & 0.01 \\
0.07 & -0.01 & 0.24 & -0.10 & -0.02 & 0.02 & -0.02 \\
0.03 &    0 & -0.10 & 0.21 & -0.11 &    0 & -0.03 \\
0.02 &    0 & -0.02 & -0.11 & 0.24 & -0.02 & -0.06 \\
-0.09 &    0 & 0.02 &    0 & -0.02 & 0.10 & -0.08 \\
-0.03 & 0.01 & -0.02 & -0.03 & -0.06 & -0.08 & 0.36
}+
\setstackgap{L}{1.1\baselineskip}
\fixTABwidth{T}
\parenMatrixstack{
   0 & 0.01 & -0.01 & -0.02 & -0.06 & -0.09 & 0.31 \\
-0.01 &    0 &    0 &    0 & -0.01 & 0.01 &    0 \\
0.01 &    0 &    0 & -0.14 & 0.17 & -0.05 & 0.07 \\
0.02 &    0 & 0.14 &    0 & -0.13 & 0.03 & 0.03 \\
0.06 & 0.01 & -0.17 & 0.13 &    0 & -0.05 & 0.02 \\
0.09 & -0.01 & 0.05 & -0.03 & 0.05 &    0 & -0.10 \\
-0.31 &    0 & -0.07 & -0.03 & -0.02 & 0.10 &    0
}i$,

$\occmat^{\uparrow\downarrow}=
\setstackgap{L}{1.1\baselineskip}
\fixTABwidth{T}
\parenMatrixstack{
0.07 &    0 & 0.02 & -0.01 & 0.03 & -0.01 & -0.04 \\
-0.01 &    0 &    0 &    0 &    0 &    0 &    0 \\
0.05 & -0.01 & 0.13 & -0.03 & -0.06 & 0.01 &    0 \\
-0.02 &    0 & 0.01 & 0.14 & -0.21 & 0.04 & -0.01 \\
0.07 & 0.01 & -0.09 & 0.12 & -0.04 & -0.03 & -0.01 \\
0.01 & -0.01 & -0.01 & 0.03 & 0.05 & 0.06 & -0.23 \\
-0.07 &    0 &    0 & -0.01 & -0.01 & 0.04 & -0.03
}+
\setstackgap{L}{1.1\baselineskip}
\fixTABwidth{T}
\parenMatrixstack{
0.03 &    0 & -0.02 & 0.01 &    0 & -0.04 & 0.06 \\
   0 &    0 &    0 &    0 &    0 &    0 & -0.01 \\
-0.01 &    0 & 0.04 & -0.12 & 0.10 & -0.03 & 0.06 \\
   0 & -0.01 & 0.21 & -0.13 & -0.01 & 0.04 & -0.02 \\
   0 &    0 & 0.06 & 0.03 & -0.13 & 0.01 & 0.06 \\
0.23 &    0 & 0.05 & 0.04 & -0.01 & -0.07 & -0.01 \\
0.04 &    0 & 0.02 & -0.01 & 0.03 & 0.01 & -0.07
}i$,

$\occmat^{\downarrow\uparrow}=
\setstackgap{L}{1.1\baselineskip}
\fixTABwidth{T}
\parenMatrixstack{
0.07 & -0.01 & 0.05 & -0.02 & 0.07 & 0.01 & -0.07 \\
   0 &    0 & -0.01 &    0 & 0.01 & -0.01 &    0 \\
0.02 &    0 & 0.13 & 0.01 & -0.09 & -0.01 &    0 \\
-0.01 &    0 & -0.03 & 0.14 & 0.12 & 0.03 & -0.01 \\
0.03 &    0 & -0.06 & -0.21 & -0.04 & 0.05 & -0.01 \\
-0.01 &    0 & 0.01 & 0.04 & -0.03 & 0.06 & 0.04 \\
-0.04 &    0 &    0 & -0.01 & -0.01 & -0.23 & -0.03
}+
\setstackgap{L}{1.1\baselineskip}
\fixTABwidth{T}
\parenMatrixstack{
-0.03 &    0 & 0.01 &    0 &    0 & -0.23 & -0.04 \\
   0 &    0 &    0 & 0.01 &    0 &    0 &    0 \\
0.02 &    0 & -0.04 & -0.21 & -0.06 & -0.05 & -0.02 \\
-0.01 &    0 & 0.12 & 0.13 & -0.03 & -0.04 & 0.01 \\
   0 &    0 & -0.10 & 0.01 & 0.13 & 0.01 & -0.03 \\
0.04 &    0 & 0.03 & -0.04 & -0.01 & 0.07 & -0.01 \\
-0.06 & 0.01 & -0.06 & 0.02 & -0.06 & 0.01 & 0.07
}i$,

$\occmat^{\downarrow\downarrow}=
\setstackgap{L}{1.1\baselineskip}
\fixTABwidth{T}
\parenMatrixstack{
0.06 &    0 & 0.02 & -0.03 &    0 & 0.03 & -0.01 \\
   0 & 0.13 &    0 &    0 &    0 &    0 &    0 \\
0.02 &    0 & 0.13 & 0.05 & -0.04 & -0.02 &    0 \\
-0.03 &    0 & 0.05 & 0.26 & 0.06 &    0 & 0.02 \\
   0 &    0 & -0.04 & 0.06 & 0.13 & 0.02 & -0.02 \\
0.03 &    0 & -0.02 &    0 & 0.02 & 0.20 & 0.02 \\
-0.01 &    0 &    0 & 0.02 & -0.02 & 0.02 & 0.06
}+
\setstackgap{L}{1.1\baselineskip}
\fixTABwidth{T}
\parenMatrixstack{
   0 &    0 & -0.01 & -0.01 & 0.02 & -0.04 & -0.02 \\
   0 &    0 &    0 &    0 &    0 &    0 &    0 \\
0.01 &    0 &    0 & -0.12 & -0.06 & -0.05 & -0.01 \\
0.01 &    0 & 0.12 &    0 & -0.12 & -0.01 & 0.02 \\
-0.02 &    0 & 0.06 & 0.12 &    0 & -0.05 &    0 \\
0.04 &    0 & 0.05 & 0.01 & 0.05 &    0 & -0.05 \\
0.02 &    0 & 0.01 & -0.02 &    0 & 0.05 &    0
}i$.

$\occmat(\textrm{U}4)$:

$\occmat^{\uparrow\uparrow}=
\setstackgap{L}{1.1\baselineskip}
\fixTABwidth{T}
\parenMatrixstack{
0.35 &    0 & 0.07 & -0.03 & 0.02 & 0.09 & -0.03 \\
   0 & 0.13 & 0.01 &    0 &    0 &    0 & -0.01 \\
0.07 & 0.01 & 0.24 & 0.10 & -0.02 & -0.02 & -0.02 \\
-0.03 &    0 & 0.10 & 0.21 & 0.11 &    0 & 0.03 \\
0.02 &    0 & -0.02 & 0.11 & 0.24 & 0.02 & -0.06 \\
0.09 &    0 & -0.02 &    0 & 0.02 & 0.10 & 0.08 \\
-0.03 & -0.01 & -0.02 & 0.03 & -0.06 & 0.08 & 0.36
}+
\setstackgap{L}{1.1\baselineskip}
\fixTABwidth{T}
\parenMatrixstack{
    0 & -0.01 & -0.01 & 0.02 & -0.06 & 0.09 & 0.32 \\
0.01 &    0 &    0 &    0 & 0.01 & 0.01 &    0 \\
0.01 &    0 &    0 & 0.14 & 0.18 & 0.05 & 0.07 \\
-0.02 &    0 & -0.14 &    0 & 0.13 & 0.02 & -0.03 \\
0.06 & -0.01 & -0.18 & -0.13 &    0 & 0.05 & 0.02 \\
-0.09 & -0.01 & -0.05 & -0.02 & -0.05 &    0 & 0.10 \\
-0.32 &    0 & -0.07 & 0.03 & -0.02 & -0.10 &    0 
}i$,

$\occmat^{\uparrow\downarrow}=
\setstackgap{L}{1.1\baselineskip}
\fixTABwidth{T}
\parenMatrixstack{
-0.06 &    0 & -0.02 & -0.01 & -0.03 & -0.01 & 0.04 \\
-0.01 &    0 &    0 &    0 &    0 &    0 &    0 \\
-0.05 & -0.01 & -0.13 & -0.03 & 0.06 & 0.01 &    0 \\
-0.02 &    0 & 0.01 & -0.14 & -0.21 & -0.04 & -0.01 \\
-0.07 & 0.01 & 0.09 & 0.12 & 0.04 & -0.03 & 0.01 \\
0.01 & 0.01 & -0.01 & -0.03 & 0.05 & -0.06 & -0.23 \\
0.07 &    0 &    0 & -0.01 & 0.01 & 0.04 & 0.03
}+
\setstackgap{L}{1.1\baselineskip}
\fixTABwidth{T}
\parenMatrixstack{
-0.03 &    0 & 0.02 & 0.01 &    0 & -0.03 & -0.06 \\
   0 &    0 &    0 &    0 &    0 &    0 & -0.01 \\
0.01 &    0 & -0.04 & -0.12 & -0.10 & -0.03 & -0.06 \\
   0 & 0.01 & 0.21 & 0.13 & -0.01 & -0.04 & -0.02 \\
   0 &    0 & -0.06 & 0.03 & 0.13 & 0.01 & -0.06 \\
0.23 &    0 & 0.05 & -0.04 & -0.01 & 0.07 & -0.01 \\
-0.04 &    0 & -0.02 & -0.01 & -0.03 & 0.01 & 0.07
}i$,

$\occmat^{\downarrow\uparrow}=
\setstackgap{L}{1.1\baselineskip}
\fixTABwidth{T}
\parenMatrixstack{
-0.06 & -0.01 & -0.05 & -0.02 & -0.07 & 0.01 & 0.07 \\
   0 &    0 & -0.01 &    0 & 0.01 & 0.01 &    0 \\
-0.02 &    0 & -0.13 & 0.01 & 0.09 & -0.01 &    0 \\
-0.01 &    0 & -0.03 & -0.14 & 0.12 & -0.03 & -0.01 \\
-0.03 &    0 & 0.06 & -0.21 & 0.04 & 0.05 & 0.01 \\
-0.01 &    0 & 0.01 & -0.04 & -0.03 & -0.06 & 0.04 \\
0.04 &    0 &    0 & -0.01 & 0.01 & -0.23 & 0.03
}+
\setstackgap{L}{1.1\baselineskip}
\fixTABwidth{T}
\parenMatrixstack{
0.03 &    0 & -0.01 &    0 &    0 & -0.23 & 0.04 \\
   0 &    0 &    0 & -0.01 &    0 &    0 &    0 \\
-0.02 &    0 & 0.04 & -0.21 & 0.06 & -0.05 & 0.02 \\
-0.01 &    0 & 0.12 & -0.13 & -0.03 & 0.04 & 0.01 \\
   0 &    0 & 0.10 & 0.01 & -0.13 & 0.01 & 0.03 \\
0.03 &    0 & 0.03 & 0.04 & -0.01 & -0.07 & -0.01 \\
0.06 & 0.01 & 0.06 & 0.02 & 0.06 & 0.01 & -0.07
}i$,

$\occmat^{\downarrow\downarrow}=
\setstackgap{L}{1.1\baselineskip}
\fixTABwidth{T}
\parenMatrixstack{
0.06 &    0 & 0.02 & 0.03 &    0 & -0.03 & -0.01 \\
   0 & 0.13 &    0 &    0 &    0 &    0 &    0 \\
0.02 &    0 & 0.13 & -0.05 & -0.04 & 0.02 &    0 \\
0.03 &    0 & -0.05 & 0.26 & -0.06 &    0 & -0.02 \\
   0 &    0 & -0.04 & -0.06 & 0.13 & -0.02 & -0.02 \\
-0.03 &    0 & 0.02 &    0 & -0.02 & 0.20 & -0.02 \\
-0.01 &    0 &    0 & -0.02 & -0.02 & -0.02 & 0.06
}+
\setstackgap{L}{1.1\baselineskip}
\fixTABwidth{T}
\parenMatrixstack{
   0 &    0 & -0.01 & 0.01 & 0.02 & 0.04 & -0.02 \\
   0 &    0 &    0 &    0 &    0 &    0 &    0 \\
0.01 &    0 &    0 & 0.12 & -0.06 & 0.04 & -0.01 \\
-0.01 &    0 & -0.12 &    0 & 0.12 & -0.01 & -0.02 \\
-0.02 &    0 & 0.06 & -0.12 &    0 & 0.05 &    0 \\
-0.04 &    0 & -0.04 & 0.01 & -0.05 &    0 & 0.05 \\
0.02 &    0 & 0.01 & 0.02 &    0 & -0.05 &    0 
}i$.

\bigskip
\noindent Relaxed crystal structure in the format of POSCAR:\\
U4 O8          \\                         
   1.00000000000000   \\  
     5.4502594375427025    0.0000052530566619    0.0000020954538445\\
     0.0000044650827639    5.4501069433499660    0.0000062034840623\\
     0.0000026533835966    0.0000056369556111    5.4505198141867979\\
   U    O \\
     4     8\\
Direct\\
  0.5000000073939900  0.0000000100853638  0.0000000023791443\\
  0.5000000075683451  0.4999999899409293  0.4999999979628626\\
 -0.0000000073498820  0.0000000099938806  0.4999999979528320\\
 -0.0000000078177981  0.4999999895210447  0.0000000023041541\\
  0.2482942322261207  0.7483076767449037  0.7516923115519583\\
  0.2482928031767960  0.2483054715778811  0.7482957256505772\\
  0.2516946216220042  0.2483104597837050  0.2482947541735445\\
  0.2517063451208649  0.7483190278296348  0.2517012305365951\\
  0.7517071967287041  0.7516945284186531  0.2517042742162016\\
  0.7517057681018967  0.2516923234785128  0.2483076882208449\\
  0.7482936545958139  0.2516809723944772  0.7482987693186320\\
  0.7483053786331446  0.7516895402310277  0.7517052457326464

\subsubsection{Properties of $\mathbb{S}_0$ (3\textbf{k} AFM, LDA+U, U=5eV)}

\begin{table*}[!h]%
\begin{ruledtabular}
\begin{tabular}{ll}
\textrm{Properties (unit)}&
\textrm{Values}\\
\colrule
Energy (eV) & -138.219 \\
Basis vectors (\AA)& $a=(
5.4647,     0,     0
)$, $b=(
    0, 5.4644,     0
)$, $c=(
    0,     0, 5.4651
)$\\
\begin{tabular}{@{}l@{}}Atom coordinates \\ (in basis vectors)\end{tabular} &
{\begin{tabular}{@{}l@{}}U1:$(0.5000 ,     0 ,     0)$, U2:$(0.5000 , 0.5000 , 0.5000)$ \\U3:$(   0 ,     0 , 0.5000)$, U4:$(  0 , 0.5000 ,     0 )$ \\O1:$(
0.2485, 0.7486, 0.7514
)$, O2:$(
0.2485, 0.2486, 0.7485
)$\\O3:$(
0.2514, 0.2486, 0.2485
)$, O4:$(
0.2515, 0.7486, 0.2515
)$\\O5:$(
0.7515, 0.7514, 0.2515
)$, O6:$(
0.7515, 0.2514, 0.2486
)$\\O7:$(
0.7485, 0.2514, 0.7485
)$, O8:$(
0.7486, 0.7514, 0.7515
)$\end{tabular}} \\ 
Strain & \begin{tabular}{@{}l@{}}$(\epsilon_{xx},\epsilon_{yy},\epsilon_{zz},\epsilon_{xy},\epsilon_{xz},\epsilon_{yz})=(
0.0001,     0, 0.0002,     0 ,     0 ,     0
)$, \\$(\epsilon_{A1g},\epsilon_{Eg.0},\epsilon_{Eg.1},\epsilon_{T2g.0},\epsilon_{T2g.1},\epsilon_{T2g.2})=(
0.0002,     0, 0.0001,     0 ,     0 ,     0
)$.\end{tabular}\\
Spin moments ($\mu_B$)& \begin{tabular}{@{}l@{}}U1:$(
 0.694, -0.647, -0.687
)$, U2:$(
-0.676,  0.674, -0.680
)$ \\ U3:$(
 0.676,  0.659,  0.694
)$, U4:$(
-0.661, -0.660,  0.707
)$\end{tabular}\\
Orbit moments ($\mu_B$)& \begin{tabular}{@{}l@{}}U1:$(
-2.062,  1.932,  2.043
)$, U2:$(
 2.013, -2.005,  2.022
)$ \\ U3:$(
-2.013, -1.965, -2.061
)$, U4:$(
 1.971,  1.969, -2.097
)$\end{tabular}\\
Total magnetic moments ($\mu_B$)& \begin{tabular}{@{}l@{}}U1:$(
-1.368, 1.285, 1.356
)$, U2:$(
1.337, -1.331, 1.342
)$ \\ U3:$(
-1.337, -1.306, -1.367
)$, U4:$(
1.310, 1.309, -1.390
)$\end{tabular}\\
Energy (eV) and lattice (\AA) of und & -138.214, 5.4642
\end{tabular}
\end{ruledtabular}
\end{table*}

\clearpage

\subsection{\label{sec:smu0fm}FM at $U=0$}

In this section, we present the FM states calculated by GGA (i.e., $U=0$) with and without SOC, as the reference state in the absence of the Hubburd $U$. 

\subsubsection{Properties of FM without SOC at $U=0$}

\begin{table*}[h]%
\begin{ruledtabular}
\begin{tabular}{ll}
\textrm{Properties (unit)}&
\textrm{Values}\\
\colrule
Energy (eV) & -30.937 \\
Basis vectors (\AA)& $a=(2.7085, 2.7085, 0.0000), b=(0.0000, 2.7085, 2.7085), c=(2.7085, 0.0000, 2.7085)$\\
\begin{tabular}{@{}l@{}}Atom coordinates \\ (in basis vectors)\end{tabular} &
{\begin{tabular}{@{}l@{}}$U1=(0.5000, 0.5000, 0.5000)$ \\$O1=(0.2500, 0.2500, 0.2500)$, $O2=(0.7500, 0.7500, 0.7500)$\end{tabular}} \\ 
Strain & \begin{tabular}{@{}l@{}}($\epsilon_{xx},\epsilon_{yy},\epsilon_{zz},\epsilon_{xy},\epsilon_{xz},\epsilon_{yz}$)=(0, 0, 0, 0, 0, 0), \\($\epsilon_{A1g},\epsilon_{Eg.0},\epsilon_{Eg.1},\epsilon_{T2g.0},\epsilon_{T2g.1},\epsilon_{T2g.2}$)=(0, 0, 0, 0, 0, 0).\end{tabular}\\
Spin moments ($\mu_B$)& (1.988, 0.000, 0.000)\\
Energy (eV) and lattice (\AA) of und & -30.937, 2.7085
\end{tabular}
\end{ruledtabular}
\end{table*}

$\occmat^\uparrow=
\setstackgap{L}{1.1\baselineskip}
\fixTABwidth{T}
\parenMatrixstack{
0.3703 &     0 & 0.0318 &     0 &     0 &     0 &     0 \\
    0 & 0.1979 &     0 &     0 &     0 &     0 &     0 \\
0.0318 &     0 & 0.3539 &     0 &     0 &     0 &     0 \\
    0 &     0 &     0 & 0.3952 &     0 & 0.0002 &     0 \\
    0 &     0 &     0 &     0 & 0.3546 &     0 & -0.0314 \\
    0 &     0 &     0 & 0.0002 &     0 & 0.3304 &     0 \\
    0 &     0 &     0 &     0 & -0.0314 &     0 & 0.3711
}$,

$\occmat^\downarrow=
\setstackgap{L}{1.1\baselineskip}
\fixTABwidth{T}
\parenMatrixstack{
0.0488 &     0 & 0.0086 &     0 &     0 &     0 &     0 \\
    0 & 0.1535 &     0 &     0 &     0 &     0 &     0 \\
0.0086 &     0 & 0.0444 &     0 &     0 &     0 &     0 \\
    0 &     0 &     0 & 0.0555 &     0 &     0 &     0 \\
    0 &     0 &     0 &     0 & 0.0444 &     0 & -0.0086 \\
    0 &     0 &     0 &     0 &     0 & 0.0376 &     0 \\
    0 &     0 &     0 &     0 & -0.0086 &     0 & 0.0488
}$.

\subsubsection{Properties of FM with SOC at $U=0$}

\begin{table*}[!h]%
\begin{ruledtabular}
\begin{tabular}{ll}
\textrm{Properties (unit)}&
\textrm{Values}\\
\colrule
Energy (eV) & -33.710 \\
Basis vectors (\AA)& $a=(
2.7098, 2.7098,     0
), b=(
0, 2.7098, 2.7193
), c=(
2.7098,     0, 2.7193
)$\\
\begin{tabular}{@{}l@{}}Atom coordinates \\ (in basis vectors)\end{tabular} &
{\begin{tabular}{@{}l@{}}$U1=(
0.5000, 0.5000, 0.5000
)$ \\$O1=(
0.2500, 0.2500, 0.2500
)$, $O2=(
0.7500, 0.7500, 0.7500
)$\end{tabular}} \\ 
Strain & \begin{tabular}{@{}l@{}}($\epsilon_{xx},\epsilon_{yy},\epsilon_{zz},\epsilon_{xy},\epsilon_{xz},\epsilon_{yz}$)=(
-0.0012, -0.0012, 0.0023,     0,     0,     0
), \\($\epsilon_{A1g},\epsilon_{Eg.0},\epsilon_{Eg.1},\epsilon_{T2g.0},\epsilon_{T2g.1},\epsilon_{T2g.2}$)=(
 0,     0, 0.0029,     0,     0,     0
).\end{tabular}\\
Spin moments ($\mu_B$)& (0, 0, 1.525)\\
Orbit moments ($\mu_B$)& (0, 0, -1.587)\\
Total magnetic moments ($\mu_B$)& (0, 0, -0.062)\\
Energy (eV) and lattice (\AA) of und & -33.709, 2.7130
\end{tabular}
\end{ruledtabular}
\end{table*}

$\occmat^{\uparrow\uparrow}=
\setstackgap{L}{1.1\baselineskip}
\setstacktabbedgap{9pt}
\fixTABwidth{T}
\parenMatrixstack{
0.34 &    0 &    0 &    0 &    0 &    0 &    0 \\
   0 & 0.20 &    0 &    0 &    0 &    0 &    0 \\
   0 &    0 & 0.33 &    0 &    0 &    0 &    0 \\
   0 &    0 &    0 & 0.30 &    0 &    0 &    0 \\
   0 &    0 &    0 &    0 & 0.33 &    0 &    0 \\
   0 &    0 &    0 &    0 &    0 & 0.30 &    0 \\
   0 &    0 &    0 &    0 &    0 &    0 & 0.34
}+$
$\setstackgap{L}{1.1\baselineskip}
\fixTABwidth{T}
\parenMatrixstack{
   0 &    0 &    0 &    0 & 0.02 &    0 & 0.23 \\
   0 &    0 &    0 &    0 &    0 & 0.05 &    0 \\
   0 &    0 &    0 &    0 & 0.10 &    0 & -0.02 \\
   0 &    0 &    0 &    0 &    0 &    0 &    0 \\
-0.02 &    0 & -0.10 &    0 &    0 &    0 &    0 \\
   0 & -0.05 &    0 &    0 &    0 &    0 &    0 \\
-0.23 &    0 & 0.02 &    0 &    0 &    0 &    0
}i$,

$\occmat^{\uparrow\downarrow}=
\setstackgap{L}{1.1\baselineskip}
\fixTABwidth{T}
\parenMatrixstack{
   0 &    0 &    0 &    0 &    0 &    0 &    0 \\
-0.04 &    0 & 0.03 &    0 &    0 &    0 &    0 \\
   0 & -0.01 &    0 &    0 &    0 &    0 &    0 \\
   0 &    0 &    0 &    0 & -0.11 &    0 & 0.01 \\
   0 &    0 &    0 & 0.07 &    0 & -0.05 &    0 \\
   0 &    0 &    0 &    0 & 0.03 &    0 & -0.04 \\
   0 &    0 &    0 & -0.01 &    0 & 0.06 &    0
}+
\setstackgap{L}{1.1\baselineskip}
\fixTABwidth{T}
\parenMatrixstack{
   0 &    0 &    0 & -0.01 &    0 & -0.06 &    0 \\
   0 &    0 &    0 &    0 & -0.03 &    0 & -0.04 \\
   0 &    0 &    0 & -0.07 &    0 & -0.05 &    0 \\
0.01 &    0 & 0.11 &    0 &    0 &    0 &    0 \\
   0 & 0.01 &    0 &    0 &    0 &    0 &    0 \\
0.04 &    0 & 0.03 &    0 &    0 &    0 &    0 \\
   0 &    0 &    0 &    0 &    0 &    0 &    0
}i$,

$\occmat^{\downarrow\uparrow}=
\setstackgap{L}{1.1\baselineskip}
\fixTABwidth{T}
\parenMatrixstack{
   0 & -0.04 &    0 &    0 &    0 &    0 &    0 \\
   0 &    0 & -0.01 &    0 &    0 &    0 &    0 \\
   0 & 0.03 &    0 &    0 &    0 &    0 &    0 \\
   0 &    0 &    0 &    0 & 0.07 &    0 & -0.01 \\
   0 &    0 &    0 & -0.11 &    0 & 0.03 &    0 \\
   0 &    0 &    0 &    0 & -0.05 &    0 & 0.06 \\
   0 &    0 &    0 & 0.01 &    0 & -0.04 &    0
}+
\setstackgap{L}{1.1\baselineskip}
\fixTABwidth{T}
\parenMatrixstack{
   0 &    0 &    0 & -0.01 &    0 & -0.04 &    0 \\
   0 &    0 &    0 &    0 & -0.01 &    0 &    0 \\
   0 &    0 &    0 & -0.11 &    0 & -0.03 &    0 \\
0.01 &    0 & 0.07 &    0 &    0 &    0 &    0 \\
   0 & 0.03 &    0 &    0 &    0 &    0 &    0 \\
0.06 &    0 & 0.05 &    0 &    0 &    0 &    0 \\
   0 & 0.04 &    0 &    0 &    0 &    0 &    0
}i$,

$\occmat^{\downarrow\downarrow}=
\setstackgap{L}{1.1\baselineskip}
\fixTABwidth{T}
\parenMatrixstack{
0.07 &    0 & 0.03 &    0 &    0 &    0 &    0 \\
   0 & 0.17 &    0 &    0 &    0 &    0 &    0 \\
0.03 &    0 & 0.08 &    0 &    0 &    0 &    0 \\
   0 &    0 &    0 & 0.12 &    0 &    0 &    0 \\
   0 &    0 &    0 &    0 & 0.08 &    0 & -0.03 \\
   0 &    0 &    0 &    0 &    0 & 0.06 &    0 \\
   0 &    0 &    0 &    0 & -0.03 &    0 & 0.07
}+
\setstackgap{L}{1.1\baselineskip}
\fixTABwidth{T}
\parenMatrixstack{
   0 &    0 &    0 &    0 & 0.01 &    0 & -0.02 \\
   0 &    0 &    0 &    0 &    0 & -0.01 &    0 \\
   0 &    0 &    0 &    0 & -0.01 &    0 & -0.01 \\
   0 &    0 &    0 &    0 &    0 &    0 &    0 \\
-0.01 &    0 & 0.01 &    0 &    0 &    0 &    0 \\
   0 & 0.01 &    0 &    0 &    0 &    0 &    0 \\
0.02 &    0 & 0.01 &    0 &    0 &    0 &    0
}i$.
\clearpage

\section{\label{sec:smphonons} phonon properties}
\vspace{5mm}
\justifying

\subsection{Phonon spectra for FM $S_0^1$ without SOC}

\begin{figure}[h]
    \centering
    \includegraphics[width=0.20\columnwidth]{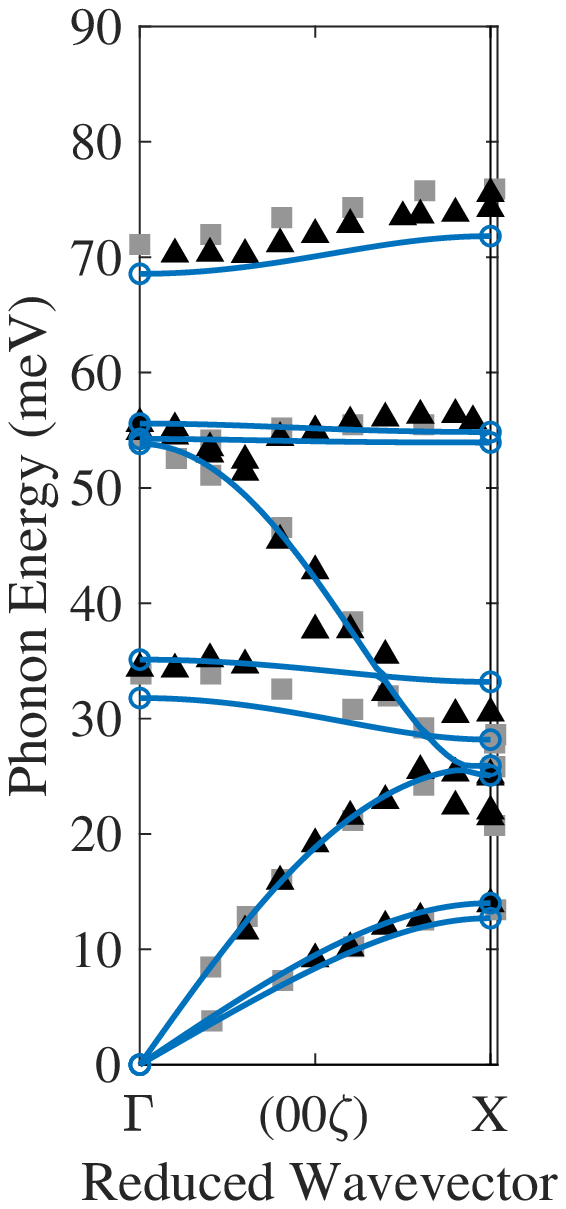}
    \caption{\label{fig:figureSM1} Phonon spectra of UO$_2$ for FM $S_0^1$, calculated using LDA+$U$ ($U=4$ eV) without SOC, and compared with inelastic neutron scattering data from Pang \emph{et al.}\cite{pang_phonon_2013} at 300 K (grey squares) and from this work at 600 K (black triangles). The hollow points are directly computed using DFT+$U$, while the corresponding lines are Fourier interpolations.}
\end{figure}

The phonon spectra of FM UO$_2$ were calculated for $S_0^1$ by using LDA+$U$ ($U=4$ eV) in order to understand the influence of the $T_{2g}$ shear strain distortion of the $S_0$ states on the phonons (see Figure~\ref{fig:figureSM1}). 
The phonon spectra are calculated along the $(100)$ direction, between the $\Gamma$ point and $X$ point, and compared with experiments from Pang \emph{et al.}\cite{pang_phonon_2013} and from this work. The TA, TO1, and TO2 branches each split into two separate branches, meaning the $T_{2g}$ shear strain distortion in $S_0^1$ breaks the symmetry of the respective phonon branches.

\subsection{Phonon spectra for 3\textbf{k} AFM $\mathbb{S}_0$}

\begin{figure}[h]
    \centering
    \includegraphics[width=0.46\columnwidth]{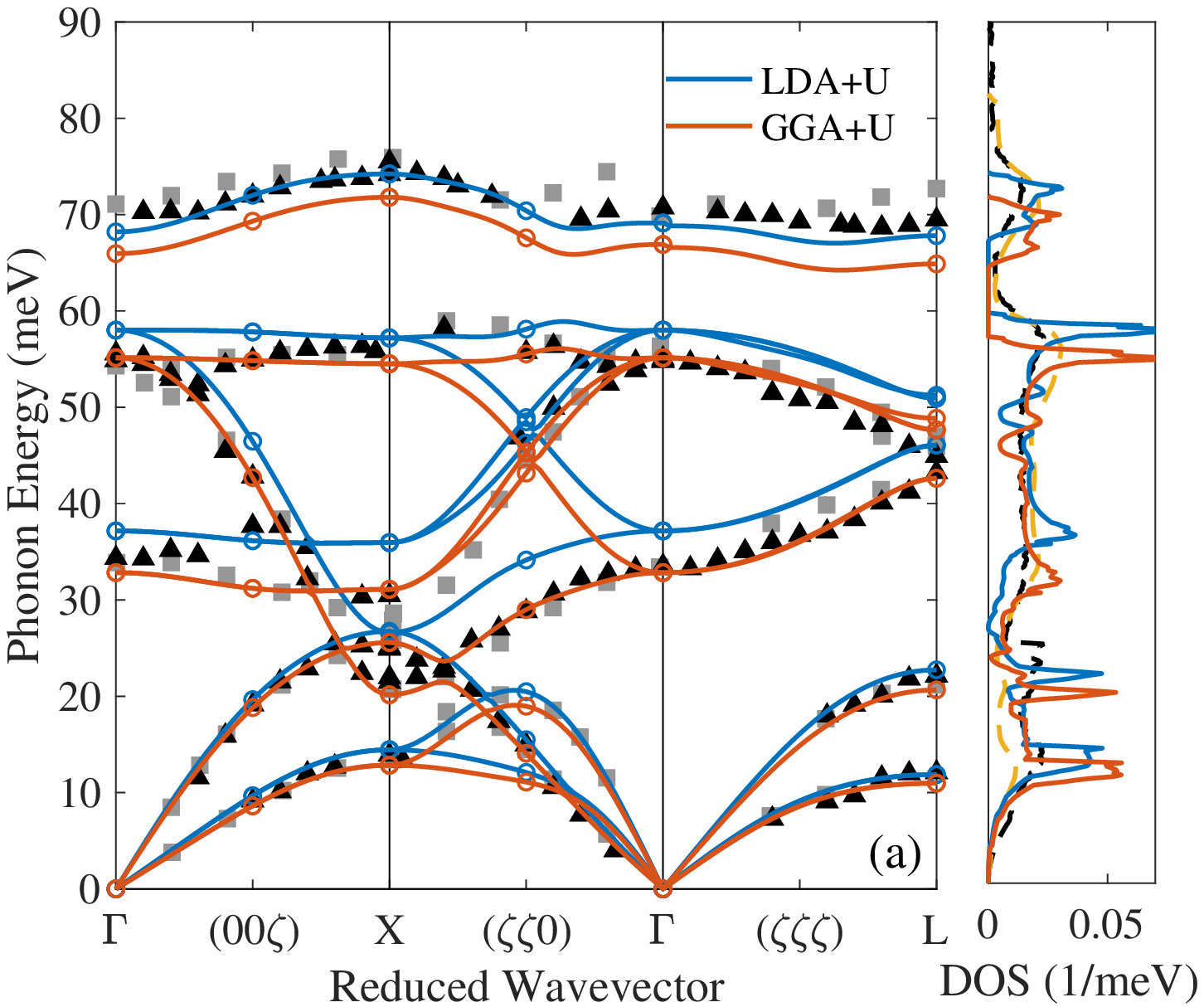}
    \includegraphics[width=0.46\columnwidth]{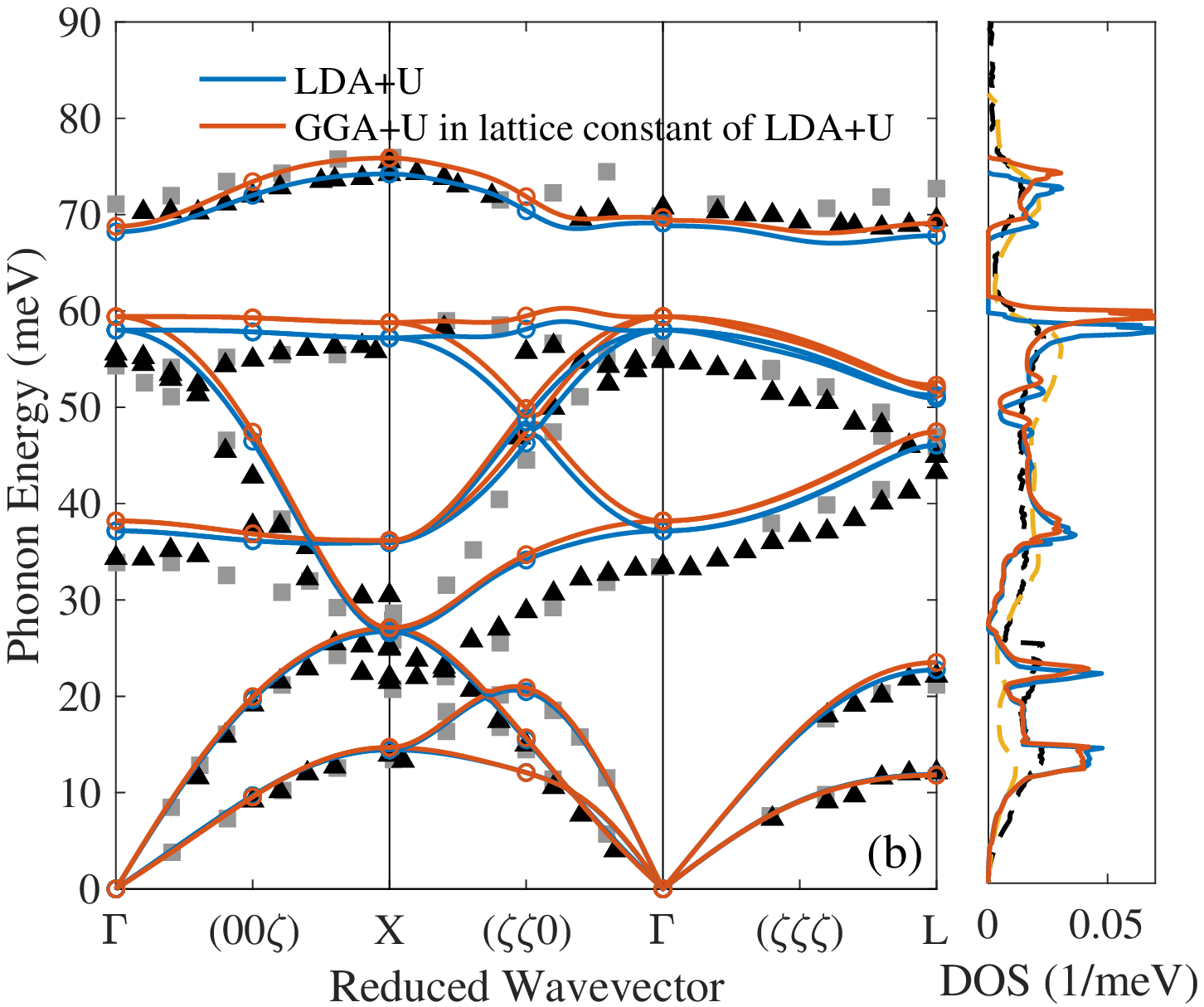}
    \caption{\label{fig:figureSM2} Unfolded 3\textbf{k} AFM UO$_2$ phonon spectra of $\mathbb{S}_0$, calculated using both LDA/GGA+$U$+SOC ($U=4$ eV) (a) in the respective relaxed unit cells of LDA/GGA+$U$+SOC, and (b) in the relaxed unit cell of LDA+$U$+SOC. Both panels are compared to the inelastic neutron scattering data from Pang \emph{et al.}\cite{pang_phonon_2013} at 300 K (grey squares) and this work at 600 K (black triangles). The hollow points are directly computed using DFT+$U$ while the corresponding lines are Fourier interpolations.}
\end{figure}

To understand the source of the difference in the phonon spectra calculated for 3\textbf{k} AFM UO$_2$ in $\mathbb{S}_0$ using LDA+$U$ and GGA+$U$ ($U=4$ eV) (see Figure~\ref{fig:figureSM2}(a)), phonon spectra were calculated using GGA+$U$ in the cell volume of LDA+$U$ (see Figure~\ref{fig:figureSM2}(b)). When using the same cell volume, LDA+$U$ and GGA+$U$ show approximately identical phonon spectra, i.e., the largest percentage error in measured phonon frequency is reduced from 24\% in Figure~\ref{fig:figureSM2}(a) to 3\% in Figure~\ref{fig:figureSM2}(b), meaning that differences in the LDA+$U$ and GGA+$U$ phonon spectra are mainly due to differences in the lattice constant.

\begin{figure}[h]
    \centering
    \includegraphics[width=0.46\columnwidth]{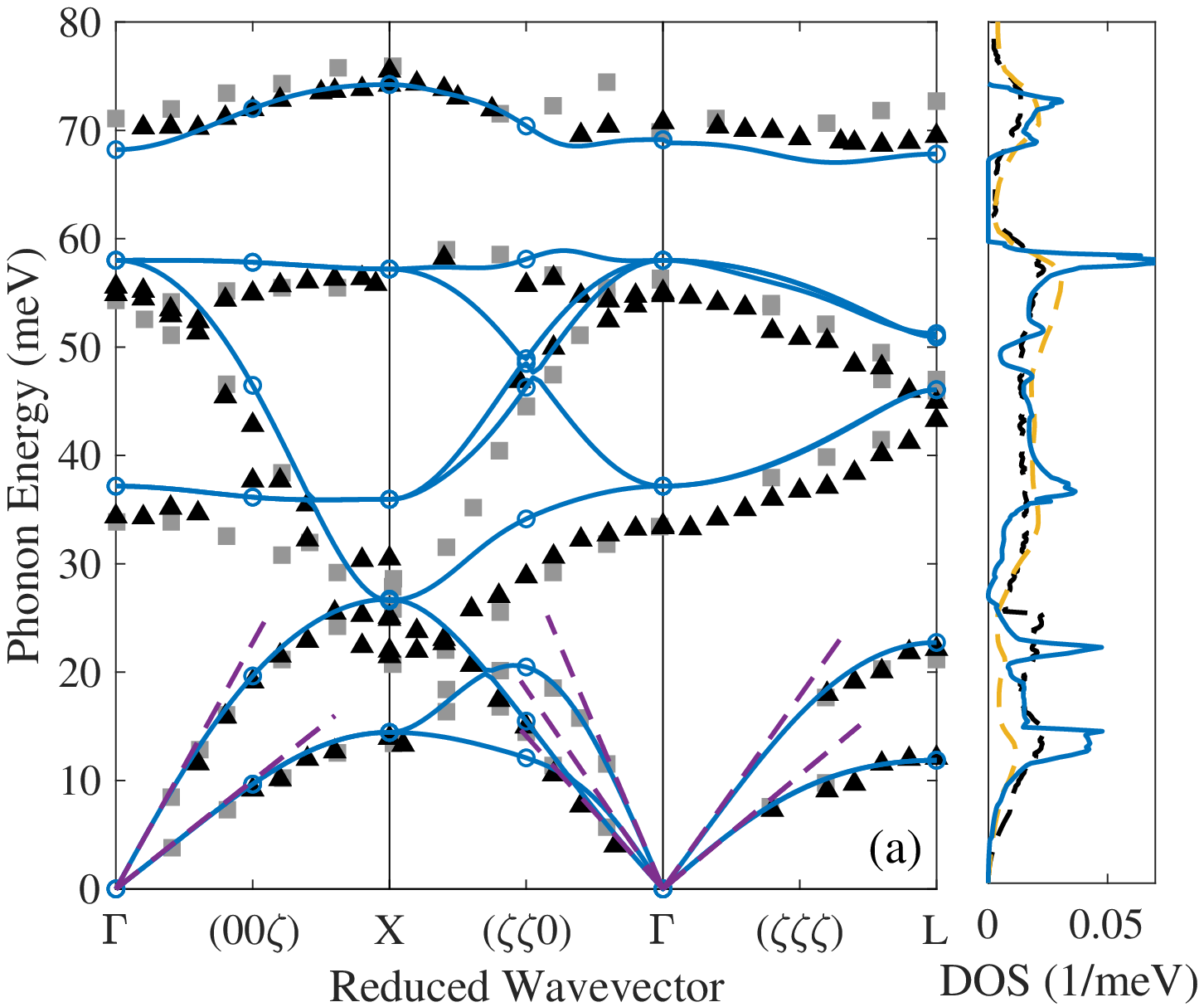}
    \includegraphics[width=0.46\columnwidth]{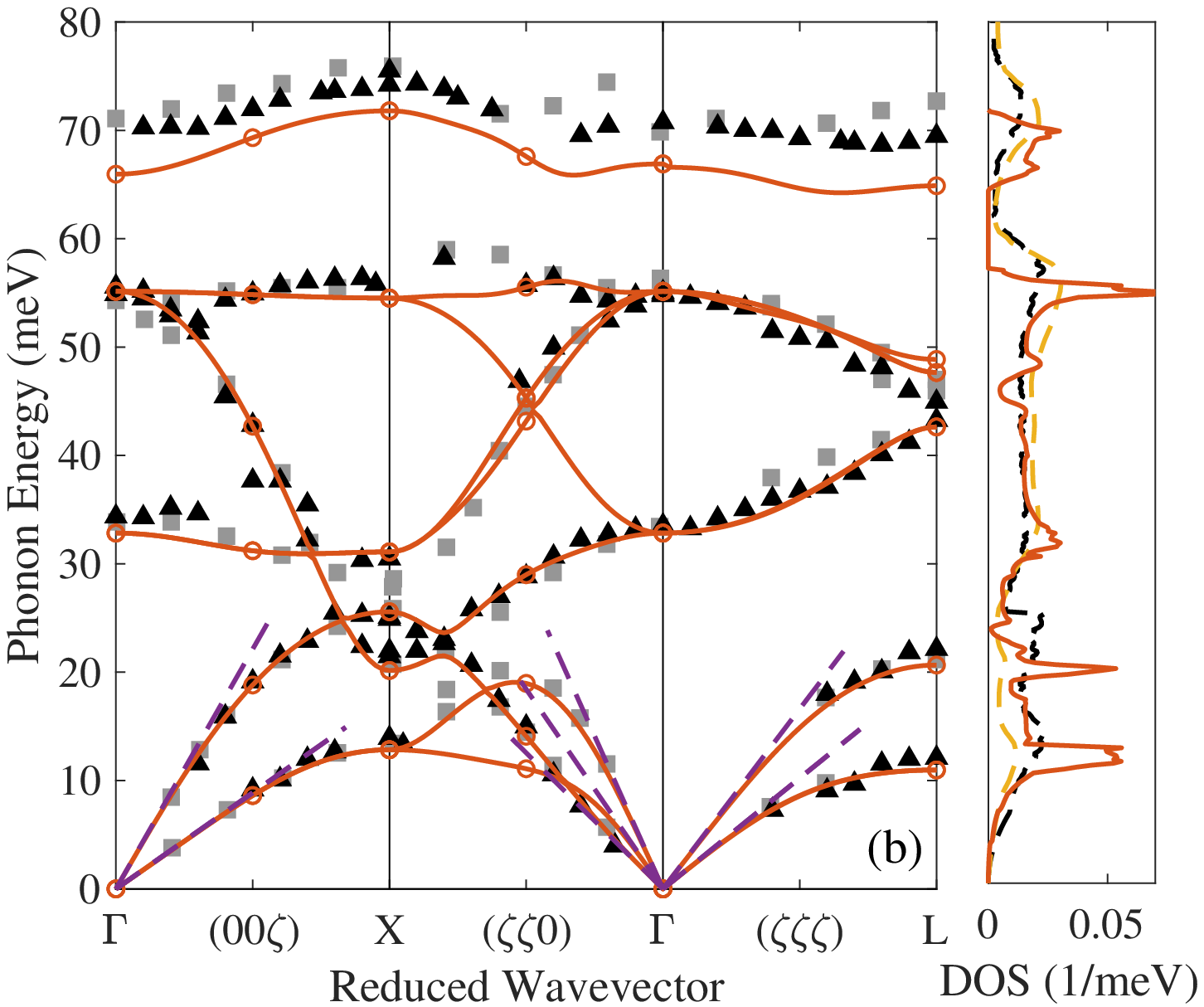}
    \caption{\label{fig:figureSM4} Unfolded 3\textbf{k} AFM UO$_2$ phonon spectra of $\mathbb{S}_0$, calculated using (a) LDA+$U$+SOC and (b) GGA+$U$+SOC ($U=4$ eV), compared with inelastic neutron scattering data from Pang \emph{et al.}\cite{pang_phonon_2013} at 300 K (grey squares), this work at 600 K (black triangles), this work at 77 K (dashed black curves), and Bryan \emph{et al.}\cite{bryan_impact_2019} at 10 K (dashed yellow curves). The hollow points are directly computed using DFT+$U$, while the corresponding lines are Fourier interpolations. The dashed purple lines at $\Gamma$ points indicate the slopes of acoustic phonon branches, as calculated from our elastic constant results. }
\end{figure}

Furthermore, we use our elastic constants (see Table III in main text) to plot the slopes of acoustic phonon branches at the $\Gamma$ points in Figure~\ref{fig:figureSM4}. The calculated slopes of acoustic phonon branches are in good agreement with the calculated phonon spectra. Therefore, our calculated acoustic phonon branches and elastic constants are consistent. 

\subsection{Phonon spectra at $U=0$ for the FM state with and without SOC}

\begin{figure}[h]
    \centering
    \includegraphics[width=0.38\columnwidth]{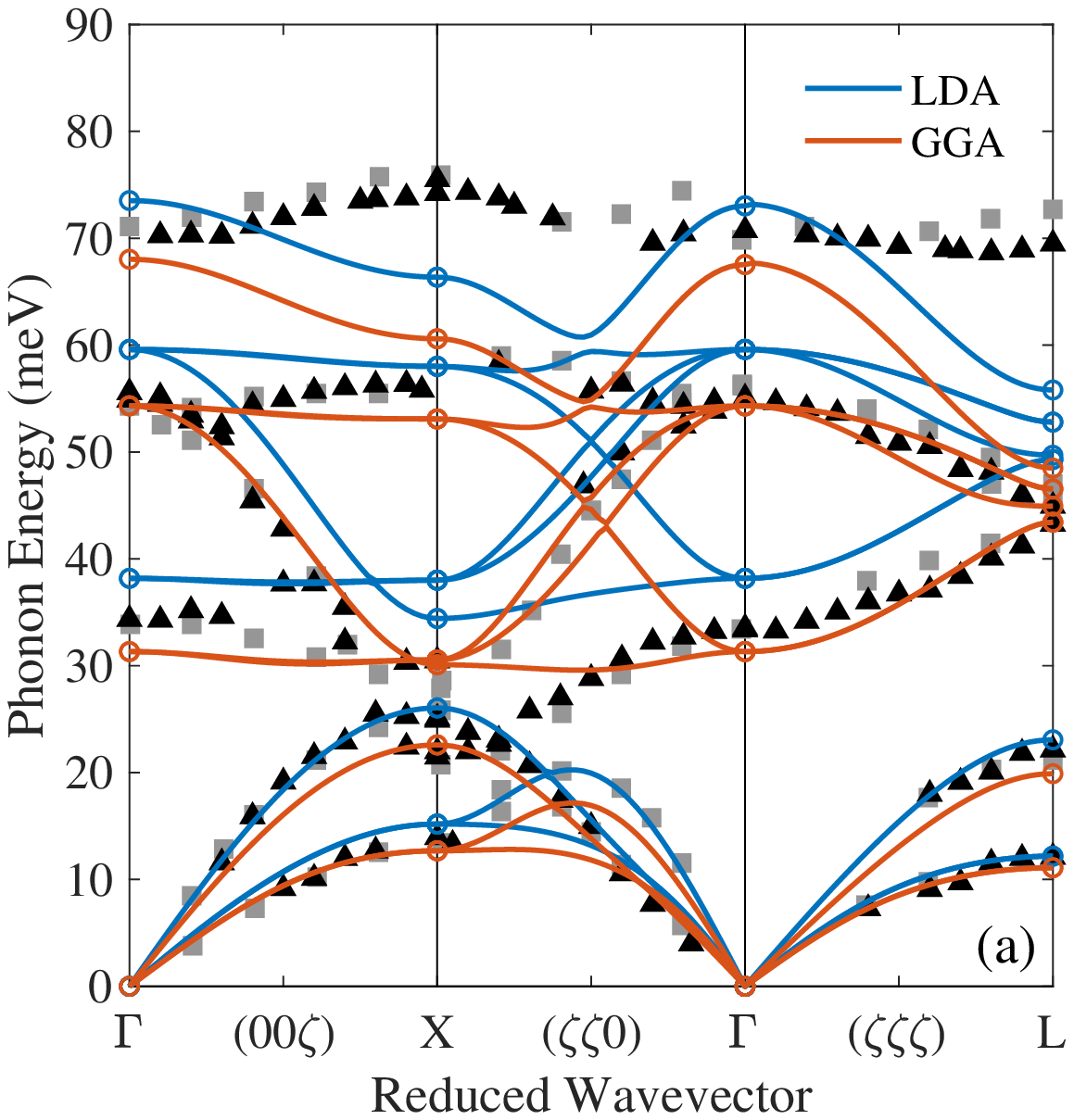}
    \includegraphics[width=0.38\columnwidth]{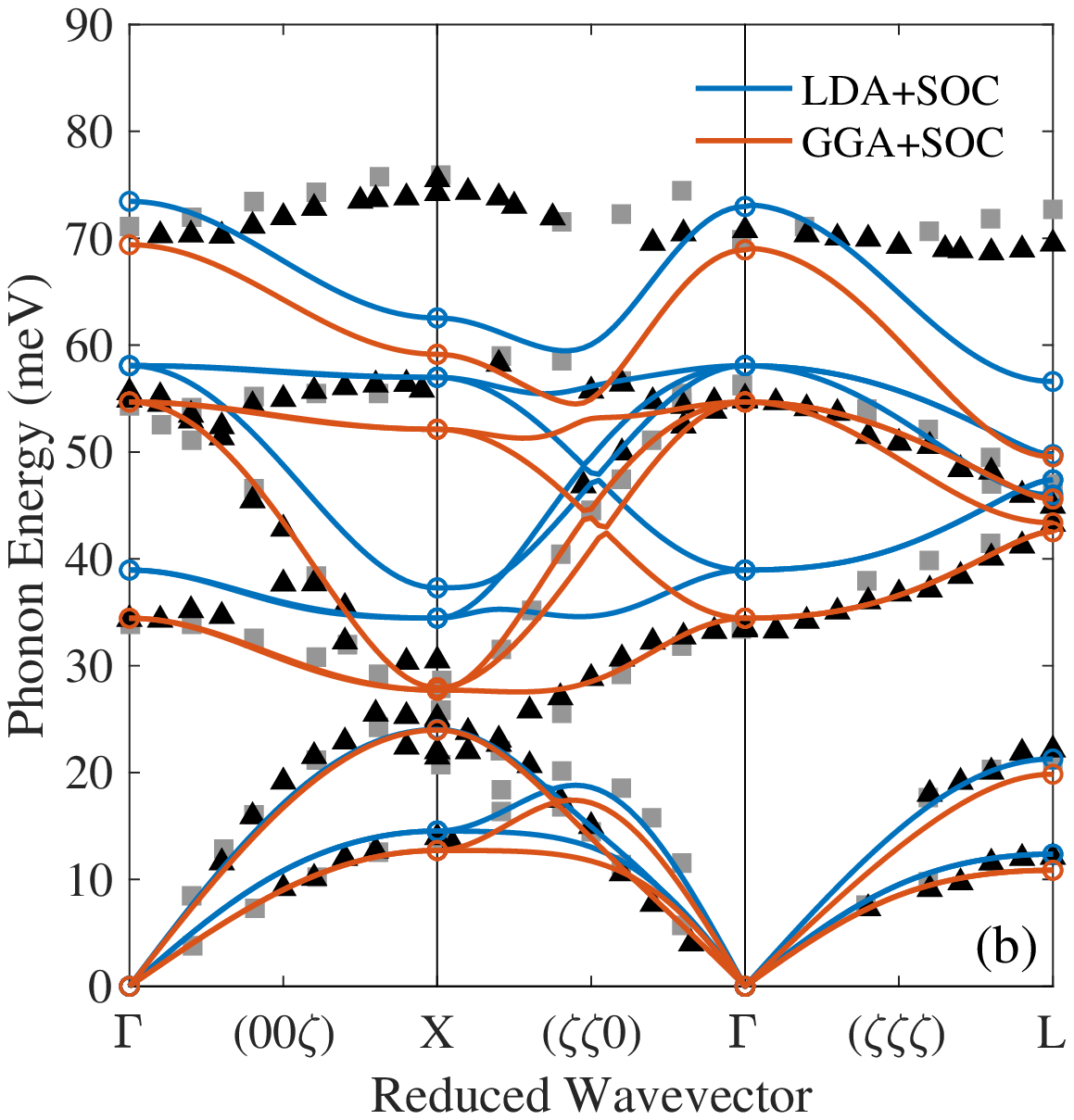}
    \caption{\label{fig:figureSM3} FM UO$_2$ phonon spectra calculated using (a) DFT ($U=0$) and (b) DFT+SOC ($U=0$), compared with inelastic neutron scattering data from Pang \emph{et al.}\cite{pang_phonon_2013} at 300 K (grey squares) and from this work at 600 K (black triangles). The hollow points are directly computed using DFT, while the corresponding lines are Fourier interpolations. }
\end{figure}

The phonon spectra for $U=0$ were calculated for FM UO$_2$ using LDA and GGA, both with and without SOC, as presented in Figure~\ref{fig:figureSM3}. The major difference between $U=4$ eV and $U=0$ is seen on the highest optical phonon branch near the $X$ and $L$ points, (e.g., the highest phonon frequency at the $X$ point decreases from 75 meV at $U$=4eV to 65 meV at $U=0$ for LDA+$U$. For $U=0$, SOC has only minor effects on phonon spectra. 

\subsection{Previously published DFT+$U$ phonon spectra}

\begin{figure}[h]
    \centering
    \includegraphics[width=0.38\columnwidth]{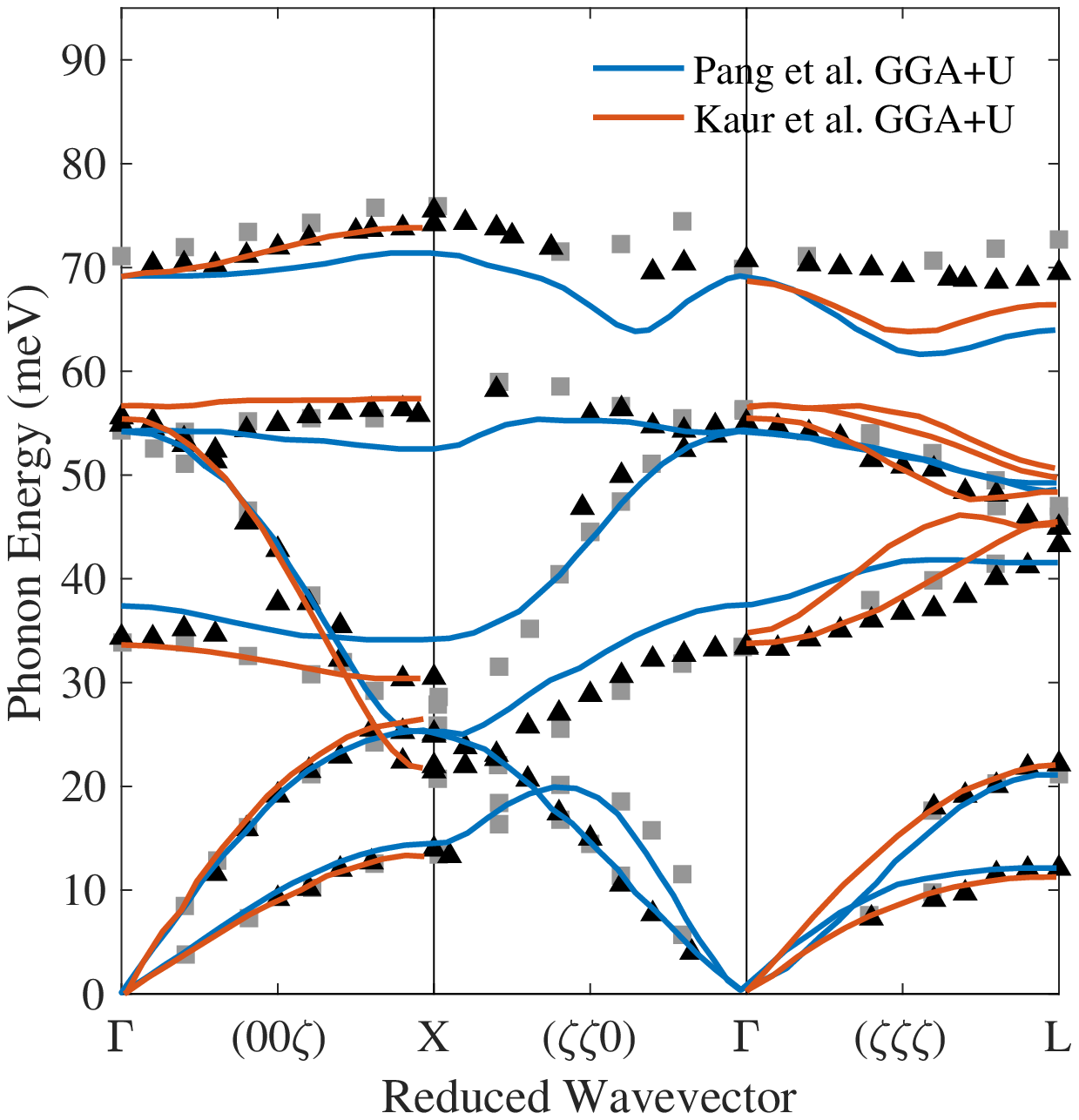}
    \includegraphics[width=0.38\columnwidth]{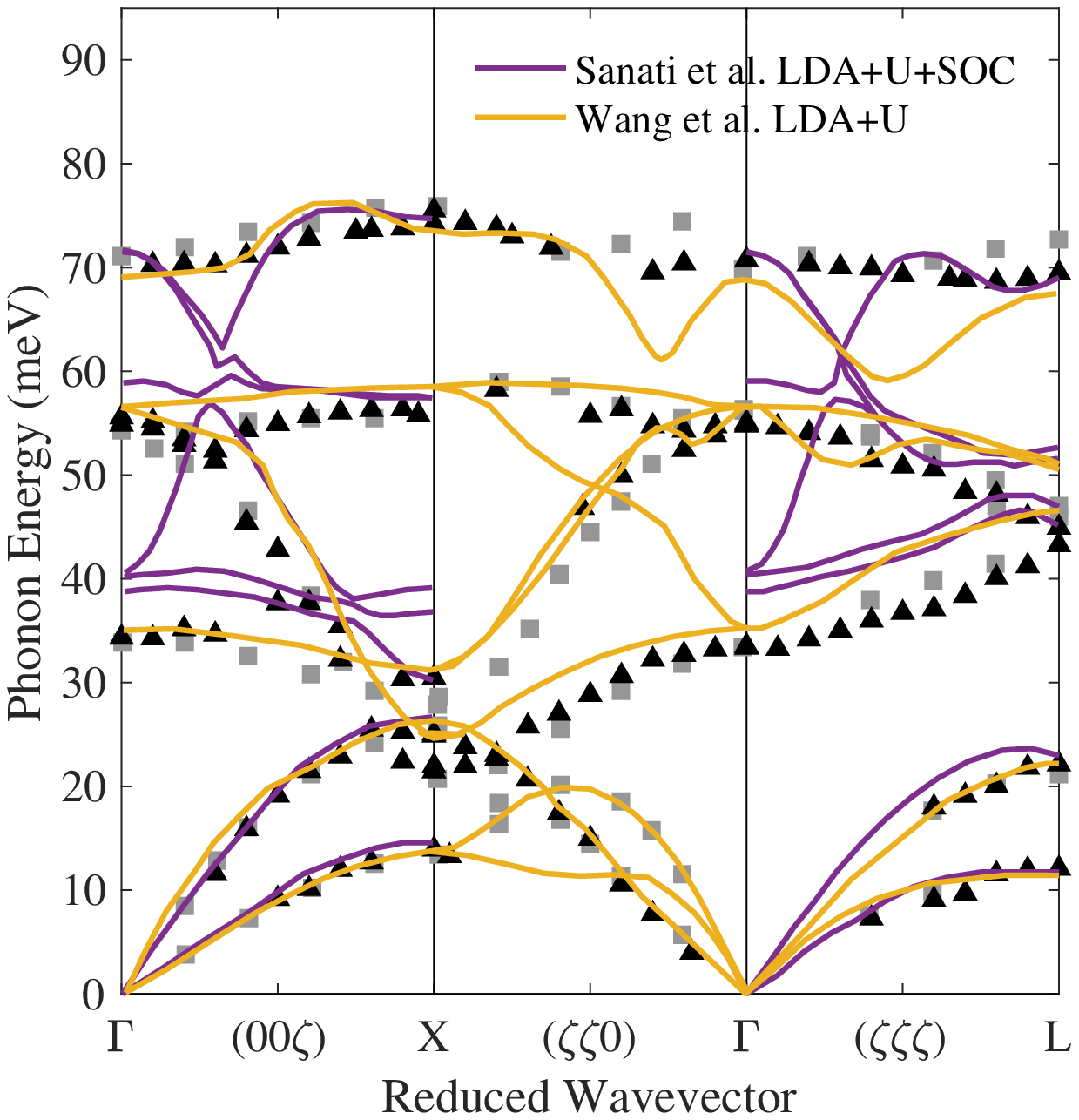}
    \caption{\label{fig:figureSMphononLit} Phonon dispersion calculations from the literature \cite{pang_phonon_2013,wang_phonon_2013,sanati_elastic_2011,kaur_thermal_2013} compared with inelastic neutron scattering data from Pang \emph{et al.}\cite{pang_phonon_2013} at 300 K (grey squares) and from this work at 600 K (black triangles).}
\end{figure}

We review the results of phonon spectra calculations using DFT+$U$, including GGA+$U$ from Pang et al.\cite{pang_phonon_2013} and Kaur et al.\cite{kaur_thermal_2013}, LDA+$U$ from Wang et al.\cite{wang_phonon_2013}, and LDA+$U$+SOC from Sanati et al.\cite{sanati_elastic_2011}. These results are compared with experimental data of inelastic neutron scattering from Pang et al.\cite{pang_phonon_2013} and this work (see Figure~\ref{fig:figureSMphononLit}). In both Pang et al.'s and Wang et al.'s results, major discrepancies occur in optical branches, especially the highest branch is not as flat as the experiments; in Sanati et al.'s results, extra error exists as they predicted the highest band to be two-fold at $\Gamma$ point; while the Kaur et al.'s results have the overall best agreements in $(001)$ direction, their results show strong symmetry breaking in $(111)$ direction. 

\subsection{Phonon unfolding}

\begin{figure}[h]
    \centering
    \includegraphics[width=0.32\columnwidth]{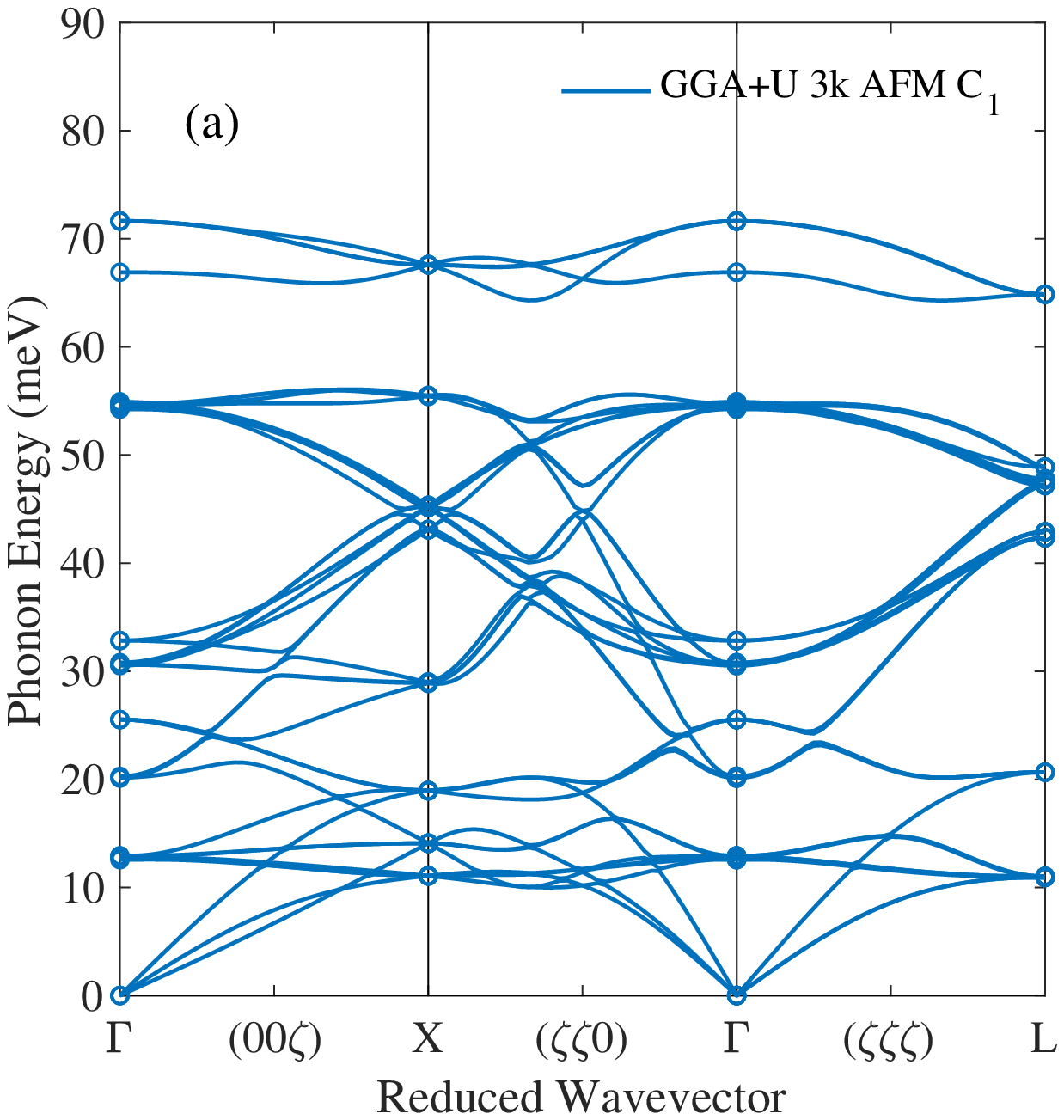}
    \includegraphics[width=0.32\columnwidth]{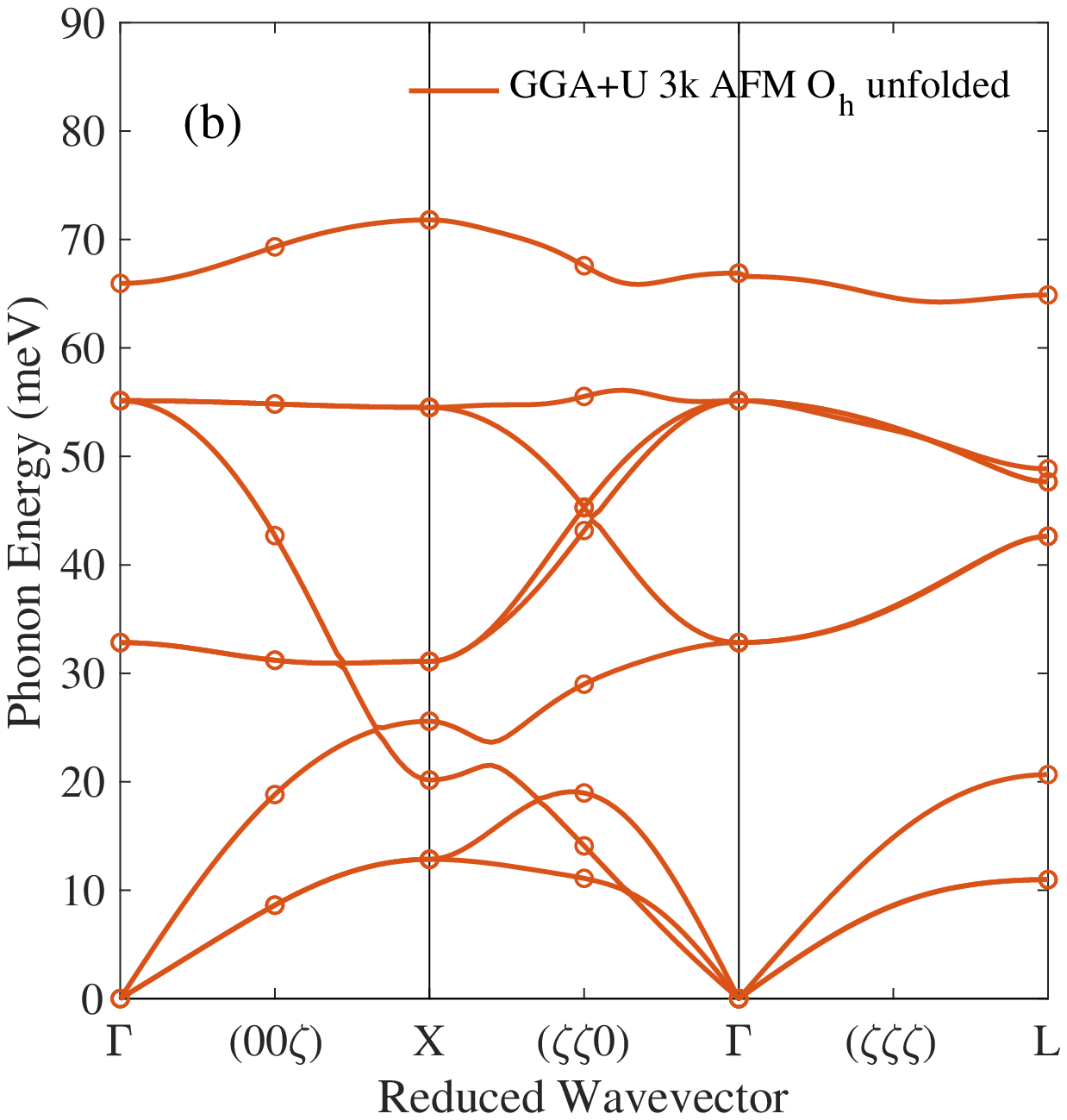}
    \includegraphics[width=0.33\columnwidth]{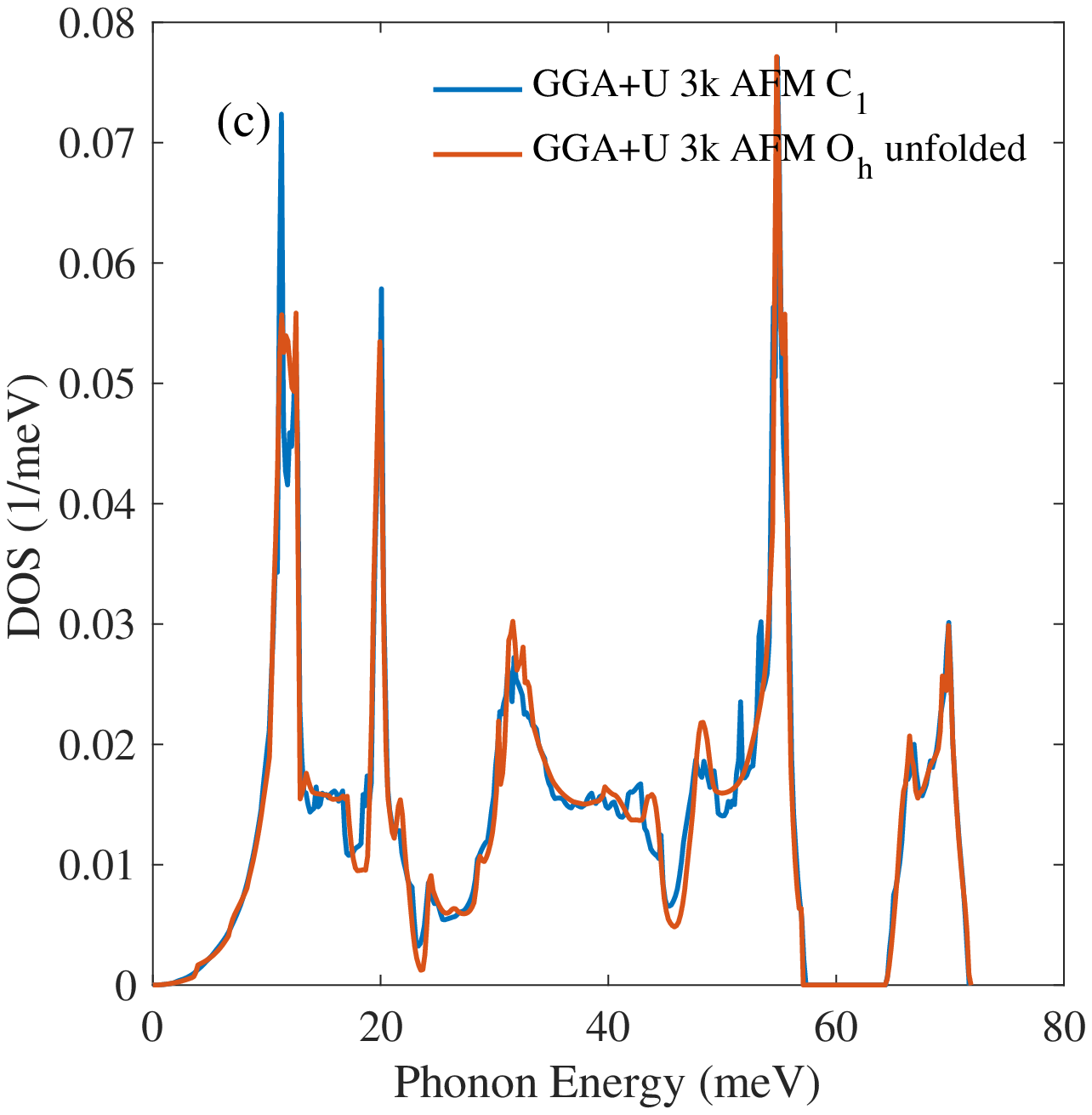}
    \caption{\label{fig:figureSMphononUnfold} Calculated (a) phonon dispersion, (b) unfolded phonon dispersion, (c) phonon density of states comparison before and after the unfolding process. DFT calculations are in 3\textbf{k} AFM $\mathbb{S}_0$ structure using GGA+$U$+SOC ($U=4$ eV) with $C_1$ symmetry.}
\end{figure}

In the case of 1\textbf{k} and 3\textbf{k} AFM, the magnetism breaks the
symmetry of the $Fm\bar{3}m$ space group, and therefore we use a primitive unit
cell of $\hat{\mathbf{S}}_{C}\hat{\mathbf{a}}$, which is commensurate with both
1\textbf{k} and 3\textbf{k}. When computing phonons for 1\textbf{k} and
3\textbf{k}, we discretize the Brillouin zone using
$\hat{\mathbf{S}}_{BZ}=2\hat{\mathbf{1}}$. In order to best compare with
experiment, we also unfold the phonon band structure back to the original primitive unit cell $\hat{\mathbf{a}}$, 
averaging any translational symmetry breaking due
to the magnetism. Here we use the phonon dispersion of the relaxed 3\textbf{k} AFM $\mathbb{S}_0$
structure calculated by GGA+$U$+SOC ($U=4$ eV) as an example.
Figure~\ref{fig:figureSMphononUnfold}(a) presents the original phonon
calculation using $C_1$ symmetry, which has been unfolded to produce
Figure~\ref{fig:figureSMphononUnfold}(b).
Figure~\ref{fig:figureSMphononUnfold}(c) compares the phonon density of states
before and after unfolding. The very small difference between the density of states
indicates that the symmetry breaking in the 3\textbf{k} magnetic state is very small.

\subsection{Phonon spectra for FM and 1\textbf{k} AFM $\mathbb{S}_0$ along different directions }

\begin{figure}[h]
    \centering
    \includegraphics[width=0.76\columnwidth]{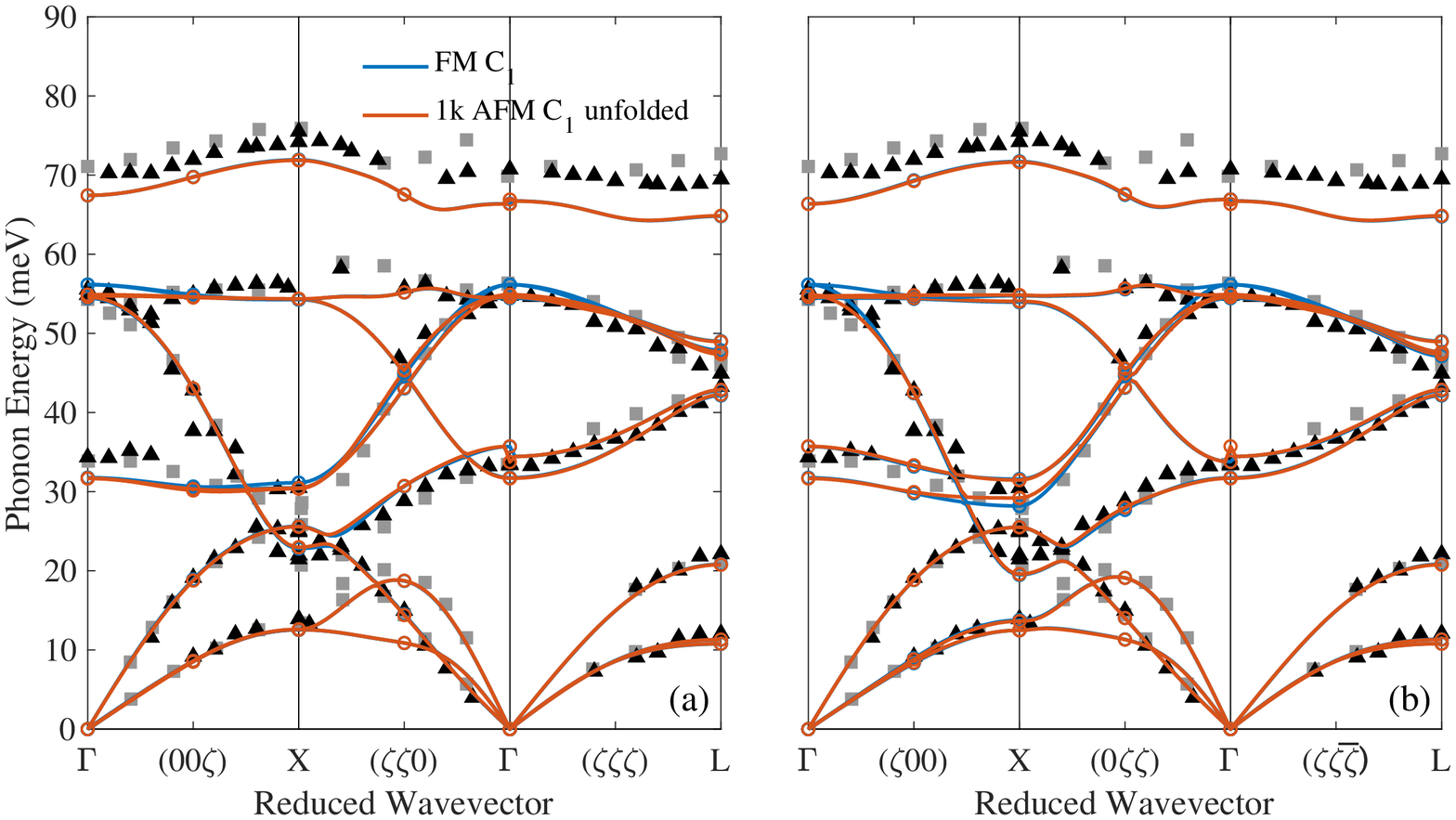}
    \includegraphics[width=0.76\columnwidth]{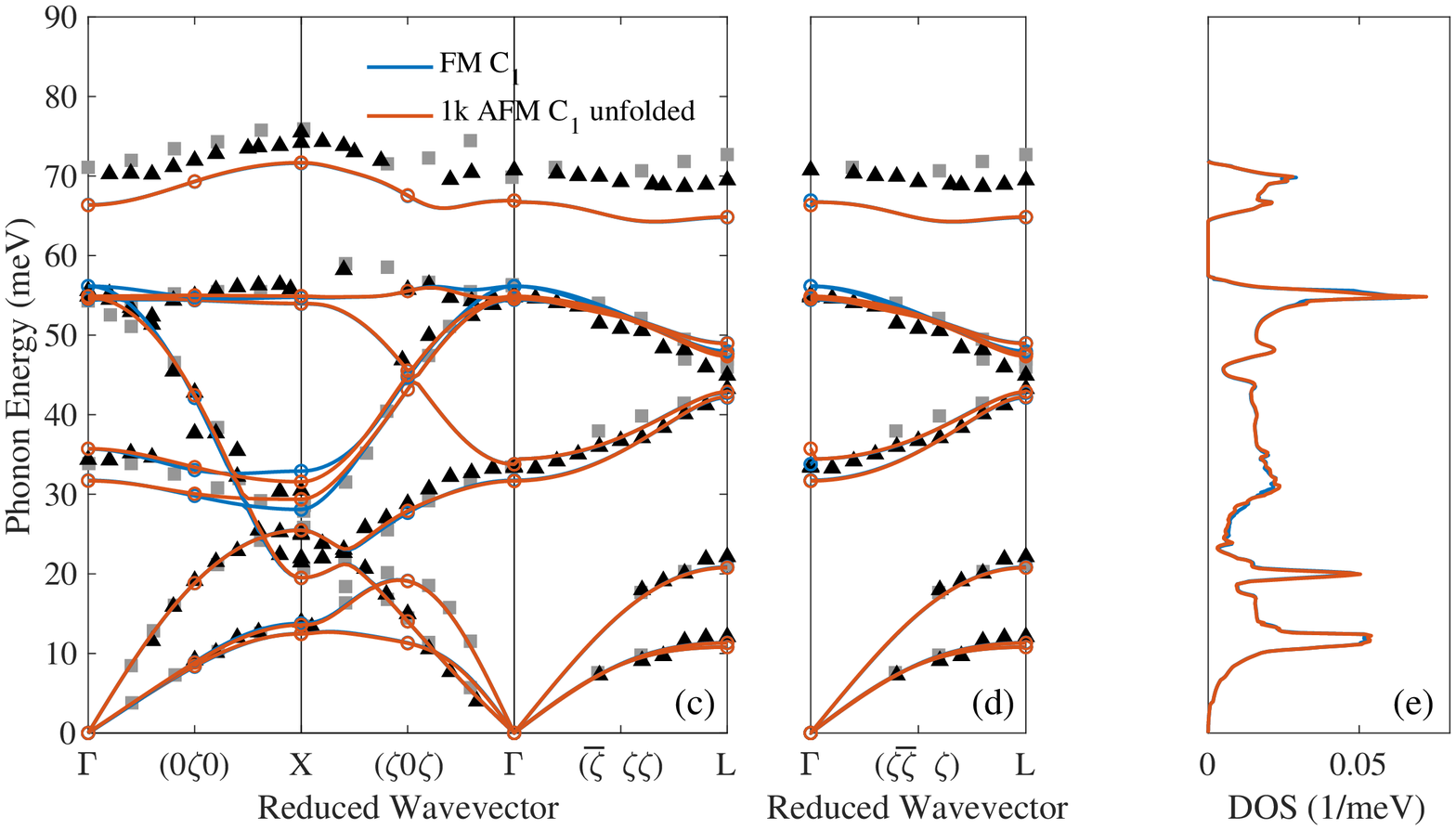}
    \caption{\label{fig:figureSMphononFmDirections} In FM and 1\textbf{k} AFM $\mathbb{S}_0$ structure (a--d) phonon dispersion along different directions and (e) phonon density of states, calculated using GGA+$U$+SOC ($U=4$ eV) with $C_1$ symmetry.}
\end{figure}

In the case of FM $\mathbb{S}_0$ and 1\textbf{k} AFM states, the structural distortions also break the symmetry of the $Fm\bar{3}m$ space group, and phonon calculations are performed using $C_1$ symmetry (and unfolded into $C_1$ symmetry for 1\textbf{k} AFM).  
For $O_h$ symmetry, there are three equivalent high symmetry points $X$ and four
equivalent $L$, which are no longer equivalent in $C_1$ symmetry. We present
the phonon dispersion along the different directions and phonon density of
states in Figure~\ref{fig:figureSMphononFmDirections}. The symmetry breaking in
the FM and 1\textbf{k} AFM states is noticeable in select cases.

\subsection{Elasticity matrix}

In this section, we present the full elasticity matrices, calculated by GGA+$U$+SOC and LDA+$U$+SOC ($U=4$ eV) in FM and AFM $\mathbb{S}_0$ structures using $C_1$ symmetry. The elasticity matrix is given in units of GPa:

$\mathbf{C}=\begin{pmatrix}
C_{11} & C_{12} & C_{13} & C_{14} & C_{15} & C_{16}\\
C_{21} & C_{22} & C_{23} & C_{24} & C_{25} & C_{26}\\
C_{31} & C_{32} & C_{33} & C_{34} & C_{35} & C_{36}\\
C_{41} & C_{42} & C_{43} & C_{44} & C_{45} & C_{46}\\
C_{51} & C_{52} & C_{53} & C_{54} & C_{55} & C_{56}\\
C_{61} & C_{62} & C_{63} & C_{64} & C_{65} & C_{66}
\end{pmatrix}$.

For GGA+$U$, in FM $\mathbb{S}_0$ structure:
$\mathbf{C}=\begin{pmatrix}
    359.68 & 109.73 & 114.63 & -0.03 &  0.05 &  0.28\\
    109.73 & 360.05 & 114.63 & -0.30 &  0.28 &  0.05\\
    114.63 & 114.63 & 355.36 & -0.03 & -0.61 & -0.28\\
     -0.03 &  -0.30 &  -0.03 & 58.21 &  0.03 & -0.30\\
      0.05 &   0.28 &  -0.61 &  0.03 & 62.17 & -0.06\\
      0.28 &   0.05 &  -0.28 & -0.30 & -0.06 & 62.06
\end{pmatrix}$.

For GGA+$U$, in  1\textbf{k} AFM $\mathbb{S}_0$ structure:
$\mathbf{C}=\begin{pmatrix}
    379.95 & 115.51 & 123.18 &  0.00 &  0.00 & -0.07\\
    115.51 & 377.81 & 121.50 &  0.00 & -0.44 &  0.00\\
    123.18 & 121.50 & 381.76 &  0.00 &  0.16 &  0.08\\
      0.00 &   0.00 &   0.00 & 60.90 &  0.00 & -0.88\\
      0.00 &  -0.44 &   0.16 &  0.00 & 60.96 &  0.52\\
     -0.07 &   0.00 &   0.08 & -0.88 &  0.52 & 63.99
\end{pmatrix}$.

For GGA+$U$, in  3\textbf{k} AFM $\mathbb{S}_0$ structure:
$\mathbf{C}=\begin{pmatrix}
    380.39 & 120.49 & 120.00 & 0.27 & -0.47 &  0.63\\
    120.49 & 380.04 & 118.75 & 0.47 & -0.99 & -0.68\\
    120.00 & 118.75 & 380.05 &  0.29 & -0.88 & -0.80\\
     0.27 &  0.47 &  0.29 & 62.61 &  0.00 & -0.30\\
     -0.47 &  -0.99 &  -0.88 &  0.00 & 62.80 & -0.30\\
      0.63 &  -0.68  & -0.80&  -0.30 & -0.30 & 62.48
\end{pmatrix}$.

For LDA+$U$, in FM $\mathbb{S}_0$ structure:
$\mathbf{C}=\begin{pmatrix}
    395.27 & 133.39 & 137.59 &  0.00 &  0.07 & 0.03\\
    133.39 & 395.28 & 137.59 &  0.00 & -0.07 & 0.00\\
    137.59 & 137.59 & 393.88 &  0.00 & -0.01& -0.01\\
      0.00 &   0.00 &   0.00 & 76.65 & -0.02 &-0.01\\
      0.07 &  -0.07 &  -0.01 & -0.02 & 79.69 & 0.16\\
      0.03 &   0.00 &  -0.01 & -0.01 &  0.16& 79.70
\end{pmatrix}$.

For LDA+$U$, in  1\textbf{k} AFM $\mathbb{S}_0$ structure:
$\mathbf{C}=\begin{pmatrix}
    400.22 & 128.15 & 134.14 &  0.00 &  0.12 &  2.24\\
    128.15 & 398.64 & 133.18 &  0.12 & -0.75 &  1.85\\
    134.14 & 133.18 & 397.65 & -0.12  & 0.12 &  1.17\\
      0.00 &   0.12 &  -0.12 & 78.33 &  1.00 & -1.32\\
      0.12 &  -0.75 &   0.12 &  1.00 & 78.66 &  1.42\\
      2.24  &  1.85 &   1.17 & -1.32 &  1.42 & 78.37
\end{pmatrix}$.

For LDA+$U$, in  3\textbf{k} AFM $\mathbb{S}_0$ structure:
$\mathbf{C}=\begin{pmatrix}
    396.72 & 132.14 & 133.14 &  0.84  & 0.87 & -2.02\\
    132.14 & 392.75 & 130.62 &  0.36  & 2.16 & -1.32\\
    133.14 & 130.62 & 393.87 & -0.59  & 0.28 &  0.19\\
      0.84 &   0.36 &  -0.59 & 77.58  & 0.00 &  0.08\\
      0.87 &   2.16 &   0.28 &  0.00  &77.51 &  0.00\\
     -2.02 &  -1.32 &   0.19 &  0.08  & 0.00 & 78.26
\end{pmatrix}$.

\clearpage 

\section{\label{sec:smpatch}VASP patch files implementing OMC}
\vspace{5mm}
\justifying

In this section, we introduce the patch files for modifying the original VASP source code, such that we  can monitor and impose the real portion of the occupation matrices during the calculations. For collinear magnetic calculations (i.e., FM and 1\textbf{k} AFM calculations without SOC), the so-called standard version of VASP is used, and its patch file is given by:

\begin{lstlisting}[language=Fortran,caption={}]
--- LDApU.F	2015-06-29 10:00:32.000000000 -0400
+++ 541_LDApU_dotF	2021-04-07 18:44:30.490196961 -0400
@@ -37,6 +37,12 @@
       REAL(q), PRIVATE, ALLOCATABLE, SAVE :: ORBMOM(:,:,:)
 
       COMPLEX(q), PRIVATE, SAVE :: V(7,7,7,7)
+
+      ! new global variable
+      REAL(q), PUBLIC, ALLOCATABLE, SAVE :: tempdm(:,:)
+      INTEGER, PRIVATE, SAVE :: L0, it
+      LOGICAL, PRIVATE, SAVE :: iffile
+      CHARACTER (24) :: denupfile, dendnfile
         
       CONTAINS
 
@@ -430,6 +436,85 @@
 
+      ! determine the iteration
+      inquire(file='iteration.out',exist=iffile)
+      if (iffile.eq. .false.)then
+        open(65,file="iteration.out")
+        write(65,*) '2'
+        close(65)
+        it=1
+      else
+        open(65,file="iteration.out")
+        read(65,*) it
+        close(65)
+        open(65,file="iteration.out")
+        write(65,*) it+1
+        close(65)
+      endif
+
+
+      ! read in the density matrix if needed
+      L0=LANG_LDAPLUSU(ITYP)
+      if (it <0)then
+        allocate (tempdm(2*L0+1,2*L0+1))
+        108  FORMAT('density_matrix_up.inp.',I1)
+        WRITE(denupfile,108) IATOM
+        109  FORMAT('density_matrix_dn.inp.',I1)
+        WRITE(dendnfile,109) IATOM
+
+        print*,'***Read in up density matrix ***',IATOM
+        inquire(file=denupfile,exist=iffile)
+        if (iffile.eq. .true.)then
+          open(65,file=denupfile)
+          tempdm=0.d0
+          open(65,file=denupfile)
+          do l1=1,(2*L0+1)
+            read(65,*) tempdm(l1,:)
+          enddo
+          close(65)
+          OCC_MAT((2*L0+4):(4*L0+4), (2*L0+4):(4*L0+4)  ,1)=tempdm(:,:)
+        else
+          print*,'File does not exist... cannot read.'
+          !CALL M_exit(); stop 
+        endif
+
+        print*,'***Read in dn density matrix ***',IATOM
+        inquire(file=dendnfile,exist=iffile)
+        if (iffile.eq. .true.)then
+          open(65,file=dendnfile)
+          tempdm=0.d0
+          open(65,file=dendnfile)
+          do l1=1,(2*L0+1)
+            read(65,*) tempdm(l1,:)
+          enddo
+          close(65)
+          OCC_MAT((2*L0+4):(4*L0+4), (2*L0+4):(4*L0+4)  ,2)=tempdm(:,:)
+        else
+          print*,'File does not exist... cannot read.'
+          !CALL M_exit(); stop 
+        endif
+        deallocate (tempdm)
+      endif
+
+
+      ! write out the density matrix
+      103  FORMAT('density_matrix_up.out.',I1)
+      WRITE(denupfile,103) IATOM
+      104  FORMAT('density_matrix_dn.out.',I1)
+      WRITE(dendnfile,104) IATOM
+      open(65,file=denupfile)
+      do l1=(2*L0+4),(4*L0+4)
+        write(65,'(7F12.6)') (real(OCC_MAT(l1,l2,1)),l2=(2*L0+4),(4*L0+4))
+      enddo
+      close(65)
+      open(65,file=dendnfile)
+      do l1=(2*L0+4),(4*L0+4)
+        write(65,'(7F12.6)') (real(OCC_MAT(l1,l2,2)),l2=(2*L0+4),(4*L0+4))
+      enddo
+      close(65)
+
+
+
       OCC_MAT_ALL(:,:,:,IATOM)=OCC_MAT(:,:,:)
       
 !==============================================================
\end{lstlisting}

The input file \emph{iteration.out} determines whether to impose the occupation matrices into a given calculation, as well as how many electronic self-consistency steps are to be imposed. To impose the occupation matrices into the first $N$ electronic self-consistency steps, \emph{iteration.out} only contains a negative number $-N$. The input files \emph{density\_matrix\_up.inp.i} and \emph{density\_matrix\_dn.inp.i} define the imposing occupation matrices of up-spin and down-spin components (i.e., $\occmat^{\uparrow}$ and $\occmat^{\downarrow}$) for the $i$th U atom, respectively. And the output files \emph{density\_matrix\_up.out.i} and \emph{density\_matrix\_dn.out.i} record the converged $\occmat^{\uparrow}$ and $\occmat^{\downarrow}$ for the $i$th U atom, respectively, during the calculation.

For noncollinear magnetic calculations (i.e., 3\textbf{k} AFM calculation without SOC, and all SOC calculations), the noncollinear version of VASP is used. Its patch file is given by:

\begin{lstlisting}[language=Fortran,caption={}]
--- LDApU.F	2015-06-29 10:00:32.000000000 -0400
+++ 541_LDApU_dotF	2021-04-07 18:44:30.490196961 -0400
@@ -37,6 +37,12 @@
       REAL(q), PRIVATE, ALLOCATABLE, SAVE :: ORBMOM(:,:,:)
 
       COMPLEX(q), PRIVATE, SAVE :: V(7,7,7,7)
+
+      ! new global variable
+      REAL(q), PUBLIC, ALLOCATABLE, SAVE :: tempdm(:,:)
+      INTEGER, PRIVATE, SAVE :: L0, it
+      LOGICAL, PRIVATE, SAVE :: iffile
+      CHARACTER (24) :: densp1file, densp2file, densp3file, densp4file
         
       CONTAINS
 
@@ -430,6 +436,85 @@
 
+      ! determine the iteration
+      inquire(file='iteration.out',exist=iffile)
+      if (iffile.eq. .false.)then
+        open(65,file="iteration.out")
+        write(65,*) '2'
+        close(65)
+        it=1
+      else
+        open(65,file="iteration.out")
+        read(65,*) it
+        close(65)
+        open(65,file="iteration.out")
+        write(65,*) it+1
+        close(65)
+      endif
+
+
+      ! read in the density matrix if needed
+      L0=LANG_LDAPLUSU(ITYP)
+      if (it <0)then
+        allocate (tempdm(2*L0+1,2*L0+1))
+        108  FORMAT('density_matrix_sp1.inp.',I1)
+        WRITE(densp1file,108) IATOM
+        109  FORMAT('density_matrix_sp2.inp.',I1)
+        WRITE(densp2file,109) IATOM
+        110  FORMAT('density_matrix_sp3.inp.',I1)
+        WRITE(densp3file,110) IATOM
+        111  FORMAT('density_matrix_sp4.inp.',I1)
+        WRITE(densp4file,111) IATOM
+
+        print*,'***Read in sp1 density matrix ***',IATOM
+        inquire(file=densp1file,exist=iffile)
+        if (iffile.eq. .true.)then
+          open(65,file=densp1file)
+          tempdm=0.d0
+          open(65,file=densp1file)
+          do l1=1,(2*L0+1)
+            read(65,*) tempdm(l1,:)
+          enddo
+          close(65)
+          OCC_MAT((L0**2+1):(L0**2+2*L0+1), (L0**2+1):(L0**2+2*L0+1)  ,1)=tempdm(:,:)
+        else
+          print*,'File does not exist... cannot read.'
+          !CALL M_exit(); stop 
+        endif
+
+        print*,'***Read in sp2 density matrix ***',IATOM
+        inquire(file=densp2file,exist=iffile)
+        if (iffile.eq. .true.)then
+          open(65,file=densp2file)
+          tempdm=0.d0
+          open(65,file=densp2file)
+          do l1=1,(2*L0+1)
+            read(65,*) tempdm(l1,:)
+          enddo
+          close(65)
+          OCC_MAT((L0**2+1):(L0**2+2*L0+1), (L0**2+1):(L0**2+2*L0+1)  ,2)=tempdm(:,:)
+        else
+          print*,'File does not exist... cannot read.'
+          !CALL M_exit(); stop 
+        endif
+
+        print*,'***Read in sp3 density matrix ***',IATOM
+        inquire(file=densp3file,exist=iffile)
+        if (iffile.eq. .true.)then
+          open(65,file=densp3file)
+          tempdm=0.d0
+          open(65,file=densp3file)
+          do l1=1,(2*L0+1)
+            read(65,*) tempdm(l1,:)
+          enddo
+          close(65)
+          OCC_MAT((L0**2+1):(L0**2+2*L0+1), (L0**2+1):(L0**2+2*L0+1)  ,3)=tempdm(:,:)
+        else
+          print*,'File does not exist... cannot read.'
+          !CALL M_exit(); stop 
+        endif
+
+        print*,'***Read in sp4 density matrix ***',IATOM
+        inquire(file=densp4file,exist=iffile)
+        if (iffile.eq. .true.)then
+          open(65,file=densp4file)
+          tempdm=0.d0
+          open(65,file=densp4file)
+          do l1=1,(2*L0+1)
+            read(65,*) tempdm(l1,:)
+          enddo
+          close(65)
+          OCC_MAT((L0**2+1):(L0**2+2*L0+1), (L0**2+1):(L0**2+2*L0+1)  ,4)=tempdm(:,:)
+        else
+          print*,'File does not exist... cannot read.'
+          !CALL M_exit(); stop 
+        endif
+        deallocate (tempdm)
+      endif
+
+
+      ! write out the density matrix
+      103  FORMAT('density_matrix_sp1.out.',I1)
+      WRITE(densp1file,103) IATOM
+      104  FORMAT('density_matrix_sp2.out.',I1)
+      WRITE(densp2file,104) IATOM
+      105  FORMAT('density_matrix_sp3.out.',I1)
+      WRITE(densp3file,105) IATOM
+      106  FORMAT('density_matrix_sp4.out.',I1)
+      WRITE(densp4file,106) IATOM
+      open(65,file=densp1file)
+      do l1=(L0**2+1),(L0**2+2*L0+1)
+        write(65,'(7F12.6)') (real(OCC_MAT(l1,l2,1)),l2=(L0**2+1),(L0**2+2*L0+1))
+      enddo
+      close(65)
+      open(65,file=densp2file)
+      do l1=(L0**2+1),(L0**2+2*L0+1)
+        write(65,'(7F12.6)') (real(OCC_MAT(l1,l2,2)),l2=(L0**2+1),(L0**2+2*L0+1))
+      enddo
+      close(65)
+      open(65,file=densp3file)
+      do l1=(L0**2+1),(L0**2+2*L0+1)
+        write(65,'(7F12.6)') (real(OCC_MAT(l1,l2,3)),l2=(L0**2+1),(L0**2+2*L0+1))
+      enddo
+      close(65)
+      open(65,file=densp4file)
+      do l1=(L0**2+1),(L0**2+2*L0+1)
+        write(65,'(7F12.6)') (real(OCC_MAT(l1,l2,4)),l2=(L0**2+1),(L0**2+2*L0+1))
+      enddo
+      close(65)
+
+
+
       OCC_MAT_ALL(:,:,:,IATOM)=OCC_MAT(:,:,:)
       
 !==============================================================
\end{lstlisting}

Compared to the standard version, the only difference in the noncollinear version is that the up-spin and down-spin components are replaced by four spinors: denoted as $sp1$, $sp2$, $sp3$, and $sp4$, corresponding to $\occmat^{\uparrow\uparrow}$, $\occmat^{\uparrow\downarrow}$, $\occmat^{\downarrow\uparrow}$, and $\occmat^{\downarrow\downarrow}$, respectively.
\clearpage

\section{Time-of-flight inelastic neutron scattering measurement on UO$_2$}
Inelastic neutron-scattering (INS) measurements of the
phonon dispersion for UO$_2$ at 600 K and PDOS for UO$_2$ at 77 K were made on the wide Angular Range Chopper Spectrometer (ARCS) at the Spallation Neutron Source \cite{abernathyDesignOperationWide2012}. An incident neutron energy of $E_i=120$ meV was used for both measurements, which is high enough to capture the phonon cutoff at around 80 meV and allow for summing over enough zones in momentum space to obtain a PDOS. Other setup and data processing details can be found elsewhere \cite{pangPhononDensityStates2014,bryan_impact_2019} and are not repeated here. 

This section gives the measured phonon frequency data at 600 K in the $(\zeta,0,0)$ (Table~\ref{tab:tablesphonon100}), $(\zeta,\zeta,0)$ (Table~\ref{tab:tablesphonon110}), and $(\zeta,\zeta,\zeta)$ directions (Table~\ref{tab:tablesphonon111}), as well as the density of state data at 77 K (Table~\ref{tab:tablesphonondos}).

\begin{table}[h]
\caption{\label{tab:tablesphonon100}
Measured UO$_2$ phonon frequency in the $(\zeta,0,0)$ direction at 600 K.
}

\begin{tabular}{lccllcc}
\hline\hline
$\zeta$           & frequency (meV)  & error (meV)       &  & $\zeta$           & frequency (meV)  & error (meV)       \\
\colrule
TA:               &                  &                   &  & LA:               &                  &                   \\
\colrule
0.5 & 9.1671 & 0.3121 &  & 0.3 & 11.5772 & 0.7505 \\
0.6 & 10.1261 & 0.2734 &  & 0.4 & 15.8962 & 0.6283 \\
0.7 & 11.9895 & 0.4727 &  & 0.5 & 19.1363 & 0.7682 \\
0.8 & 12.6863 & 0.6755 &  & 0.6 & 21.5178 & 0.3640 \\
1.0               & 13.9307 & 0.7945 &  & 0.7 & 22.8865 & 1.1725  \\
                  &                  &                   &  & 0.9 & 25.2697 & 0.7814 \\
                  &                  &                   &  & 1.0               & 24.9141 & 0.7045 \\
\\
TO1:              &                  &                   &  & LO1:              &                  &                   \\
\colrule
0                 & 34.3567 & 0.2223 &  & 0                 & 55.5753 & 0.3591 \\
0.1 & 34.2838 & 0.2991 &  & 0.1 & 55.1916 & 0.3528 \\
0.2 & 35.1755 & 0.3689 &  & 0.2 & 52.9257 & 0.3876 \\
0.3 & 34.6423 & 0.6637 &  & 0.3 & 51.3413 & 0.4171 \\
0.5 & 37.6713 & 0.6039 &  & 0.4 & 45.4522 & 0.5993 \\
0.7 & 35.4691 & 1.2993 &  & 0.5 & 42.8027 & 0.4614 \\
0.9 & 30.3478 & 0.2273 &  & 0.6 & 37.6789 & 0.5295 \\
1.0                 & 30.4801 & 1.2655  &  & 0.7 & 32.2373 & 0.4363 \\
                  &                  &                   &  & 0.8 & 25.4852 & 0.2879 \\
                  &                  &                   &  & 0.9 & 22.4016 & 0.1990 \\
                  &                  &                   &  & 1.0                 & 21.4728 & 0.2198 \\
                  \\
TO2:              &                  &                   &  & LO2:              &                  &                   \\
\colrule
0                 & 54.8792 & 0.2334 &  & 0.1 & 70.2667 & 0.8892 \\
0.1 & 54.4859 & 0.2135 &  & 0.2 & 70.3262 & 0.4290 \\
0.2 & 53.4574 & 0.2696 &  & 0.3 & 70.2251 & 0.6242 \\
0.3 & 52.3708 & 0.2729 &  & 0.4 & 71.1486 & 0.4630 \\
0.4 & 54.3544 & 1.3971 &  & 0.5 & 71.9566 & 0.3386 \\
0.5 & 54.9363 & 1.0158 &  & 0.6 & 72.8155 & 0.2275 \\
0.6 & 55.6976 & 0.5017 &  & 0.8 & 73.4891 & 0.3567 \\
0.7 & 56.0451 & 0.5137 &  & 0.8 & 73.6372 & 0.4617 \\
0.8 & 56.2867 & 0.5647 &  & 0.9 & 73.7968 & 0.3709 \\
0.9 & 56.3545 & 0.8727 &  & 1.0                 & 74.2015 & 0.6906 \\
1.0                 & 55.8099 & 0.3545 &  &                   &                  &                   \\
\hline\hline
\end{tabular}

\end{table}

\begin{table}[h]
\caption{\label{tab:tablesphonon110}
Measured UO$_2$ phonon frequency in the $(\zeta,\zeta,0)$ direction at 600 K.
}

\begin{tabular}{lccllcc}
\hline\hline
$\zeta$           & frequency (meV)  & error (meV)       &  & $\zeta$           & frequency (meV)  & error (meV)       \\
\colrule
TA:               &                  &                   &  &  LO1:            &                  &                   \\
\colrule
0.95 & 13.3000 & 0.4211 &  &     0 & 55.0614 &  0.3045   \\
 &  &  &  &     0.1 & 53.8267 &  1.1137   \\
  &  &  &  &     0.2 & 52.4418 &  0.9746   \\
   &  &  &  &     0.4 & 49.9379 &  0.9051   \\
    &  &  &  &     0.525 & 46.8626 &  1.0193   \\
                  &                  &                   &  &                   &                  &                   \\
LA:               &                  &                   &  & TO1:              &                  &                   \\
\colrule
0.175 & 3.9939 & 0.3702 &  & 0                 & 33.5350 & 0.2791 \\
0.3 & 7.6936 & 0.2138 &  & 0.1 & 33.2261 & 0.2224 \\
0.4 & 10.5601 & 0.1884 &  & 0.2 & 32.7013 & 0.2277 \\
0.5 & 14.9738 & 0.4602 &  & 0.3 & 32.2231 & 0.2023 \\
0.6 & 17.4222 & 0.4405 &  & 0.4 & 30.6529 & 0.2644 \\
0.7 & 20.6661 & 0.3133 &  & 0.5 & 28.8257 & 0.2121 \\
0.8 & 23.0523 & 0.4391 &  & 0.6 & 27.0225 & 0.2953 \\
0.9 & 23.7848 & 0.5687 &  & 0.7 & 25.7884 & 0.2920 \\
1                 & 25.1044 & 0.4624 &  & 0.8 & 22.6643 & 0.5068 \\
                  &                  &                   &  & 0.9 & 21.9699 & 0.3265 \\
                  &                  &                   &  & 1                 & 21.9783 & 0.4089 \\
                  &                  &                   &  &                   &                  &                   \\
TO2:              &                  &                   &  & LO2:              &                  &                   \\
\colrule
0                 & 54.8975 & 0.5170 &  & 0                 & 70.7392 & 0.8429 \\
0.1 & 54.7225 & 0.6134 &  & 0.2 & 70.4241 & 0.5566 \\
0.2 & 54.2925 & 0.8693 &  & 0.3 & 69.5506 & 0.7862 \\
0.3 & 54.7220 & 0.6406 &  & 0.625 & 71.9323 & 0.9053 \\
0.4 & 56.3601 & 1.4859  &  & 0.75 & 73.0211 & 0.8919 \\
0.5 & 55.7216 & 1.2747  &  & 0.8 & 73.7998 & 0.4012 \\
0.8 & 58.2236 & 0.6034 &  & 0.9 & 74.3109 & 0.5391 \\
                  &                  &                   &  & 1                 & 75.5080 & 0.5152 \\
\hline\hline
\end{tabular}

\end{table}

\begin{table}[h]
\caption{\label{tab:tablesphonon111}
Measured UO$_2$ phonon frequency in the $(\zeta,\zeta,\zeta)$ direction at 600 K.}

\begin{tabular}{lccllcc}
\hline\hline
$\zeta$           & frequency (meV)  & error (meV)       &  & $\zeta$           & frequency (meV)  & error (meV)       \\
\colrule
TA:               &                  &                   &  & LA:               &                  &                   \\
\colrule
0.3  & 9.0804 & 0.3228 &  & 0.2  & 7.2788 & 0.0375 \\
0.35  & 9.7025 & 0.3876 &  & 0.3  & 17.9888 & 0.6133  \\
0.4  & 11.5468 & 0.2706 &  & 0.35  & 19.1084 & 0.4392  \\
0.45  & 11.9859 & 0.3181 &  & 0.4  & 20.0660 & 0.3748  \\
0.5  & 12.0910 & 0.6167 &  & 0.45  & 21.8306 & 1.1388   \\
                   &                  &                   &  & 0.5  & 22.1334 & 0.4236  \\
                   &                  &                   &  &                    &                  &                    \\
TO1:               &                  &                   &  & LO1:               &                  &                    \\
\colrule
0                  & 33.3759 & 0.1666 &  & 0                  & 54.7776 & 0.2539  \\
0.05 & 33.2763 & 0.1975 &  & 0.05 & 54.6347 & 0.3000  \\
0.1  & 34.1746 & 0.1708 &  & 0.1  & 54.0615 & 0.4509  \\
0.15  & 35.0347 & 0.1879 &  & 0.15  & 53.6299 & 0.2052  \\
0.2  & 35.9997 & 0.2336 &  & 0.2  & 51.4734 & 0.5552  \\
0.25  & 36.7348 & 0.2668 &  & 0.25  & 50.8322 & 0.2953  \\
0.35  & 38.3816 & 0.3368 &  & 0.3  & 50.5464 & 0.3554  \\
0.4  & 40.0969 & 0.5767 &  & 0.35  & 48.3810 & 0.5290  \\
0.45  & 41.2284 & 0.5288 &  & 0.4  & 48.1316 & 0.3977  \\
0.5  & 43.2628 & 0.6820 &  & 0.45  & 45.9698 & 0.5279  \\
                   &                  &                   &  & 0.5  & 44.9379 & 0.4690  \\
                   &                  &                   &  &                    &                  &                    \\
LO2:               &                  &                   &  &                    &                  &                    \\
\colrule
0.1  & 70.3385 & 0.6441 &  &                    &                  &                    \\
0.15  & 70.0344 & 0.6350 &  &                    &                  &                    \\
0.2  & 69.9371 & 0.7446 &  &                    &                  &                    \\
0.25  & 69.2741 & 0.7689 &  &                    &                  &                    \\
0.3250  & 68.9561 & 0.5485 &  &                    &                  &                    \\
0.35  & 68.8439 & 0.5456 &  &                    &                  &                    \\
0.4  & 68.6587 & 0.5626 &  &                    &                  &                    \\
0.45  & 68.9236 & 0.3539 &  &                    &                  &                    \\
0.5  & 69.4832 & 0.4330 &  &                    &                  &                    \\
\hline\hline
\end{tabular}

\end{table}
\clearpage

\begin{longtable}{llllllllll}
\caption{Measured UO$_2$ phonon density of states at 77 K.}\\
\label{tab:tablesphonondos}\\
\hline\hline

$\omega$ (meV) & DOS (1/meV) & $\omega$ (meV) & DOS (1/meV) & $\omega$ (meV) & DOS (1/meV) & $\omega$ (meV) & DOS (1/meV) & $\omega$ (meV) & DOS (1/meV) \\
\hline
0.25           & 0.0000      & 20.25          & 0.0184      & 40.25          & 0.0141      & 60.25          & 0.0092      & 80.25          & 0.0018      \\
0.5            & 0.0001      & 20.5           & 0.0185      & 40.5           & 0.0144      & 60.5           & 0.0073      & 80.5           & 0.0016      \\
0.75           & 0.0001      & 20.75          & 0.0189      & 40.75          & 0.0133      & 60.75          & 0.0062      & 80.75          & 0.0015      \\
1              & 0.0002      & 21             & 0.0182      & 41             & 0.0135      & 61             & 0.0060      & 81             & 0.0015      \\
1.25           & 0.0003      & 21.25          & 0.0201      & 41.25          & 0.0138      & 61.25          & 0.0046      & 81.25          & 0.0015      \\
1.5            & 0.0005      & 21.5           & 0.0207      & 41.5           & 0.0140      & 61.5           & 0.0044      & 81.5           & 0.0023      \\
1.75           & 0.0006      & 21.75          & 0.0183      & 41.75          & 0.0142      & 61.75          & 0.0042      & 81.75          & 0.0012      \\
2              & 0.0008      & 22             & 0.0185      & 42             & 0.0147      & 62             & 0.0021      & 82             & 0.0011      \\
2.25           & 0.0011      & 22.25          & 0.0193      & 42.25          & 0.0135      & 62.25          & 0.0028      & 82.25          & 0.0016      \\
2.5            & 0.0013      & 22.5           & 0.0184      & 42.5           & 0.0131      & 62.5           & 0.0033      & 82.5           & 0.0016      \\
2.75           & 0.0016      & 22.75          & 0.0200      & 42.75          & 0.0139      & 62.75          & 0.0028      & 82.75          & 0.0017      \\
3              & 0.0019      & 23             & 0.0223      & 43             & 0.0154      & 63             & 0.0029      & 83             & 0.0016      \\
3.25           & 0.0022      & 23.25          & 0.0211      & 43.25          & 0.0133      & 63.25          & 0.0034      & 83.25          & 0.0017      \\
3.5            & 0.0025      & 23.5           & 0.0184      & 43.5           & 0.0142      & 63.5           & 0.0029      & 83.5           & 0.0010      \\
3.75           & 0.0029      & 23.75          & 0.0212      & 43.75          & 0.0141      & 63.75          & 0.0028      & 83.75          & 0.0014      \\
4              & 0.0033      & 24             & 0.0223      & 44             & 0.0141      & 64             & 0.0028      & 84             & 0.0009      \\
4.25           & 0.0037      & 24.25          & 0.0224      & 44.25          & 0.0133      & 64.25          & 0.0029      & 84.25          & 0.0021      \\
4.5            & 0.0042      & 24.5           & 0.0219      & 44.5           & 0.0136      & 64.5           & 0.0029      & 84.5           & 0.0013      \\
4.75           & 0.0047      & 24.75          & 0.0215      & 44.75          & 0.0134      & 64.75          & 0.0029      & 84.75          & 0.0015      \\
5              & 0.0052      & 25             & 0.0233      & 45             & 0.0139      & 65             & 0.0033      & 85             & 0.0018      \\
5.25           & 0.0057      & 25.25          & 0.0064      & 45.25          & 0.0146      & 65.25          & 0.0032      & 85.25          & 0.0020      \\
5.5            & 0.0063      & 25.5           & 0.0073      & 45.5           & 0.0136      & 65.5           & 0.0025      & 85.5           & 0.0018      \\
5.75           & 0.0069      & 25.75          & 0.0070      & 45.75          & 0.0140      & 65.75          & 0.0040      & 85.75          & 0.0017      \\
6              & 0.0075      & 26             & 0.0074      & 46             & 0.0133      & 66             & 0.0032      & 86             & 0.0020      \\
6.25           & 0.0081      & 26.25          & 0.0074      & 46.25          & 0.0136      & 66.25          & 0.0038      & 86.25          & 0.0013      \\
6.5            & 0.0088      & 26.5           & 0.0079      & 46.5           & 0.0141      & 66.5           & 0.0039      & 86.5           & 0.0009      \\
6.75           & 0.0091      & 26.75          & 0.0076      & 46.75          & 0.0132      & 66.75          & 0.0040      & 86.75          & 0.0010      \\
7              & 0.0091      & 27             & 0.0086      & 47             & 0.0131      & 67             & 0.0036      & 87             & 0.0008      \\
7.25           & 0.0092      & 27.25          & 0.0092      & 47.25          & 0.0143      & 67.25          & 0.0044      & 87.25          & 0.0008      \\
7.5            & 0.0099      & 27.5           & 0.0092      & 47.5           & 0.0136      & 67.5           & 0.0049      & 87.5           & 0.0008      \\
7.75           & 0.0096      & 27.75          & 0.0089      & 47.75          & 0.0131      & 67.75          & 0.0056      & 87.75          & 0.0012      \\
8              & 0.0101      & 28             & 0.0093      & 48             & 0.0132      & 68             & 0.0066      & 88             & 0.0010      \\
8.25           & 0.0104      & 28.25          & 0.0106      & 48.25          & 0.0133      & 68.25          & 0.0065      & 88.25          & 0.0014      \\
8.5            & 0.0107      & 28.5           & 0.0111      & 48.5           & 0.0131      & 68.5           & 0.0071      & 88.5           & 0.0014      \\
8.75           & 0.0112      & 28.75          & 0.0104      & 48.75          & 0.0136      & 68.75          & 0.0075      & 88.75          & 0.0016      \\
9              & 0.0114      & 29             & 0.0123      & 49             & 0.0138      & 69             & 0.0087      & 89             & 0.0009      \\
9.25           & 0.0126      & 29.25          & 0.0123      & 49.25          & 0.0142      & 69.25          & 0.0083      & 89.25          & 0.0004      \\
9.5            & 0.0127      & 29.5           & 0.0121      & 49.5           & 0.0146      & 69.5           & 0.0104      & 89.5           & 0.0011      \\
9.75           & 0.0136      & 29.75          & 0.0133      & 49.75          & 0.0139      & 69.75          & 0.0094      & 89.75          & 0.0012      \\
10             & 0.0141      & 30             & 0.0126      & 50             & 0.0138      & 70             & 0.0104      & 90             & 0.0011      \\
10.25          & 0.0150      & 30.25          & 0.0134      & 50.25          & 0.0143      & 70.25          & 0.0122      & 90.25          & 0.0015      \\
10.5           & 0.0153      & 30.5           & 0.0133      & 50.5           & 0.0148      & 70.5           & 0.0130      & 90.5           & 0.0012      \\
10.75          & 0.0159      & 30.75          & 0.0143      & 50.75          & 0.0143      & 70.75          & 0.0123      & 90.75          & 0.0003      \\
11             & 0.0176      & 31             & 0.0141      & 51             & 0.0150      & 71             & 0.0126      & 91             & 0.0011      \\
11.25          & 0.0175      & 31.25          & 0.0150      & 51.25          & 0.0157      & 71.25          & 0.0133      & 91.25          & 0.0015      \\
11.5           & 0.0176      & 31.5           & 0.0144      & 51.5           & 0.0153      & 71.5           & 0.0134      & 91.5           & 0.0013      \\
11.75          & 0.0203      & 31.75          & 0.0158      & 51.75          & 0.0156      & 71.75          & 0.0136      & 91.75          & 0.0013      \\
12             & 0.0191      & 32             & 0.0160      & 52             & 0.0152      & 72             & 0.0137      & 92             & 0.0008      \\
12.25          & 0.0210      & 32.25          & 0.0148      & 52.25          & 0.0157      & 72.25          & 0.0138      & 92.25          & 0.0013      \\
12.5           & 0.0221      & 32.5           & 0.0163      & 52.5           & 0.0170      & 72.5           & 0.0150      & 92.5           & 0.0003      \\
12.75          & 0.0223      & 32.75          & 0.0152      & 52.75          & 0.0178      & 72.75          & 0.0136      & 92.75          & 0.0003      \\
13             & 0.0227      & 33             & 0.0146      & 53             & 0.0169      & 73             & 0.0135      & 93             & 0.0002      \\
13.25          & 0.0214      & 33.25          & 0.0154      & 53.25          & 0.0175      & 73.25          & 0.0135      & 93.25          & 0.0009      \\
13.5           & 0.0223      & 33.5           & 0.0154      & 53.5           & 0.0179      & 73.5           & 0.0139      & 93.5           & 0.0006      \\
13.75          & 0.0228      & 33.75          & 0.0162      & 53.75          & 0.0187      & 73.75          & 0.0128      & 93.75          & 0.0011      \\
14             & 0.0217      & 34             & 0.0158      & 54             & 0.0188      & 74             & 0.0110      & 94             & 0.0004      \\
14.25          & 0.0229      & 34.25          & 0.0147      & 54.25          & 0.0179      & 74.25          & 0.0116      & 94.25          & 0.0008      \\
14.5           & 0.0210      & 34.5           & 0.0160      & 54.5           & 0.0189      & 74.5           & 0.0107      & 94.5           & 0.0002      \\
14.75          & 0.0215      & 34.75          & 0.0159      & 54.75          & 0.0198      & 74.75          & 0.0098      & 94.75          & 0.0008      \\
15             & 0.0200      & 35             & 0.0158      & 55             & 0.0202      & 75             & 0.0083      & 95             & 0.0006      \\
15.25          & 0.0188      & 35.25          & 0.0157      & 55.25          & 0.0207      & 75.25          & 0.0080      & 95.25          & 0.0007      \\
15.5           & 0.0181      & 35.5           & 0.0161      & 55.5           & 0.0203      & 75.5           & 0.0065      & 95.5           & 0.0002      \\
15.75          & 0.0171      & 35.75          & 0.0157      & 55.75          & 0.0209      & 75.75          & 0.0066      & 95.75          & 0.0000      \\
16             & 0.0146      & 36             & 0.0154      & 56             & 0.0214      & 76             & 0.0049      & 96             & 0.0009      \\
16.25          & 0.0156      & 36.25          & 0.0152      & 56.25          & 0.0217      & 76.25          & 0.0044      & 96.25          & 0.0003      \\
16.5           & 0.0141      & 36.5           & 0.0157      & 56.5           & 0.0208      & 76.5           & 0.0034      & 96.5           & 0.0010      \\
16.75          & 0.0148      & 36.75          & 0.0157      & 56.75          & 0.0227      & 76.75          & 0.0042      & 96.75          & 0.0005      \\
17             & 0.0139      & 37             & 0.0150      & 57             & 0.0233      & 77             & 0.0031      & 97             & 0.0004      \\
17.25          & 0.0148      & 37.25          & 0.0150      & 57.25          & 0.0226      & 77.25          & 0.0031      & 97.25          & 0.0006      \\
17.5           & 0.0139      & 37.5           & 0.0152      & 57.5           & 0.0199      & 77.5           & 0.0024      & 97.5           & 0.0006      \\
17.75          & 0.0135      & 37.75          & 0.0150      & 57.75          & 0.0212      & 77.75          & 0.0023      & 97.75          & 0.0006      \\
18             & 0.0138      & 38             & 0.0147      & 58             & 0.0209      & 78             & 0.0024      & 98             & 0.0007      \\
18.25          & 0.0152      & 38.25          & 0.0141      & 58.25          & 0.0199      & 78.25          & 0.0023      & 98.25          & 0.0004      \\
18.5           & 0.0129      & 38.5           & 0.0156      & 58.5           & 0.0186      & 78.5           & 0.0024      & 98.5           & 0.0004      \\
18.75          & 0.0143      & 38.75          & 0.0138      & 58.75          & 0.0176      & 78.75          & 0.0015      & 98.75          & 0.0007      \\
19             & 0.0155      & 39             & 0.0139      & 59             & 0.0147      & 79             & 0.0015      & 99             & 0.0000      \\
19.25          & 0.0144      & 39.25          & 0.0142      & 59.25          & 0.0142      & 79.25          & 0.0021      & 99.25          & 0.0006      \\
19.5           & 0.0172      & 39.5           & 0.0145      & 59.5           & 0.0128      & 79.5           & 0.0024      & 99.5           & 0.0015      \\
19.75          & 0.0164      & 39.75          & 0.0137      & 59.75          & 0.0113      & 79.75          & 0.0018      & 99.75          & 0.0012      \\
20             & 0.0175      & 40             & 0.0143      & 60             & 0.0093      & 80             & 0.0018      & 100            & 0.0014           \\
\hline\hline

\end{longtable}
 
\clearpage 

\section{Electronic density of states}

\begin{figure}[h]
    \centering
    \includegraphics[width=0.30\columnwidth]{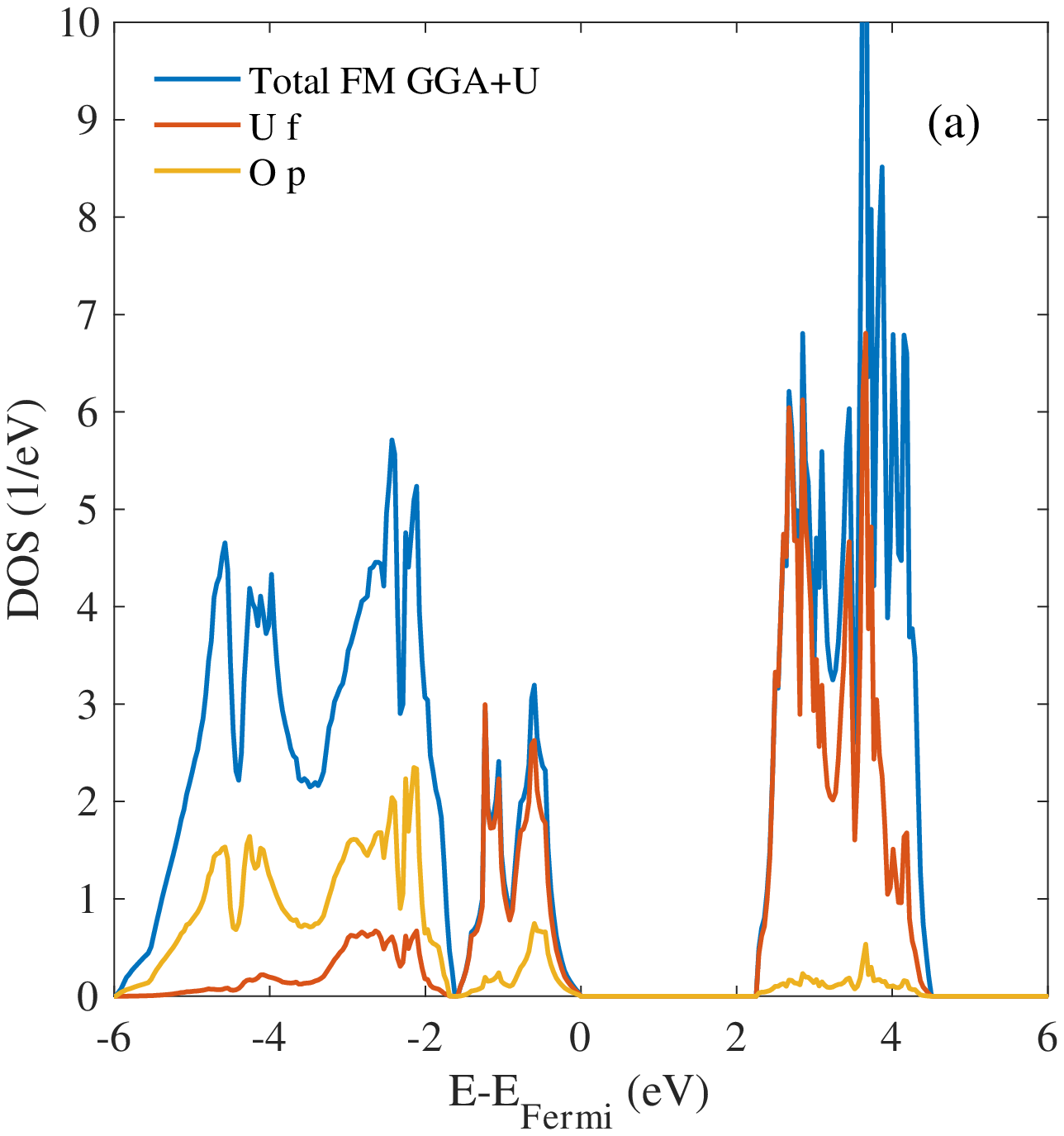}
    \includegraphics[width=0.30\columnwidth]{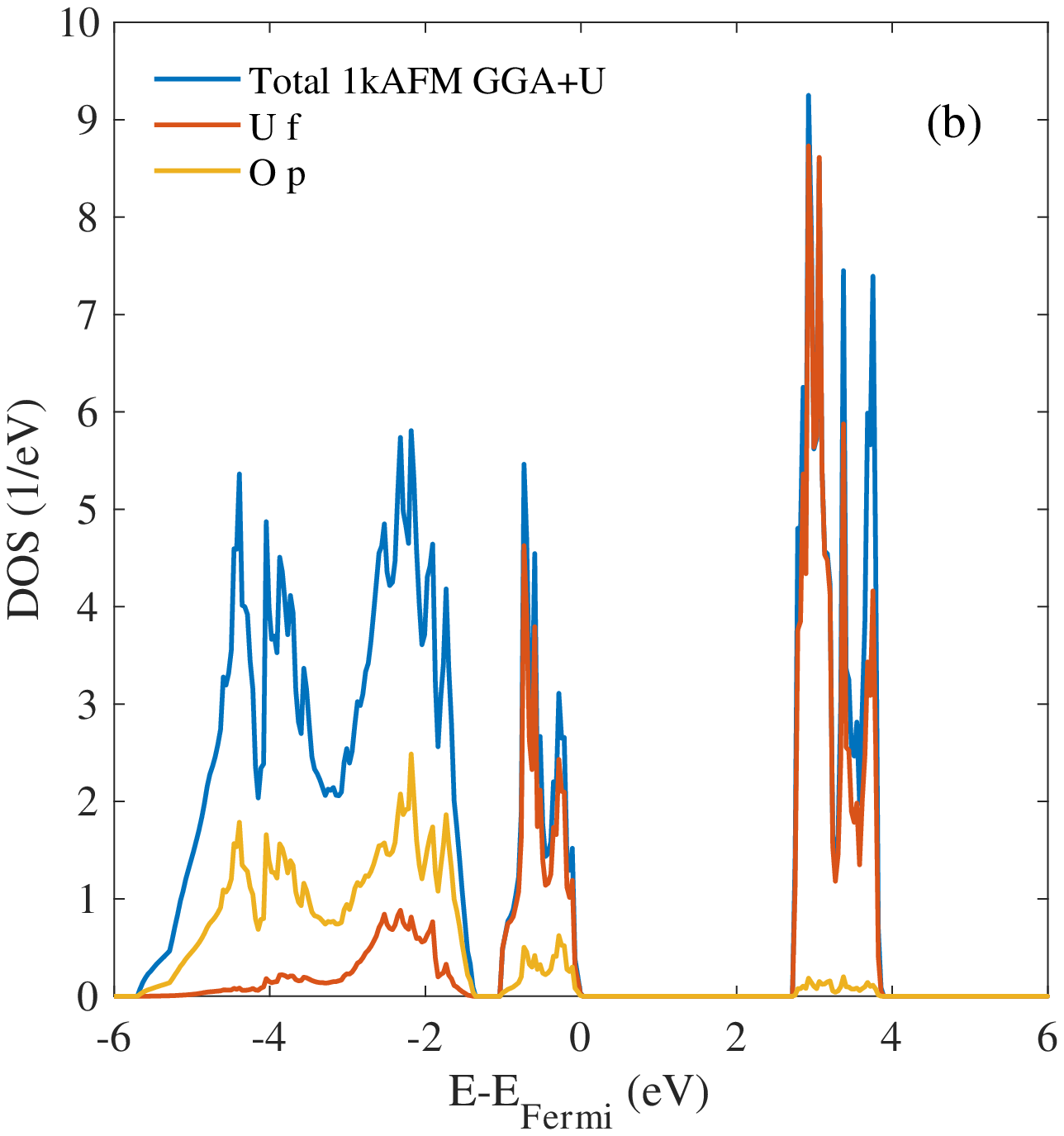}
    \includegraphics[width=0.30\columnwidth]{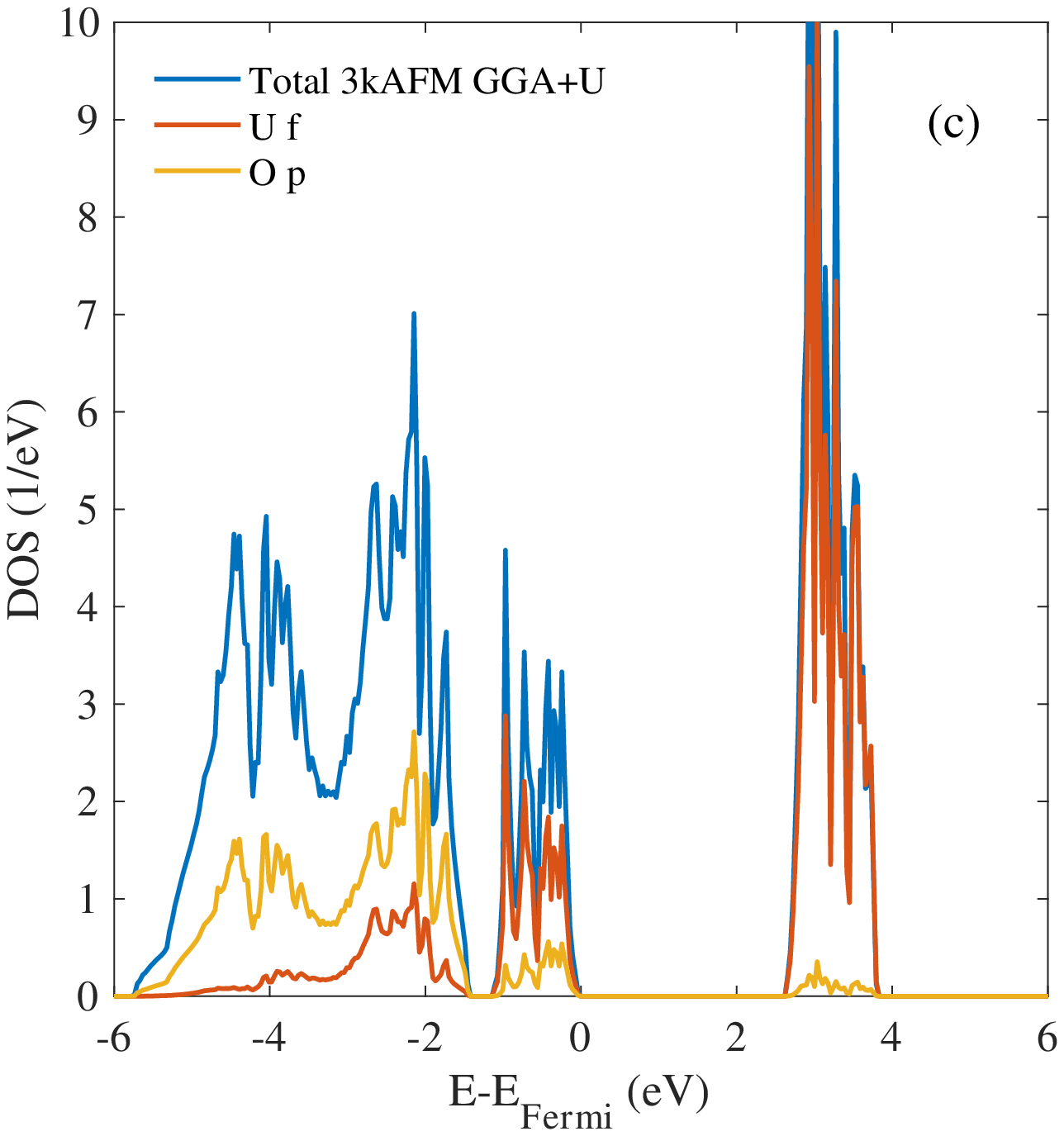}
    \includegraphics[width=0.30\columnwidth]{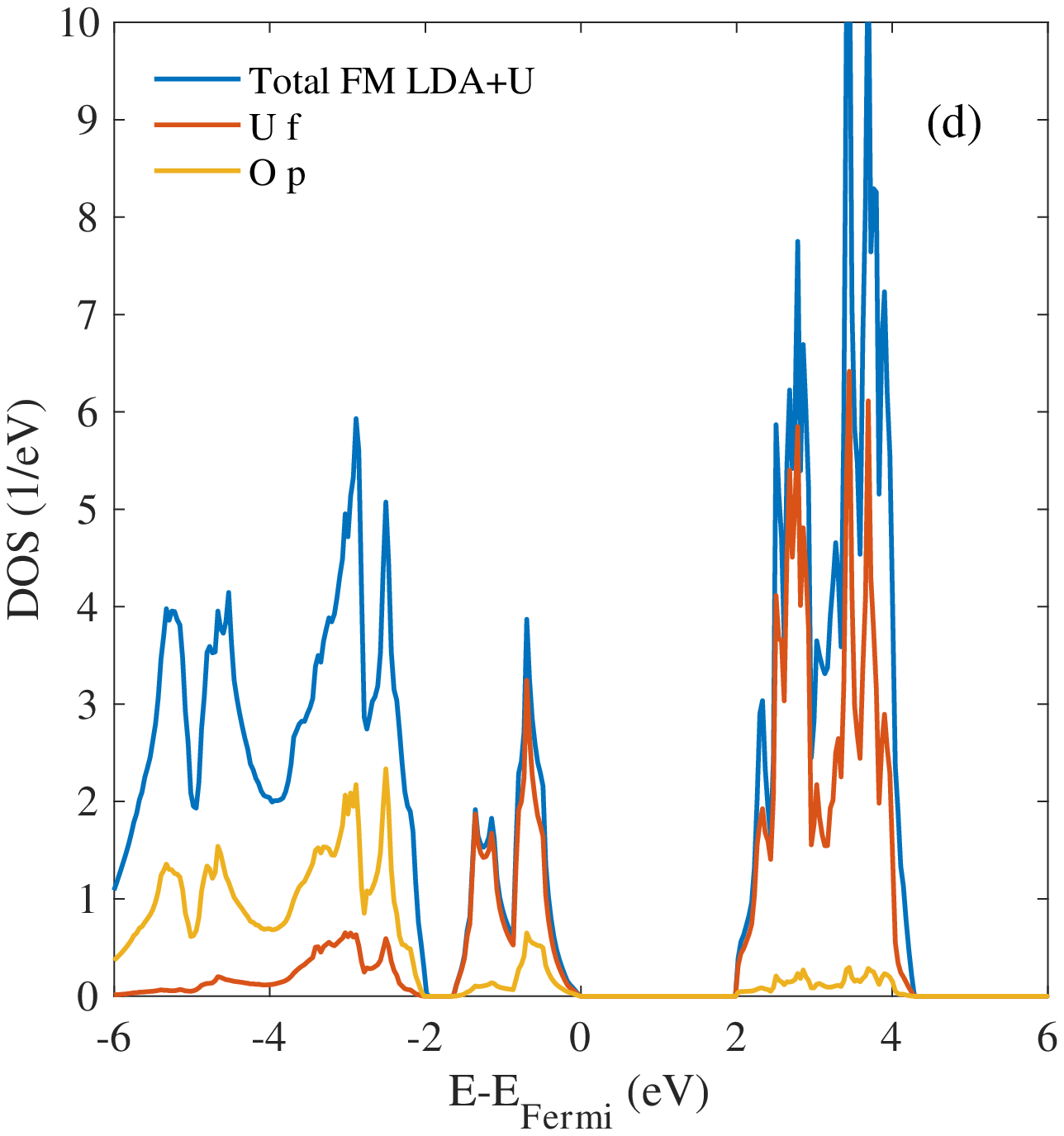}
    \includegraphics[width=0.30\columnwidth]{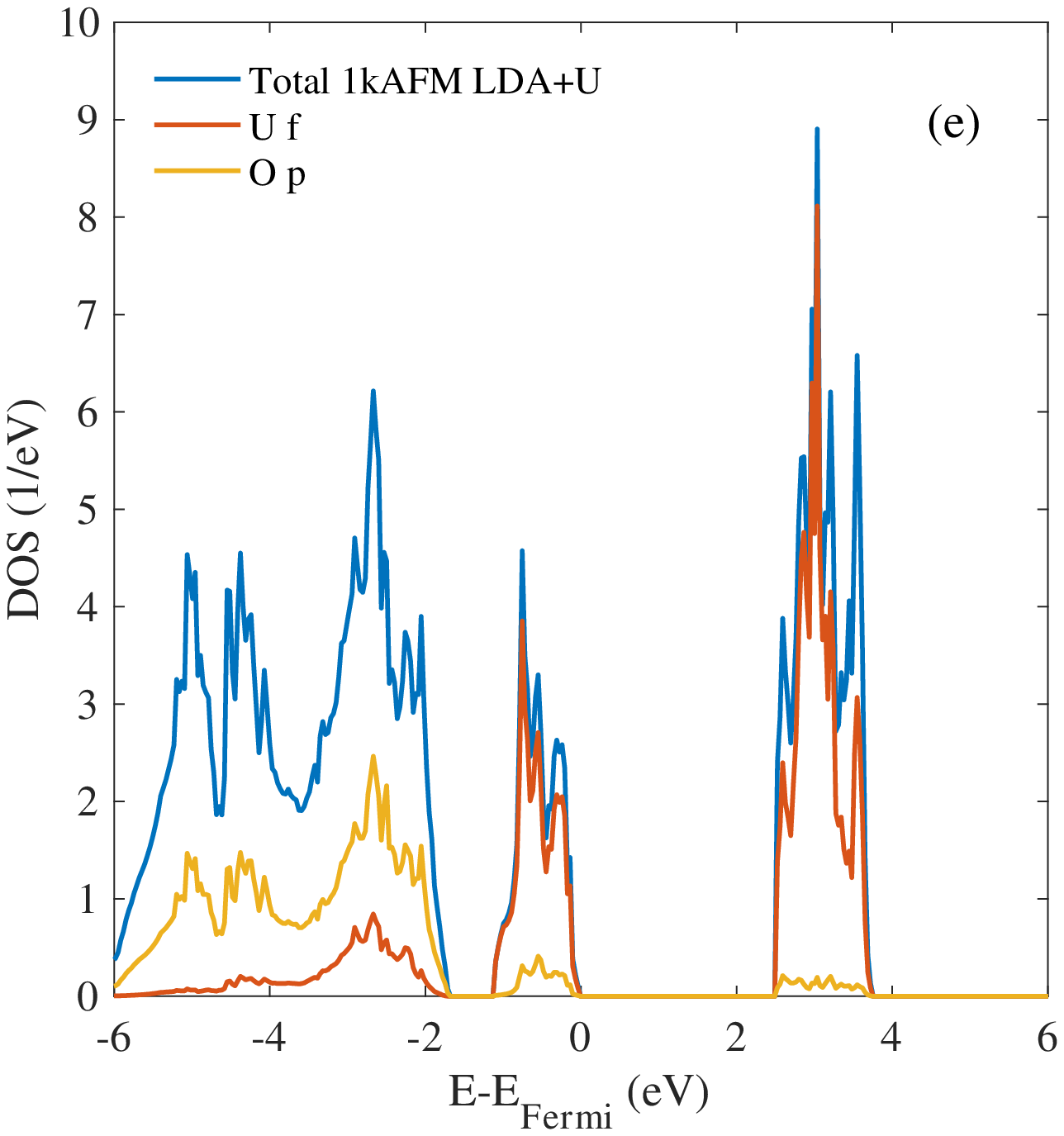}
    \includegraphics[width=0.30\columnwidth]{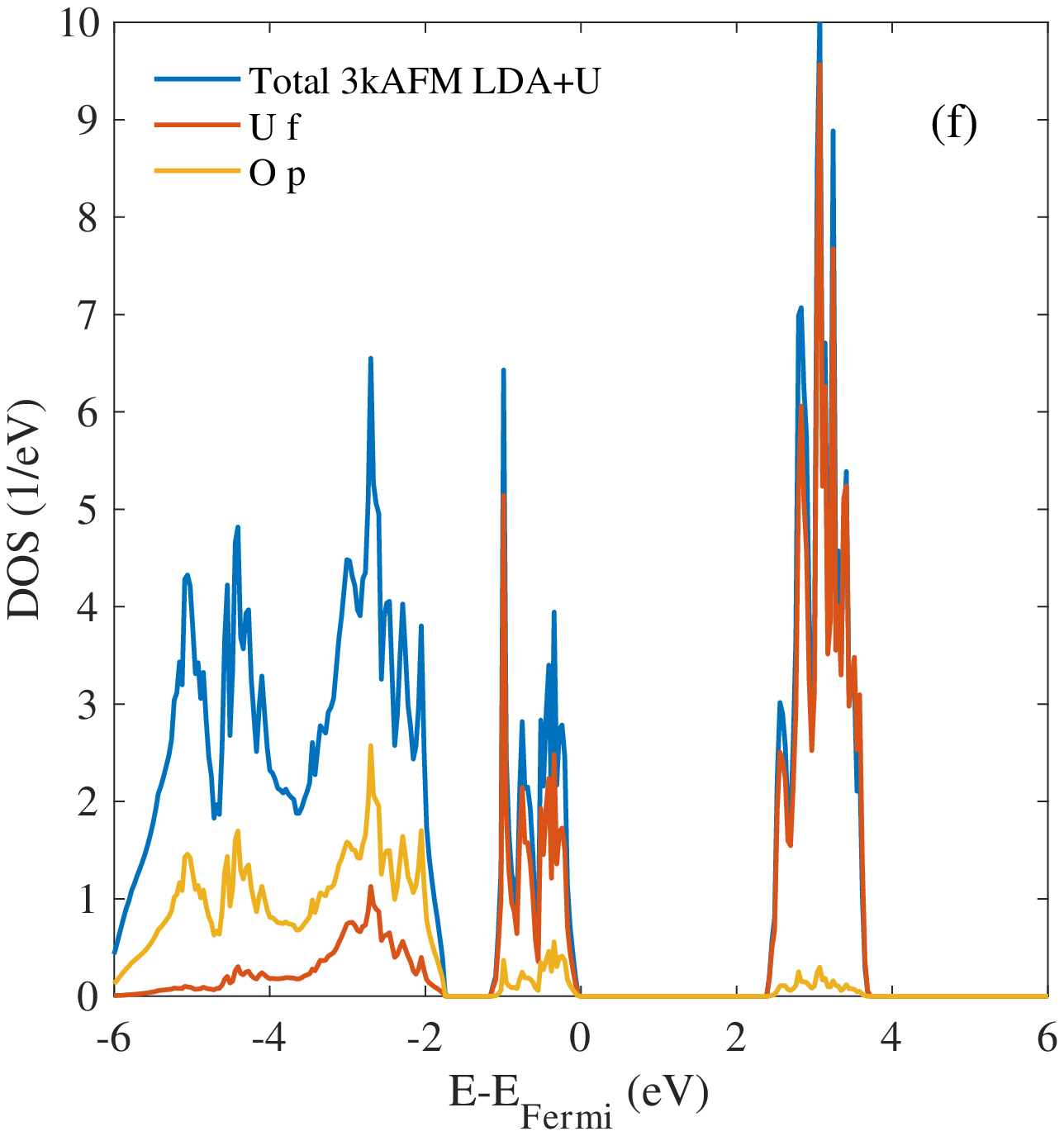}
    \caption{\label{fig:figureSMdos} The total and projected electronic density of states calculated from the $\mathbb{S}_0$ state in the FM, 1k AFM, and 3k AFM structures, using GGA+$U$+SOC and LDA+$U$+SOC ($U=4$ eV). }
\end{figure}

The total and projected electronic density of states (DOS) of $\mathbb{S}_0$ in FM, 1\textbf{k} AFM, and 3\textbf{k} AFM structures, calculated by both GGA+$U$+SOC and LDA+$U$+SOC ($U=4$ eV), are presented in Figure~\ref{fig:figureSMdos}. The band gaps are 2.25, 2.72, and 2.62 eV for FM, 1\textbf{k} AFM, and 3\textbf{k} AFM structures in GGA+$U$, respectively, and are 1.98, 2.49, and 2.39 eV for FM, 1\textbf{k} AFM, and 3\textbf{k} AFM structures in LDA+$U$, respectively. 

\clearpage
\section{DFT+$U$ convergence tests and influence of $J$ }

\subsection{Energy difference convergence of cut-off energy and K-point mesh}

\begin{figure}[h]
    \centering
    \includegraphics[width=0.40\columnwidth]{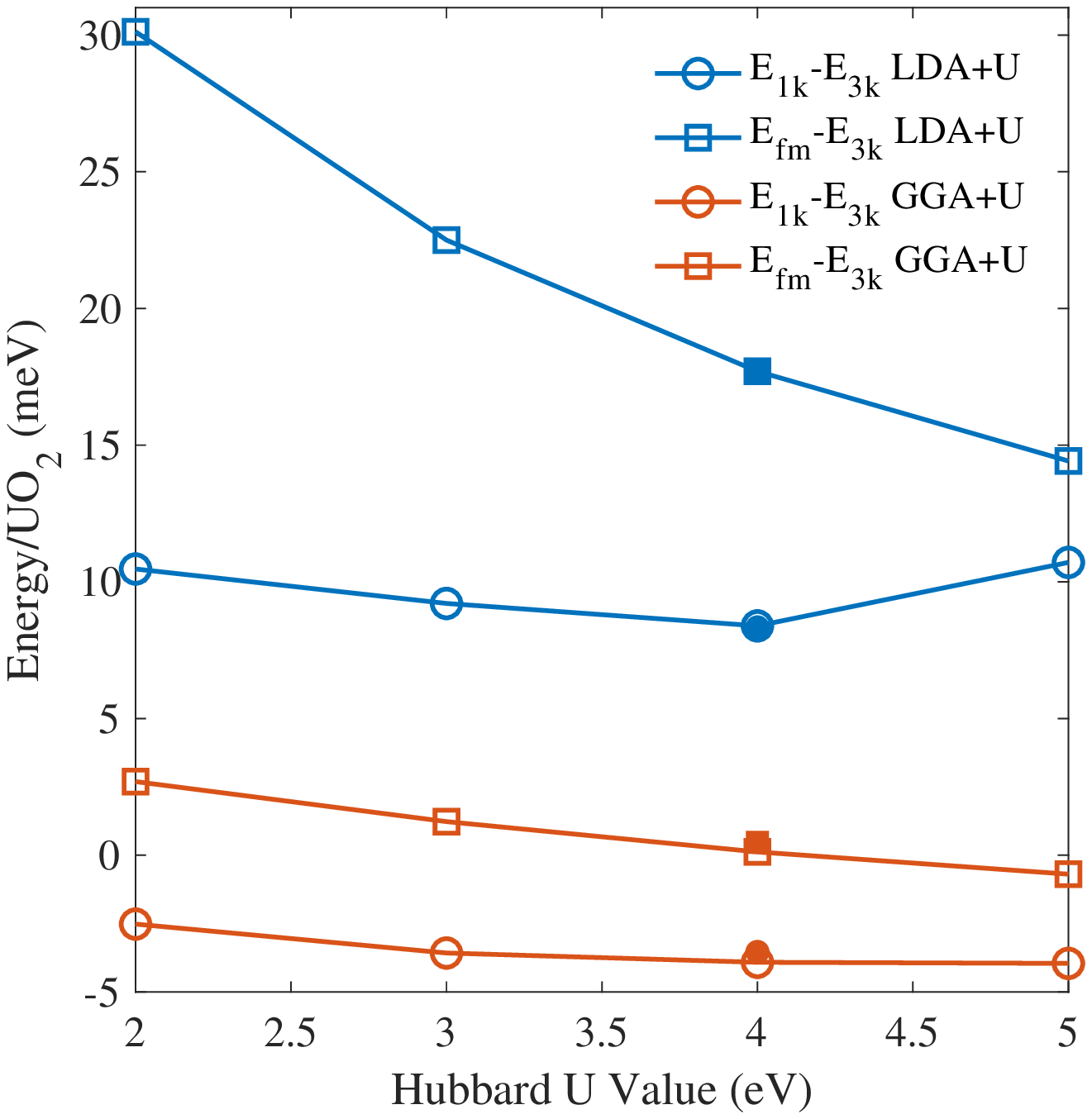}
    \caption{\label{fig:SMconvergence} Convergence tests for the energy difference between magnetic structures, calculated using a cut-off energy of 550 eV, a K-point mesh of $13\times13\times13$ for FM and $7\times7\times7$ for AFM (open symbols), and a cut-off energy of 650 eV, a K-point mesh of $20\times20\times20$ for FM and $10\times10\times10$ for AFM (solid symbols at $U=4$ eV).  }
\end{figure}

 The settings of our calculations are: the cut-off energy is 550 eV, the convergence criteria is $1\times10^{-8}$ eV for energy, the K-point mesh is $13\times13\times13$ for FM calculations in an unitcell and $7\times7\times7$ for AFM calculations in a conventional cell, respectively. As in Figure 1 of our main text, the energy differences are on the order of several meV, so here we test the energy difference convergence using a higher cut-off energy and denser K-point mesh ( see Figure~\ref{fig:SMconvergence}). We re-calculate the energies of the different magnetic structures of GGA+$U$ and LDA+$U$ ($U=4$ eV) by using a cut-off energy 650 eV and k-point mesh $20\times20\times20$ for FM calculations and $10\times10\times10$ for AFM calculations, respectively. The new calculations are added as solid symbols. The difference between the results of the different settings are smaller than 0.4 meV, showing good convergence.

\subsection{Results of LDAUTYPE=1 using $J$}

\begin{figure}[h]
    \centering
    \includegraphics[width=0.40\columnwidth]{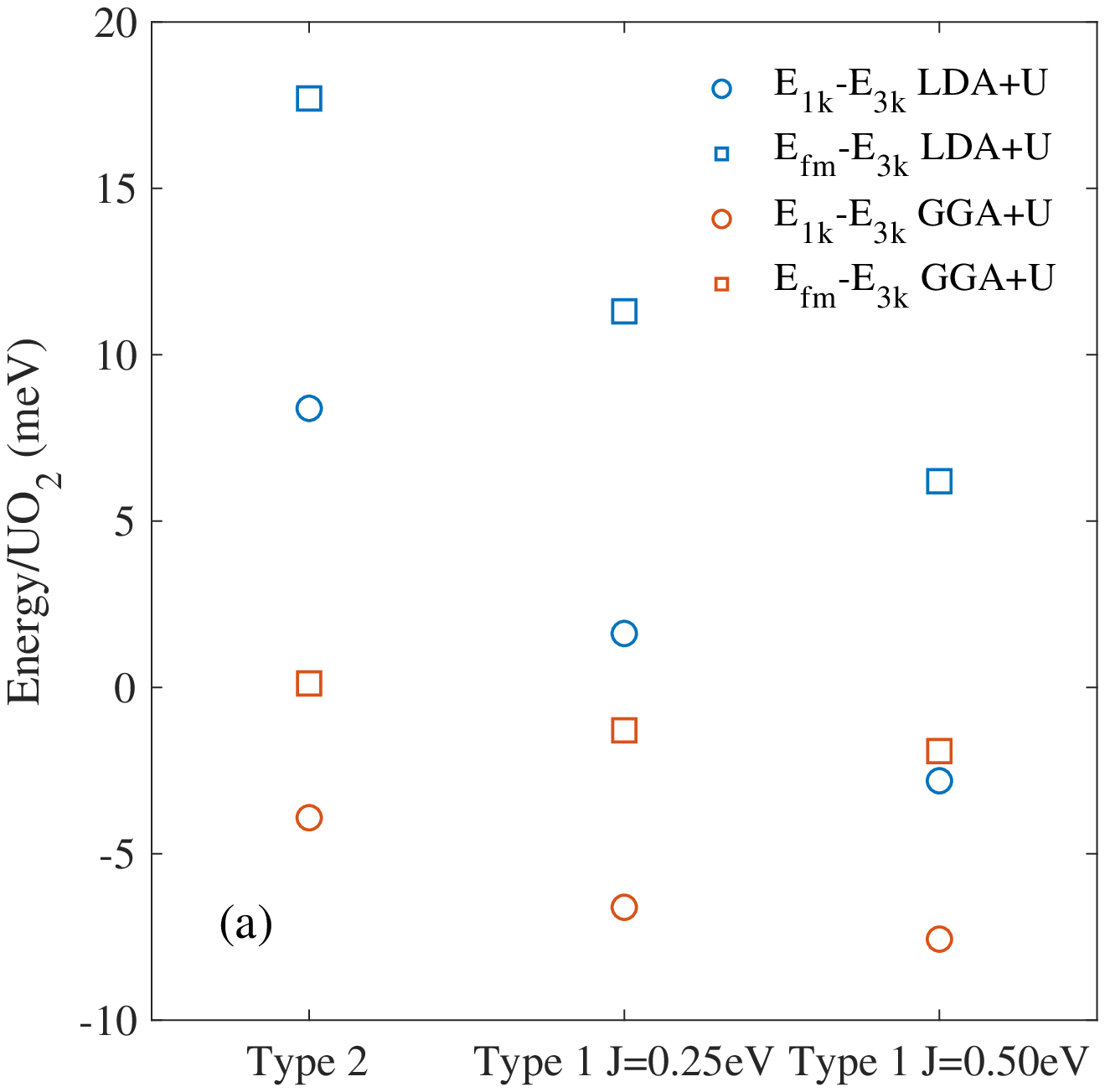}
    \includegraphics[width=0.40\columnwidth]{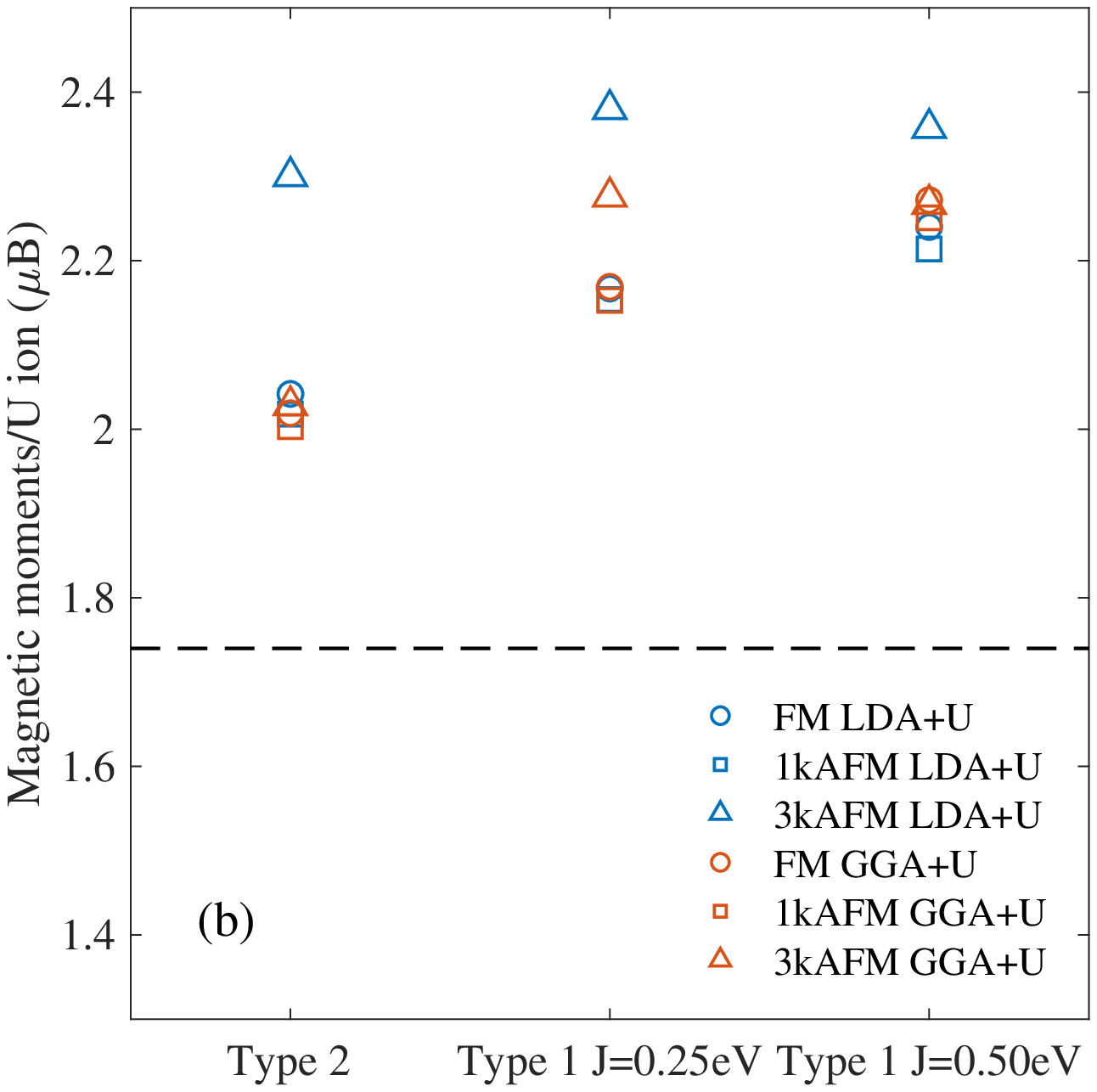}
    \caption{\label{fig:type1} The calculated (a) energy difference between different magnetic structures and (b) magnetic moments of U ion  using GGA+$U$+SOC and LDA+$U$+SOC calculations ($U=4$ eV). Type 1 refers the rotationally invariant DFT+$U$ approach by Liechtenstein et al.\cite{liechtenstein_density-functional_1995}, and Type 2 refers to the simplified rotationally invariant approach by Dudarev et al.\cite{dudarev_electron-energy-loss_1998}. The black dashed horizontal line in (b) corresponds to the experimental value 1.74 $\mu B$ from Faber et al.\cite{faberNeutronDiffractionStudyMathrmO1975}.}
\end{figure}

We re-calculate the results of Figure 1, the energy difference between
different magnetic structures, using the rotationally invariant DFT+$U$
approach by Liechtenstein et al.\cite{liechtenstein_density-functional_1995}
($LDAUTYPE=1$, referred as Type 1). We use the electronic charge density 
from $LDAUTYPE=2$ (referred to as Type 2) to initiate the Type 1
calculations, i.e., we do not perform our ground state search process for Type
1. Type 1 and Type 2 are equivalent when $J=0$.  We test two nonzero $J$ values
in our Type 1 calculations: $J=0.25$ and $0.5$ eV, with $U=4$ eV. The
effect of $J$ within Type 1 is illustrated in
Figure~\ref{fig:type1}. 
In Figure~\ref{fig:type1}(a), 
$E_{fm}-E_{3k}$ and $E_{1k}-E_{3k}$ are both decreasing with increasing $J$; meaning that
$J$ prefers collinear magnetic structures like FM and
1k AFM rather than noncollinear structures like 3k AFM. 
In
Figure~\ref{fig:type1}(b), increasing $J$ increases the magnetic moments of
U atoms within FM and 1\textbf{k}, while 3\textbf{k} increases and then slightly decreases.
Interestingly, the $J$ is pushing both the energetics and the magnetic moments in the wrong direction relative to experiment.

\bibliographystyle{apsrev4-1} 